%% file: nr.5e.tex
\documentclass[a4paper,final,12pt]{article}
\usepackage{amssymb,amsfonts,mathpple,setspace,floatrow,paralist,amsmath,graphicx,marvosym,nicefrac,accents}
\usepackage{footmisc,currfile,arydshln,mathtools,pifont,textgreek,calc,geometry,siunitx}
\usepackage[displaymath,mathlines,pagewise]{lineno}
\usepackage[notref,notcite]{showkeys}
\usepackage{LaTexPaperStyle,cleveref}
\usepackage{tabu,array,anyfontsize}
\usepackage[center]{subfigure}
\usepackage[export]{adjustbox}
\usepackage[makeroom]{cancel}
\usepackage[percent]{overpic}
\usepackage[round]{natbib}
\usepackage[alpine]{ifsym}
\usepackage{hyperref}

\def\I{1}
\def\figs{\I}
\def\Rcode{\I}
\def\appndx{\I}
\def\spce{1.23}
\def\submission{1}
\def\figspace{-15mm}
\def\appendixOnline{1}
\def\figtabsname{_figstabs.5e.tex}
\def\appndixname{_appendix.5e.tex}
\def\abstrctname{_abstract.5e.tex}

\newcommand{\onlineappendix}{http://www.danielbuncic.com/pdf/ESTAR_online_appendix.pdf}
\newcommandx\intheappendix[1][1=\appendixOnline]{\ifthenelse{\equal{1}{#1}}{\href{\onlineappendix}{Online Appendix}\xspace}{\hyperlink{sec:appendix}{Appendix}\xspace}}
\graphicspath{{./graphics/}}
\IfFileExists{C:/localtex/tex/latex/misc/tcilatex.tex}{\input{tcilatex}}{}
\defcitealias{bis:2006}{QIS5}
\fulllink
\bibpunct{(}{)}{;}{a}{,}{,}
\newcolumntype{d}{D{.}{.}{5.4}}
\renewcommand{\Espace}{-1mm}
\renewcommand{\Authorspace}{\vspace{-1.5mm}}
\renewcommand{\Titlefont}{\fontsize{21}{21}\selectfont}

\definecolor{HeaderColor}{rgb}{.0,.0,.0}
\definecolor{VersionColor}{rgb}{.0,.0,.0}
\allowdisplaybreaks

\DeclareMathAlphabet{\mathitbf}{OML}{cmm}{b}{sf}
\renewcommand{\dsum}{\ensuremath \mathop {\scalebox {1.11}{$\displaystyle \sum$}}}
\DeclareNewFloatType{pdftable}{placement=htbp,name=Table}

\renewcommand{\captionfont}{\small\sf}
\renewcommand{\aref}[1]{\hyperref[#1]{Appendix}}
\let\oldCref\Cref\renewcommand{\Cref}[1]{\hyperref[#1]{\oldCref{#1}}}
\let\oldcref\cref\renewcommand{\cref}[1]{\hyperref[#1]{\oldcref{#1}}}

\newenvironment{enum}{ \begin{list}{\color{myblu}{(\emph{\roman{enumi}}\hspace{1.4pt})}}{            \leftmargin=23pt\partopsep=4pt\topsep=1pt\itemsep=1pt\parsep=4pt\usecounter{enumi}}}{\end{list}}
\newenvironment{enuma}{\begin{list}{\color{myblu}{(\hspace{.5pt}\emph{\alph{enumi}}\hspace{.5pt})}} {\leftmargin=20pt\partopsep=4pt\topsep=1pt\itemsep=1pt\parsep=4pt\usecounter{enumi}}}{\end{list}}
\newenvironment{enumI}{\begin{list}{\color{myblu}{ (\emph{\Roman{enumi}}\hspace{1.0pt})}}{\leftmargin=20pt\partopsep=4pt\topsep=1pt\itemsep=1pt\parsep=4pt\usecounter{enumi}}}{\end{list}}

\makeatletter
\renewcommand{\@fnsymbol}[1]{\ensuremath{ \ifcase#1\or  \star\or \ast\or \text{\Gentsroom}\or {\natural}\or
 \mathsection\or \mathparagraph\or \|\or \#\or \text{\VarIceMountain}\or \text{\ding{73}}\or \blacklozenge\or
 \text{\ding{167}}\or \text{\Football}\or \star\or {\star\star} \or {\flat} \or {\sharp} \or {\imath}\or
 \text{\ding{118}}\or \text{\ding{50}} \or \text{\ding{68}}\or \text{\StoneMan}
 \else\@ctrerr\fi}}
\renewcommand\@makefntext[1]{\noindent\makebox[1em][r]{\@makefnmark}#1}
\makeatother
\def\MYTITLE{Econometric issues with Laubach and Williams' \\
estimates of the natural rate of interest}
\input{./graphics/RCode_style.sty}
\begin{document}

\title{%
\MYTITLE%
\thanks{%
Without implications, I am grateful to James Stock, Adrian Pagan, Neil
Ericsson, Paolo Giordani, Jesper Lind\'{e}, Michael Kiley, Glen Rudebusch,
Luc Bauwens, Francesco Ravazollo, Simon van Norden, Fabio Canova, Eric
Leeper, Georgi Krustev, Bernd Schwaab, Alessandro Galesi, Claus Brand,
Wolfgang Lemke, and Eric Renault for helpful comments and discussions. I
thank Lorand Abos for excellent research assistance.}\\
\vspace{05mm}\setcounter{footnote}{0}}
\author{ \href{http://www.danielbuncic.com}{{Daniel Buncic}}%
\setcounter{footnote}{2}\thanks{%
Corresponding author: Stockholm Business School, Stockholm University,
SE-103 37, Stockholm, Sweden. Email: \emailto{daniel.buncic@sbs.su.se}. Web: %
\url{http://www.danielbuncic.com}.} 
\affiliation{Stockholm University} }
\date{\vspace{-7mm}\version{November 22, 2019}{(\currfilebase)}{\today}\\
\vspace{-10mm}}
\maketitle

\begin{abstract}
\input{\abstrctname} \thispagestyle{empty} \ifthenelse{\equal{1}{%
\submission}}{\thispagestyle{empty}}{}\newpage
\end{abstract}


\thispagestyle{empty}

\ifthenelse{\equal{1}{\submission}}{
 \secondtitle[2.166][\SecondTitlespace]{\MYTITLE}
 {\vspace{24.7mm}}
\begin{abstract}
\IfFileExists{\abstrctname}{\input{\abstrctname}} {} \setcounter{page}{1}
\thispagestyle{empty}
\newpage
\end{abstract}}

\ifthenelse{\equal{0}{\submission}}{}

\newpage\newpage

\let\oldref\ref \renewcommand{\ref}[1]{(\oldref{#1})}%

\setcounter{page}{1} 
\setstretch{\spce} 

\section{Introduction \label{sec:intro}}

Since the global financial crisis, nominal interest rates have declined
substantially to levels last witnessed in the early 1940s following the
Great Depression. The academic as well as policy literature has attributed
this decline in nominal interest rates to a decline in the natural rate of
interest; namely, the rate of interest consistent with employment at full
capacity and inflation at its target. In this literature, \citeauthor*{%
holston.etal:2017}' (\citeyear{holston.etal:2017}) estimates of the natural
rate have become particularly influential and are widely regarded as a
benchmark. The Federal Reserve Bank of New York (FRBNY) maintains an entire
website dedicated to providing updates to \cites{holston.etal:2017}
estimates of the natural rate, not only for the United States (U.S.), but
also for the Euro Area, Canada and the United Kingdom (U.K.) (see %
\url{https://www.newyorkfed.org/research/policy/rstar}).

In \cites{holston.etal:2017} model, the natural rate of interest is defined
as the sum of trend growth of output $g_{t}$ and `\emph{other factor}' $%
z_{t} $. This `\emph{other factor}' $z_{t}$ is meant to capture various
underlying structural factors such as savings/investment imbalances,
demographic changes, and fiscal imbalances that influence the natural rate,
but which are not captured by trend growth $g_{t}$. In \autoref{fig:HLW_zf}
below, I show filtered (as well as smoothed) estimates of %
\cites{holston.etal:2017} `\emph{other factor}' $z_{t}$.\footnote{\cite%
{holston.etal:2017} do not show a plot of `\emph{other factor}' $z_{t}$ on
the FRBNY website (as of 22$^{nd}$ of June, 2020).}

\begin{figure}[h]
\centering
\includegraphics[width=1\textwidth,rotate=00,trim={0 0 0
0},clip]{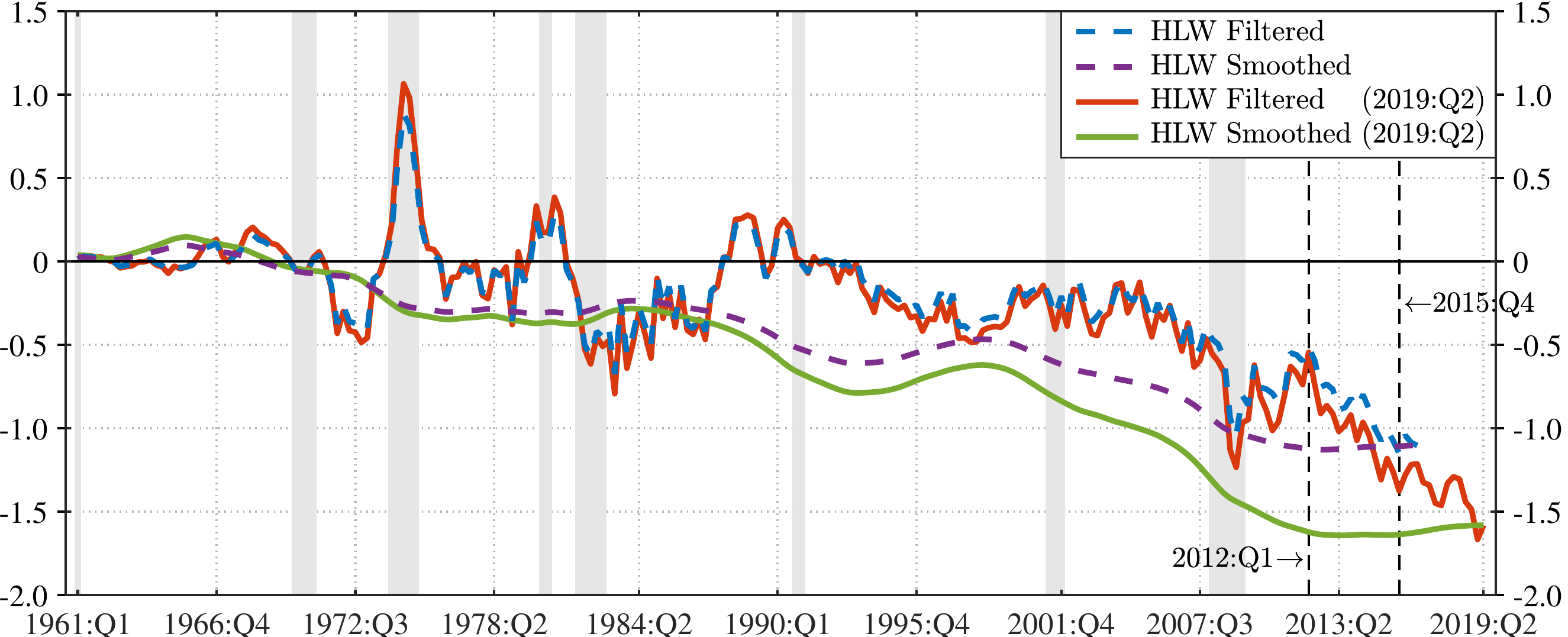} \vspace{-06mm}
\caption{Filtered and smoothed estimates of \cites{holston.etal:2017} \emph{%
`other factor'} $z_{t}$.}
\label{fig:HLW_zf}
\end{figure}
\vsp[-2]

\noindent The dashed lines in \autoref{fig:HLW_zf} show estimates obtained
with data ending in 2017:Q1, while the solid lines are estimates based on
data extended to 2019:Q2. The strong and persistent downward trending
behaviour of `\emph{other factor}' $z_{t}$ is striking from \autoref%
{fig:HLW_zf}, particularly from 2012:Q1 onwards. The two (black) dashed
vertical lines mark the periods 2012:Q1 and 2015:Q4. In 2015:Q4, the Federal
Reserve started the tightening cycle and raised nominal interest rates by 25
basis points. In 2012:Q1, real rates began to rise due to a (mild)
deterioration in inflation expectations.\footnote{%
See panel (a) of \autoref{fig:HLW_factors}, which shows plots of the federal
funds rate, the real interest rate, as well as inflation and inflation
expectations.} Both led to an increase in the real rate. Yet, %
\cites{holston.etal:2017} estimates of `\emph{other factor}' $z_{t}$
declined by about 50 basis points from 2012:Q1 to 2015:Q4, and then another
50 basis points from 2015:Q4 to 2019:Q2, reaching a value of $-1.58$ in
2019:Q2. Because $z_{t}$ evolves as a driftless random walk in the model,
the only parameter that \emph{`controls'} the influence of $z_{t}$ on the
natural rate is the `\emph{signal-to-noise ratio}' $\lambda _{z}$.\footnote{%
This description is somewhat imprecise to avoid cumbersome language. Since $%
z_{t}$ evolves as $z_{t}=z_{t-1}+\sigma _{z}\epsilon _{t}$, with $\epsilon
_{t}$ being standard normal, it is the standard deviation $\sigma _{z}$ that
is the only parameter that influences the evolution of $z_{t}$. However,
\cite{holston.etal:2017} determine $\sigma _{z}$ indirectly through the `%
\emph{signal-to-noise ratio}' $\lambda _{z}$, so it is the size of $\lambda
_{z}$ that matters for the evolution of $z_{t}$.} Thus, how exactly this
parameter is estimated is of fundamental importance for the determination of
the natural rate of interest.

In this paper, I show that \cites{holston.etal:2017} implementation of %
\cites{stock.watson:1998} Median Unbiased Estimation (MUE) is unsound. It
cannot recover the ratios of interest $\lambda _{g}=\sigma _{g}/\sigma
_{y^{\ast }}$ and $\lambda _{z}=a_{r}\sigma _{z}/\sigma _{\tilde{y}}$ from
Stages 1 and 2 of their three stage procedure needed for the estimation of
the full structural model. The implementation of MUE of $\lambda _{z}$ in
Stage~2 is particularly problematic, as \cites{holston.etal:2017} procedure
is based on an \emph{`unnecessarily'} misspecified Stage~2 model. This
misspecified Stage 2 model not only fails to identify the ratio of interest $%
\lambda _{z}=a_{r}\sigma _{z}/\sigma _{\tilde{y}}$, but moreover, due to the
way \cite{holston.etal:2017} implement MUE in Stage 2, leads to spuriously
large and excessively amplified estimates of $\lambda _{z}$. Since the
magnitude of $\lambda _{z}$ determines and drives the downward trending
behaviour of `\emph{other factor}' $z_{t}$, this misspecification is
consequential. Correcting their Stage 2 model and the MUE\ implementation
results in a substantial quantitative reduction in the point estimate of $%
\lambda _{z}$, and hence also $\sigma _{z}$. For instance, using data ending
in 2017:Q1, \cites{holston.etal:2017} estimate of $\lambda _{z}$ is $%
0.030217 $ and yields an implied value of $0.150021$ for $\sigma _{z}$.
After the correction, $\lambda _{z}$ is estimated to be $0.000754$ with an
implied value for $\sigma _{z}$ of $0.003746$.\footnote{%
These are my replicated estimates using data up to 2017:Q1, but they are
effectively identical to those listed in Table 1, column 1 for the U.S. on
page S60 in \cite{holston.etal:2017}.} The resulting filtered (and smoothed)
estimates of $z_{t}$ are markedly different, with the one from the correct
Stage 2 implementation not only being very close to zero, but also highly
insignificant statistically. The $p-$values corresponding to the structural
break statistics from which $\lambda _{z}$ is estimated are of an order of
magnitude of 0.5. These results highlight that there is no evidence of \emph{%
`other factor'} $z_{t}$ in this model. The large and persistent downward
trend in \cites{holston.etal:2017} estimates thus appears to be spurious.

In \Sref{sec:S2}, I outline in detail the Stage 2 model and the MUE
procedure that \cite{holston.etal:2017} implement to estimate $\lambda _{z}$%
. I show that their Stage 2 model is misspecified and that due to this,
their MUE procedure cannot identify the ratio of interest $a_{r}\sigma
_{z}/\sigma _{\tilde{y}}$ from $\lambda _{z}$. Instead, it recovers $\lambda
_{z}=a_{r}\sigma _{z}/(\sigma _{\tilde{y}}+0.5a_{g}\sigma _{g})$ if $%
(a_{g}+4a_{r})=0$. If $(a_{g}+4a_{r})\neq 0$, then additional parameters
enter the denominator of $\lambda _{z}$, making it more intricate to recover
$\sigma _{z}$ from $\lambda _{z}$, as it will be necessary to make
additional assumptions about the time series properties of the nominal
interest rate which is not explicitly modelled by \cite{holston.etal:2017},
but rather added as an exogenous variable. The terms $a_{r}$ and $a_{g}$ are
the parameters on the lagged real interest rate and lagged trend growth in
the Stage 2 model of the output gap equation (see \Sref{sec:S2} for more
details). In the full model, these are restricted so that $a_{g}=-4a_{r}$.
In their specification of the Stage 2 model, \cite{holston.etal:2017} do not
impose this restriction. Moreover, they include only one lag of trend growth
$g_{t}$ in the output gap equation and, curiously, further add an intercept
term to the specification that is not present in the full model (see
equation \ref{S2:ytilde}). Since \cites{stock.watson:1998} MUE relies upon
\cite{chow:1960} type structural break tests to estimate $\lambda _{z}$,
these differences in the output gap specification lead to substantially
larger $F$ statistics (see \autoref{fig:seqaF} for a visual presentation)
and therefore estimates of $\lambda _{z}$. To demonstrate that their
misspecified Stage 2 model and MUE procedure leads to spurious and
excessively large estimates of $\lambda _{z}$ when the true value is zero, I
implement a simulation experiment in \Sref{sec:S2}. This simulation
experiment shows that the mean estimate of $\lambda _{z}$ can be as high as $%
0.028842$, with a $45.7\%$ probability (relative frequency) of observing a
value larger than estimated from the empirical data, when computed from
simulated data which were generated from a model with the true $\lambda
_{z}=0$. These simulation results are concerning, because they suggest that
it is \cites{holston.etal:2017} MUE procedure itself that leads to the
excessively large estimates of $\lambda _{z}$, rather than the size of the
true $\lambda _{z}$ in the data.

Although \Sref{sec:S2} describes the core problem with %
\cites{holston.etal:2017} estimation procedure, there are other issues with
the model and how it is estimated. Some of these are outlined in %
\Sref{sec:other}.\ For instance, \cites{holston.etal:2017} estimates of the
natural rate, trend growth, `\emph{other factor}' $z_{t}$ and the output gap
are extremely sensitive to the starting date of the sample used to estimate
the model. Estimating the model with data beginning in 1972:Q1 (or 1967:Q1)
leads to negative estimates of the natural rate of interest toward the end
of the sample period. These negative estimates are again driven purely by
the exaggerated downward trending behaviour of `\emph{other factor}' $z_{t}$%
. The 1972:Q1 sample start was chosen to match the starting date used in the
estimation of this model for the Euro Area. Out of the four countries that %
\cites{holston.etal:2017} model is fitted to, only the Euro Area estimates
of the natural rate turn negative in 2013.\footnote{%
Only the Euro Area estimates are based on a sample that starts in 1972:Q1,
while the estimates for the U.K., Canada and the U.S. are based on samples
starting in 1961:Q1.} The fact that it is also possible to generate such
negative estimates of the natural rate from \cites{holston.etal:2017} model
for the U.S. by simply adjusting the start of the estimation period to match
that of the Euro Area data suggests that the model is far from robust, and
therefore inappropriate for use in policy analysis. Furthermore, because
Kalman Filtered estimates of the natural rate of interest will be moving
averages of the observed variables that enter the state-space model, a
circular or confounding relationship between the natural rate and the
(nominal)\ policy rate will arise, because any central bank induced change
in the policy target will be mechanically transferred to the natural rate
via the Kalman Filtered estimate of the state vector. This makes it
impossible to address \textit{`causal'} questions regarding the relationship
between natural rates and policy rates.

Median Unbiased Estimation is neither well known nor widely used at policy
institutions. To give some background on the methodology, and to be able to
understand why \cites{holston.etal:2017} implementation of MUE\ in Stage 2
is unsound, I provide a concise but important and informative overview of
the methodology in \Sref{sec:MUE}.\ This section is essential for readers
unfamiliar with the estimator. It reviews and summarises the conditions when
it is likely to encounter \emph{`pile-up'} at zero problems with Maximum
Likelihood Estimation (MLE) of such models. Namely, MLE is likely to
generate higher \emph{`pile-up'} at zero frequencies than MUE when the
initial conditions of the state vector are unknown and need to be estimated,
and when the true `\emph{signal-to-noise ratio}' is very small (close to
zero). Since \cite{holston.etal:2017} do not estimate the initial conditions
of the state vector, but instead use very tightly specified prior values,
and because their MUEs of the `\emph{signal-to-noise ratio}' are everything
else but very small in the context of MUE, it seems highly unlikely a priori
that MLE should generate higher \emph{`pile-up'} at zero probabilities than
MUE. From \cites{stock.watson:1998} simulation results we know that MLE
(with a diffuse prior) is substantially more efficient than MUE when the `%
\emph{signal-to-noise ratio}' is not extremely small. MLE should thus be
preferred as an estimator.

For reasons of completeness, I provide a comprehensive description of %
\cites{holston.etal:2017} Stage 1 model and their first stage MUE
implementation in \Sref{sec:S1}. As in the Stage 2 model, I show
algebraically that their MUE procedure cannot recover the ratio $\sigma
_{g}/\sigma _{y^{\ast }}$ from $\lambda _{g}$ because the error term in the
first difference of the constructed trend variable $y_{t}^{\ast }$ in the
first stage model depends on the real interest rate, as well as `\emph{other
factor}' $z_{t}$ and trend growth $g_{t}$. This means that when the long-run
standard deviation from the MUE\ procedure is constructed, it will not only
equal $\sigma _{y^{\ast }}$ as required, but also depend on $\sigma _{z}$, $%
\sigma _{g}$, as well as the long-run standard deviation of the real rate.
Rewriting a simpler version of the Stage 1 model in local level model form
also fails to identify the ratio of interest $\sigma _{g}/\sigma _{y^{\ast }}
$ from MUE of $\lambda _{g}$. The inability to recover the ratio $\sigma
_{g}/\sigma _{y^{\ast }}$ from the first stage model thus appears to be a
broader issue highlighting the unsuitability of MUE in this context. This
section also illustrates that it is empirically unnecessary to use MUE\ to
estimate $\sigma _{g}$ in the first stage model since MLE does not lead to
\emph{`pile-up'} at zero problems with $\sigma _{g}$, neither in the local
level model nor in the local linear trend (or unobserved component) model
form. Estimating $\sigma _{g}$ directly by MLE\ in the second and third
stages confirms this result, yielding in fact larger point estimates than
implied by the first stage MUE of $\lambda _{g}$ obtained from %
\cites{holston.etal:2017} procedure. Readers not interested in the
computational intricacies and nuances of the Stage 1 model may skip this
section entirely, and only refer back to it as needed for clarification of
later results. The key contribution of this paper relates to the correct
estimation of $\lambda _{z}$ in \cites{holston.etal:2017} Stage 2 model and
its impact on the natural rate of interest through `\emph{other factor}' $%
z_{t}$.

MUE of $\lambda _{z}$ based on the correctly specified Stage 2 model
suggests that there is no role for `\emph{other factor}' $z_{t}$ in this
model and given this data.\footnote{%
This result is inline with the MLE based estimates of $\sigma _{z}$.
Furthermore, these results also carry over to the Euro Area, Canadian and
U.K. estimates of $z_{t}$ which are not reported here, but will be made
available on the author's webpage.} This brings the focus back to (the
estimates of) trend growth in this model. \cites{holston.etal:2017}
estimates give the impression that trend growth has markedly slowed since
the global financial crisis, particularly in the immediate aftermath of the
crisis. In panels (b) and (c) of \autoref{fig:HLW_factors}, I\ show plots of %
\cites{holston.etal:2017} estimates of $g_{t}$ together with a few simple
alternative ones (annualized GDP growth is superimposed in panel (b)). Trend
growth is severely underestimated from 2009:Q3 onwards. From the robust
(median) estimates of average GDP growth over the various expansion periods
shown in \autoref{tab:sumstatGDP}, trend growth is only approximately 25
basis points lower at 2.25\% since 2009:Q3 than over the pre financial
crisis expansion from 2002:Q1 to 2007:Q4.\footnote{%
GDP\ growth is close to being serially uncorrelated over the last two
expansion periods, with low variances.} Survey based 10 year-ahead
expectations of annualized real GDP\ growth plotted in \autoref%
{Afig:SPF_GDP_growth} and \autoref{Afig:giglio_GDP_growth} also suggest that
trend growth remained stable (these plots are discussed further in %
\Sref{sec:other}). The key point to take away from this discussion is that %
\cites{holston.etal:2017} (one sided) Kalman Filter based estimate of
$g_{t}$ is excessively \emph{`pulled down'} by the large decline in GDP
during the financial crisis, and this strongly and adversely effects the
estimate of trend growth for many periods \emph{after} the crisis.

The rest of the paper is organised as follows. In \Sref{sec:model}, %
\cites{holston.etal:2017} structural model of the natural rate of interest
is described. \Sref{sec:MUE} gives a concise background to %
\cites{stock.watson:1998} Median Unbiased Estimation. In \Sref{sec:HLW}, I
provide a detailed description of the Stage 1 and Stage 2 models, and report
the results of the full Stage 3 model estimates. Some additional issues with
the model are discussed in \Sref{sec:other}, and \Sref{sec:conclusion}
concludes the study.

\section{Holston, Laubach and Williams' (2017) Model \label{sec:model}}

\citeallauthors{holston.etal:2017} use the following \textit{`structural'}
model to estimate the natural rate of interest:\footnote{%
In what follows, I use the same notation as in \cite{holston.etal:2017} (see
equations 3 to 9 on pages S61 to S63) to facilitate a direct comparison.
Also note that this model builds on an earlier specification of \cite%
{laubach.williams:2003}, where trend growth $g_{t}$ is scaled by another
parameter $c$, and where also a stationary AR(2) process for the \emph{%
`other factor'} $z_{t}$ was considered in addition to the $I(1)$
specification in \ref{z}.} \bsq\label{eq:hlw}\vsp[-0]
\begin{align}
\text{Output}& \text{:} & y_{t}& =y_{t}^{\ast }+\tilde{y}_{t}  \label{gdp} \\
\text{Inflation}& \text{:} & \pi _{t}& =b_{\pi }\pi _{t-1}+(1-b_{\pi })\pi
_{t-2,4}+b_{y}\tilde{y}_{t-1}+\varepsilon _{t}^{\pi }  \label{AS} \\
\text{Output gap}& \text{:} & \tilde{y}_{t}& =a_{y,1}\tilde{y}_{t-1}+a_{y,2}%
\tilde{y}_{t-2}+\tfrac{a_{r}}{2}[\left( r_{t-1}-r_{t-1}^{\ast }\right)
+\left( r_{t-2}-r_{t-2}^{\ast }\right) ]+\varepsilon _{t}^{\tilde{y}}
\label{IS} \\
\text{Output trend}& \text{:} & y_{t}^{\ast }& =y_{t-1}^{\ast
}+g_{t-1}+\varepsilon _{t}^{y^{\ast }}  \label{y*} \\
\text{Trend growth}& \text{:} & g_{t}& =g_{t-1}+\varepsilon _{t}^{g}
\label{g} \\
\text{Other factor}& \text{:} & z_{t}& =z_{t-1}+\varepsilon _{t}^{z},
\label{z}
\end{align}%
\esq where $y_{t}$ is 100 times the (natural) log of real GDP, $y_{t}^{\ast
} $ is the permanent or trend component of GDP, $\tilde{y}_{t}$ is its
cyclical component, $\pi _{t}$ is annualized quarter-on-quarter PCE
inflation, and $\pi _{t-2,4}=\left( \pi _{t-2}+\pi _{t-3}+\pi _{t-4}\right)
/3$. The real interest rate $r_{t}$ is computed as:
\begin{equation}
r_{t}=i_{t}-\pi _{t}^{e},  \label{r}
\end{equation}%
where expected inflation is constructed as:%
\begin{equation}
\pi _{t}^{e}=(\pi _{t}+\pi _{t-1}+\pi _{t-2}+\pi _{t-3})/4  \label{pi}
\end{equation}%
and $i_{t}$ is the \emph{exogenously} determined nominal interest rate, the
federal funds rate.

The natural rate of interest $r_{t}^{\ast }$ is computed as the sum of trend
growth $g_{t}$ and \emph{`other factor'} $z_{t}$, both of which are $I(1)$
processes. The real interest rate gap is defined as $\tilde{r}%
_{t}=(r_{t}-r_{t}^{\ast })$. The error terms $\varepsilon _{t}^{\ell
},\forall \ell =\{\pi ,\tilde{y},y^{\ast }\hsp[-1],g,z\}$ are assumed to be $%
i.i.d$ normal distributed, mutually uncorrelated, and with time-invariant
variances denoted by $\sigma _{\ell }^{2}$. Notice from \ref{AS} that
inflation is restricted to follow an integrated AR(4) process. From the
description of the data, we can see that the nominal interest rate $i_{t}$
as well as inflation $\pi _{t}$ are defined in annual or annualized terms,
while output, and hence the output gap, trend and trend growth in output are
defined at a quarterly rate. Due to this measurement mismatch, \cite%
{holston.etal:2017} adjust the calculation of the natural rate in their code
so that trend growth $g_{t}$ is scaled by 4 whenever it enters equations
that relate it to annualized variables. The natural rate is thus factually
computed as $r_{t}^{\ast }=4g_{t}+z_{t}$.\footnote{%
This generates some confusion when working with the model, as it is not
clear whether the estimated $z_{t}$ factor is to be interpreted at an annual
or quarterly rate.} In the descriptions that follow, I\ will use the
annualized $4g_{t}$ trend growth rate whenever it is important to highlight
a result or in some of the algebraic derivations, and will leave the
equations in \ref{eq:hlw} as in \cite{holston.etal:2017} otherwise for ease
of comparability.

\cite{holston.etal:2017} argue that due to \textit{`pile-up'} at zero
problems with Maximum Likelihood (ML) estimation of the variances of the
innovation terms $\varepsilon _{t}^{g}$ and $\varepsilon _{t}^{z}$ in \ref%
{eq:hlw}, estimates of $\sigma _{g}^{2}$ and $\sigma _{z}^{2}$ are
\textquotedblleft \textit{likely to be biased towards zero}%
\textquotedblright\ (page S64). To avoid such \textit{`pile-up'} at zero
problems, they employ Median Unbiased Estimation (MUE)\ of \cite%
{stock.watson:1998} in two preliminary steps --- Stage 1 and Stage 2 --- to
get estimates of what they refer to as \textit{`}\emph{signal-to-noise
ratios'} defined as $\lambda _{g}=\sigma _{g}/\sigma _{_{y^{\ast }}}$ and $%
\lambda _{z}=a_{r}\sigma _{z}/\sigma _{\tilde{y}}$. In Stage 3, the
remaining parameters of the full model in \ref{eq:hlw} are estimated,
conditional on the median unbiased estimates $\hat{\lambda}_{g}$ and $\hat{%
\lambda}_{z}$ obtained in Stages 1 and 2, respectively.

In the above description, I\ intentionally differentiate between the \textit{%
`}\emph{signal-to-noise ratio' }terminology of \cite{holston.etal:2017} and
the one used in \cite{harvey:1989} and in the broader literature on
state-space models and exponential smoothing, where the signal-to-noise
ratio would be defined as $\sigma _{y^{\ast }}/\sigma _{\tilde{y}}$ or $%
\left( \sigma _{g}/\sigma _{\tilde{y}}\right) $ from the relations in \ref%
{eq:hlw}.\footnote{%
As noted on page 337 in \cite{harvey:2006}, the signal-to-noise ratio
\textquotedblleft \emph{plays the key role in determining how observations
should be weighted for prediction and signal extraction.}\textquotedblright}
To be more explicit, in the context of the classic local level model of \cite%
{muth:1960}:\bsq\label{eq:ll}%
\begin{align}
y_{t}& =\mu _{t}+\varepsilon _{t} \\
\mu _{t}& =\mu _{t-1}+\eta _{t},
\end{align}%
\esq the signal-to-noise ratio is computed as $\sigma _{\eta }/\sigma
_{\varepsilon }$. In the extended version of the model in \ref{eq:ll} known
as the local linear trend model: \bsq\label{eq:llt}
\begin{align}
y_{t}& =\mu _{t}+\varepsilon _{t} \\
\mu _{t}& =\mu _{t-1}+\tau _{t-1}+\eta _{t} \\
\tau _{t}& =\tau _{t-1}+\zeta _{t},
\end{align}%
\esq two signal-to-noise ratios, namely $\sigma _{\eta }/\sigma
_{\varepsilon }$ and $\sigma _{\zeta }/\sigma _{\varepsilon }$, can be
formed.\footnote{%
The processes $\varepsilon _{t},\eta _{t}$ and $\zeta _{t}$ are uncorrelated
white noise. These two state-space formulations are described in more detail
in Chapters 2 and 4 of \cite{harvey:1989}. \cite{harvey:1989} also shows how
to derive their relation to simple and double exponential smoothing models.}
Note here that the model of \cite{holston.etal:2017} in \ref{eq:hlw} is
essentially an extended and more flexible version of the local linear trend
model in \ref{eq:llt}. Referring to $\lambda _{g}=\sigma _{g}/\sigma
_{_{y^{\ast }}}$ as a signal-to-noise ratio as \cite{holston.etal:2017} do
is thus rather misleading, since it would correspond to $\sigma _{\zeta
}/\sigma _{\eta }$ in the local linear trend model in \ref{eq:llt}, which
has no relation to the traditional signal-to-noise ratio terminology of \cite%
{harvey:1989} and others in this literature.\footnote{%
Readers familiar with the \cite{hodrick.prescott:1997} (HP) filter will
recognize that the local linear trend model in \ref{eq:llt} --- with the
extra \emph{`smoothness'} restriction $\sigma _{\eta }=0$ --- defines the
state-space model representation of the HP\ filter, where the square of the
inverse of the signal-to-noise ratio ($\sigma _{\varepsilon }^{2}/\sigma
_{\zeta }^{2}$ in \ref{eq:llt} or equivalently $\sigma _{\tilde{y}%
}^{2}/\sigma _{g}^{2}$ in \ref{eq:hlw}) is the HP\ filter smoothing
parameter that is frequently set to $1600$ in applications involving
quarterly GDP data.}

Before the three stage procedure of \cite{holston.etal:2017} is described, I
outline in detail how \cites{stock.watson:1998} median unbiased estimator is
implemented, what normalization assumptions it imposes, and how look-up
tables for the construction of the estimator are computed. I also include a
replication of \cites{stock.watson:1998} empirical estimation of trend
growth of U.S. real GDP per capita. Although the section that follows below
may seem excessively detailed, long, and perhaps unnecessary, the intention
here is to provide the reader with an overview of how median unbiased
estimation is implemented, what it is intended for, and when one can expect
to encounter \emph{`pile-up'} at zero problems to materialize. Most
importantly, it should highlight that \emph{`pile-up'} at zero is not a
problem in the general sense of the word, but rather only a nuisance in
situations when it is necessary to distinguish between \emph{very} small
variances and ones that are zero.

\section{\cites{stock.watson:1998} Median Unbiased Estimation \label{sec:MUE}%
}

\cite{stock.watson:1998} proposed Median Unbiased Estimation (MUE) in the
general setting of Time Varying Parameter (TVP) models. TVP models are
commonly specified in a way that allows their parameters to change gradually
or smoothly over time. This is achieved by defining the parameters to evolve
as driftless random walks (RWs), with the variances of the innovation terms
in the RW equations assumed to be small. One issue with Kalman Filter based
ML estimation of such models is that estimates of these variances can
frequently \emph{`pile-up'} at zero when the true error variances are \emph{%
`very'} small, but nevertheless, non-zero.\footnote{%
See the discussion in Section 1 of \cite{stock.watson:1998} for additional
motivation and explanations. As the title of \cites{stock.watson:1998} paper
suggests, MUE was introduced for \textquotedblleft \textit{coefficient
variance estimation in TVP models\textquotedblright } when this variance is
expected to be small.}

\cite{stock.watson:1998} show simulation evidence of \emph{`pile-up'} at
zero problems with Kalman Filter based ML estimation in Table 1 on page 353
of their paper. In their simulation set-up, they consider the following data
generating process for the series $GY_{t}$:\footnote{%
See their GAUSS\ files \texttt{TESTCDF.GSS} and \texttt{ESTLAM.GSS} for
details on the data generating process, which are available from Mark
Watson's homepage at \url{http://www.princeton.edu/~mwatson/ddisk/tvpci.zip}.%
}\bsq\label{eq:tvp_sim}%
\begin{align}
GY_{t}& =\beta _{t}+\varepsilon _{t} \\
\beta _{t}& =\beta _{t-1}+(\lambda /T)\eta _{t},
\end{align}%
\esq where $\varepsilon _{t}$ and $\eta _{t}$ are drawn from $i.i.d.$
standard normal distributions, $\beta _{00}$ is initialized at 0, and the
sample size is held fixed at $T=500$ observations, using $5000$
replications. The $\lambda $ values that determine the size of the variance
of $\Delta \beta _{t}$ are generated over a grid from 0 to 30, with unit
increments.\footnote{%
To be precise, in their GAUSS\ code, \cite{stock.watson:1998} use a range
from 0 to 80 for $\lambda $, with finer step sizes for lower $\lambda $
values (see, for instance, the file \texttt{TESTCDF.GSS}). That is, $\lambda
$ is a sequence between 0 to 30 with increments of 0.25, then 0.5 unit
increments from 30 to 60, and unit increments from 60 to 80. In Tables 1 to
3 of their paper, results are reported for $\lambda $ values from 0 up to $%
30 $ only, with unit increments.} Four median unbiased estimators relying on
four different structural break test statistics are compared to two ML
estimators. The first ML estimator, referred to as the maximum profile
likelihood estimator (MPLE), treats the initial state vector as an unknown
parameter to be estimated. The second estimator, the maximum marginal
likelihood estimator (MMLE), treats the initial state vector as a Gaussian
random variable with a given mean and variance. When the variance of the
integrated part of the initial state vector goes to infinity, MMLE produces
a likelihood with a diffuse prior.

How one treats the initial condition in the Kalman Filter recursions matters
substantially for the \emph{`pile-up'} at zero problem with MLE. This fact
has been known, at least, since the work of \cite{shephard.harvey:1990}.%
\footnote{%
On page 340, \cite{shephard.harvey:1990} write to this: \emph{%
\textquotedblleft \ldots we show that the results for the fixed and known
start-up and the diffuse prior are not too different. However, in Section 4
we demonstrate that the sampling distribution of the ML estimator will
change dramatically when we specify a fixed but unknown start-up
procedure.\textquotedblright } Their Tables II and III quantify how much
worse the ML estimator that attempts to estimate the initial condition in
the local level model performs compared to MLE with a diffuse prior.} The
simulation results reported in Table 1 on page 353 in \cite%
{stock.watson:1998} show that \emph{`pile-up'} at zero frequencies are \emph{%
considerably} lower when MMLE with a diffuse prior is used than for MPLE,
which estimates the initial state vector. For instance, for the smallest
considered non-zero population value of $\lambda =1$, which implies a
standard deviation of $\Delta \beta _{t}$ ($\sigma _{\Delta \beta }$
henceforth)\ of $\lambda /T=1/500=0.002$, MMLE produces an at most $14\ $%
percentage points higher \emph{`pile-up'} at zero frequency than MUE (ie., $%
0.60$ or $60\%$ for MMLE versus $0.46$ or $46\%$ for MUE based on the \cite%
{quandt:1960} Likelihood Ratio, henceforth QLR, structural break test
statistic).\footnote{%
The four different MUEs based on the different structural break tests appear
to perform equally well.} For MPLE, this frequency is $45$ percentage points
higher at $0.91$ ($91\%$). At $\lambda =5$ ($\sigma _{\Delta \beta }=0.01$)
and $\lambda =10$ ($\sigma _{\Delta \beta }=0.02$), these differences in the
\emph{`pile-up'} at zero frequencies reduce to $11$ and $4$ percentage
points, respectively, for MMLE, but remain still sizeable for MPLE. At $%
\lambda =20$ ($\sigma _{\Delta \beta }=0.04$), the \emph{`pile-up'} at zero
problem disappears nearly entirely for MMLE\ and MUE, with \emph{`pile-up'}
frequencies dropping to $2$ and $1$ percentage points, respectively, for
these two estimators, staying somewhat higher at $7$ percentage points for
MPLE.

Using MUE instead of MLE to mitigate \emph{`pile-up'} at zero problems
comes, nevertheless, at a cost; that is, a loss in estimator efficiency
whenever $\lambda $ (or $\sigma _{\Delta \beta }$) is not \emph{very} small.
From Table 2 on page 353 in \cite{stock.watson:1998}, which shows the
asymptotic relative efficiency of MUE (and MPLE) relative to MMLE, it is
evident that for true $\lambda $ values of $10$ or greater $(\sigma _{\Delta
\beta }\geq 0.02)$, the 4 different MUEs yield asymptotic relative
efficiencies (AREs) as low as 0.65 (see the results under the $L$ and
\textrm{MW} columns in Table 2).\footnote{%
The QLR structural break test seems to be the most efficient among the MUEs,
yielding the highest AREs across the various MUE implementations.} This
means that MMLE only needs $65\%$ of MUE's sample size to achieve the same
probability of falling into a given null set. Only for very small values of $%
\lambda \leq 4$ $(\sigma _{\Delta \beta }\leq 0.008)$ are the AREs of MUE\
and MMLE of a similar magnitude, ie., close to 1, suggesting that both
estimators achieve approximately the same precision.

Three important points are to be taken away from this review of the
simulation results reported in \cite{stock.watson:1998}. First, with MLE,
\emph{`pile-up'} at zero frequencies are substantially smaller when the
initial state vector is treated as a known fixed quantity or when a diffuse
prior is used, which is the case with MMLE (but not with MPLE). Second,
\emph{`pile-up'} at zero frequencies of MMLE\ are at most $4$ percentage
points higher than those of MUE once $\lambda \geq 10$ ($\sigma _{\Delta
\beta }=0.02$). Third, MUE can be considerably less efficient than MMLE, in
particular for \emph{`larger'} values of $\lambda \geq 10$ ($\sigma _{\Delta
\beta }=0.02$). This suggests that MLE\ with a diffuse prior should be
preferred whenever MUE\ based estimates of $\lambda $ (or $\sigma _{\Delta
\beta }$) are \emph{`large'} enough to indicate that \emph{`pile-up'} at
zero problems are unlikely to materialize.

To provide the reader with an illustration of how MUE is implemented, and
how its estimates compare to the two maximum likelihood based procedures
(MPLE and MMLE), I replicate the empirical example in Section 4 of \cite%
{stock.watson:1998} which provides estimates of trend growth of U.S. real
GDP per capita over the period from 1947:Q2 to 1995:Q4. Note that trend
growth in GDP is one of the two components that make up the real natural
rate $r_{t}^{\ast }$ in \cite{holston.etal:2017}. It is thus beneficial to
illustrate the implementation of MUE in this specific context, rather than
in the more general setting of time varying parameter models.

\subsection{Median unbiased estimation of U.S. trend growth \label%
{subsec:MUE}}

\cite{stock.watson:1998} use the following specification to model the
evolution of annualized trend growth in real per capita GDP for the U.S.,
denoted by $GY_{t}$ below:\footnote{%
That is, $GY_{t}=400\Delta \ln (\text{real per capita GDP}_{t})$, where $%
\Delta $ is the first difference operator (see Section 4 on page 354 in \cite%
{stock.watson:1998}). I again follow their notation as closely as possible
for comparability reasons.} \bsq\label{eq:sw98}
\begin{align}
GY_{t}& =\beta _{t}+u_{t}  \label{eq:sw1} \\
\Delta \beta _{t}& =(\lambda /T)\eta _{t}  \label{eq:swRW} \\
a(L)u_{t}& =\varepsilon _{t},  \label{eq:sw3}
\end{align}%
\esq where $a(L)$ is a (\emph{`stationary'}) lag polynomial with all roots
outside the unit circle, $\lambda $ is the parameter of interest, $T$ is the
sample size, and $\eta _{t}$ and $\varepsilon _{t}$ are two uncorrelated
disturbance terms, with variances $\sigma _{\eta }^{2}$ and $\sigma
_{\varepsilon }^{2}$, respectively. The growth rate of per capita GDP is
thus composed of a stationary component $u_{t}$ and a random walk component $%
\beta _{t}$ for trend growth. \cite{stock.watson:1998} set $a(L)$ to a $%
4^{th}$ order lag-polynomial, so that $u_{t}$ follows an AR(4) process. The
model in \ref{eq:sw98} can be recognized as the local level model of \cite%
{muth:1960} defined earlier in \ref{eq:ll}, albeit with the generalisation
that $u_{t}$ follows an AR(4)\ process, rather than white noise. Being in
the class of local level models means that the estimate of trend growth will
be an exponentially weighted moving average (EWMA) of $GY_{t}$.\footnote{%
\cite{stock.watson:1998} offer a discussion of the rationale behind the
random walk specification of trend growth in $GY_{t}$ in the second
paragraph on the left of page 355. Without wanting to get into a technical
discussion, one might want to view the random walk specification of trend
growth $\beta _{t}$ as a purely statistical tool to allow for a slowly
changing mean, rather than interpreting trend growth as an $I(1)$ process.}

It is important to highlight here that \cites{stock.watson:1998} discussion
of the theoretical results of the estimator in Sections $2.2-2.3$ of their
paper emphasizes that MUE of $\lambda $ in the model in \ref{eq:sw98} is
only possible with the \textquotedblleft \textit{normalisation }$\mathbf{D}%
=1 $\textquotedblright . They write at the top of page 351\ (right column):
\textquotedblleft \textit{Henceforth, when }$k=1$\textit{, we thus set }$%
\mathbf{D}=1$\textit{. When }$\mathbf{X}_{t}=1$\textit{, under this
normalization, }$\lambda $\textit{\ is }$T$\textit{\ times the ratio of the
long-run standard deviation of }$\Delta \beta _{t}$\textit{\ to the long run
standard deviation of }$u_{t}$\textit{.}\textquotedblright \footnote{%
The parameter $k$ here refers to the column dimension of regressor vector $%
\mathbf{X}_{t}$. When $k=1$, then only a model with an intercept is fitted,
ie., $\mathbf{X}_{t}$ contains only a unit constant and no other regressors.}
Denoting the long-run standard deviation of a stochastic process by $\bar{%
\sigma}(\cdot )$, this means that%
\begin{equation}
\lambda =T\frac{\bar{\sigma}(\Delta \beta _{t})}{\bar{\sigma}(u_{t})}=T\frac{%
\sigma _{\Delta \beta }}{\sigma _{\varepsilon }/a(1)},  \label{eq:lambda}
\end{equation}%
or alternatively, expressed in signal-to-noise ratio form as used by \cite%
{holston.etal:2017}:%
\begin{equation}
\frac{\lambda }{T}=\frac{\bar{\sigma}(\Delta \beta _{t})}{\bar{\sigma}(u_{t})%
}=\frac{\sigma _{\Delta \beta }}{\sigma _{\varepsilon }/a(1)},
\label{eq:s2n}
\end{equation}%
where $\bar{\sigma}(u_{t})=\sigma _{\varepsilon }/a(1)$ since $u_{t}$
follows a stationary AR(4) process, $a(1)=(1-\sum_{i=1}^{4}a_{i})$, and $%
\bar{\sigma}(\Delta \beta _{t})=\sigma _{\Delta \beta }$ due to $\eta _{t}$
being $i.i.d.$, yielding further the relation $\sigma _{\Delta \beta
}=(\lambda /T)\sigma _{\eta }$. As a result of the identifying
\textquotedblleft \textit{normalization }$\mathbf{D}=1$\textquotedblright\
of MUE, \ref{eq:s2n} implies that $\sigma _{\eta }=\sigma _{\varepsilon
}/a(1)$. That is, the long-run standard deviation of the stationary
component $u_{t}$ is equal to the standard deviation of the trend growth
innovations $\eta _{t}$.

\cite{stock.watson:1998} write on page 354: \textquotedblleft \textit{Table
3 is a lookup table that permits computing median unbiased estimates, given
a value of the test statistic. The normalization used in Table 3 is that $%
\mathbf{D}=1$, and users of this lookup table must be sure to impose this
normalization when using the resulting estimator of $\lambda $.}%
\textquotedblright\ Moreover, the numerical results that are reported in\
Section 3, which is appropriately labelled \textquotedblleft \textit{%
Numerical Results for the univariate Local-Level Model}\textquotedblright ,
are obtained from simulations that employ the local level model of \ref%
{eq:tvp_sim} as the data generating process (see \cites{stock.watson:1998}
GAUSS\ programs \texttt{ESTLAM.GSS}, \texttt{TESTCDF.GSS}, and \texttt{%
LOOKUP.GSS} in the \texttt{tvpci.zip} file that accompanies their paper).
These numerical results do not only include the simulations regarding \emph{%
`pile-up'} at zero frequencies reported in Table 1, asymptotic power
functions plotted in Figure 1, or the AREs provided in Table 2 of \cite%
{stock.watson:1998}, but also the look-up tables for the construction of the
median unbiased estimator of $\lambda $ in Table 3. It must therefore be
kept in mind that these look-up table values are valid only for the
univariate local level model, or for models that can be (re-)written in
\textit{\textquotedblleft local level form\textquotedblright }.

\autoref{tab:sw98_T4} below reports the replication results of Tables 4 and
5 in \cite{stock.watson:1998}.\footnote{%
All computations are implemented in Matlab, using their GDP growth data
provided in the file \texttt{DYPC.ASC}. Note that I also obtained look-up
table values based on a finer grid of $\lambda $ values from their original
GAUSS file \texttt{LOOKUP.GSS} (commenting out the lines \texttt{if
(lamdat[i,1] .<= 30) .and (lamdat[i,1]-floor(lamdat[i,1]) .== 0);} in
\texttt{LOOKUP.GSS} to list look-up values for the entire grid of $\lambda $%
's considered), rather than those listed in Table 3 on page 354 of their
paper, where the grid is based on unit increments in $\lambda $ from 0 to
30. I further changed the settings in the tolerance on the gradient in their
maximum likelihood (maxlik) library routine to \texttt{\_max\_GradTol = 1e-08%
} and used the printing option \texttt{format /rd 14,14} for a more precise
printing of all results up to 14 decimal points. Lastly, there is a small
error in the construction of the lag matrix in the estimation of the AR(4)
model in file \texttt{TST\_GDP1.GSS} (see lines 40 to 47). The first column
in the \texttt{w} matrix is the first lag of the demeaned per capita trend
growth series, while columns 2 to 4 are the second to fourth lags of the
raw, that is, not demeaned per capita trend growth series. Correcting this
leads to mildly higher, yet still insignificant, point estimates of all $%
\sigma_{\Delta\beta}$. For instance, the point estimate of $%
\sigma_{\Delta\beta}$ based on \cites{nyblom:1989} $L$ statistic yields
0.1501, rather than 0.1303, but remains still statistically insignificant,
with the lower value of the confidence interval being 0. To exactly
replicate the results in \cites{stock.watson:1998}, I compute the lag matrix
as they do.} Columns one and two in the top half of \autoref{tab:sw98_T4}
show test statistics and $p-$values of the four structural break tests that
are considered: $i)$ \cites{nyblom:1989} $L$ test, $ii)$ %
\cites{andrews.ploberger:1994} mean Wald (MW) test, $iii)$ %
\cites{andrews.ploberger:1994} exponential Wald (EW) test, and $iv)$ %
\cites{quandt:1960} Likelihood ratio (QLR) test, together with corresponding
$p-$values.

As a reminder, the MW, EW and QLR tests are \cite{chow:1960} type structural
break tests, which test for a structural break in the unconditional mean of
a series at a given or known point in time. \cite{chow:1960} break tests
require a partitioning of the data into two sub-periods. When the break date
is unknown, these tests are implemented by rolling through the sample. To be
more concrete, denote by $\mathcal{Y}_{t}$ the series to be tested for a
structural break in the unconditional mean. Let the dummy variable $%
D_{t}(\tau )=1$ if $t>\tau ,$ and $0$ otherwise, where $\tau =\{\tau
_{0},\tau _{0}+1,\tau _{0}+2,\ldots ,\tau _{1}\}$ is an index (or sequence)
of grid points between endpoints $\tau _{0}$ and $\tau _{1}$. As is common
in this literature, \cite{stock.watson:1998} set these endpoints at the $%
15^{th}$ and $85^{th}$ percentiles of the sample size $T$, that is, $\tau
_{0}=0.15T$ and $\tau _{1}=0.85T$.\footnote{%
To be precise, $\tau _{0}$ is computed as $\mathtt{floor(0.15\ast T)}$ and $%
\tau _{1}$ as $T-\tau _{0}$. Also, it is standard practice in the structural
break literature to trim out some upper/lower percentiles of the search
variable to avoid having too few observations at the beginning or at the end
of the sample in the 0 and 1 dummy regimes created by $D_{t}(\tau )$. In
fact, the large sample approximation of the distribution of the QLR test
statistic depends on $\tau _{0}$ and $\tau _{1}$. \cite{stock.watson:2011}
write to this on page 558: \textquotedblleft \emph{For the large-sample
approximation to the distribution of the QLR statistic to be a good one, the
sub-sample endpoints, }$\tau _{0}$ \emph{and} $\tau _{1}$\emph{, cannot be
too close to the beginning or the end of the sample}.\textquotedblright\
Employing endpoints other than the $15^{th}$ upper/lower percentile values
used by \cite{stock.watson:1998} in the simulation of the look-up table for $%
\lambda $ is thus likely to affect the values provided in Table 3 of \cite%
{stock.watson:1998}, due to the endpoints' influence on the distribution of
the structural break test statistics.} For each $\tau \in \lbrack \tau
_{0},\tau _{1}]$, the following regression of $\mathcal{Y}_{t}$ on an
intercept term and $D_{t}(\tau )$ is estimated:%
\begin{equation}
\mathcal{Y}_{t}=\zeta _{0}+\zeta _{1}D_{t}(\tau )+\epsilon _{t},  \label{Zt}
\end{equation}%
and the $F$ statistic (the square of the $t-$statistic) on the point
estimate $\hat{\zeta}_{1}$ is constructed. The sequence $\{F(\tau )\}_{\tau
=\tau _{0}}^{\tau _{1}}$ of $F$ statistics is then utilized to compute the
MW, EW and QLR structural break test statistics needed in the implementation
of MUE. These are calculated as:\bsq\label{eq:breakTests}%
\begin{align}
\mathrm{MW}& =\frac{1}{N_{\tau }}\sum\limits_{\tau =\tau _{0}}^{\tau
_{1}}F(\tau ) \\
\mathrm{EW}& =\ln \left( \frac{1}{N_{\tau }}\sum_{\tau =\tau _{0}}^{\tau
_{1}}\exp \left\{ \frac{1}{2}F(\tau )\right\} \right)  \label{EW} \\
\mathrm{QLR}& =\max_{\tau \in \lbrack \tau _{0},\tau _{1}]}\{F(\tau
)\}_{\tau =\tau _{0}}^{\tau _{1}},
\end{align}%
\esq where $N_{\tau }$ denotes the number of grid points in $\tau $. %
\cites{nyblom:1989} $L$ test statistic is computed without sequentially
partitioning the data via the sum of squared cumulative sums of $\mathcal{Y}%
_{t}$. More specifically, let $\hat{\mu}_{\mathcal{Y}}$ denote the sample
mean of $\mathcal{Y}_{t}$, $\hat{\sigma}_{\mathcal{Y}}^{2}$ the sample
variance of $\mathcal{Y}_{t}$, and $\mathcal{\tilde{Y}}_{t}=\mathcal{Y}_{t}-%
\hat{\mu}_{\mathcal{Y}}$ the demeaned $\mathcal{Y}_{t}$ process. %
\cites{nyblom:1989} $L$ statistic is then constructed as:%
\begin{equation}
L=T^{-1}\sum_{t=1}^{T}\vartheta _{t}^{2}/\hat{\sigma}_{\mathcal{Y}}^{2},
\label{eqL}
\end{equation}%
where $\vartheta _{t}$ is the scaled cumulative sum of $\mathcal{\tilde{Y}}%
_{t}$, ie., $\vartheta _{t}=T^{-1/2}\sum_{s=1}^{t}\mathcal{\tilde{Y}}_{s} $.

Median unbiased estimates of $\lambda $ based on \cites{stock.watson:1998}
look-up tables are reported in column 3 of \autoref{tab:sw98_T4}, followed
by respective 90\% confidence intervals (CIs) in square brackets. The last
two columns show estimates of $\sigma _{\Delta \beta }$ computed as $\hat{%
\sigma}_{\Delta \beta }=\hat{\lambda}/T\times \hat{\sigma}_{\varepsilon }/%
\hat{a}(1)$, with 90\% CIs also in square brackets.\ In the bottom half of %
\autoref{tab:sw98_T4}, MLE\ and MUE based parameter estimates of the model
in \ref{eq:sw98} are reported. The columns under the MPLE and MMLE headings
show, respectively, MLE based results when the initial state vector is
estimated and when a diffuse prior is used. The diffuse prior for the $I(1)$
element of the state vector is centered at 0 with a variance of $10^{6}$.
The next two columns under the headings MUE(0.13) and MUE(0.62) report
parameter estimates of the model in \ref{eq:sw98} with $\sigma _{\Delta
\beta }$ held fixed at its MUE$\ $point estimate of $0.13$ and upper 90\% CI
value of $0.62$, respectively.
The last column under the heading SW.GAUSS lists the corresponding MUE(0.13)
estimates obtained from running \cites{stock.watson:1998} GAUSS\ code as
reference values.\footnote{%
See the results reported in Table 5 on page 354 in \cite{stock.watson:1998},
where nevertheless only two decimal points are reported. MPLE and MMLE are
also replicated accurately to 6 decimal points.} 

As can be seen from the results in \autoref{tab:sw98_T4}, consistent with
the \emph{`pile-up'} at zero problem documented in the simulations in \cite%
{stock.watson:1998} (and also \cite{shephard.harvey:1990}), the MPLE\
estimate of $\sigma _{\Delta \beta }$ goes numerically to zero (up to 11
decimal points), while MMLE produces a \textit{`sizeable'} point estimate of
$\sigma _{\Delta \beta }$ of $0.044$. Although \cite{stock.watson:1998} (and
also\ I) do not report a standard error for $\hat{\sigma}_{\Delta \beta }$
in the tables containing the estimation results, the estimate of $\mathrm{%
stderr}(\hat{\sigma}_{\Delta \beta })$ is $0.1520$, suggesting that $\hat{%
\sigma}_{\Delta \beta }$ is very imprecisely estimated.\footnote{\cite%
{stock.watson:1998} compute standard errors for the remaining MMLE\
parameters (see column three in the upper part of Table 5 on page 354 in
their paper. They write in the notes to Table 5: \textquotedblleft \textit{%
Because of the nonnormal distribution of the MLE of }$\lambda $, \textit{the
standard error for }$\sigma _{\Delta \beta }$ \textit{is not reported}%
.\textquotedblright\ Evidently, \textit{`testing'} the null hypothesis of $%
\sigma _{\Delta \beta }=0$ using a standard $t-$ratio does not make any
sense statistically. Nevertheless, $\hat{\sigma}_{\Delta \beta }$ is very
imprecisely estimated, and highly likely to be \emph{`very'} close to zero.
The MMLE log-likelihood function with the restriction $\sigma _{\Delta \beta
}=0$ is $-547.5781$, while the (unrestricted) MMLE is $-547.4805$, with the
difference between the two being very small of about $0.10$.} From the MUE\
results reported in the first column of the top half of \autoref{tab:sw98_T4}
it is evident that all 4 structural break tests yield confidence intervals
for $\lambda $ and hence also $\sigma _{\Delta \beta }$ that include zero.
Thus, even when using MUE as the \textit{`preferred'} estimator, one would
conclude that $\hat{\lambda}$ and $\hat{\sigma}_{\Delta \beta }$ are \textit{%
not} statistically different from zero.

An evident practical problem with the use of \cites{stock.watson:1998} MUE\
is that the 4 different structural break tests can produce vastly different
point estimates of $\lambda $. This is clearly visible from \autoref%
{tab:sw98_T4}, where the 4 tests yield $\lambda $ estimates with an implied $%
\hat{\sigma}_{\Delta \beta }$ range between $0.0250$ (for QLR)$\ $and $%
0.1303 $ (for $L$). From the simulation results in \cite{stock.watson:1998}
we know that all 4 tests seem to behave equally well in the \emph{`pile-up'}
at zero frequency simulations (see Table 1 in \cite{stock.watson:1998}).
However, the QLR test performed \emph{`best'} in the efficiency results,
producing the largest (closes to 1)\ asymptotic relative efficiencies in
Table 2 of \cite{stock.watson:1998}. Analysing these results in the context
of the empirical estimation of trend growth, the most accurate MUE estimator
based on the QLR\ structural break test produces an estimate of $\sigma
_{\Delta \beta }$ that is 5 times \textit{smaller} than the largest one
based on the $L$ structural break test, with the MMLE\ estimate of $\sigma
_{\Delta \beta } $ being approximately double the size of the QLR estimate.

To provide a visual feel of how different the MLE\ and MUE based estimates
of U.S. trend growth are, I show plots of the \emph{smoothed} estimates in %
\autoref{fig:sw98_F4} (these correspond to Figures 4 and 3 in \cite%
{stock.watson:1998}). The top panel displays the MPLE, MMLE, MUE(0.13), and
MUE(0.62) estimates together with a 90\%\ CI of the MMLE estimate (shaded
area), as well as a dashed yellow line that shows \cites{stock.watson:1998}
GAUSS code based MUE(0.13) estimate for reference.\ The plot in the bottom
panel of \autoref{fig:sw98_F4} superimposes the actual $GY_{t}$ series to
portray the variability in the trend growth estimates relative to the
variation in the data from which these were extracted.\footnote{%
Notice from the top panel of \autoref{tab:sw98_T4} that there are four
different estimates of $\lambda $, and thus four $\hat{\sigma}_{\Delta \beta
}$. Rather then showing smoothed trend estimates for all four of these, I
follow \cite{stock.watson:1998} and only show estimates based on %
\cites{nyblom:1989} $L$ statistic, which has the largest $\lambda ~$%
estimate, and hence also $\sigma _{\Delta \beta }$.} The $y-$axis range is
set as in Figures 4 and 3 in \cite{stock.watson:1998}. As can be seen from %
\autoref{fig:sw98_F4}, there is only little variability in the MLE based
trend growth estimates, with somewhat more variation from MUE(0.13).
Nonetheless, all three trend growth point estimates stay within the 90\%
error bands of MMLE. Moreover, the plots in \autoref{fig:sw98_F4} confirm
the lack of precision of MUE. Trend growth could be anywhere between a
constant value of about $1.8\%$ ($\hat{\beta}_{00}$ from MPLE), which is a
flat line graphically when $\sigma _{\Delta \beta }$ is held fixed at its
lower $90\%$\ CI value of 0, and a rather volatile series which produces a
range between nearly $4.5\%$ in 1950 and less than $0.5\%$ in 1980 when $%
\sigma _{\Delta \beta }$ is set at its upper $90\%$\ CI value of 0.62.

Given the previous results and discussion, one could argue that the
statistical evidence in support of any important time varying trend growth
in real U.S.\ GDP per capita is rather weak in this model and data set. As a
robustness check and in the context of a broader replication of the time
varying trend growth estimates of \cite{stock.watson:1998}, I obtain real
GDP per capita data from the Federal Reserve Economic Data (FRED2) database
and re-estimate model. These results are reported in \autoref{tab:sw98_T4_2}%
, which is arranged in the same way as \autoref{tab:sw98_T4} (only the last
column with heading SW.GAUSS is removed). The sample period is again from
1947:Q2 to 1995:Q4, using an AR(4)\ model to approximate $u_{t}$ in \ref%
{eq:sw1}.\footnote{%
The results using an ARMA($2,2$) model for $u_{t}$ instead are qualitatively
the same.} From \autoref{tab:sw98_T4_2} it is clear that not only do the two
MLE based estimates of $\sigma _{\Delta \beta }$ yield point estimates that
are numerically equal to zero, but so do all 4 MUEs. Hence, trend growth may
well be constant. More importantly, it demonstrates that MUE\ can also lead
to zero estimates of $\sigma _{\Delta \beta }$ and that there is nothing
unusual about that.\footnote{%
I show later that the Stage 2 MUE procedure of \cite{holston.etal:2017} is
incorrectly implemented and based on a misspecified Stage 2 model. Once this
is corrected, the Stage 2 $\lambda _{z}$ that one obtains is very close to
zero, resulting in the full model MLE\ and MUE\ estimates being very similar.%
}

Before I proceed to describe how the three stage procedure of \cite%
{holston.etal:2017} is implemented, a brief procedural description of %
\cites{stock.watson:1998} MUE\ that lists the main steps needed to replicate
the results reported in \autoref{tab:sw98_T4} and \autoref{fig:sw98_F4} is
provided below.

\begin{enum}
\item Fit an AR(4)\ model to $GY_{t}$, construct $\hat{a}(L)$ from the
estimated AR(4) coefficients $\left\{ \hat{a}_{j}\right\} _{j=1}^{4}$, and
filter the series to remove the AR(4) serial dependence. Let $\widetilde{GY}%
_{t}=\hat{a}(L)GY_{t}$ denote the AR(4)\ filtered series.\footnote{%
This is the generalized least squares (GLS) step in the original TVP model
description on page 350 in \cite{stock.watson:1998}.} Use the residuals $%
\hat{\varepsilon}_{t}$ from the fitted AR(4)\ model for $GY_{t}$ to compute
an estimate of the standard deviation of $\varepsilon _{t}$ and denote it by
$\hat{\sigma}_{\varepsilon }$. Also, let $\hat{a}(1)=\big(1-\sum_{j=1}^{4}%
\hat{a}_{j}\big)$.

\item Test for a structural break in the unconditional mean of the AR(4)
filtered series $\widetilde{GY}_{t}$ using the four structural break tests
described above. That is, replace $\mathcal{Y}_{t}$ in \ref{Zt} with $%
\widetilde{GY}_{t}$, run the dummy variable regression in \ref{Zt}, and
compute the structural break statistics as defined in \ref{eq:breakTests}
and \ref{eqL}.

\item Given these structural break test statistics, use the look-up values
provided in Table 3 on page 354 in \cite{stock.watson:1998} to find the
corresponding $\lambda $ value by interpolation. Once an estimate of $%
\lambda $ is available, compute $\hat{\sigma}_{\Delta \beta }=T^{-1}\hat{%
\lambda}\hat{\sigma}_{\varepsilon }/\hat{a}(1)$, where $\hat{\sigma}%
_{\varepsilon }$ and $\hat{a}(1)$ are obtained from Step $(i)$.

\item With $\sigma _{\Delta \beta }$ held fixed at its median unbiased
estimate obtained in Step $(iii)$, estimate the remaining parameters of the
model in \ref{eq:sw98} using the Kalman Filter and MLE, namely, MPLE, where
the initial value is estimated as well. Finally, using the estimates of the
full set of parameters of the model in \ref{eq:sw98}, apply the Kalman
Smoother to extract an estimate of annualized trend growth of U.S. real per
capita GDP.
\end{enum}

\section{Three stage estimation procedure of \protect\cite{holston.etal:2017}
\label{sec:HLW}}

\cite{holston.etal:2017} employ MUE in two preliminary stages that are based
on restricted versions of the full model in \ref{eq:hlw} to obtain estimates
of the \emph{`signal-to-noise ratios'} $\lambda _{g}=\sigma _{g}/\sigma
_{_{y^{\ast }}}$ and $\lambda _{z}=a_{r}\sigma _{z}/\sigma _{\tilde{y}}$.
These ratios are then held fixed in Stage 3 of their procedure, which
produces estimates of the remaining parameters of the model in \ref{eq:hlw}.
In order to conserve space in the main text, I provide all algebraic details
needed for the replication of the three individual stages in the %
\aref{appendix}, which includes also some additional discussion as well as
R-Code extracts to show the exact computations. In the results that are
reported in this section, I have used their R-Code from the file \href{https://www.newyorkfed.org/medialibrary/media/research/economists/williams/data/HLW_Code.zip}%
{\texttt{HLW\_Code.zip}} made available on Willams' website at the New York
Fed to (numerically) accurately reproduce their results.\footnote{%
Williams' website at the Federal Reserve Bank of New York is at: %
\url{https://www.newyorkfed.org/research/economists/williams/pub}. Their
R-Code is available from the website: %
\url{https://www.newyorkfed.org/medialibrary/media/research/economists/williams/data/HLW_Code.zip}%
. The weblink to the file with their real time estimates is: %
\url{https://www.newyorkfed.org/medialibrary/media/research/economists/williams/data/Holston_Laubach_Williams_real_time_estimates.xlsx}%
. Note here that all my results exactly match their estimates provided in
the \href{https://www.newyorkfed.org/medialibrary/media/research/economists/williams/data/Holston_Laubach_Williams_real_time_estimates.xlsx}%
{\texttt{Holston\_Laubach\_Williams\_real\_time\_estimates.xlsx}} file in
Sheet 2017Q1.} The sample period that I cover ends in 2017:Q1.\ The
beginning of the sample is the same as in \cite{holston.etal:2017}. That is,
it starts in 1960:Q1, where the first 4 quarters are used for initialisation
of the state vector, while the estimation period starts in 1961:Q1.

\cite{holston.etal:2017} adopt the general state-space model (SSM) notation
of \cite{hamilton:1994} in their three stage procedure. The SSM is
formulated as follows:\footnote{%
The state-space form that they use is described on pages 9 to 11 of their
online appendix that is included with the R-Code \texttt{HLW\_Code.zip} file
from Williams' website at the New York Fed. Note that I use exactly the same
state-space notation to facilitate the comparison to \cite{holston.etal:2017}%
, with the only exception being that I include one extra selection matrix
term $\mathbf{S}$ in front of $\boldsymbol{\epsilon }_{t}$ in \ref{eq:RQ} as
is common in the literature to match the dimension of the state vector to $%
\boldsymbol{\epsilon }_{t}$ when there are identities due to lagged values.
I also prefer not to transpose the system matrices $\mathbf{A}$\textbf{\ }and%
\textbf{\ }$\mathbf{H}$ in \ref{eq:RQ}, as it is not necessary and does not
improve the readability.}%
\begin{equation}
\begin{array}{l}
\mathbf{y}_{t}=\mathbf{Ax}_{t}+\mathbf{H}\boldsymbol{\xi }_{t}+\boldsymbol{%
\nu }_{t} \\
\boldsymbol{\xi }_{t}=\mathbf{F}\boldsymbol{\xi }_{t-1}+\mathbf{S}%
\boldsymbol{\varepsilon }_{t}%
\end{array}%
\text{, \ \ where }%
\begin{bmatrix}
\boldsymbol{\nu }_{t} \\
\boldsymbol{\varepsilon }_{t}%
\end{bmatrix}%
\sim \mathsf{MNorm}\left(
\begin{bmatrix}
\boldsymbol{0} \\
\boldsymbol{0}%
\end{bmatrix}%
,%
\begin{bmatrix}
\mathbf{R} & \boldsymbol{0} \\
\boldsymbol{0} & \mathbf{W}%
\end{bmatrix}%
\right) ,  \label{eq:RQ}
\end{equation}%
where we can define $\boldsymbol{\epsilon }_{t}=\mathbf{S}\boldsymbol{%
\varepsilon }_{t}$, so that $\mathrm{Var}(\boldsymbol{\epsilon }_{t})=%
\mathrm{Var}(\mathbf{S}\boldsymbol{\varepsilon }_{t})=\mathbf{SWS}^{\prime }=%
\mathbf{Q}$ to make it consistent with the notation used in \cite%
{holston.etal:2017}. The (observed) measurement vector is denoted by $%
\mathbf{y}_{t}$ in \ref{eq:RQ}, $\mathbf{x}_{t}$ is a vector of exogenous
variables, $\mathbf{A}$\textbf{, }$\mathbf{H}$ and $\mathbf{F}$ are
conformable system matrices, $\boldsymbol{\xi }_{t}$ is the latent state
vector, $\mathbf{S}$ is a selection matrix, and the notation $\mathsf{MNorm}%
\left( \boldsymbol{\mu },\boldsymbol{\Sigma }\right) $ denotes a
multivariate normal random variable with mean vector $\boldsymbol{\mu }$ and
covariance matrix $\boldsymbol{\Sigma }$. The disturbance terms $\boldsymbol{%
\nu }_{t}$ and $\boldsymbol{\varepsilon }_{t}$ are serially uncorrelated,
and the (individual) covariance matrices $\mathbf{R}$ and $\mathbf{W}$ are
assumed to be diagonal matrices, implying zero correlation between the
elements of the measurement and state vector disturbance terms. The
measurement vector $\mathbf{y}_{t}$ in \ref{eq:RQ} is the same for all three
stages and is defined as $\mathbf{y}_{t}=[y_{t},~\pi _{t}]^{\prime }$, where
$y_{t}$ and $\pi _{t}$ are the log of real GDP\ and annualized PCE
inflation, respectively, as defined in \Sref{sec:model}. The exact form of
the remaining components of the SSM\ in \ref{eq:RQ} changes with the
estimation stage that is considered, and is described in detail either in
the text below or in the \aref{appendix}.

As I have emphasized in the description of MUE in \Sref{sec:MUE}, the
simulation results of \cite{stock.watson:1998} show that \emph{`pile-up'} at
zero frequencies for MLE are not only a function of the size of the variance
of $\Delta \beta _{t}=(\lambda /T)\eta _{t}$ (or alternatively $\lambda $),
but also depend critically on whether the initial condition of the state
vector is estimated or not. Now \cite{holston.etal:2017} \emph{do not}
estimate the initial condition of the state vector in any of the three
stages that are implemented. Instead, they apply the HP filter to log GDP
data with the smoothing parameter set to $36000$ to get a preliminary
estimate of $y_{t}^{\ast }$ and trend growth $g_{t}$ (computed as the first
difference of the HP\ filter estimate of $y_{t}^{\ast }$) using data from
1960:Q1 onwards. \textit{`Other factor' }$z_{t}$ is initialized at 0.%
\footnote{%
See the listing in \coderef{R:stage3}{1} in the \hyperref[Rcode]{A.6~R-Code
Snippets} section of the \aref{appendix}, which shows the first 122 lines of
their R-file \texttt{rstar.stage3.R}. Line 30 shows the construction of the
initial state vector as $\boldsymbol{\xi }_{00}=[y_{0}^{\ast },y_{-1}^{\ast
},y_{-2}^{\ast },g_{-1},g_{-2},z_{-1},z_{-2}]^{\prime }$ where subscripts $%
[0,-1,-2]$ refer to the time periods 1960:Q4, 1960:Q3, and 1960:Q2,
respectively. In terms of their R-Code, we have: \texttt{xi.00 <- c(100*%
\texttt{g.pot}[3:1],100*g.pot.diff[2:1],0,0)}, where \texttt{g.pot} is the
HP filtered trend and \texttt{g.pot.diff} is its first difference, ie.,
trend growth, with the two zeros at the end being the initialisation of $%
z_{t}$. This yields the following numerical values: [806.45, 805.29, 804.12,
1.1604, 1.1603, 0, 0]. The same strategy is also used in the first two
stages (see their R-files \texttt{rstar.stage1.R} and \texttt{rstar.stage2.R}%
).\label{fn:1}} This means that $\boldsymbol{\xi }_{00}$ has known and fixed
quantities in all three stages.\ Given the simulation evidence provided in
Table 1 on page 353 in \cite{stock.watson:1998}, one may thus expect a
priori \emph{`pile-up'} at zero frequencies of MLE (without estimation of
the initial conditions) to be only marginally larger than those of MUE,
especially for everything but very small values of $\lambda $.

Also, \cite{holston.etal:2017} determined the covariance matrix of the
initial state vector in an unorthodox way. Even though every element of the
state vector $\boldsymbol{\xi }_{t}$ in all three estimation stages is an $%
I(1)$ variable, they do not employ a diffuse prior on the state vector.
Instead, the covariance matrix is determined with a call to the function
\texttt{calculate.covariance.R} (see the code snippet in \coderef{R:covar}{2}
for details on this function, and also lines 66, 84, and 88, respectively,
in their R-files \texttt{rstar.stage1.R}, \texttt{rstar.stage2.R}, and
\texttt{rstar.stage3.R}, with line 88 in \texttt{rstar.stage3.R }also shown
on the second page of the code snippet in \coderef{R:stage3}{1}). To
summarize what this function does, consider the Stage 1 model, which is
estimated with a call to \texttt{rstar.stage1.R}. The function \texttt{%
calculate.covariance.R }first sets the initial covariance matrix to $0.2$
times a three dimensional identity matrix $\mathbf{I}_{3}$. Their procedure
then continues by using data from 1961:Q1 to the end of the sample to get an
estimate of $\sigma _{y^{\ast }}^{2}$ from the Stage 1 model. Lastly, the
initial covariance matrix $\mathbf{P}_{00}$ to be used in the \textit{`}%
\emph{final}\textit{'} estimation of the Stage 1 model is then computed as:%
\bsq\label{eq:P00S1}%
\begin{align}
\mathbf{P}_{00}& =\mathbf{F}\,\mathrm{\mathrm{diag}}([0.2,~0.2,~0.2])\,%
\mathbf{F}^{\prime }+\mathbf{\hat{Q}}  \label{eq:P00S1a} \\
& =%
\begin{bmatrix}
1 & 0 & 0 \\
1 & 0 & 0 \\
0 & 1 & 0%
\end{bmatrix}%
\begin{bmatrix}
0.2 & 0 & 0 \\
0 & 0.2 & 0 \\
0 & 0 & 0.2%
\end{bmatrix}%
\begin{bmatrix}
1 & 0 & 0 \\
1 & 0 & 0 \\
0 & 1 & 0%
\end{bmatrix}%
^{\prime }+%
\begin{bmatrix}
\hat{\sigma}_{y^{\ast }}^{2} & 0 & 0 \\
0 & 0 & 0 \\
0 & 0 & 0%
\end{bmatrix}
\label{eq:P00S1b} \\[2mm]
& =%
\begin{bmatrix}
0.4711 & 0.2 & 0.0 \\
0.2 & 0.2 & 0.0 \\
0.0 & 0.0 & 0.2%
\end{bmatrix}%
,  \label{eq:P00S1c}
\end{align}%
\esq with $\mathbf{\hat{Q}}$ a $(3\times 3)$ dimensional zero matrix with
element $(1,1)$ set to $\hat{\sigma}_{y^{\ast }}^{2}=0.27113455739$ from the
initial run of the Stage 1 model. What this procedure effectively does is to
set $\mathbf{P}_{00}$ to the first time period's predicted state covariance
matrix, given an initial state covariance matrix of $0.2\times \mathbf{I}%
_{3} $ and the estimate $\hat{\sigma}_{y^{\ast }}^{2}$, where $\hat{\sigma}%
_{y^{\ast }}^{2}$ was obtained by MLE\ and the Kalman Filter using $%
0.2\times \mathbf{I}_{3}$ as the initial state covariance. This way of
initialising $\mathbf{P}_{00}$ is rather circular, as it fundamentally
presets $\mathbf{P}_{00}$ at $0.2\times \mathbf{I}_{3}$.\footnote{%
In footnote 6 on page S64 in \cite{holston.etal:2017} (and also in the
description of the \texttt{calculate.covariance.R} file), they write:
\textquotedblleft \emph{We compute the covariance matrix of these states
from the gradients of the likelihood function.}\textquotedblright\ Given the
contents of the R-Code, it is unclear how and if this was implemented.}

When the state vector contains $I(1)$ variables, it is not only standard
practice to use a diffuse prior, but it is highly recommended. For instance,
\cite{harvey:1989} writes to this on the bottom of page 121:
\textquotedblleft \textit{When the transition equation is non-stationary,
the unconditional distribution of the state vector is not defined. Unless
genuine prior information is available, therefore, the initial distribution
of }$\boldsymbol{\alpha }_{0}$\textit{\ \textbf{must} be specified in terms
of a diffuse or non-informative prior}.\textquotedblright\ (emphasis added, $%
\boldsymbol{\alpha }_{t}$ is the state-vector in Harvey's notation). It is
not clear why \cite{holston.etal:2017} do not use a diffuse prior.\footnote{%
In an earlier paper using a similar model for the NAIRU, \cite{laubach:2001}
discusses the use of diffuse priors. \cite{laubach:2001} writes on page
222:\ \textquotedblleft \emph{The most commonly used approach in the
presence of a nonstationary state is to integrate the initial value out of
the likelihood by specifying an (approximately) diffuse
prior.\textquotedblright\ }He then proceeds to describe an alternative
procedure that can be implemented by using: \textquotedblleft \emph{a few
initial observations to estimate the initial state by GLS, and use the
covariance matrix of the estimator as initial value for the conditional
covariance matrix of the state.}\textquotedblright\ The discussion is then
closed with the statement: \textquotedblleft \emph{This is the first
approach considered here. Because this estimate of the initial state and its
covariance matrix are functions of the model parameters, under certain
parameter choices the covariance matrix may be ill conditioned. The routines
then choose the diffuse prior described above as default}.\textquotedblright%
\ Thus even here, the diffuse prior is the "\emph{safe}" default option.
Note that their current procedure does not use: \textquotedblleft \emph{a
few initial observations to estimate the initial state}\textquotedblright ,
but the same sample of data that are used in the final model, ie., with data
beginning in 1961:Q1.} However, one may conjecture that it could be due to
their preference for reporting Kalman Filtered (one-sided) rather than the
more efficient Kalman Smoothed (two-sided) estimates of the latent state
vector $\boldsymbol{\xi }_{t}$ which includes trend growth $g_{t}$ and \emph{%
`other factor'} $z_{t}$ needed to construct $r_{t}^{\ast }$.\footnote{%
Note that Filtered estimates of $g_{t}$, $z_{t}$ and thus also $r_{t}^{\ast
} $ are very volatile at the beginning of the sample period (until about
1970) when $\mathbf{P}_{00}$ is initialized with a diffuse prior.}

As a final point in relation to the probability of \emph{`pile-up'} at zero
problems arising due to small variances of the state innovations, and hence
the rationale for employing MUE rather than MLE\ in the first place, one can
observe from the size of the $\sigma _{g}$ and $\sigma _{z}$ estimates for
the U.S. reported in Table 1 on page S60 in \cite{holston.etal:2017} that
these are rather \emph{`large'} at $0.122$ and $0.150$, respectively. The
simulation results in Table 1 in \cite{stock.watson:1998} show that \emph{%
`pile-up'} at zero frequencies drop to $0.01$ for both, MMLE\ and MUE, when
the true population value of $\lambda $ is $30$ ($\sigma _{\Delta \beta
}=0.06$). Given the fact that \cite{holston.etal:2017} do not estimate the
initial value of the state vector, and that their median unbiased estimates
are about two times larger than $0.06$, it seems highly implausible that
\emph{`pile-up'} at zero problems should materialize with a higher
probability for MLE than for MUE.

\subsection{Stage 1 model \label{sec:S1}}

\cites{holston.etal:2017} first stage model takes the following restricted
form of the full model presented in equation \ref{eq:hlw}:\footnote{%
See \hyperref[sec:AS1]{Section A.1} in the \hyperref[appendix]{Appendix} for
the exact matrix expressions and expansions of the first stage SSM. Note
that one key difference of \cites{holston.etal:2017} SSM specification
described in equations \ref{AS1:m} and \ref{AS1:s} in the \hyperref[appendix]%
{Appendix} is that the expansion of the system matrices for the Stage 1
model does not include the drift term $g$ in the trend specification in \ref%
{S1d}, so that $y_{t}^{\ast }$ follows a random walk \emph{without} drift.
Evidently, such a specification cannot match the upward trend in the GDP
data. To resolve this mismatch, \cite{holston.etal:2017} `\textit{detrend'}
output $y_{t}$ in the estimation (see \hyperref[sec:AS1]{Section A.1} in the
\hyperref[appendix]{Appendix} which describes how this is done and also
shows snippets of their R-Code).}\bsq\label{eq:stag1}%
\begin{align}
y_{t}& =y_{t}^{\ast }+\tilde{y}_{t}  \label{S1a} \\
\pi _{t}& =b_{\pi }\pi _{t-1}+\left( 1-b_{\pi }\right) \pi _{t-2,4}+b_{y}%
\tilde{y}_{t-1}+\varepsilon _{t}^{\pi }  \label{S1b} \\
\tilde{y}_{t}& =a_{y,1}\tilde{y}_{t-1}+a_{y,2}\tilde{y}_{t-2}+\mathring{%
\varepsilon}_{t}^{\tilde{y}}  \label{S1c} \\
y_{t}^{\ast }& =g+y_{t-1}^{\ast }+\mathring{\varepsilon}_{t}^{y^{\ast }}\!\!,
\label{S1d}
\end{align}%
\esq where the vector of Stage 1 parameters to be estimated is:%
\begin{equation}
\boldsymbol{\theta }_{1}=[a_{y,1},~a_{y,2},~b_{\pi },~b_{y},~g,~\sigma _{%
\tilde{y}},~\sigma _{\pi },~\sigma _{y^{\ast }}]^{\prime }.  \label{S1theta1}
\end{equation}%
To be able to distinguish the disturbance terms of the full model in \ref%
{eq:hlw} from the ones in the restricted Stage 1 model in \ref{eq:stag1}
above, I have placed a ring $(\mathring{\phantom{y}})$ symbol on the error
terms in \ref{S1c} and \ref{S1d}. These two disturbance terms from the
restricted model are defined as:
\begin{equation}
\mathring{\varepsilon}_{t}^{y^{\ast }}=g_{t-1}-g+\varepsilon _{t}^{y^{\ast }}
\label{S1eps_ystar0}
\end{equation}%
and
\begin{equation}
\mathring{\varepsilon}_{t}^{\tilde{y}}=\tfrac{a_{r}}{2}[\left(
r_{t-1}-4g_{t-1}-z_{t-1}\right) +\left( r_{t-2}-4g_{t-2}-z_{t-2}\right)
]+\varepsilon _{t}^{\tilde{y}}.  \label{S1eps_ytilde0}
\end{equation}%
From the relations in \ref{S1eps_ystar0} and \ref{S1eps_ytilde0} it is clear
that, due to the restrictions in the Stage 1 model, the error terms $%
\mathring{\varepsilon}_{t}^{\tilde{y}}$ and $\mathring{\varepsilon}%
_{t}^{y^{\ast }}$ in \ref{eq:stag1} will not be uncorrelated anymore, since $%
\mathrm{Cov}(\mathring{\varepsilon}_{t}^{\tilde{y}},\mathring{\varepsilon}%
_{t}^{y^{\ast }})=-\tfrac{a_{r}}{2}4\sigma _{g}^{2}$ given the assumptions
of the full model in \ref{eq:hlw}.\ The separation of trend and cycle shocks
in this formulation of the Stage 1 model is thus more intricate, as both
shocks will respond to one common factor, the missing $g_{t-1}$.

In the implementation of the Stage 1 model, \cite{holston.etal:2017} make
two important modelling choices that have a substantial impact on the $%
\boldsymbol{\theta }_{1}$ parameter estimates, and thus also the estimate of
the `\emph{signal-to-noise ratio}' $\lambda _{g}$ used in the later stages.
The first is the tight specification of the prior variance of the initial
state vector $\mathbf{P}_{00}$ discussed in the introduction of this
section. The second is a lower bound restriction on $b_{y}$ in the inflation
equation in \ref{S1b} ($b_{y}\geq 0.025$ in the estimation). The effect of
these two choices on the estimates of the Stage 1 model parameters are shown
in \autoref{tab:Stage1} below. The left block of the estimates in \autoref%
{tab:Stage1} (under the heading `HLW Prior') reports four sets of results
where the state vector was initialized using their values for $\boldsymbol{%
\xi }_{00}$ and $\mathbf{P}_{00}$. The first column of this block
(HLW.R-File) reports estimates from running \cites{holston.etal:2017} R-Code
for the first stage model. These are reported as reference values. The
second column ($b_{y}\geq 0.025$) shows my replication of %
\cites{holston.etal:2017} results using the same initial values for
parameter vector $\boldsymbol{\theta }_{1}$ in the optimisation routine and
also the same lower bound constraint on $b_{y}$. The third column
(Alt.Init.Vals) displays the results I obtain when a different initial value
for $b_{y}$ is used, with the lower bound restriction $b_{y}\geq 0.025$
still in place. The fourth column ($b_{y} $ Free)\ reports results when the
lower bound constraint on $b_{y}$ is removed.\footnote{\label{FN:initVals}
To find the initial values for $\boldsymbol{\theta }_{1}$, \cite%
{holston.etal:2017} apply the HP filter to GDP\ to obtain an initial
estimate of the cycle and trend components of GDP. These estimates are then
used to find initial values for (some of) the components of parameter vector
$\boldsymbol{\theta }_{1}$ by running OLS\ regressions of the HP cycle
estimate on two of its own lags (an AR(2) essentially), and by running
regressions of inflation on its own lags and one lag of the HP cycle.
Interestingly, although readily available, rather than taking the
coefficient on the lagged value of the HP cycle in the initialization of $%
b_{y}$, which yields a value of $0.0921$, \cite{holston.etal:2017} use the
lower bound value of $0.025$ for $b_{y}$ as the initial value. In the
optimisation, this has the effect that the estimate for $b_{y}$ is
effectively stuck at $0.025$, although it is not the global optimum in the
restricted model, which is at $b_{y}=0.097185$ (see also the values of the
log-likelihood function reported in the last row of \autoref{tab:Stage1}).}
The right block in \autoref{tab:Stage1} shows parameter estimates when a
diffuse prior for $\boldsymbol{\xi }_{t}$ is used, where $\mathbf{P}_{00}$
is set to $10^{6}$ times a three dimensional identity matrix, with the left
and right columns showing, respectively, the estimates with and without the
lower bound restriction on $b_{y}$ imposed. 

Notice initially from the first two columns in the left block of \autoref%
{tab:Stage1} that their numerical results are accurately replicated up to 6
decimal points. From these results we also see that the lower bound
restriction on $b_{y}$ is binding. \cite{holston.etal:2017} set the initial
value for $b_{y}$ at $0.025$, and there is no movement away from this value
in the numerical routine. Specifying an alternative initial value for $b_{y}$%
, which is determined in the same way as for the remaining parameters in $%
\boldsymbol{\theta }_{1}$, leads to markedly different estimates, while
removing the lower bound restriction on $b_{y}$ all together results in the
ML estimate of $b_{y}$ to converge to zero. Evidently, these three scenarios
yield also noticeably different values for ${\hat{\sigma}}_{y^{\ast }}$,
that is, values between $0.4190$ and $0.6177$. The diffuse prior based
results (with and without the lower bound restriction) in the right block of %
\autoref{tab:Stage1} show somewhat less variability in ${\hat{\sigma}}%
_{y^{\ast }}$, but affect the persistence of the cycle variable $\tilde{y}%
_{t}$ in the model, with the smallest AR(2) lag polynomial root being 1.1190
when $b_{y}\geq 0.025$ is imposed, while it is only 1.0251 and thus closer
to the unit circle when $b_{y}$ is left unrestricted.

As a final comment, there is only little variation in the likelihoods of the
different estimates that are reported in the respective left and right
blocks of \autoref{tab:Stage1}. For instance, the largest difference in
log-likelihoods is obtained from the diffuse prior results shown in the
right block of \autoref{tab:Stage1}. If we treat the lower bound as a
restriction, a Likelihood Ratio (LR) test of the null hypothesis of the
difference in these likelihoods being zero yields $%
-2(-536.9803-(-535.9596))=2.0414$, which, with one degree of freedom has a $%
p-$value of $0.1531$ and cannot be rejected at conventional significance
levels. Hence, there is only limited information in the data to compute a
precise estimate of $b_{y}$. This empirical fact is known in the literature
as a \emph{`flat Phillips curve'}.\footnote{%
That the output gap is nearly uninformative for inflation (forecasting) once
structural break information is conditioned upon --- regardless of what
measure of the output gap is used or whether it is combined as an ensemble
from multiple measures --- is shown in \cite{buncic.muller:2017} for the
U.S. and for Switzerland.}

Given the Stage 1 estimate $\skew{0}\boldsymbol{\hat{\theta}}_{1}$, \cite%
{holston.etal:2017} use the following steps to implement median unbiased
estimation of their `\emph{signal-to-noise ratio}' $\lambda _{g}=\sigma
_{g}/\sigma _{y^{\ast }}$.

\begin{enuma}
\item Use the Stage 1 model to extract an estimate of $y_{t}^{\ast }$ from
the Kalman Smoother and construct annualised trend growth as $\Delta \hat{y}%
_{t|T}^{\ast }=400(\hat{y}_{t|T}^{\ast }-\hat{y}_{t-1|T}^{\ast })$, where $%
\hat{y}_{t|T}^{\ast }$ here denotes the Kalman Smoothed estimate of $%
y_{t}^{\ast }$.\footnote{%
Note that, although the series is annualised (scaled by 400), this does not
have an impact on the magnitude of the structural break tests. The numerical
values that one obtains for $\lambda _{g}$ are identical if scaled by 100
instead.}

\item Apply the three structural break tests described in \ref{eq:breakTests}
to the $\Delta \hat{y}_{t|T}^{\ast }$ series. Specifically, replace $Y_{t}$
in \ref{Zt} with the constructed $\Delta \hat{y}_{t|T}^{\ast }$ series, run
the dummy variable regression in \ref{Zt}, and compute the structural break
statistics as defined in \ref{eq:breakTests} and \ref{eqL}. Note that \cite%
{holston.etal:2017} specify the endpoint values of the search-grid over $%
\tau $ at $\tau _{0}=4$ and $\tau _{1}=T-4$.\footnote{%
This effectively tests for a structural break in nearly every time period in
the sample. Interestingly, adjusting the $\tau $ grid to cover the $15^{th}$
upper/lower percentiles of $T$ as in \cite{stock.watson:1998} leads to no
important differences in the structural break test statistics, or the size
of the $\lambda $ estimates that one obtains in Stage 1. Nevertheless, it
should be kept in mind that it is not clear what critical values the
structural break test statistics should be compared to and also what $%
\lambda $ values for MUE are the appropriate ones to use with such endpoint
values. Also, \cite{holston.etal:2017} do not compute \cites{nyblom:1989} $L$
statistic.}

\item Given the structural break test statistics computed in Step ($b$),
find the corresponding $\lambda $ values in the look-up table of \cite%
{stock.watson:1998}. Return the ratio ${\lambda }/T=\sigma _{g}/\sigma
_{y^{\ast }}$ which \cite{holston.etal:2017} denoted by $\lambda _{g}$,
where their preferred estimate of $\lambda $ is based on the EW statistic of
\cite{andrews.ploberger:1994}.
\end{enuma}

\autoref{tab:Stage1_lambda_g} shows the range of $\lambda _{g}$ estimates
computed from the five sets of $\skew{0}\boldsymbol{\hat{\theta}}_{1}$
values reported in \autoref{tab:Stage1}, using all four structural break
tests of \cite{stock.watson:1998}. \autoref{tab:Stage1_lambda_g} is arranged
in the same format as \autoref{tab:Stage1}, again showing %
\cites{holston.etal:2017} estimates of $\lambda _{g}$ obtained from running
their R-Code in the first column of the left block for reference. As can be
seen from \autoref{tab:Stage1_lambda_g}, the range of $\hat{\lambda}_{g}$
values one obtains from \cites{holston.etal:2017} MUE procedure is between $%
0 $ to $0.08945$ (if only the three structural break tests implemented by
\cite{holston.etal:2017} are considered, and up to $0.09419$ if the $L$
statistic is computed as well. Note that this range is not due to
statistical uncertainty, but simply due to the choice of structural break
test, which prior for $\mathbf{P}_{00}$ is used, and whether the lower bound
constraint on $b_{y}$ is imposed. Since these estimates determine the
relative variation in trend growth through the magnitude of $\sigma
_{y^{\ast }}$, they have a direct impact not only on the variation in the
permanent component of GDP, but also on the natural rate of interest through
the ratio $\lambda _{g}=\sigma _{g}/\sigma _{y^{\ast }}$ utilized in the
later stages of the three step procedure of \cite{holston.etal:2017}. %

\subsubsection{\cites{holston.etal:2017} rational for MUE in Stage 1 \label%
{S1MUE}}

Comparing the MUE procedure that \cite{holston.etal:2017} implement to the
one by \cite{stock.watson:1998}, it is evident that they are fundamentally
different. Instead of rewriting the true model of interest in local level
form to make it compatible with \cites{stock.watson:1998} look-up tables,
\cite{holston.etal:2017} instead formulate a restricted Stage 1 model that
not only sets $a_{r}$ in the output gap equation to zero, but also makes the
awkward assumption that trend growth is constant when computing the \emph{%
`preliminary'} estimate of $y_{t}^{\ast }$.

The rationale behind \cites{holston.etal:2017} implementation of MUE is as
follows. Suppose we observe trend $y_{t}^{\ast }$. Then, a local level model
for $\Delta y_{t}^{\ast }$ can be formulated as:\bsq\label{eq:hlwLL1}%
\begin{align}
\Delta y_{t}^{\ast }& =g_{t}+\varepsilon _{t}^{y^{\ast }}  \label{hlw_ll1a}
\\
\Delta g_{t}& =\varepsilon _{t}^{g},  \label{hlw_ll1b}
\end{align}%
\esq where $\Delta y_{t}^{\ast }$, $g_{t}$ and $\varepsilon _{t}^{y^{\ast }}$
are the analogues to $GY_{t},\beta _{t}$ and $u_{t}$, respectively, in %
\cites{stock.watson:1998} MUE in \ref{eq:sw98}, with $\varepsilon
_{t}^{y^{\ast }}$ in \ref{hlw_ll1a}, nonetheless, assumed to be $i.i.d.$
rather than an autocorrelated AR(4) process as $u_{t}$ in \ref{eq:sw1}.
Under \cites{stock.watson:1998} assumptions, MUE of the local level model in %
\ref{eq:hlwLL1} yields $\lambda _{g}=\lambda /T$ defined as:%
\begin{equation}
\frac{\lambda }{T}=\frac{\bar{\sigma}(\varepsilon _{t}^{g})}{\bar{\sigma}%
(\varepsilon _{t}^{y^{\ast }})}=\frac{\sigma _{g}}{\sigma _{y^{\ast }}},
\label{hlw_rationale1}
\end{equation}%
where $\bar{\sigma}(\cdot )$ denotes again the long-run standard deviation,
and the last equality in \ref{hlw_rationale1} follows due to $\varepsilon
_{t}^{y^{\ast }}$ and $\varepsilon _{t}^{g}$ assumed to be uncorrelated
white noise processes.

Since $\Delta y_{t}^{\ast }$ is not observed, \cite{holston.etal:2017}
replace it with the Kalman Smoother based estimate $\Delta \hat{y}%
_{t|T}^{\ast }$ obtained from the restricted Stage 1 model in \ref{eq:stag1}%
. To illustrate what impact this has on their MUE procedure, let $%
a_{y}(L)=(1-a_{y,1}L-a_{y,2}L^{2})$ and $a_{r}(L)=\tfrac{a_{r}}{2}(L+L^{2})$
denote two lag polynomials that capture the dynamics in the output gap $%
\tilde{y}_{t}$ and the real rate cycle $\tilde{r}_{t}=(r_{t}-r_{t}^{\ast
})=(r_{t}-4g_{t}-z_{t})$, respectively. Also, define $\psi
(L)=a_{y}(L)^{-1}a_{r}(L)$ and $\psi (1)=a_{r}/(1-a_{y,1}-a_{y,2})$. The
output gap equation of the true (full) model in \ref{eq:hlw} can then be
written compactly as:%
\begin{align}
a_{y}(L)\tilde{y}_{t}& =a_{r}(L)\tilde{r}_{t}+\varepsilon _{t}^{\tilde{y}},
\label{s2dya} \\
\intxt{or in differenced form and solved for $\Delta \tilde{y}_t$ as:}\Delta
\tilde{y}_{t}& =a_{y}(L)^{-1}\left[ a_{r}(L)\Delta \tilde{r}_{t}+\Delta
\varepsilon _{t}^{\tilde{y}}\right] .  \label{s2dyb}
\end{align}%
Observed output, and trend and cycle are related by the identity%
\begin{align}
y_{t}& =y_{t}^{\ast }+\tilde{y}_{t}  \notag \\
\therefore \Delta y_{t}& =\Delta y_{t}^{\ast }+\Delta \tilde{y}_{t}.
\label{trndCycl}
\end{align}%
This relation, together with \ref{hlw_ll1a} and \ref{s2dyb}, can be written
as:
\begin{align}
\Delta y_{t}-\Delta \tilde{y}_{t}& =\Delta y_{t}^{\ast }  \notag \\
\Delta y_{t}-\underbrace{a_{y}(L)^{-1}\left[ a_{r}(L)\Delta \tilde{r}%
_{t}+\Delta \varepsilon _{t}^{\tilde{y}}\right] }_{\Delta \tilde{y}_{t}}& =%
\underbrace{g_{t}+\varepsilon _{t}^{y^{\ast }}}_{_{\Delta y_{t}^{\ast }}}.
\label{eq:mis1}
\end{align}%
\qquad

Because the data $\Delta y_{t}$ are fixed, any restriction imposed on the $%
\Delta \tilde{y}_{t}$ process translates directly into a misspecification of
the right hand side of \ref{eq:mis1}; the $\Delta y_{t}^{\ast }$ term. In
the Stage 1 model, $a_{r}$ is restricted to zero. For the relation in \ref%
{eq:mis1} to balance, $\Delta y_{t}^{\ast }$ effectively becomes:\footnote{%
Note that we need to formulate a local level model for trend growth as in %
\ref{eq:hlwLL1} to be able to apply the MUE\ framework of \cite%
{stock.watson:1998}. To arrive at \ref{hlw_ll2a}, add $%
[a_{y}(L)^{-1}a_{r}(L)\Delta \tilde{r}_{t}]$ to both sides of \ref{eq:mis1}.
The ring $(\mathring{\phantom{y}})$ symbol on $\mathring{\nu}_{t}^{y\ast }$
highlights again that it is obtained from the restricted model.
} \bsq\label{S1LLfalse}%
\begin{align}
\Delta y_{t}^{\ast }& =g_{t}+\mathring{\nu}_{t}^{y\ast }  \label{hlw_ll2a} \\
\Delta g_{t}& =\varepsilon _{t}^{g},  \label{hlw_ll2b}
\end{align}%
\esq where
\begin{equation}
\mathring{\nu}_{t}^{y\ast }=\varepsilon _{t}^{y^{\ast }}+\psi (L)\Delta
\tilde{r}_{t}.
\end{equation}%
\cites{holston.etal:2017} implementation of MUE relies on the (constructed)
local level model relations from the restricted Stage 1 model in \ref%
{S1LLfalse} and requires us to evaluate the ratio of the long-run standard
deviations of $\varepsilon _{t}^{g}$ and $\mathring{\nu}_{t}^{y\ast }$:
\begin{equation}
\frac{\bar{\sigma}(\varepsilon _{t}^{g})}{\bar{\sigma}(\mathring{\nu}%
_{t}^{y\ast })}.  \label{s2n_s1}
\end{equation}%
Evidently, $\varepsilon _{t}^{g}$ in \ref{hlw_ll2b} has not changed, so the
numerator of the \emph{`signal-to-noise ratio'} in \ref{s2n_s1} is still $%
\bar{\sigma}(\varepsilon _{t}^{g})=\sigma _{g}\ $, due to $\varepsilon
_{t}^{g}$ being an $i.i.d.$ process. However, the term $\mathring{\nu}%
_{t}^{y\ast }$ in \ref{hlw_ll2a} is not uncorrelated white noise anymore.
Moreover, the long-run standard deviation $\bar{\sigma}(\mathring{\nu}%
_{t}^{y\ast })$ in the denominator of \ref{s2n_s1} now also depends on the
(long-run) standard deviation of $\psi (L)\Delta \tilde{r}_{t}$, and will be
equal to $\sigma _{y^{\ast }}$ if and only if $a_{r}=0$ in the \textit{%
empirical data}.\footnote{%
If monetary policy is believed to be effective in cyclical aggregate demand
management, then $a_{r}$ cannot be 0 and one would not have formulated the
main model of interest assuming that $a_{r}$ is different from zero (viz,
negative). Also, this restriction cannot be enforced in the data.}

To see what the long-run standard deviation of $\mathring{\nu}_{t}^{y\ast }$
looks like, assume for simplicity that $\varepsilon _{t}^{y^{\ast }}$ and $%
\Delta \tilde{r}_{t}$ are uncorrelated, so that the long-run standard
deviation calculation of $\mathring{\nu}_{t}^{y\ast }$ can be broken up into
a part involving $\varepsilon _{t}^{y^{\ast }}$ and another part involving $%
\psi (L)\Delta \tilde{r}_{t}$, where the latter decomposes as:%
\begin{align*}
\psi (L)\Delta \tilde{r}_{t}& =\psi (L)[\Delta r_{t}-4\Delta g_{t}-\Delta
z_{t}] \\
& =\psi (L)[\Delta r_{t}-4\varepsilon _{t}^{g}-\varepsilon _{t}^{z}].
\end{align*}%
Assuming that the shocks $\{\varepsilon _{t}^{g},\varepsilon _{t}^{z}\}$ are
uncorrelated with the (change in the) real rate $\Delta r_{t}$, the long-run
standard deviation of $\psi (L)\Delta \tilde{r}_{t}$ can be evaluated as:%
\begin{align}
\bar{\sigma}\left( \psi (L)\Delta \tilde{r}_{t}\right) & =\bar{\sigma}\left(
\psi (L)\Delta r_{t}\right) +\bar{\sigma}\left( \psi (L)4\varepsilon
_{t}^{g}\right) +\bar{\sigma}\left( \psi (L)\varepsilon _{t}^{z}\right)
\notag \\
& =\bar{\sigma}\left( \psi (L)\Delta r_{t}\right) +\psi (1)\left[ 4\sigma
_{g}+\sigma _{z}\right] ,  \label{LRv}
\end{align}%
since $\varepsilon _{t}^{g}$ and $\varepsilon _{t}^{z}$ are uncorrelated in
the model. Because the nominal rate $i_{t}$ is exogenous, it will not be
possible to say more about the first term on the right hand side of \ref{LRv}
unless we assume some time series process for $\Delta r_{t}$. Suppose that $%
r_{t}$ follows a random walk, so that $\Delta r_{t}=\varepsilon _{t}^{r}$,
with $\mathrm{Var}(\varepsilon _{t}^{r})=\sigma _{r}^{2}$. Then $\bar{\sigma}%
\left( \psi (L)\Delta \tilde{r}_{t}\right) =a_{r}/(1-a_{y,1}-a_{y,2})\left[
\sigma _{r}+4\sigma _{g}+\sigma _{z}\right] $, and we obtain $\bar{\sigma}(%
\mathring{\nu}_{t}^{y\ast })=\sigma _{y^{\ast }}+a_{r}/(1-a_{y,1}-a_{y,2})%
\left[ \sigma _{r}+4\sigma _{g}+\sigma _{z}\right] $. The MUE ratio in \ref%
{s2n_s1} based on the restricted Stage 1 model yields:%
\begin{equation}
\frac{\bar{\sigma}(\varepsilon _{t}^{g})}{\bar{\sigma}(\mathring{\nu}%
_{t}^{y\ast })}=\frac{\sigma _{g}}{\sigma _{y^{\ast
}}+a_{r}/(1-a_{y,1}-a_{y,2})\left[ \sigma _{r}+4\sigma _{g}+\sigma _{z}%
\right] }\neq \frac{\sigma _{g}}{\sigma _{y^{\ast }}}.  \label{S1_Lg0}
\end{equation}%
Thus, \cites{holston.etal:2017} implementation of MUE in Stage 1 cannot
recover the \textit{`}\emph{signal-to-noise ratio'} of interest $\frac{%
\sigma _{g}}{\sigma _{y^{\ast }}}$ from $\lambda _{g}$.

Note here that the autocorrelation pattern in $\mathring{\nu}_{t}^{y\ast }$
is also reflected in the $\Delta \hat{y}_{t|T}^{\ast }$ series which is used
as the observable counterpart to $\Delta y_{t}^{\ast }$ in \ref{hlw_ll2a}.
That is, $\Delta \hat{y}_{t|T}^{\ast }$ has a significant and sizeable AR(1)
coefficient of $-0.2320$ (standard error $\approx 0.0649$). Inline with Step
$(i)$ of \cites{stock.watson:1998} implementation of MUE (the GLS\ step),
one would thus need to AR(1)\ filter the constructed $\Delta \hat{y}%
_{t|T}^{\ast }$ series used in the local level model \emph{before}
implementing the structural break tests. Accounting for this autocorrelation
patter in $\Delta \hat{y}_{t|T}^{\ast }$ leads to very different $\lambda
_{g}$ point estimates (see \autoref{tab:stage1_MUE_AR1}, which is arranged
in the same way as the top half of \autoref{tab:sw98_T4}, with the last
column showing $\lambda _{g}=\lambda /T$ rather than $\sigma _{g}$ to be
able to compare these to column one of \autoref{tab:Stage1_lambda_g}).

\subsubsection{Rewriting the Stage 1 model in local level model form}

One nuisance with the Stage 1 model formulation of \cite{holston.etal:2017}
in \ref{eq:stag1} is that trend growth is initially assumed to be constant
to compute a first estimate of $y_{t}^{\ast }$. This estimate is then used
to construct the empirical counterpart of $\Delta y_{t}^{\ast }$ to which
MUE is applied.

A more coherent way to implement MUE in the context of the Stage 1 model is
to rewrite the local linear trend model in local level form. To see how this
could be done, we can simplify the Stage 1 model by excluding the inflation
equation \ref{S1b} and replacing the constant trend growth equation in \ref%
{S1d} with the original trend and trend growth equations in \ref{y*} and \ref%
{g}. Since the specification of the full model in \ref{eq:hlw} assumes that
the error terms $\varepsilon _{t}^{\ell },\forall \ell =\{\pi ,\tilde{y}%
,y^{\ast }\hsp[-1],g,z\}$ are $i.i.d.$ Normal and mutually uncorrelated, and
$\hat{b}_{y}\approx 0$ in the unrestricted Stage 1 model (see the results
under the heading `$b_{y}$ Free' in \autoref{tab:Stage1}), this
simplification is unlikely to induce any additional misspecification into
the model.

The modified Stage 1 model we can work with thus takes the following form:%
\bsq\label{Stage1:mod}
\begin{align}
y_{t}& =y_{t}^{\ast }+\tilde{y}_{t}  \label{S1M1a} \\
a_{y}(L)\tilde{y}_{t}& =\mathring{\varepsilon}_{t}^{\tilde{y}}  \label{S1M1b}
\\
y_{t}^{\ast }& =y_{t-1}^{\ast }+g_{t-1}+\varepsilon _{t}^{y^{\ast }}
\label{S1M1c} \\
g_{t}& =g_{t-1}+\varepsilon _{t}^{g},  \label{S1M1d}
\end{align}%
\esq where $\mathring{\varepsilon}_{t}^{\tilde{y}}=a_{r}(L)\tilde{r}%
_{t}+\varepsilon _{t}^{\tilde{y}}$ again due to the restriction of the
output gap equation of the full model in \ref{eq:hlw}.\footnote{%
If the disturbance term $\mathring{\varepsilon}_{t}^{\tilde{y}}$ is $i.i.d.$%
, then the model in \ref{Stage1:mod} can be recognized as \cites{clark:1987}
Unobserved Component (UC) model. However, $\mathring{\varepsilon}_{t}^{%
\tilde{y}}$ is not $i.i.d.$ and instead follows a general ARMA\ process with
non-zero autocovariances, which are functions of $\sigma _{g}^{2}$, $\sigma
_{z}^{2}$, the autocovariances of inflation $\pi _{t}$, as well as the
exogenously specified interest rate $i_{t}$. To see this, recall from %
\Sref{sec:model} that the real interest rate gap $\tilde{r}_{t}$ is defied
as $\tilde{r}_{t}=\left[ i_{t}-\delta (L)\pi _{t}-4g_{t}-z_{t}\right] $,
where expected inflation $\pi _{t}^{e}=\delta (L)\pi _{t}$ and $\delta (L)=%
\tfrac{1}{4}\left( 1+L+L^{2}+L^{3}\right) $, so that we can re-express $%
\mathring{\varepsilon}_{t}^{\tilde{y}}$ as:%
\begin{equation}
\mathring{\varepsilon}_{t}^{\tilde{y}}=a_{r}(L)\left[ i_{t}-\delta (L)\pi
_{t}-4g_{t}-z_{t}\right] +\varepsilon _{t}^{\tilde{y}}.  \label{eps_tilde}
\end{equation}%
The product of the two lag polynomials $a_{r}(L)\delta (L)$ in \ref%
{eps_tilde} yields a $5^{th}$ order lag polynomial for inflation. If $i_{t}$
and $\pi _{t}$ were uncorrelated white noise processes (which they are
clearly not), then we would obtain an MA(5)\ process for $\mathring{%
\varepsilon}_{t}^{\tilde{y}}$ when $a_{r}$ is non-zero. Since $\pi _{t}$ is
modelled as an integrated AR(4), the implied process for $\mathring{%
\varepsilon}_{t}^{\tilde{y}}$ is a higher order ARMA\ process, the exact
order of which depends on the assumptions one places on the exogenously
specified interest rate $i_{t}$. To determine this process exactly is of no
material interest here. However, the important point to take away from this
is that $\mathring{\varepsilon}_{t}^{ \tilde{y}}$ is autocorrelated and
follows a higher order ARMA\ process. Moreover, if $i_{t},\pi _{t},g_{t}$
and $z_{t}$ do not co-integrate, then $\mathring{\varepsilon}_{t}^{\tilde{y}%
} $ will be an $I(1)$ process.} The local linear trend model in \ref%
{Stage1:mod} can now be rewritten in local level model form by differencing %
\ref{S1M1a} and \ref{S1M1b}, and bringing $y_{t-1}^{\ast }$ to the left side
of \ref{S1M1c} to give the relations:\bsq\label{Stage1:mod2}
\begin{align}
\Delta y_{t}& =\Delta y_{t}^{\ast }+\Delta \tilde{y}_{t}  \label{S1M2a} \\
a_{y}(L)\Delta \tilde{y}_{t}& =\Delta \mathring{\varepsilon}_{t}^{\tilde{y}}
\label{S1M2b} \\
\Delta y_{t}^{\ast }& =g_{t-1}+\varepsilon _{t}^{y^{\ast }}  \label{S1M2c} \\
g_{t}& =g_{t-1}+\varepsilon _{t}^{g}.  \notag
\end{align}%
\esq

Substituting \ref{S1M2b} and \ref{S1M2c} into \ref{S1M2a} yields the local
level model:%
\begin{align}
\Delta y_{t}& =g_{t-1}+u_{t}  \label{S1Mod2b} \\
\Delta g_{t}& =\varepsilon _{t}^{g},  \label{S1Mod2c}
\end{align}%
where $u_{t}$ is defined as:%
\begin{align}
u_{t}& =\varepsilon _{t}^{y^{\ast }}+a_{y}(L)^{-1}\Delta \mathring{%
\varepsilon}_{t}^{\tilde{y}}  \notag \\
\underbrace{a_{y}(L)u_{t}}_{\text{AR(2)}}& =\underbrace{a_{y}(L)\varepsilon
_{t}^{y^{\ast }}}_{\text{MA(2)}}+\Delta \mathring{\varepsilon}_{t}^{\tilde{y}%
}  \notag \\
a_{y}(L)u_{t}& =b(L)\varepsilon _{t},  \label{eq:ut_arma}
\end{align}%
with $b(L)\varepsilon _{t}=a_{y}(L)\varepsilon _{t}^{y^{\ast }}+\Delta
\mathring{\varepsilon}_{t}^{\tilde{y}}$ on the right hand side of \ref%
{eq:ut_arma} denoting a general MA process. The $u_{t}$ term in \ref%
{eq:ut_arma} thus follows a higher order ARMA model. If $a_{r}=0$, then $%
\mathring{\varepsilon}_{t}^{\tilde{y}}=\varepsilon _{t}^{\tilde{y}}$ in \ref%
{eps_tilde} and $\Delta \mathring{\varepsilon}_{t}^{\tilde{y}}=\Delta
\varepsilon _{t}^{\tilde{y}}$, which is an integrated MA(1)\ process, so
that the right hand side would be the sum of an MA(2)\ and an MA(1),
yielding an overall MA(2) for $b(L)\varepsilon _{t}$. With $a_{y}(L)$ being
an AR(2) lag polynomial for the cycle component, we would then get an ARMA$%
(2,2)$ for $u_{t}$ in \ref{eq:ut_arma}. If $a_{r}\neq 0$, then $\Delta
\mathring{\varepsilon}_{t}^{\tilde{y}}$ follows a higher order ARMA process.
In the empirical implementation of MUE, I follow \cite{stock.watson:1998},
and use an AR(4) as an approximating model for $u_{t}$.\footnote{%
They also considered an ARMA($2,3$) model (see page 355 in their paper). It
is well known that higher order ARMA\ models can be difficult to estimate
numerically due to potential root cancellations in the AR and MA lag
polynomials. Inspection of the autocorrelation and partial autocorrelation
functions of $\Delta y_{t}$ indicate that an AR(4) model is more than
adequate to capture the time series dynamics of $\Delta y_{t}$. I have also
estimated an ARMA$(2,2)$\ model for $\Delta y_{t}\,$, with the overall
qualitative conclusions being the same and the quantitative results very
similar.}

The relations in \ref{S1Mod2b} to \ref{eq:ut_arma} are now in local level
model form to which MUE\ can be applied to as outlined in equations \ref%
{eq:sw98} to \ref{eq:s2n} in \Sref{subsec:MUE}.\footnote{%
I am grateful to James Stock for his email correspondence on this point.} To
examine if we can recover the \textit{`signal-to-noise ratio'} of interest $%
\sigma _{g}/\sigma _{y^{\ast }}$ from this MUE procedure, we need to evaluate%
\begin{equation}
\frac{\bar{\sigma}(\varepsilon _{t}^{g})}{\bar{\sigma}(u_{t})}.
\label{eq:S1Lambda}
\end{equation}%
In the numerator of \ref{eq:S1Lambda}, the term $\bar{\sigma}(\varepsilon
_{t}^{g})=\sigma _{g}$ as before. Nevertheless, the denominator term $\bar{%
\sigma}(u_{t})=\bar{\sigma}(\varepsilon _{t}^{y^{\ast }}+a_{y}(L)^{-1}\Delta
\mathring{\varepsilon}_{t}^{\tilde{y}})\neq \sigma _{y^{\ast }}$. With $%
\mathring{\varepsilon}_{t}^{\tilde{y}}=a_{r}(L)\tilde{r}_{t}+\varepsilon
_{t}^{\tilde{y}}$, we have:\
\begin{align}
\bar{\sigma}(u_{t})& =\bar{\sigma}(\varepsilon _{t}^{y^{\ast
}}+a_{y}(L)^{-1}\Delta \mathring{\varepsilon}_{t}^{\tilde{y}})  \notag \\
& =\bar{\sigma}(\varepsilon _{t}^{y^{\ast }}+\psi (L)\Delta \tilde{r}%
_{t}+a_{y}(L)^{-1}\Delta \varepsilon _{t}^{\tilde{y}}),  \label{LRv2}
\end{align}%
where the middle part in \ref{LRv2} (ie., $\psi (L)\Delta \tilde{r}_{t}$)
will again be as before in \ref{LRv} and therefore depend on $\Delta r_{t}$,
$\varepsilon _{t}^{g}$ and $\varepsilon _{t}^{z}$. Notice here also that
even if we knew $a_{r}=0$, so that the middle part in \ref{LRv2} is $0$,
there is no mechanism to enforce a zero correlation between $\varepsilon
_{t}^{y^{\ast }}$ and $\varepsilon _{t}^{\tilde{y}}$ in the data, because $%
u_{t}$ appears in reduced form in the local level model. We would thus need
the empirical correlation between $\varepsilon _{t}^{y^{\ast }}$ and $%
\varepsilon _{t}^{\tilde{y}}$ to be zero for the long-run standard deviation
$\bar{\sigma}(u_{t})$ to equal $\sigma _{y^{\ast }}$ even when the true $%
a_{r}=0$. Estimates from the existing business cycle literature suggest that
trend and cycle shocks are negatively correlated (see for instance Table 3
in \cite{morley.etal:2003}, who estimate this correlation to be $-0.9062$,
or Table 1 in the more recent study by \cite{grant.chan:2017a} whose
estimate is $-0.87$). I obtain an estimate of $-0.9426$ (see \autoref%
{tab:clarkddUC} below).

For completeness, parameter estimates of MUE applied to the local level
transformed Stage 1 model defined in \ref{Stage1:mod} are reported in %
\autoref{tab:MUE_S1}. This table is arranged in the same way as \autoref%
{tab:sw98_T4}, with all computations performed in exactly the same way as
before. The MUE results in the last two columns of the bottom part of the
table are based on the exponential Wald (EW) structural break test as used
in \cite{holston.etal:2017}. Overall, these estimates are very similar to %
\cites{stock.watson:1998} estimates, despite different time periods and GDP
data being used. The $\lambda $ (and also $\sigma _{g}$) estimates are not
statistically different from 0, and the MMLE $\hat{\sigma}_{g}$ of $0.1062$
is rather sizeable and quite close to the one implied by MUE.

\subsubsection{Estimating the local linear trend version of the Stage 1 model%
}

So far, \textit{`pile-up'} at zero problems were examined in the local level
model form which is compatible with MUE. As a last exercise, I estimate the
modified Stage 1 model in \ref{Stage1:mod} in local linear trend model form.
Two different specifications of the model are estimated. The first assumes
all error terms to be uncorrelated. This version is referred to as %
\cites{clark:1987} UC0 model. The second allows for a non-zero correlation
between $\varepsilon _{t}^{y^{\ast }}$ and $\mathring{\varepsilon}_{t}^{%
\tilde{y}}$. This version is labelled \cites{clark:1987} UC model. The aim
here is to not only examine empirically how valid the zero correlation
assumption is and to quantify its magnitude, but also to investigate whether
\textit{`pile-up'} at zero problems materialize more generally in UC\
models. In \autoref{tab:clarkddUC}, the parameter estimates of the two UC
models are reported, together with standard errors of the parameter
estimates (these are listed under the columns with the heading Std.error).

As can be seen from the estimates in \autoref{tab:clarkddUC}, there exists
no evidence of \textit{`pile-up'} at zero problems with MLE in either of
these two UC models.\footnote{%
I use a diffuse prior on the initial state vector in the estimation of both
UC models, and do not estimate the initial value. This is analogous to MMLE
in \cite{stock.watson:1998}. The input data are $100$ times the log of real
GDP.} The estimates of $\sigma _{g}$ from the two UC models are $0.0463$ and
$0.0322$, respectively, and are based on quarterly data. Expressed at an
annualized rate, they amount to approximately $0.1852$ and $0.1288$, and
hence are similar in magnitude to the corresponding MUE based estimates
obtained from the transformed model in \autoref{tab:MUE_S1}. Notice also
that the correlation between $\mathring{\varepsilon}_{t}^{\tilde{y}}$ and $%
\varepsilon _{t}^{y^{\ast }}$ (denoted by $\mathrm{Corr}(\mathring{%
\varepsilon}_{t}^{\tilde{y}},\varepsilon _{t}^{y^{\ast }})$ in \autoref%
{tab:clarkddUC}) is estimated to be $-0.9426$ ($t-$statistic is
approximately $-10$). The magnitude of the $\hat{\sigma}_{y^{\ast }}$ and $%
\hat{\sigma}_{\tilde{y}}$ coefficients nearly doubles when an allowance for
a non-zero correlation between $\mathring{\varepsilon}_{t}^{\tilde{y}}$ and $%
\varepsilon _{t}^{y^{\ast }}$ is made.\footnote{%
As is common with UC models, the improvement in the log-likelihood due to
the addition of the extra correlation parameter is rather small. Although it
is important to empirically capture the correlation between $\mathring{%
\varepsilon}_{t}^{\tilde{y}}$ and $\varepsilon _{t}^{y^{\ast }}$ as it
affects the trend growth estimate (see \autoref{fig:MUE_S1}), the overall
level of information contained in the data appears to be limited and
therefore makes it difficult to decisively distinguish one model over the
other statistically. Also, one other aspect of the empirical GDP data that
both models fail to capture is the global financial crisis. The level of GDP
dropped substantially and in an unprecedented manner. Simply \emph{%
`smoothing'} the data to extract a trend as the UC models implicity do may
thus not adequately capture this drop in the level of the series.}

\autoref{fig:MUE_S1} shows plots of the various trend growth estimates from
the modified Stage 1 models reported in \autoref{tab:MUE_S1} and \autoref%
{tab:clarkddUC}. The plots are presented in the same way as in \autoref%
{fig:sw98_F4} earlier, with the (annualized) trend growth estimates from the
two UC models superimposed. Analogous to the results in \cite%
{stock.watson:1998}, the variation in the MUE based estimates is once again
large. Trend growth can be a flat line when the lower 90\% CI\ of MUE is
considered or rather variable when the upper CI bound is used.
Interestingly, the MMLE, Clark UC model (with non-zero $\mathrm{Corr}(%
\mathring{\varepsilon}_{t}^{\tilde{y}},\varepsilon _{t}^{y^{\ast }})$) and
MUE$(\hat{\lambda}_{EW})$ trend growth estimates are very similar visually.
More importantly, the effect of restricting $\mathrm{Corr}(\mathring{%
\varepsilon}_{t}^{\tilde{y}},\varepsilon _{t}^{y^{\ast }})$ to zero on the
trend growth estimate can be directly seen in \autoref{fig:MUE_S1}. The UC0
model produces a noticeably more variable trend growth estimate than the UC
model.

Two conclusions can be drawn from this section. Firstly, %
\cites{holston.etal:2017} implementation of MUE in Stage 1 and the resulting
$\lambda _{g}$ estimate cannot recover the \textit{`signal-to-noise ratio'}
of interest $\sigma _{g}/\sigma _{y^{\ast }}$. Secondly, there is no
evidence of \textit{`pile-up'} at zero problems materializing when
estimating $\sigma _{g}$ directly by MLE. Replacing $\sigma _{g}$ in $%
\mathbf{Q}$ by $\hat{\lambda}_{g}\sigma _{y^{\ast }}$ in the Stage 2 and
full model log-likelihood functions (see \ref{S2Q} and \ref{Q3}) where $\hat{%
\lambda}_{g}$ was obtained from MUE applied to the Stage 1 model is not only
unsound but empirically entirely unnecessary.

\subsection{Stage 2 Model\label{sec:S2}}

The second stage model of \cite{holston.etal:2017} consists of the following
system of equations, which are again a restricted version of the full model
in \ref{eq:hlw}:\bsq\label{eq:stag2}%
\begin{align}
y_{t}& =y_{t}^{\ast }+\tilde{y}_{t}  \label{S2:y} \\
\pi _{t}& =b_{\pi }\pi _{t-1}+\left( 1-b_{\pi }\right) \pi _{t-2,4}+b_{y}%
\tilde{y}_{t-1}+\varepsilon _{t}^{\pi }  \label{S2:pi} \\
a_{y}(L)\tilde{y}_{t}& =a_{0}+\tfrac{a_{r}}{2}(r_{t-1}+r_{t-2})+a_{g}g_{t-1}+%
\mathring{\varepsilon}_{t}^{\tilde{y}}  \label{S2:ytilde} \\
y_{t}^{\ast }& =y_{t-1}^{\ast }+g_{t-2}+\mathring{\varepsilon}_{t}^{y^{\ast
}}  \label{S2:ystar} \\
g_{t-1}& =g_{t-2}+\varepsilon _{t-1}^{g}.  \label{S2:g}
\end{align}%
\esq Given the estimate of $\lambda _{g}$ from Stage 1, the vector of Stage
2 parameters to be estimated by MLE is:\footnote{%
See \hyperref[sec:AS2]{Section A.2} in the \hyperref[appendix]{Appendix} for
the exact matrix expressions and expansions of the SSM of Stage 2. In the $%
\mathbf{Q}$ matrix, $\sigma _{g}$ is replaced by $\hat{\lambda}_{g}\sigma
_{y^{\ast }}$, where $\hat{\lambda}_{g}$ is the estimate from the first
stage model (see \ref{S2Q}). The state vector $\boldsymbol{\xi }_{t}$ is
initialized using the same procedure as outlined in \ref{eq:P00S1a} and %
\fnref{fn:1}, with the numerical values of $\boldsymbol{\xi }_{00}$ and $%
\mathbf{P}_{00}$ given in \ref{AS2:xi00} and \ref{AS2:P00}.}%
\begin{equation}
\boldsymbol{\theta }_{2}=[a_{y,1},~a_{y,2},~a_{r},~a_{0},~a_{g},~b_{\pi
},~b_{y},~\sigma _{\tilde{y}},~\sigma _{\pi },~\sigma _{y^{\ast }}]^{\prime
}.  \label{S2theta2}
\end{equation}%
As in the first stage model in \ref{eq:stag1}, I again use the ring symbol $(%
\mathring{\phantom{y}})$ on the disturbance terms in \ref{S2:ytilde} and \ref%
{S2:ystar} to distinguish them from the $i.i.d.$ error terms of the full
model in \ref{eq:hlw}.

Examining the formulation of the Stage 2 model in \ref{eq:stag2} and
comparing it to the full model in \ref{eq:hlw}, it is evident that \cite%
{holston.etal:2017} make two \emph{`misspecification'} choices that are
important to highlight. First, they include $g_{t-2}$ instead of $g_{t-1}$
in the trend equation in \ref{S2:ystar}, so that the $\mathring{\varepsilon}%
_{t}^{y^{\ast }}$ error term is in fact:\footnote{\cite{holston.etal:2017}
only report the $\mathbf{\mathbf{Q}}$ matrix in their documentation, which
is a diagonal matrix and takes the form given in \ref{S2Q}. In \hyperref[sec:AS2]%
{Section A.2} of the \hyperref[appendix]{Appendix}, I\ show how this matrix
is obtained. In \hyperref[sec:AS21]{Section A.2.1}, the correct Stage 2
model state-space form is provided, applying the same \emph{`trick'} as used
in the Stage 3 state-space model specification. The two $\mathbf{\mathbf{Q}}$
matrices are listed in \ref{S2Q} and \ref{S2Qcorrect}.}%
\begin{align}
\mathring{\varepsilon}_{t}^{y^{\ast }}& =\varepsilon _{t}^{y^{\ast }}+%
\overbrace{g_{t-1}-g_{t-2}}^{\varepsilon _{t-1}^{g}\text{ from \ref{S2:g}}}
\notag \\
& =\varepsilon _{t}^{y^{\ast }}+\varepsilon _{t-1}^{g}.  \label{e_ystar}
\end{align}%
As a result of this, $\mathring{\varepsilon}_{t}^{y^{\ast }}$ in \ref%
{e_ystar} follows an MA(1)\ process, instead of white noise as $\varepsilon
_{t}^{y^{\ast }}$ in \ref{y*}. Moreover, due to the $\varepsilon _{t-1}^{g}$
term in \ref{e_ystar}, the covariance between the two error terms in \ref%
{S2:ystar} and \ref{S2:g} is no longer zero, but rather $\sigma _{g}^{2}$.
Thus, treating $\mathbf{W}$ in \ref{eq:RQ} as a diagonal variance-covariance
matrix in the estimation of the second stage model is incorrect.

Second, \cite{holston.etal:2017} do not only add an (unnecessary) intercept
term $a_{0}$ to the output gap equation in \ref{S2:ytilde}, but they also
account for only one lag in trend growth $g_{t}$, and further fail to impose
the $a_{g}=-4a_{r}$ restriction in the estimation of $a_{g}$. Due to this,
the error term $\mathring{\varepsilon}_{t}^{\tilde{y}}$ in \ref{S2:ytilde}
can be seen to consist of the following two components:%
\begin{align}
\mathring{\varepsilon}_{t}^{\tilde{y}}& =\overbrace{%
-a_{r}(L)4g_{t}-a_{r}(L)z_{t}+\varepsilon _{t}^{\tilde{y}}}^{\text{missing
true model part}}-\overbrace{(a_{0}+a_{g}g_{t-1})}^{\text{added Stage 2 part}%
}  \notag \\
& =\underbrace{-a_{r}(L)z_{t}+\varepsilon _{t}^{\tilde{y}}}_{\text{desired
terms}}-\underbrace{\left[ a_{0}+a_{g}g_{t-1}+a_{r}(L)4g_{t}\right] }_{\text{%
unnecessary terms}},  \label{S2:eps_ytilde}
\end{align}%
where the \emph{`desired terms'} on the right-hand side of \ref%
{S2:eps_ytilde} are needed for \cites{holston.etal:2017} implementation of
MUE in the second stage, whose logic I will explain momentarily, while the
\emph{`unnecessary terms'} are purely due to the \emph{ad hoc} addition of
an intercept term, changing lag structure on $g_{t}$ and failure to impose
the $a_{g}=-4a_{r}$ restriction.

To be consistent with the full model specification in \ref{eq:hlw}, the
relations in \ref{S2:ytilde} and \ref{S2:ystar} should have been formulated
as:%
\bsq\label{eq:stag2a}
\begin{align}
a_{y}(L)\tilde{y}_{t}& =a_{r}(L)[r_{t}-4g_{t}]+\mathring{\varepsilon}_{t}^{%
\tilde{y}}  \label{S2a:ytilde} \\
y_{t}^{\ast }& =y_{t-1}^{\ast }+g_{t-1}+\varepsilon _{t}^{y^{\ast }}
\label{S2a:ystar}
\end{align}%
\esq so that only the two missing lags of $z_{t}$ from \ref{S2a:ytilde}
appear in the error term $\mathring{\varepsilon}_{t}^{\tilde{y}}$,
specifically:
\begin{equation}
\mathring{\varepsilon}_{t}^{\tilde{y}}=-a_{r}(L)z_{t}+\varepsilon _{t}^{%
\tilde{y}}.  \label{e_ytilde_true}
\end{equation}%
Such a specification could have been easily obtained from the full Stage 3
state-space model form described in \hyperref[sec:AS3]{Section A.3} in the
\hyperref[appendix]{Appendix}, by simply removing the last two row entries
of the state vector $\boldsymbol{\xi }_{t}$ in \ref{AS3:xi}, and adjusting
the $\mathbf{H}$, $\mathbf{F}$, and $\mathbf{S}$ matrices in the state and
measurement equations to be conformable with this state vector. This is
illustrated in \hyperref[sec:AS21]{Section A.2.1} in the \hyperref[appendix]{%
Appendix}. The \emph{`correctly specified'} Stage 2 model should thus have
been:\bsq\label{S2full0}%
\begin{align}
y_{t}& =y_{t}^{\ast }+\tilde{y}_{t} \\
\pi _{t}& =b_{\pi }\pi _{t-1}+\left( 1-b_{\pi }\right) \pi _{t-2,4}+b_{y}%
\tilde{y}_{t-1}+\varepsilon _{t}^{\pi } \\
a_{y}(L)\tilde{y}_{t}& =a_{r}(L)[r_{t}-4g_{t}]+\mathring{\varepsilon}_{t}^{%
\tilde{y}}  \label{S2_ytilde0} \\
y_{t}^{\ast }& =y_{t-1}^{\ast }+g_{t-1}+\varepsilon _{t}^{y^{\ast }} \\
g_{t-1}& =g_{t-2}+\varepsilon _{t-1}^{g}.
\end{align}%
\esq

To see why this matters, let us examine how one would implement MUE\ in the
Stage 2 model, following again \cites{holston.etal:2017} logic as applied in
Stage 1. That is, one would first need to define a local level model
involving $z_{t}$ to be in the same format as in \ref{eq:sw98}. If we assume
for the moment that the true state variables $\tilde{y}_{t}$ and $g_{t}$, as
well as parameters $a_{y,1},$ $a_{y,2}$ and $a_{r}$ are known, and we ignore
the econometric issues that arise when these are replaced by estimates, then
the following local level model from the \emph{`correctly specified'} Stage
2 model in \ref{S2_ytilde0} can be formed:\bsq\label{MUE2}%
\begin{align}
\overbrace{a_{y}(L)\tilde{y}_{t}-a_{r}(L)[r_{t}-4g_{t}]}^{\text{analogue to }%
GY_{t}\text{ in \ref{eq:sw1}}}& =\underbrace{\overbrace{-a_{r}(L)z_{t}}^{%
\mathclap{\substack{\hsp[13]\text{analogue to } \beta _{t} \text{ in
\ref{eq:sw1}}}}}+\varepsilon _{t}^{\tilde{y}}}_{\mathring{\varepsilon}_{t}^{%
\tilde{y}}\text{ in \ref{e_ytilde_true}}}  \label{MUE2a} \\
\underbrace{-a_{r}(L)\Delta z_{t}}_{\mathclap{\substack{\text{analogue to }
\\[1pt] \Delta\beta _{t} \text{ in \ref{eq:swRW}}}}}& =\underbrace{%
-a_{r}(L)\varepsilon _{t}^{z}}_{\mathclap{\substack{\text{analogue to }
\\[1pt] (\lambda /T)\eta _{t} \text{ in \ref{eq:swRW}}}}},  \label{MUE2b}
\end{align}%
\esq where $a_{y}(L)\tilde{y}_{t}-a_{r}(L)[r_{t}-4g_{t}]$ and $%
-a_{r}(L)z_{t} $ in \ref{MUE2a} are the analogues to $GY_{t}$ and $\beta
_{t} $ in \ref{eq:sw1}, $\varepsilon _{t}^{\tilde{y}}$ corresponds to $u_{t}$
(but is $i.i.d.$ from the full model assumptions in \ref{eq:hlw} rather than
an autocorrelated time series process as $u_{t}$ in \ref{eq:sw1}), and $%
-a_{r}(L)\Delta z_{t}$ and $-a_{r}(L)\varepsilon _{t}^{z}$ are the
counterparts to $\Delta \beta _{t}$ and $(\lambda /T)\eta _{t}$ in the state
equation in \ref{eq:swRW}.\footnote{%
To arrive at \ref{MUE2b}, simply multiply \ref{z} in the full model by $%
-a_{r}(L)$.}

The equations in \ref{MUE2} are now in local level model form suitable for
MUE. The Stage 2 MUE procedure implemented on this constructed $%
GY_{t}=a_{y}(L)\tilde{y}_{t}-a_{r}(L)[r_{t}-4g_{t}]$ series produces the $%
\lambda _{z}=\lambda /T$ ratio corresponding to \ref{eq:s2n}, that is:%
\footnote{%
To make this clear, MUE returns an estimate of $\lambda $ by using the
look-up table on page 354 in \cite{stock.watson:1998} to find the closest
matching value of one of the four structural break test statistics defined
in \ref{eq:breakTests} and \ref{eqL} which test for a structural break in
the unconditional mean of the constructed $GY_{t}$ series by running a dummy
variable regression of the form defined in \ref{Zt}.}%
\begin{equation}
\frac{\lambda }{T}=\frac{\bar{\sigma}(\Delta \beta _{t})}{\bar{\sigma}%
(\varepsilon _{t}^{\tilde{y}})}=\frac{\bar{\sigma}(-a_{r}(L)\Delta z_{t})}{%
\sigma _{\tilde{y}}}=\frac{a_{r}(1)\sigma _{z}}{\sigma _{\tilde{y}}}=\frac{%
a_{r}\sigma _{z}}{\sigma _{\tilde{y}}}.  \label{mue2_ratio}
\end{equation}%
The last two steps in \ref{mue2_ratio} follow due to $a_{r}(1)=\frac{a_{r}}{2%
}(1+1^{2})=a_{r}$ and $\bar{\sigma}(\varepsilon _{t}^{\tilde{y}})=\sigma _{%
\tilde{y}}$, with $\bar{\sigma}(\cdot )$ denoting again the long-run
standard deviation. The final term in \ref{mue2_ratio} gives %
\cites{holston.etal:2017} ratio $\lambda _{z}=a_{r}\sigma _{z}/\sigma _{%
\tilde{y}}$.\footnote{%
In \cite{laubach.williams:2003}, $\lambda _{z}$ is curiously defined as the
ratio $a_{r}\sigma _{z}/(\sigma _{\tilde{y}}\sqrt{2})$ (see page 1064,
second paragraph on the right). It is not clear where the extra $\sqrt{2}$
term comes from.} This is the logic behind \cites{holston.etal:2017}
implementation of MUE in Stage 2.

However, because \cite{holston.etal:2017} define the Stage 2 model in
\textit{`misspecified'} form in \ref{eq:stag2}, $\mathring{\varepsilon}_{t}^{%
\tilde{y}}$ is no longer simply equal to $-a_{r}(L)z_{t}+\varepsilon _{t}^{%
\tilde{y}}$ as needed for the right-hand side of \ref{MUE2a}, but now also
includes the \emph{`unnecessary terms'} $\left[
a_{0}+a_{g}g_{t-1}+a_{r}(L)4g_{t}\right] $ (see the decomposition in \ref%
{S2:eps_ytilde}). What effect this has on the Stage 2 MUE\ procedure can be
seen by first rewriting $a_{g}g_{t-1}$ as:%
\begin{align}
a_{g}g_{t-1}& =\tfrac{a_{g}}{2}(g_{t-1}+g_{t-1})  \notag \\
& =\tfrac{a_{g}}{2}(g_{t-1}+\underbrace{g_{t-2}+\varepsilon _{t-1}^{g}}%
_{g_{t-1}\text{ from \textrm{\ref{S2:g}}}})  \notag \\
& =a_{g}(L)g_{t}+\tfrac{a_{g}}{2}\varepsilon _{t-1}^{g},  \label{ag}
\end{align}%
where $a_{g}(L)=\frac{a_{g}}{2}(L+L^{2})$. The additional \emph{`unnecessary
terms'} on the right-hand side of \ref{S2:eps_ytilde} become:\vsp[-2]%
\begin{align}
-\left[ a_{0}+a_{g}g_{t-1}+a_{r}(L)4g_{t}\right] & =-[a_{0}+\overbrace{%
a_{g}(L)g_{t}+\tfrac{a_{g}}{2}\varepsilon _{t-1}^{g}}^{a_{g}g_{t-1}\text{
from \ref{ag}}}+a_{r}(L)4g_{t}]  \notag \\
& =-[a_{0}+\tfrac{(a_{g}+4a_{r})}{2}(g_{t-1}+g_{t-2})+\tfrac{a_{g}}{2}%
\varepsilon _{t-1}^{g}].  \label{unTerms}
\end{align}

In \cites{holston.etal:2017} Stage 2 model in \ref{eq:stag2}, the
constructed local level model takes then the form:\bsq\label{S2wrong}%
\begin{align}
\overbrace{a_{y}(L)\tilde{y}_{t}-a_{0}-a_{r}(L)r_{t}-a_{g}g_{t-1}}^{\text{%
misspecified analogue to }GY_{t}\text{ in \ref{MUE2a}}}& =\overbrace{%
-a_{r}(L)z_{t}}^{\mathclap{~\text{analogue to } \beta _{t} }}+\mathring{\nu}%
_{t}^{\tilde{y}}  \label{S2wrong_a} \\
\underbrace{-a_{r}(L)\Delta z_{t}}_{\mathclap{\substack{\text{analogue}
\\[1pt] \text{to } \Delta\beta _{t}}}}& =\underbrace{-a_{r}(L)\varepsilon
_{t}^{z}}_{\mathclap{\substack{\text{analogue} \\[1pt] \text{to } (\lambda
/T)\eta _{t} }}},  \label{S2wrong_b}
\end{align}%
\bigskip \esq where $\mathring{\nu}_{t}^{\tilde{y}}$ in \ref{S2wrong_a} is
the misspecified analogue to $\varepsilon _{t}^{\tilde{y}}$ in \ref{MUE2a}
and is defined as:
\begin{equation}
\mathring{\nu}_{t}^{\tilde{y}}=\varepsilon _{t}^{\tilde{y}}-[a_{0}+\tfrac{%
(a_{g}+4a_{r})}{2}(g_{t-1}+g_{t-2})+\tfrac{a_{g}}{2}\varepsilon _{t-1}^{g}].
\label{S2_nu_ring}
\end{equation}%
As can be seen, the error term $\mathring{\nu}_{t}^{\tilde{y}}$ in \ref%
{S2_nu_ring} will not be white noise. Moreover, forming the MUE\ $\lambda /T$
ratio from the model in \ref{S2wrong} in the same way as in \ref{mue2_ratio}
leads to:
\begin{equation}
\frac{\lambda }{T}=\frac{\bar{\sigma}(-a_{r}(L)\Delta z_{t})}{\bar{\sigma}(%
\mathring{\nu}_{t}^{\tilde{y}})}=\frac{a_{r}(1)\sigma _{z}}{\bar{\sigma}(%
\mathring{\nu}_{t}^{\tilde{y}})}=\frac{a_{r}\sigma _{z}}{\bar{\sigma}(%
\mathring{\nu}_{t}^{\tilde{y}})},  \label{Lambda_z_correct}
\end{equation}%
and now requires the evaluation of the long-run standard deviation of $%
\mathring{\nu}_{t}^{\tilde{y}}$ in the denominator, which will not be equal
to $\sigma _{\tilde{y}}$ as from the \emph{`correctly'} specified Stage 2
model defined in \ref{S2full0}. Note here that, even in the unlikely
scenario that $(a_{g}+4a_{r})=0$ in the data, the long-run standard
deviation of $\mathring{\nu}_{t}^{\tilde{y}}$ will also depend on $\tfrac{%
a_{g}}{2}\sigma _{g}$ because of the $\tfrac{a_{g}}{2}\varepsilon _{t-1}^{g}$
term in $\mathring{\nu}_{t}^{\tilde{y}}$, so that one obtains:
\begin{equation}
\lambda _{z}=\frac{\lambda }{T}=\frac{a_{r}\sigma _{z}}{(\sigma _{\tilde{y}%
}+a_{g}\sigma _{g}/2)}.  \label{Lz00}
\end{equation}%
Thus, MUE applied to \cites{holston.etal:2017} \textit{`misspecified' }Stage
2 model as defined in \ref{eq:stag2} cannot recover the ratio of interest $%
\lambda _{z}=a_{r}\sigma _{z}/\sigma _{\tilde{y}}$.\footnote{%
If $(a_{g}+4a_{r})\neq 0$, additional $\sigma _{g}$ terms enter the long-run
standard deviation in the denominator of $\lambda _{z}$.}

Before I\ discuss in the next section what effect\ the \emph{%
`misspecification'} of the Stage 2 model has on \cites{holston.etal:2017}
median unbiased estimates of $\lambda _{z}$, I\ report the estimates of the
two different Stage 2 models in \autoref{tab:Stage2}. The first and second
columns show replicated results which are based on \cites{holston.etal:2017}
R-Code as well as my own implementation and serve as reference values. In
the third column under the heading `MLE$(\sigma _{g})$', $\sigma _{g}$ is
estimated directly by MLE together with the other parameters of the model
without using $\hat{\lambda}_{g}$ from Stage 1.\footnote{%
I use the same initial values for the parameter and the state vector (mean
and variance) as in the exact replication of \cite{holston.etal:2017}. Using
a diffuse prior instead leads to only minor differences in the numerical
values. The implied $\lambda _{g}$ and $\sigma _{g}$ estimates are shown in
brackets and were computed from the \emph{`signal-to-noise ratio'} relation $%
\lambda _{g}=\sigma _{g}/\sigma _{y^{\ast }}.$} The last column under the
heading `MLE$(\sigma _{g}).\mathcal{M}_{0}$' reports estimates obtained from
the \emph{`correctly specified'} Stage 2 model defined in \ref{S2full0},
where $\sigma _{g}$ is once again estimated directly by MLE.

The results in \autoref{tab:Stage2} can be summarized as follows. First,
there exists no evidence of \emph{`pile-up'} at zero problems materializing
when estimating $\sigma _{g}$ directly by MLE; not in the \textit{%
`misspecified' }Stage 2 model, nor in the \emph{`correctly specified'}%
\textit{\ }one. This finding is consistent with the earlier results from the
first stage. The Stage 2 MLE of $\sigma _{g}$ is in fact nearly $50\%$
\textit{larger} than the estimate implied by $\hat{\lambda}_{g}$ from MUE in
Stage 1. MUE in Stage 1 thus seems to be redundant. Second, the estimate of $%
a_{g}$ is about eight times the magnitude of $-a_{r}$, so that $%
(a_{g}+4a_{r})\approx 0.3132\neq 0$. Therefore, the ratio in \ref{Lz00} will
have additional $\sigma _{g}$ terms in the denominator, making the
evaluation of this quantity more intricate. And third, despite the different
Stage 2 model specifications, the resulting parameter estimates as well as
the log-likelihood values across the three different models in columns two
to four of \autoref{tab:Stage2} are very similar. This suggests that,
overall, the data are uninformative about the model parameters.\footnote{%
These findings also hold when using data for the Euro Area, the U.K., and
Canada, but are not reported here.}

Note here that, although the results in \autoref{tab:Stage2} indicate that
\emph{`misspecifying'} the Stage 2 model does not have an important impact
on the parameter estimates that are obtained, I show below that it
substantially and spuriously amplifies the size of the $\lambda _{z}$
estimate.%

\subsubsection{\cites{holston.etal:2017} implementation of MUE in Stage 2
\label{sec:MUE2}}

Recall again conceptually how MUE in Stage 2 would need to be implemented
following the same logic as in Stage 1 before.\ First, one needs to
construct an observable counterpart to $GY_{t}$ as given in \ref{MUE2a} from
the Stage 2 model estimates. Then, the four structural break tests described
in \Sref{subsec:MUE} are applied to test for a break in the unconditional
mean of (the AR filtered) $GY_{t}$ series. This corresponds to Step $(ii)$
in \cites{stock.watson:1998} procedural description. Constructing a local
level model of the form described in \ref{MUE2} enables us to implement MUE\
to yield the ratio $\lambda /T=\bar{\sigma}(\Delta \beta _{t})/\bar{\sigma}%
(\varepsilon _{t}^{\tilde{y}})$ as defined in \ref{mue2_ratio}.

\cites{holston.etal:2017} implementation of MUE in Stage 2, nonetheless,
departs from this description in two important ways. First, instead of using
the \emph{`correctly specified'} Stage 2 model defined in \ref{S2full0},
they work with the \emph{`misspecified'} model given in \ref{eq:stag2}.
Second, rather than leaving\ the $a_{y,1},$ $a_{y,2},$ $a_{r},$ $a_{g}$ and $%
a_{0}$ parameters fixed at their Stage 2 estimates and constructing the
observable counterpart to $GY_{t}$ in \ref{S2wrong_a} only once outside the
dummy variable regression loop, \cite{holston.etal:2017} essentially \emph{%
`re-estimate'} these parameters by including the vector $\boldsymbol{%
\mathcal{X}}_{t}$ defined in \ref{XX} below as a regressor in the structural
break regression in \ref{eqS2regs}. For the \emph{`misspecified'} Stage 2
model, this has the effect of substantially increasing the size and
variability of not only the dummy variable coefficients $\hat{\zeta}_{1}$ in %
\ref{eqS2regs}, but also the corresponding $F$ statistics used in the
computation of the \textrm{MW}, \textrm{EW}, and \textrm{QLR} structural
break tests needed for MUE$\ $of $\lambda _{z}$.

To illustrate how \cite{holston.etal:2017} implement MUE in the second
stage, I list below the main steps that they follow to compute $\lambda _{z}$%
.

\begin{enumI}
\item Given the Stage 2 estimate $\skew{0}\boldsymbol{\hat{\theta}}_{2}$
from the model in \ref{eq:stag2}, use the Kalman Smoother to obtain
(smoothed) estimates of the latent state vector $\boldsymbol{\xi }%
_{t}=[y_{t}^{\ast },~y_{t-1}^{\ast },~y_{t-2}^{\ast },~g_{t-1}]^{\prime }$.
Then form estimates of the cycle variable and its lags as $\hat{\tilde{y}}%
_{t-i|T}=(y_{t-i}-\hat{y}_{t-i|T}^{\ast }),\forall i=0,1,2$.

\item Construct
\begin{equation}
\mathcal{Y}_{t}=\hat{\tilde{y}}_{t|T}  \label{YY}
\end{equation}%
and the $(1\times 5)$ vector
\begin{equation}
\boldsymbol{\mathcal{X}}_{t}=[\hat{\tilde{y}}_{t-1|T},~\hat{\tilde{y}}%
_{t-2|T},~(r_{t-1}+r_{t-2})/2,~\hat{g}_{t-1|T},~1],  \label{XX}
\end{equation}%
where $r_{t}$ is the real interest rate, $\hat{g}_{t-1|T}$ is the Kalman
Smoothed estimate of $g_{t-1}$ and $1$ is a scalar to capture the constant $%
a_{0}$ (intercept term).

\item For each $\tau \in \lbrack \tau _{0},\tau _{1}]$, run the following
dummy variable regression analogous to \ref{Zt}:%
\begin{equation}
\mathcal{Y}_{t}=\boldsymbol{\mathcal{X}}_{t}\boldsymbol{\phi }+\zeta
_{1}D_{t}(\tau )+\epsilon _{t},  \label{eqS2regs}
\end{equation}%
where $\boldsymbol{\mathcal{X}}_{t}$ is as defined in \ref{XX} and $%
\boldsymbol{\phi \ }$is a $(5\times 1)$ parameter vector. The structural
break dummy variable $D_{t}(\tau )$ takes the value $1$ if $t>\tau $ and $0$
otherwise, and $\tau =\{\tau _{0},\ldots ,\tau _{1}\}$ is an index of grid
points between $\tau _{0}=4$ and $\tau _{1}=T-4$. Use the sequence of $F$
statistics $\{F(\tau )\}_{\tau =\tau _{0}}^{\tau _{1}}$ on the dummy
variable coefficients to compute the \textrm{MW}, \textrm{EW}, and \textrm{%
QLR} structural break test statistics needed for MUE.

\item Given the structural break test statistics computed in Step (\emph{III}%
\hsp[.3]), find the corresponding $\lambda $ values in look-up Table 3 of
\cite{stock.watson:1998} and return the ratio ${\lambda }/T=\lambda _{z}$,
where the preferred estimate of $\lambda $ is again based on the \textrm{EW}
structural break statistic defined in \ref{EW} as in the Stage 1 MUE.
\end{enumI}

In the top and bottom panels of \autoref{fig:seqaF} I show plots of the
sequences of $F$ statistics $\{F(\tau )\}_{\tau =\tau _{0}}^{\tau _{1}}$
computed from \cites{holston.etal:2017} \emph{`misspecified}\textit{'} Stage
2 model and the \emph{`correctly specified'} Stage 2 model defined in \ref%
{S2full0}, respectively. Two sets of sequences are drawn in each panel.%
\footnote{%
The same sequence computed from an updated data series up to 2019:Q2 is
shown in \autoref{fig:seqaF_2019Q2} in the \hyperref[appendix]{Appendix}.}
The first sequence, which I refer to as \emph{`time varying} $\boldsymbol{%
\phi }$\textit{'} (drawn as a red line in \autoref{fig:seqaF}) is
constructed by following
\cites{holston.etal:2017}
implementation outlined in Steps (\emph{I}\hsp[.3]) to (\emph{III}\hsp[.3])
above. I call this the \emph{`time varying} $\boldsymbol{\phi }$\textit{'}
sequence because the $a_{y,1},$ $a_{y,2},$ $a_{r},$ $a_{g}$ and $a_{0}$
parameters needed to \textit{`construct'} the observable counterpart to $%
GY_{t}$ in \ref{S2wrong_a} are effectively \textit{`re-estimated'} for each $%
\tau \in \lbrack \tau _{0},\tau _{1}]$ in the dummy variable regression loop
due to the inclusion of the extra $\boldsymbol{\mathcal{X}}_{t}\boldsymbol{%
\phi }$ term in \ref{eqS2regs}. For the \emph{`correctly specified}\textit{'}
Stage 2 model in \ref{S2full0}, $\boldsymbol{\mathcal{X}}_{t}$ in \ref{XX}
is replaced by the $(1\times 3)$ vector $[\hat{\tilde{y}}_{t-1|T},~\hat{%
\tilde{y}}_{t-2|T},~(r_{t-1}+r_{t-2}-4\{\hat{g}_{t-1|T}+\hat{g}%
_{t-2|T}\})/2] $.

In the second sequence, labelled \emph{`constant} $\boldsymbol{\phi }$%
\textit{' }in \autoref{fig:seqaF} and drawn as a blue line, the observable
counterpart to $GY_{t}$ is computed only once outside the structural break
regression loop, with the dummy variable regression performed without the
extra $\boldsymbol{\mathcal{X}}_{t}\boldsymbol{\phi }$ term in \ref{eqS2regs}%
, ie., it is computed in its \textit{`original'} form as given in \ref{Zt}.%
\footnote{%
Note that \cites{stock.watson:1998} MUE look-up table values for $\lambda $
were constructed by simulation with the structural break test testing the
unconditional mean of the $GY_{t}$ series for a break, without any other
variables being included in the regression. This form of the structural
break regression is thus compatible with
\cites{stock.watson:1998}
look-up table values.} More specifically, for the \emph{`misspecified}%
\textit{'} and \emph{`correctly specified}\textit{' }Stage 2 models, the
observable counterparts to the $GY_{t}$ series are constructed as:%
\begin{align}
GY_{t}& =\hat{\tilde{y}}_{t|T}-\hat{a}_{y,1}\hat{\tilde{y}}_{t-1|T}-\hat{a}%
_{y,2}\hat{\tilde{y}}_{t-2|T}-\hat{a}_{r}(r_{t-1}+r_{t-2})/2-\hat{a}_{g}\hat{%
g}_{t-1|T}-\hat{a}_{0},  \label{GY_HLW} \\
\intxt{and}GY_{t}& =\hat{\tilde{y}}_{t|T}-\hat{a}_{y,1}\hat{\tilde{y}}%
_{t-1|T}-\hat{a}_{y,2}\hat{\tilde{y}}_{t-2|T}-\hat{a}_{r}(r_{t-1}+r_{t-2}-4\{%
\hat{g}_{t-1|T}+\hat{g}_{t-2|T}\})/2,  \label{GYcorr1}
\end{align}%
respectively. The $\hat{a}_{y,1},\hat{a}_{y,2},\hat{a}_{r},\hat{a}_{g},$ and
$\hat{a}_{0}$ coefficients are the (full sample) estimates reported in
columns 2 and 4 of \autoref{tab:Stage2} under the headings `Replicated' and
`MLE$(\sigma _{g}).\mathcal{M}_{0}$', with the corresponding latent state
estimates from the respective models.\footnote{%
For instance, $\hat{g}_{t-1|T}$ in \ref{GY_HLW} is the Kalman Smoothed
estimate of trend growth from \cites{holston.etal:2017} \emph{`misspecified}%
\textit{'} Stage 2 model, while trend growth $\hat{g}_{t-1|T}$ in \ref%
{GYcorr1} is the corresponding estimate from the \emph{`correctly specified}%
\textit{' }Stage 2 model.}

As can be seen from \autoref{fig:seqaF}, the $\{F(\tau )\}_{\tau =\tau
_{0}}^{\tau _{1}}$ sequences from the \emph{`correctly specified}\textit{'}
Stage 2 models shown in the bottom panel are not only smaller overall, but
they are nearly unaffected by
\cites{holston.etal:2017}
approach to \emph{`re-estimate'} the parameters in the structural break
loop. Both, the \emph{`constant} $\boldsymbol{\phi }$\textit{' }and the
\emph{`time varying} $\boldsymbol{\phi }$\textit{'} versions generate $%
\{F(\tau )\}_{\tau =\tau _{0}}^{\tau _{1}}$ sequences that are overall very
similar, with their maximum values being around 4.5. For the \emph{%
`misspecified}\textit{'} Stage 2 model shown in the top panel, this is not
the case. The variation as well as the magnitude of $\{F(\tau )\}_{\tau
=\tau _{0}}^{\tau _{1}}$ from the \emph{`time varying} $\boldsymbol{\phi }$%
\textit{'} and \emph{`constant} $\boldsymbol{\phi }$\textit{' }%
implementations are vastly different, with the former having a much higher
mean and maximum value.

These large differences in the $\{F(\tau )\}_{\tau =\tau _{0}}^{\tau _{1}}$
sequences from the \emph{`misspecified}\textit{'} Stage 2 models also lead
to very different estimates of $\lambda _{z}$. This can be seen from \autoref%
{tab:Stage2_lambda_z}, which shows the resulting $\lambda _{z}$ estimates in
the top part with the corresponding $L$, \textrm{MW}, \textrm{EW}, and
\textrm{QLR} structural break test statistics in the bottom part. \autoref%
{tab:Stage2_lambda_z} is arranged further into a left and a right column
block, referring to the \emph{`time varying} $\boldsymbol{\phi }$\textit{' }%
and the \emph{`constant} $\boldsymbol{\phi }$\textit{' }MUE implementations
for the three different models reported in \ref{tab:Stage2}. `Replicated'
refers to the baseline replicated results, `MLE$(\sigma _{g})$' corresponds
to the \emph{`misspecified}\textit{'} Stage 2 model but with $\sigma _{g}$
estimated by MLE, and `MLE$(\sigma _{g}).\mathcal{M}_{0}$' is from the \emph{%
`correctly specified}\textit{'} Stage 2 model with $\sigma _{g}$ again
estimated by MLE. The `HLW.R-File' column lists the results from
\cites{holston.etal:2017}
R-Code. Note that \cite{holston.etal:2017} do not report estimates based on %
\cites{nyblom:1989} $L$ statistic. The entries in the $L$ rows in \autoref%
{tab:Stage2_lambda_z} under `HLW.R-File' thus simply list `---'. 90\%
confidence intervals for $\lambda _{z}$ and $p-$values for the structural
break tests are reported in square and round brackets, respectively.%
\footnote{%
As in the replication of
\cites{stock.watson:1998}
results reported in \ref{tab:sw98_T4}, these were again obtained from their
GAUSS files.}

Consistent with the visual findings from \autoref{fig:seqaF}, the structural
break statistics from the \emph{`misspecified}\textit{'} Stage 2 model shown
under the 'Replicated' heading for the \emph{`time varying} $\boldsymbol{%
\phi }$\textit{'} and \emph{`constant} $\boldsymbol{\phi }$\textit{'}
settings are very different. The \textrm{MW}, \textrm{EW}, and \textrm{QLR}
statistics are approximately 4 to 5 times larger under the \emph{`time
varying} $\boldsymbol{\phi }$\textit{' }setting than under the \emph{%
`constant} $\boldsymbol{\phi }$\textit{'} scenario. Because %
\cites{nyblom:1989} $L$ statistic is constructed as the scaled cumulative
sum of the demeaned `$GY_{t}$' series and thus does not require the
partitioning of data, creation of dummy variables, or looping through
potential break dates, it is not affected by this choice, yielding the same
test statistic of about $0.05$ under both settings.

Under the \emph{`time varying} $\boldsymbol{\phi }$\textit{' }setting, the
\textrm{MW}, \textrm{EW}, and \textrm{QLR} statistics and \cites{nyblom:1989}
$L$ statistic generate vastly different $\lambda _{z}$ estimates. %
\cites{nyblom:1989} $L$ statistic is highly insignificant with a $p-$value
of $0.87$, resulting in a $\lambda _{z}$ estimate of exactly $0$ (%
\cites{nyblom:1989} $L$ statistic is less than $0.118$, the smallest value
in
\cites{stock.watson:1998}
look-up Table 3 which corresponds to $\lambda =0$). The \textrm{MW}, \textrm{%
EW}, and \textrm{QLR} structural break statistics on the other hand are
either weakly significant or marginally insignificant, with $p-$values
between $0.045$ and $0.13$. These borderline significant structural break
statistics generate sizable $\lambda _{z}$ point estimates between $0.025$
and $0.034$. The resulting $90\%$ confidence intervals for $\lambda _{z}$
are, nonetheless, rather wide with $0$ as the lower bound, suggesting that
these point estimates are not significantly different from zero.\footnote{%
Given the earlier discussion in \Sref{subsec:MUE} and the ARE results in
Table 2 of \cite{stock.watson:1998}, we know that MUE\ can be a very
inefficient estimator.} Under the \emph{`constant} $\boldsymbol{\phi }$%
\textit{'} setting, the four structural break statistics and the resulting $%
\lambda _{z}$ estimates tell a consistent story (see the `Replicated'
heading in the right column block). All structural break statistics are
highly insignificant, with their respective $\lambda _{z}$ point estimates
being equal to zero.

For the \emph{`correctly specified}\textit{'} Stage 2 models shown under the
headings `MLE$(\sigma _{g}).\mathcal{M}_{0}$' in \autoref%
{tab:Stage2_lambda_z}, the \emph{`time varying} $\boldsymbol{\phi }$\textit{'%
} and the \emph{`constant} $\boldsymbol{\phi }$\textit{'} estimates of $%
\lambda _{z}$ reflect the visual similarity of the $\{F(\tau )\}_{\tau =\tau
_{0}}^{\tau _{1}}$ sequences shown in the bottom panel of \autoref{fig:seqaF}%
. The $\lambda _{z}$ point estimates are of the same order of magnitude,
very close to zero (they are exactly equal to zero for \cites{nyblom:1989} $%
L $ statistic and \textrm{MW}\ under the \emph{`constant} $\boldsymbol{\phi }
$\textit{'} setting), and most importantly, substantially smaller than those
constructed from
\cites{holston.etal:2017}
\textit{`}\emph{misspecified}\textit{'} Stage 2 model.\footnote{%
In \autoref{Atab:Stage2_lambda_z_2019} in the \hyperref[appendix]{Appendix},
I present these Stage 2 MUE results for data that was updated to 2019:Q2.
The conclusion is the same.}

What is causing this large difference in the $\{F(\tau )\}_{\tau =\tau
_{0}}^{\tau _{1}}$ sequences between the\ \emph{`misspecified}\textit{'} and
\emph{`correctly specified}\textit{'} Stage 2 models in the \emph{`time
varying} $\boldsymbol{\phi }$\textit{'} setting? There are two components.
First, the Kalman Smoothed estimates of the output gap (cycle) $\hat{\tilde{y%
}}_{t|T}\ $and of (annualized) trend growth $\hat{g}_{t|T}$ can be quite
different from these two models, despite the parameter estimates and values
of the log-likelihoods being very similar. This difference is more
pronounced for the cycle estimate $\hat{\tilde{y}}_{t|T}$, particulary
towards the end of the sample period than for the trend growth estimate $%
\hat{g}_{t|T}$ (see \autoref{fig:MUE_comp_input} in the \hyperref[appendix]{%
Appendix} which shows a comparison of $\hat{\tilde{y}}_{t|T}\ $and $\hat{g}%
_{t|T}$ from the\ \emph{`misspecified}\textit{'} and \emph{`correctly
specified}\textit{'} Stage 2 models).

Second, the parameter restriction $(a_{g}+4a_{r})$ on the relationship
between the real rate and trend growth matters. More specifically, when
conditioning on $\boldsymbol{\mathcal{X}}_{t}$ in \ref{eqS2regs}, it is the
restriction $(r_{t-1}-4\hat{g}_{t-1|T})$ in $\boldsymbol{\mathcal{X}}_{t}$
that makes the largest difference to the $\{F(\tau )\}_{\tau =\tau
_{0}}^{\tau _{1}}$ sequence. To see this, I\ show plots of the $\{F(\tau
)\}_{\tau =\tau _{0}}^{\tau _{1}}$ sequences from various $\boldsymbol{%
\mathcal{X}}_{t}$ constructs corresponding to the different Stage 2 model
specifications in \autoref{fig:MUE_comp} in the \hyperref[appendix]{Appendix}%
. I use the \emph{`correctly specified}\textit{'} Stage 2 model's $\{\hat{%
\tilde{y}}_{t-i|T}\}_{i=1}^{2}$ and $\hat{g}_{t-1|T}$ estimates to form
three sets of $\boldsymbol{\mathcal{X}}_{t}$ vectors for the dummy variable
regressions in \ref{eqS2regs}. These are:\bsq\label{S2:compis}
\begin{align}
\boldsymbol{\mathcal{X}}_{t}& =[\hat{\tilde{y}}_{t-1|T},~\hat{\tilde{y}}%
_{t-2|T},~(r_{t-1}+r_{t-2})/2,~\hat{g}_{t-1|T},~1]  \label{c1} \\
\boldsymbol{\mathcal{X}}_{t}& =[\hat{\tilde{y}}_{t-1|T},~\hat{\tilde{y}}%
_{t-2|T},~r_{t-1},~\hat{g}_{t-1|T},~1]  \label{c2} \\
\boldsymbol{\mathcal{X}}_{t}& =[\hat{\tilde{y}}_{t-1|T},~\hat{\tilde{y}}%
_{t-2|T},~(r_{t-1}-4\hat{g}_{t-1|T})],  \label{c3}
\end{align}%
\esq and are labelled accordingly in \autoref{fig:MUE_comp} (the preceding
`MLE$(\sigma _{g}).\mathcal{M}_{0}$' signifies that these were constructed
using the $\{\hat{\tilde{y}}_{t-i|T}\}_{i=1}^{2}$ and $\hat{g}_{t-1|T}$
estimates from the \emph{`correctly specified}\textit{'} Stage 2 model). The
corresponding $\mathcal{Y}_{t}$ dependent variable for these structural
break regressions also uses the \emph{`correctly specified'} Stage 2 model's
output gap estimate $\hat{\tilde{y}}_{t|T}$. The $\{F(\tau )\}_{\tau =\tau
_{0}}^{\tau _{1}}$ sequences from
\cites{holston.etal:2017}
\emph{`misspecified}\textit{'} and the \emph{`correctly specified}\textit{'}
Stage 2 models are superimposed as reference values and are denoted by `HLW'
and `MLE$(\sigma _{g}).\mathcal{M}_{0}$'.

The plot corresponding to \ref{c1} (orange dashed line in \autoref%
{fig:MUE_comp}) shows a rather small difference relative to the `HLW'
benchmark (blue solid line). Thus, exchanging $\{\hat{\tilde{y}}%
_{t-i|T}\}_{i=1}^{2}$ and $\hat{g}_{t-1|T}$ from
\cites{holston.etal:2017}
\emph{`misspecified}\textit{'} Stage 2 model for those from the \emph{%
`correctly specified}\textit{'} one only has a small impact on the $\{F(\tau
)\}_{\tau =\tau _{0}}^{\tau _{1}}$ sequence and is most visible over the
1994 to 2000 period. Dropping the second lag in $r_{t}$ from $\boldsymbol{%
\mathcal{X}}_{t}$ in \ref{c2} (see the cyan dotted line in \autoref%
{fig:MUE_comp}) also has only a small impact on the $\{F(\tau )\}_{\tau
=\tau _{0}}^{\tau _{1}}$ sequence. The biggest effect on $\{F(\tau )\}_{\tau
=\tau _{0}}^{\tau _{1}}$ has the restriction $(r_{t-1}-4\hat{g}_{t-1|T})$ as
imposed in \ref{c3} (green dashed-dotted line \autoref{fig:MUE_comp}). This
is evident from the near overlapping with the red solid line corresponding
to the \emph{correctly specified}\textit{'} Stage 2 model's $\{F(\tau
)\}_{\tau =\tau _{0}}^{\tau _{1}}$ sequence. Recall that the only difference
between these two is that an extra lag of $(r_{t-1}-4\hat{g}_{t-1|T})$ is
added to $\boldsymbol{\mathcal{X}}_{t}$, and that these enter as an average,
viz, $\boldsymbol{\mathcal{X}}_{t}=[\hat{\tilde{y}}_{t-1|T},~\hat{\tilde{y}}%
_{t-2|T},~(r_{t-1}+r_{t-2}-4\{\hat{g}_{t-1|T}+\hat{g}_{t-2|T}\})/2]$.

\subsubsection{What does \cites{holston.etal:2017} Stage 2 MUE procedure
recover?}

\cites{holston.etal:2017}
Stage 2 MUE procedure implemented on the \emph{`misspecified}\textit{'}
Stage 2 model leads to spuriously large estimates of $\lambda _{z}$ when the
true value is zero. To show this, I\ perform two simple simulation
experiments.

In the first experiment, I\ simulate data from the full structural model in %
\ref{eq:hlw} using the Stage 3 parameter estimates of \cite%
{holston.etal:2017} reported in column one of \autoref{tab:Stage3} as the
true values that generate the data, but with \textit{`other factor'} $z_{t}$
set to zero for all $t$. The natural rate $r_{t}^{\ast }$ in the output gap
equation in \ref{IS} is thus solely determined by (annualized) trend growth,
that is, $r_{t}^{\ast }=4g_{t}$, which implies that $\lambda _{z}$ is zero
in the simulated data.\footnote{%
To implement the simulations from the full Stage 3 model, I\ need to define
a process for the exogenously determined interest rate in %
\cites{holston.etal:2017} model. For simplicity, I estimate a parsimonious,
but well fitting, ARMA($2,1$) model for the real interest rate series, and
then use the ARMA($2,1$) coefficients to generate a sequence of 229
simulated observations for $r_{t}$. Recall that \cite{holston.etal:2017} use
data from 1960:Q1, where the first 4 quarters are used for initialisation of
the state vector, so that in total $4+225=T$ observations are available. The
remaining series are simulated from the Stage 3 model given in \ref{eq:hlw}.
To get a realistic simulation path from the Stage 3 model, I\ initialize the
first four data points for the simulated inflation series at their observed
empirical values. For the $y_{t}^{\ast }$ series, the HP-filter based trend
estimates of GDP (also utilized in the initialisation of the State vector in
Stage 1) are used to set the first four observations. The cycle variable $%
\tilde{y}_{t}$ is initialized at zero, while trend growth $g_{t}$ is
initialized at $0.75$, which corresponds to an annualized rate of $3$
percent. In the analysis that requires a simulated path of `\emph{other
factor}' $z_{t}$, ie., when the natural rate is generated from $r_{t}^{\ast
}=4g_{t}+z_{t}$, the first four entries in $z_{t}$ are initialized at zero.
A total of $S=1000$ sequences are simulated with a total sample size of $229$
observations, where the first four entries are discarded in later analysis.}
I then implement \cites{holston.etal:2017} Stage 2 MUE procedure on the
simulated data following steps (\emph{I}\hsp[.3]) to (\emph{IV}\hsp[.3])
outlined in \Sref{sec:MUE2}\ above to yield a sequence of $S=1000$ estimates
of $\lambda _{z}$ $\left( \{\hat{\lambda}_{z}^{s}\}_{s=1}^{S}\right) $.

I use two different scenarios for $\boldsymbol{\theta }_{2}$ in the Kalman
Smoother recursions described in Step $(I)$ to extract the latent cycle as
well as trend growth series needed for the construction of $\mathcal{Y}_{t}$
and $\boldsymbol{\mathcal{X}}_{t}$ in the dummy variable regression in \ref%
{eqS2regs}. The first scenario simply takes \cites{holston.etal:2017}
empirical Stage 2 estimate $\skew{0}\boldsymbol{\hat{\theta}}_{2}$ as
reported in column one of \autoref{tab:Stage2}, and keeps these values fixed
for all 1000 generated data sequences when applying the Kalman Smoother. In
the second scenario, I re-estimate the Stage 2 parameters for each simulated
sequence to obtain new estimates $\skew{0}\boldsymbol{\hat{\theta}}%
_{2}^{s},\forall s=1,\ldots ,S$. I then apply the Kalman Smoother using
these estimates to generate the $\mathcal{Y}_{t}$ and $\boldsymbol{\mathcal{X%
}}_{t}$ sequences for the regression in \ref{eqS2regs}.

Finally, I repeat the above computations on data that were generated from
the full model in \ref{eq:hlw} with the natural rate of interest determined
by both factors, namely, $r_{t}^{\ast }=4g_{t}+z_{t}$, where $z_{t}$ was
simulated as a pure random walk. The standard deviation of $z_{t}$ was set
at the implied value from the Stage 2 estimate of $\lambda _{z}$ and the
Stage 3 estimates of $\sigma _{\tilde{y}}$ and $a_{r}$, ie., at $\sigma
_{z}=\lambda _{z}\sigma _{\tilde{y}}/a_{r}\approx 0.15$ (see row $\sigma
_{z} $ (implied) of column one in \autoref{tab:Stage3}). The objective here
is to provide a comparison of the magnitudes of the $\lambda _{z}$ estimates
that are obtained when implementing \cites{holston.etal:2017} Stage 2 MUE
procedure on data that were generate with and without \textit{`other factor'}
$z_{t}$ in the natural rate.

In \autoref{tab:Stage2_lambda_z_o}, summary statistics of $\hat{\lambda}%
_{z}^{s}$ from the two different data generating processes (DGPs) are
reported. The left column block shows results for the two different DGPs
when the Stage 2 parameter vector $\boldsymbol{\theta }_{2}$ is held fixed
at the estimates reported in column one of \autoref{tab:Stage2}. The right
column block shows corresponding results when $\boldsymbol{\theta }_{2}$ is
re-estimated for each simulated data series. The summary statistics are the
minimum, maximum, standard deviation, mean, and median of $\hat{\lambda}%
_{z}^{s}$, as well as the relative frequency of obtaining a value larger
than the empirical point estimate of \cite{holston.etal:2017}.\ This point
estimate and the corresponding relative frequency are denoted by $\hat{%
\lambda}_{z}^{\mathrm{HLW}}$ and $\Pr (\hat{\lambda}_{z}^{s}>\hat{\lambda}%
_{z}^{\mathrm{HLW}})$, respectively. To complement the summary statistics in %
\autoref{tab:Stage2_lambda_z_o}, histograms of $\hat{\lambda}_{z}^{s}$ are
shown in \autoref{fig:S2Lam_z_sim} to provide visual information about its
sampling distribution.

From the summary statistics in \autoref{tab:Stage2_lambda_z_o} as well as
the histograms in \autoref{fig:S2Lam_z_sim} we can see how similar the $\hat{%
\lambda}_{z}^{s}$ coefficients from these two different DGPs are. For
instance, when the data were simulated without \textit{`other factor'} $%
z_{t} $ (ie., $\lambda _{z}=0$), the sample mean of $\hat{\lambda}_{z}^{s}$
is $0.028842$. When the data were generated from the full model with $%
r_{t}^{\ast }=4g_{t}+z_{t}$, the sample mean of $\hat{\lambda}_{z}^{s}$ is
only $6.53\%$ higher at $0.030726$. Similarly, the relative frequencies $\Pr
(\hat{\lambda}_{z}^{s}>\hat{\lambda}_{z}^{\mathrm{HLW}})$ for these two DGPs
are $45.70\%$ and $49\%$, respectively. The inclusion of \textit{`other
factor'} $z_{t}$ in the DGP\ of the natural rate thus results in only a $3.3$
percentage points higher $\Pr (\hat{\lambda}_{z}^{s}>\hat{\lambda}_{z}^{%
\mathrm{HLW}})$.\footnote{%
When the Stage 2 parameter vector $\boldsymbol{\theta }_{2}$ is re-estimated
for each simulated sequence shown in the right column block in \autoref%
{tab:Stage2_lambda_z_o}, the sample means as well as the relative frequency $%
\Pr (\hat{\lambda}_{z}^{s}>\hat{\lambda}_{z}^{\mathrm{HLW}})$ are somewhat
lower at $0.025103$ and $0.027462$, and 33.90\% and 39.30\%, respectively.}
The histograms in \autoref{fig:S2Lam_z_sim} paint the same overall picture.
As can be seen, the Stage 2 MUE implementation has difficulties to
discriminate between these two DGPs. Moreover, it seems that it is
\cites{holston.etal:2017}
procedure itself that leads to the spuriously amplified estimates of $%
\lambda _{z}$, regardless of the data.

In a second experiment I\ simulate DGPs from entirely unrelated univariate
ARMA\ processes of the individual components of the $\mathcal{Y}_{t}$ and $%
\boldsymbol{\mathcal{X}}_{t}$ series needed for the regressions in \ref%
{eqS2regs}. To match the time series properties of the $\mathcal{Y}_{t}$ and
$\boldsymbol{\mathcal{X}}_{t}$ elements given in \ref{YY} and \ref{XX}, I
fit simple low-order ARMA\ models to $\hat{\tilde{y}}_{t|T},$ $r_{t}$ and $%
\hat{g}_{t|T}$, and then use these ARMA\ estimates to simulate artificial
data.\footnote{%
I use 4 different time series processes for $\hat{g}_{t|T}$ in these
simulations. Complete details of the simulation design are given in
\hyperref[sec:AS4]{Section A.4} of the \hyperref[appendix]{Appendix}.}
Finally, I apply \cites{holston.etal:2017} Stage 2 MUE procedure to the
simulated data as before, nevertheless starting from Step (\emph{II}\hsp[.3]%
), and thereby skipping the Kalman Smoother step. The full results from the
second experiment are reported in \autoref{tab:MUE2_Sim_extra} and \autoref%
{fig:MUE2_Sim_extra} in the \hyperref[appendix]{Appendix}. These yield
magnitudes of $\hat{\lambda}_{z}^{s}$ that are similar to those from the
first simulation experiment, with mean estimates being between $0.026117$
and $0.031798$, and relative frequencies corresponding to $\Pr (\hat{\lambda}%
_{z}^{s}>\hat{\lambda}_{z}^{\mathrm{HLW}})$ being between $38.40\%$ and $%
49.80\%$.

\subsection{Stage 3 Model \label{sec:S3}}

The analysis so far has demonstrated that the ratios of interest $\lambda
_{g}=\sigma _{g}/\sigma _{_{y^{\ast }}}$ and $\lambda _{z}=a_{r}\sigma
_{z}/\sigma _{\tilde{y}}$ required for the estimation of the full structural
model in \ref{eq:hlw} cannot be recovered from \cites{holston.etal:2017}
MUE\ procedure implemented in Stages 1 and 2. Moreover, since their
procedure is based on the \emph{`misspecified}\textit{'} Stage 2 model in %
\ref{eq:stag2}, it results in a substantially larger estimate of $\lambda
_{z}$ than when implemented on the \emph{`correctly specified}\textit{'}
Stage 2 model in \ref{S2full0}. This substantially larger estimate of $%
\lambda _{z}$ in turn leads to a greatly amplified and strongly downward
trending \emph{`other factor'} $z_{t}$. To show the impact of this on %
\cites{holston.etal:2017} estimate of the natural rate of interest, I
initially report parameter estimates of the full Stage 3 model in \autoref%
{tab:Stage3}, followed by plots of filtered estimates of the natural rate $%
r_{t}^{\ast }$, trend growth $g_{t}$, \emph{`other factor'} $z_{t}$, and the
output gap (cycle) variable $\tilde{y}_{t}$ in \autoref{fig:2017KF}.%
\footnote{%
Smoothed estimates are shown in \autoref{fig:2017KS}. In \hyperref[sec:AS3]{%
Section A.3} in the \hyperref[appendix]{Appendix}, the expansion of the
system matrices are reported as for the earlier Stage 1 and Stage 2 models.
These are in line with the full model reported in \ref{eq:hlw}. As before,
the state vector $\boldsymbol{\xi }_{t}$ is initialized using the same
procedure as outlined in \ref{eq:P00S1a} and \fnref{fn:1}, with the
numerical values of $\boldsymbol{\xi }_{00}$ and $\mathbf{P}_{00}$ given in %
\ref{AS3:xi00} and \ref{AS3:P00}.}

Given estimates of the ratios $\lambda _{g}=\sigma _{g}/\sigma _{_{y^{\ast
}}}$ and $\lambda _{z}=a_{r}\sigma _{z}/\sigma _{\tilde{y}}$ from the
previous two stages, the vector of Stage 3 parameters to be computed by MLE
is:%
\begin{equation}
\boldsymbol{\theta }_{3}=[a_{y,1},~a_{y,2},~a_{r},~b_{\pi },~b_{y},~\sigma _{%
\tilde{y}},~\sigma _{\pi },~\sigma _{y^{\ast }}]^{\prime }.
\label{eq:theta3}
\end{equation}%
In \autoref{tab:Stage3}, estimates of $\boldsymbol{\theta }_{3}$ are
presented following the same format as in \autoref{tab:Stage1} and \autoref%
{tab:Stage2} previously. Since I also estimate $\sigma _{g}$ and $\sigma
_{z} $ directly together with the other parameters by MLE without using the
Stage 1 and Stage 2 estimates of $\lambda _{g}$ and $\lambda _{z}$,
additional rows are inserted, with the values in brackets denoting implied
estimates. The first two columns in \autoref{tab:Stage3} show estimates of $%
\boldsymbol{\theta }_{3}$ obtained from running \cites{holston.etal:2017}
R-Code and my replication. The third and fourth columns (under headings `MLE(%
$\sigma _{g}|\hat{\lambda}_{z}^{\mathrm{HLW}}$)' and `MLE($\sigma
_{g}|\lambda _{z}^{\mathcal{M}_{0}}$)', respectively) report estimates when $%
\sigma _{g}$ is estimated freely by MLE, while $\lambda _{z}$ is held fixed
at either $\hat{\lambda}_{z}^{\mathrm{HLW}}=0.030217$ obtained from %
\cites{holston.etal:2017} \emph{`misspecified}\textit{'} Stage 2 model under
their \emph{`time varying} $\boldsymbol{\phi }$\textit{'} approach, or at $%
\hat{\lambda}_{z}^{\mathcal{M}_{0}}=0.000754$ computed from the \emph{%
`correctly specified}\textit{'} Stage 2 model in \ref{S2full0} with \emph{%
`constant} $\boldsymbol{\phi }$\textit{'}. The last column of \autoref%
{tab:Stage3} under heading `MLE($\sigma _{g},\sigma _{z}$)' lists the
estimates of $\boldsymbol{\theta }_{3}$ when $\sigma _{g}$ and $\sigma _{z}$
are computed directly by MLE, with the implied values of $\lambda _{g}$ and $%
\lambda _{z}$ reported in brackets.

The Stage 3 results in \autoref{tab:Stage3} can be summarized as follows.\
The MLE of $\sigma _{g}$ does not \emph{`pile-up'} at zero and is again
approximately $50\%$ larger than the estimate implied by the Stage 1 MUE\ of
$\lambda _{g}$. That is, $\hat{\sigma}_{g}\approx $ $0.045$ in the last
three columns of \autoref{tab:Stage3}, and thus very similar in size to the
Stage 2 estimates of $0.044$ and $0.045$ shown in the last two columns of %
\autoref{tab:Stage2}. Computing $\sigma _{z}$ directly by MLE leads to a
point estimate that shrinks numerically to zero, while the estimates of the
other parameters remain largely unchanged. Notice again that the
log-likelihood values of the last three models in \autoref{tab:Stage3} are
very similar, ie., between $-514.8307$ and $-514.2899$. Yet, the
corresponding estimates of $\sigma _{z}$ are either very small at $0$ or
comparatively large at $0.1371$ when implied from the \emph{`misspecified}%
\textit{'} Stage 2 model's $\hat{\lambda}_{z}^{\mathrm{HLW}}$ estimate. The $%
\hat{\sigma}_{z}$ coefficient from the \emph{`correctly specified}\textit{'}
Stage 2 model is $0.0037$ and thereby nearly 40 times smaller than from the
\emph{`misspecified}\textit{'} Stage 2 model.

The findings from \autoref{tab:Stage3} are mirrored in the filtered
estimates of $r_{t}^{\ast }$, $g_{t}$, $z_{t}$ and $\tilde{y}_{t}$ plotted
in \autoref{fig:2017KF}. The `MLE($\sigma _{g}|\lambda _{z}^{\mathcal{M}%
_{0}} $)' and `MLE($\sigma _{g},\sigma _{z}$)' estimates are visually
indistinguishable. Unsurprisingly, out of the four estimates, \emph{`other
factor'} $z_{t}$ is overall most strongly affected by the two different $%
\lambda _{z}$ values that are conditioned upon, showing either\ vary large
variability and a pronounced downward trend in $z_{t}$, or being close to
zero with very little variation (see panel (c) in \autoref{fig:2017KF}). The
effect on the estimate of the natural rate is largest in the immediate
aftermath of the global financial crisis, namely, from 2010 onwards.
Interestingly, the output gap estimates shown in panel (d) of \autoref%
{fig:2017KF} are quite similar, with the largest divergence occurring after
2012. The three trend growth estimates in panel (b) of \autoref{fig:2017KF}
which estimate $\sigma _{g}$ directly by MLE are visually indistinguishable,
despite having very different ${\sigma}_{z}$ values, namely, between $0$ and
$0.1371$ (see the lines corresponding to `MLE($\sigma _{g}|\lambda _{z}^{%
\mathrm{HLW}}$)', `MLE($\sigma _{g}|\lambda _{z}^{\mathcal{M}_{0}}$)' and
`MLE($\sigma _{g},\sigma _{z}$)'). Trend growth estimated from %
\cites{holston.etal:2017} \ Stage 1 MUE of $\lambda _{g}$ is noticeably
larger from 2009 to 2014. In comparison to the plots shown in panel (c) of %
\autoref{fig:HLW_factors}, the drop in all four trend growth estimates
following the financial crisis seems exaggerated. The pure backward looking
nature of the Kalman Filtered $g_{t}$ series exacerbates the effect of the
decline in GDP during the financial crisis on trend growth estimates after
the crisis.

A final point I would like to make here --- and without the intention to
engage in repetitive and unnecessary discussion --- is that extending the
sample period to 2019:Q2 produces the interesting empirical result that
estimating $\sigma _{z}$ directly by MLE does not lead to any \emph{`pile-up'%
} at zero problems. Moreover, the ML estimate of $\sigma _{z}$ is very
similar to the one implied from the \emph{`correctly specified}\textit{'}
Stage 2 model's $\lambda _{z}$, and thereby again in stark contrast to the
oversized estimate obtained from
\cites{holston.etal:2017}
\emph{`misspecified}\textit{'} Stage 2 model's $\lambda _{z}$.\footnote{%
These estimation results using data up to 2019:Q2 together with
corresponding plots of filtered (and smoothed) estimates are reported in %
\autoref{Atab:S3_2019}, \autoref{Afig:2019KF} and \autoref{Afig:2019KS} in
\hyperref[sec:AS3]{Section A.3} of the \hyperref[appendix]{Appendix}.} Even
so, despite the fact that the point estimate of $\sigma _{z}$ does not
shrink to zero, it is highly insignificant, which suggests that there is
little evidence in the data of \textit{`other factor'} $z_{t}$ being
relevant for the model.

\section{Other issues\label{sec:other}}

There are other issues with
\cites{holston.etal:2017}
structural model in \ref{eq:hlw} that make it unsuitable for policy
analysis. For instance, the interest rate $i_{t}$ is included as an
exogenous variable, so that the model essentially tries to find the best
fitting natural rate $r_{t}^{\ast }$ for it. With $r_{t}^{\ast
}=4g_{t}+z_{t} $, and \textit{`other factor'} $z_{t}$ the \textit{`free'}
variable due to $g_{t}$ being driven by GDP, $z_{t}$ effectively matches the
\textit{`leftover' }movements in the interest rate to make it compatible
with trend growth in the model. Since the central bank has full control over
the (fed funds) interest rate, it can set $i_{t}$ to any desired level and
the model will produce a natural rate through \textit{`other factor'} $z_{t}$
that will match it. Also, there is nothing in the structural model of \ref%
{eq:hlw} that makes the system stable. For the output gap relation in \ref%
{IS} to be stationary, the real rate cycle $r_{t}-r_{t}^{\ast }=(i_{t}-\pi
_{t}^{e})-(4g_{t}+z_{t})$ must be $I(0)$, yet there is no co-integrating
relation imposed anywhere in the system to ensure that this holds in the
model.\footnote{%
This insight is not new and has been discussed in, for instance, \cite%
{pagan.wickens:2019} (see pages $21-23$).} When trying to simulate from such
a model, with $\pi _{t}$ being integrated of order 1, the simulated paths of
the real rate $r_{t}=i_{t}-\pi _{t}^{e}$ can frequently diverge to very
large values, even with samples of size $T=229$ observations, which is the
empirical sample size.

A broader concern for policy analysis is the fact that the filtered
estimates of the state vector $\boldsymbol{\xi }_{t}$ will be (weighted
combinations of the) one-sided moving averages of the three observed
variables that enter the state-space model; namely, $i_{t}$, $y_{t}$, and $%
\pi _{t}$.\footnote{%
Smoothed estimates will be (weighted combinations of the) two-sided moving
averages of the observables. See also \cite{durbin.koopman:2012}, who write
to this on page 104: \textit{"}\emph{It follows that these conditional means
are weighted sums of past (filtering), of past and present (contemporaneous
filtering) and of all (smoothing) observations. It is of interest to study
these weights to gain a better understanding of the properties of the
estimators as is argued in Koopman and Harvey (2003). ... . In effect, the
weights can be regarded as what are known as kernel functions in the field
of nonparametric regression; ... ."}} This can be seen by writing out the
Kalman Filtered estimate of the state vector as:\footnote{%
I again follow the notation in \cite{hamilton:1994}, see pages 394-395, with
the matrices $\mathbf{A}$ and $\mathbf{H}$ however not transposed to be
consistent with my earlier notation.}%
\begin{align}
\skew{0}\boldsymbol{\hat{\xi}}_{t|t}& =\skew{0}\boldsymbol{\hat{\xi}}%
_{t|t-1}+\underbrace{\mathbf{P}_{t|t-1}\mathbf{H}^{\prime }(\mathbf{HP}%
_{t|t-1}^{\prime }\mathbf{H}^{\prime }+\mathbf{R})^{-1}}_{\mathbf{G}_{t}}(%
\mathbf{y}_{t}-\mathbf{Ax}_{t}-\mathbf{H}\skew{0}\boldsymbol{\hat{\xi}}%
_{t|t-1})  \notag \\
& =\skew{0}\boldsymbol{\hat{\xi}}_{t|t-1}+\mathbf{G}_{t}(\mathbf{y}_{t}-%
\mathbf{Ax}_{t}-\mathbf{H}\skew{0}\boldsymbol{\hat{\xi}}_{t|t-1})  \notag \\
& =(\mathbf{I}-\mathbf{G}_{t}\mathbf{H)}\skew{0}\boldsymbol{\hat{\xi}}%
_{t|t-1}+\mathbf{G}_{t}(\mathbf{y}_{t}-\mathbf{Ax}_{t})  \notag \\
& =\underbrace{(\mathbf{I}-\mathbf{G}_{t}\mathbf{H)F}}_{\mathbf{\Phi }_{t}}%
\skew{0}\boldsymbol{\hat{\xi}}_{t-1|t-1}+\mathbf{G}_{t}\underbrace{(\mathbf{y%
}_{t}-\mathbf{Ax}_{t})}_{\mathbf{\bar{y}}_{t}}  \notag \\
& =\mathbf{\Phi }_{t}\skew{0}\boldsymbol{\hat{\xi}}_{t-1|t-1}+\mathbf{G}_{t}%
\mathbf{\bar{y}}_{t},  \notag \\
\intxt{which is a (linear) recursion in
$\skew{0}\boldsymbol{\hat{\xi}}_{t|t}$ and can be thus rewritten as:}& =%
\boldsymbol{\Psi }_{t}\boldsymbol{\xi }_{0|0}+\sum_{i=0}^{t-1}\underbrace{%
\boldsymbol{\Psi }_{i}\mathbf{G}_{t-i}}_{\boldsymbol{\omega }_{ti}}\mathbf{%
\bar{y}}_{t-i}  \notag \\
& =\boldsymbol{\Psi }_{t}\boldsymbol{\xi }_{0|0}+\sum_{i=0}^{t-1}\boldsymbol{%
\omega }_{ti}\mathbf{\bar{y}}_{t-i},  \label{xirec}
\end{align}%
where $\boldsymbol{\Psi }_{i}=\prod_{n=0}^{i-1}\mathbf{\Phi }_{t-n},\forall
i=1,2,\ldots ,$ $\boldsymbol{\Psi }_{0}=\mathbf{I,}$ $\mathbf{I}$ is the
identity matrix, $\skew{0}\boldsymbol{\hat{\xi}}_{t|t-1}=\mathbf{F}\skew{0}%
\boldsymbol{\hat{\xi}}_{t-1|t-1}$ is the predicted state vector, $%
\boldsymbol{\xi }_{0|0}$ is the prior mean, $\mathbf{P}_{t|t-1}=\mathbf{FP}%
_{t-1|t-1}\mathbf{F}+\mathbf{Q}$ is the predicted state variance, $%
\boldsymbol{\omega }_{ti}=\boldsymbol{\Psi }_{i}\mathbf{G}_{t-i}$ is a time
varying weight matrix, and $\mathbf{\bar{y}}_{t}$ consists of the observed
variables $y_{t}$, $\pi _{t}$, and $i_{t}$.\footnote{%
To understand what is driving the downward trend in `\emph{other factor}' $%
z_{t}$ since the early 2000s in the model, one could examine the weight
matrix $\boldsymbol{\omega }_{ti}$ in \ref{xirec} more closely to see how it
interacts with the observable vector $\mathbf{\bar{y}}_{t}=\mathbf{y}_{t}-%
\mathbf{Ax}_{t}=[a(L)y_{t}-a_{r}(L)r_{t};~b_{\pi }(L)\pi
_{t}-b_{y}y_{t}]^{\prime }$, where $b_{\pi }(L)=1-b_{\pi }L-\frac{1}{3}%
(1-b_{\pi })(L^{2}+L^{3}+L^{4})$ is the lag polynomial capturing the
dynamics of inflation. Alternatively, the steady-state $\mathbf{P}$ matrix
could be computed recursively as in equation 13.5.3 in \cite{hamilton:1994}
to replace $\mathbf{P}_{t|t-1}$ in the recursions for $\skew{0}\boldsymbol{%
\hat{\xi}}_{t|t}$. The relation in \ref{xirec} would then yield $\skew{0}%
\boldsymbol{\hat{\xi}}_{t|t}=\Phi ^{t}\boldsymbol{\xi }_{0|0}+%
\sum_{i=0}^{t-1}\mathbf{\Phi }^{i}\mathbf{G\bar{y}}_{t-i}$, where $\mathbf{%
\Phi =(I-GH)F}$ and $\mathbf{G=PH}^{\prime }(\mathbf{HP}^{\prime }\mathbf{H}%
^{\prime }+\mathbf{R})^{-1}$ would be the steady-state analogue to $\mathbf{%
\Phi }_{t}$ and $\mathbf{G}_{t}$, with $\mathbf{P}_{t|t-1}$ replaced by $%
\mathbf{P}$ from the steady-state $\mathbf{P}$ matrix.}

This creates the following two issues. First, since the nominal interest
rate $i_{t}$ is directly controlled by the central bank, and the natural
rate is constructed from the filtered estimate of state vector $\boldsymbol{%
\xi }_{t}$, which itself is computed as a moving average of $i_{t}$ (and the
other observable variables), a circular relationship can be seen to evolve.
Any central bank induced change in the policy rate $i_{t}$ is mechanically
transferred to the natural rate $r_{t}^{\ast }$ via the Kalman Filtered
estimate of the state vector $\skew{0}\boldsymbol{\hat{\xi}}_{t|t}$ in \ref%
{xirec}. A confounding effect between $r_{t}^{\ast }$ and $i_{t}$ will
arise, making it impossible to answer questions of interest such as:
\textquotedblleft \textit{Is the natural rate low because }$i_{t}$\ \textit{%
is low, or is }$i_{t}$\textit{\ low because the natural rate is
low?\textquotedblright } with this model, as one will follow as a direct
consequence from the other.

Second, because of the one-sided moving average nature of the Kalman
Filtered estimates of the state vector, any outliers, structural breaks or
otherwise \emph{`extreme'} observations at the beginning (or end) of the
sample period can have a strong impact on these filtered estimates. For the
(two-sided) \cite{hodrick.prescott:1997} filter, such problems (and other
ones) are well known and have been discussed extensively in the literature
before.\footnote{%
There exists a large literature on the HP filter and its problems (one of
the more recent papers is by \cite{hamilton:2018}), and it is not the goal
to review or list them here. However, the study by \cite{phillips.jin:2015}
is interesting to single out, in particular the introduction section on
pages 2 to 9, as it highlights the recent public debates by James Bullard,
Paul Krugman, Tim Duy and others on the use (and misuse) of the HP filter
for the construction of output gaps for policy analysis. \cite%
{phillips.jin:2015} show also that the HP filter fails to recover the
underlying trend asymptotically in models with breaks (see section 4 in
their paper), and they further propose alternative filtering/smoothing
methods. In an earlier study, \cite{schlicht:2008} describes how to deal
with structural breaks and missing data.} However, (one-sided) Kalman Filter
based estimates will also be affected. This can be easily demonstrated here
by re-estimating the model using four different starting dates, while
keeping the end of the sample period the same at 2019:Q2. In \autoref%
{fig:T0KF} I show filtered estimates of $r_{t}^{\ast }$, $g_{t}$, $z_{t}$
and $\tilde{y}_{t}$ for the starting dates 1967:Q1, 1972:Q1, 1952:Q2 and
1947:Q1 (smoothed estimates are shown in \autoref{fig:T0KS}), together with %
\cites{holston.etal:2017} estimates using 1961:Q1 as the starting date.%
\footnote{%
In all computations, I use \cites{holston.etal:2017} R-Code and follow
exactly their three stage procedure as before to estimate the factors of
interest.}

Why are these starting dates chosen? The period following the April 1960 to
February 1961 recession was marked by temporarily unusually (and perhaps
misleadingly) high GDP\ growth, yielding an annualized mean of $6.07\%$
(median $6.47\%$), with a low standard deviation of $2.67\%$ from 1961:Q2 to
1966:Q1 (see panel (b) of \autoref{fig:HLW_factors}). Having such \textit{%
`excessive'} growth at the beginning of the sample period has an unduly
strong impact on the filtered (less so on the smoothed) estimate of trend
growth $g_{t}$ in the model. Since both $g_{t}$ and $z_{t}$ enter the
natural rate, this affects the estimate of $r_{t}^{\ast }$. To illustrate
the sensitivity of these estimates to this time period, I\ re-estimate the
model with data starting 6 years later in 1967:Q1. Also, %
\cites{holston.etal:2017} Euro Area estimates of $r_{t}^{\ast }$ are
negative from around 2013 onwards (see the bottom panel of Figure 3 on page
S63 of their paper).\footnote{%
This negative estimate in $r_{t}^{\ast }$ is driven by an excessively large
and volatile estimate of \textit{`other factor'} $z_{t}$. Some commentators
have attributed the larger decline in the natural rate to a stronger
manifestation of \textit{`}\emph{secular stagnation}\textit{'} in the Euro
Area than in the U.S.} To show that we can get the same negative estimates
of $r_{t}^{\ast }$ for the U.S., I re-estimate the model with data starting
in 1972:Q1 to match the sample period of the Euro Area in \cite%
{holston.etal:2017}. Lastly, I\ extend \cites{holston.etal:2017} data back
to 1947:Q1 to have estimates from a very long sample, using total PCE\
inflation prior to 1959:Q2 in place of Core PCE\ inflation and the Federal
Reserve Bank of New York discount rate from 1965:Q1 back to 1947:Q1 as a
proxy for the Federal Funds rate as was done in \cite{laubach.williams:2003}.%
\footnote{%
Note that from the quarterly CORE PCE data it will be possible to construct
annualized inflation only from 1947:Q2 onwards. To have an inflation data
point for 1947:Q1, annual core PCE\ data (BEA\ Series ID: DPCCRG3A086NBEA)
that extends back to 1929 was interpolated to a quarterly frequency and
subsequently used to compute (annualized) quarterly inflation data for
1947:Q1. Since \cites{holston.etal:2017} R-Code requires 4 quarters of GDP\
data prior to 1947:Q1 as initial values, annual GDP\ (BEA Series ID: GDPCA)
was interpolated to quarterly data for the period 1946:Q1 1946:Q4.} Since
inflation was rather volatile from 1947 to 1952, I also re-estimate the
model with data beginning in 1952:Q2 to exclude this volatile inflation
period from the sample.

Panel (a) in \autoref{fig:T0KF} shows how sensitive the natural rate
estimates to the different starting dates are, particularly at the beginning
of \cites{holston.etal:2017} sample, namely, from 1961 until about 1980, and
at the end of the sample from 2009 onwards. Negative natural rate estimates
are now also obtained for the U.S. when the beginning of the sample period
is aligned with that of the Euro Area in 1972:Q1 (or 1967:Q1 which excludes
the high GDP growth period). Since the natural rate is defined as the sum of
trend growth $g_{t}$ and \textit{`}\emph{other factor}\textit{'} $z_{t}$, we
can separately examine the contribution of each of these factors to $%
r_{t}^{\ast }$. From panel (b) in \autoref{fig:T0KF} it is evident that the
filtered trend growth estimates\ are the primary driver of the excessive
sensitivity in $r_{t}^{\ast }$ over the 1961 to 1980 period. For instance,
in 1961:Q1, these estimates can be as high as 6 percent, or as low as 3
percent, depending on the starting date of the sample. Also, the differences
in these estimates stay sizeable until 1972:Q1, before converging to more
comparable magnitudes from approximately 1981 onwards. Apart from the
estimate using the very long sample beginning in 1947:Q1 (see the blue line
in panel (b) of \autoref{fig:T0KF}), the other four remain surprisingly
similar, even during and after the financial crisis period, that is, from
mid 2007 to the end of the sample in 2019:Q2. Thus, the \textit{`front-end'}
variability of the natural rate estimates are driven by the \textit{%
`front-end' }variability in the estimates of trend growth $g_{t}$.

In panel (b) of \autoref{fig:T0KF}, I also superimpose MUE and MMLE
(smoothed) estimates of trend growth from \cites{stock.watson:1998} model in %
\ref{eq:tvp_sim}, as well as (smoothed) estimates from the (correlated)\ UC
model in \ref{Stage1:mod} to provide long-sample benchmarks of trend growth
from simple univariate models which can be compared to %
\cites{holston.etal:2017} estimates. These are the same estimates that are
plotted in panels (b) and (c) of \autoref{fig:HLW_factors}. To avoid
cluttering the plot with additional lines, I do not plot the mean and median
estimates computed over the more recent expansion periods as was done in %
\autoref{fig:HLW_factors}. Note, however, that the MUE estimate overlaps
with the mean and median of GDP\ growth from 2009:Q3 onwards and can thus be
used as a representative for these model free \textit{`average'} estimates
of GDP growth since the end of the financial crisis. Comparing the Kalman
Filter based estimates from the various starting dates to the MUE, MMLE, and
UC (smoothed) ones shows how different these are, particularly, from 2009:Q3
until the end of the sample.\ In the immediate post-crisis period, the
(one-sided) filter based estimates are \emph{`pulled down'} excessively by
the sharp decline in GDP and \textit{`converge'} only slowly at the very end
of the sample period towards the three long-sample benchmarks. Trend growth
is severely underestimated from 2009:Q3 onwards, and this is reflected in
the estimate of $r_{t}^{\ast }$.

In \autoref{Afig:rec_mean} in the \hyperref[appendix]{Appendix}, I show
plots of (real) GDP\ growth and the recursively estimated mean of GDP growth
over the post financial crisis period from 2009:Q3 to 2019:Q2. Trend growth
stays rather stable between $2\%$ and $3\%$ over nearly the entire period,
settling at around $2.25\%$ in 2014:Q2 and remaining at that level.
Moreover, it is never close to the filtered estimate of \cite%
{holston.etal:2017} from 2009:Q3 to 2014:Q3. In \autoref{Afig:SPF_GDP_growth}%
, I plot the mean as well as median 10 year ahead annual-average (real) GDP
growth forecasts from the Survey of Professional Forecasters\ (SPF) from
1992 to 2020.\footnote{%
The data were downloaded from: %
\url{https://www.philadelphiafed.org/research-and-data/real-time-center/survey-of-professional-forecasters/data-files/rgdp10}
(accessed on the $27^{th}$ of July, 2020).} These forecasts also remain
fairly stable between $2\%$ and $3\%$ from 2008 until 2017, and drift only
marginally lower towards the very end of the sample. In \autoref%
{Afig:giglio_GDP_growth}, Vanguard investor survey based 3 year and 10 year
ahead expectations of (real) GDP growth from February 2017 to April 2020 are
plotted. These are taken from Figure II on page 5 in \cite{giglio.etal:2020}%
. The 10 year expected growth rate shown in the right panel of \autoref%
{Afig:giglio_GDP_growth} fluctuates (mainly) between $2.8\%$ and $3.2\%$
(the 3 year expected growth rate on the left is somewhat lower). All three
plots suggest that following the financial crisis, trend growth in GDP\ is
unlikely to have dropped to the 1.3\% estimate of \cite{holston.etal:2017}.

Looking at the estimates of \emph{`other factor'} $z_{t}$ in panel (c) of %
\autoref{fig:T0KF}, we can see that it is the end of the sample, namely,
from 2009:Q1 to 2019:Q2, that is most strongly affected by the different
starting dates.\footnote{%
There is also some variability beginning in the 70s until the 80s, but this
variation seems to be largely due to the noisier nature of the filtered
estimates and is not visible from the more efficient smoothed estimates
shown in panel\ (c) of \autoref{fig:T0KS}. The differentiation here is not
important. The point to take away from this discussion is that the period
following the financial crisis yields very different estimates from the two
shorter samples, irrespective of whether smoothed or filtered estimates are
used in the construction of the natural rate.} In particular the two $z_{t}$
estimates that are based on the shorter samples starting in 1967:Q1 and
1972:Q1, which exclude the \textit{`excessive'} GDP growth period at the
beginning of \cites{holston.etal:2017} sample, generate substantially more
negative $z_{t}$ estimates. For instance, in 2009:Q1, the 1972:Q1 based $%
z_{t}$ estimate is $-2.87$ while \cites{holston.etal:2017} is $-1.22$. Also,
the $z_{t}$ estimates from the shorter samples are well below $-2$ over
nearly the entire 2014:Q4 to 2019:Q2 period.\footnote{%
This is even more pronounced in the smoothed estimates of $z_{t}$ shown in
panel (c)\ of \autoref{fig:T0KS}.} What is particularly interesting to
highlight here is how stable (and very close to zero) the estimates of $%
z_{t} $ are from the four earlier sample starts from 1947:Q1 to about
1971:Q3. Given the rapid change in demographics and population growth, as
well as factors related to savings and investment following the end of World
War II, one would expect $z_{t}$ to capture this change. Even if we look at
the period until 1990:Q1, apart from the noise in the estimates, no apparent
upward or downward trend in $z_{t}$ is visible from panel (c) of \autoref%
{fig:T0KF}. Thus, the Baby Boomer generation entering the workforce shows no
effect on $z_{t}$. Only from 1990:Q2 onwards is a decisive downward trend in
the estimates of $z_{t}$ visible.

\cite{holston.etal:2017} initialize the state vector at zero for the $z_{t}$
elements of $\boldsymbol{\xi }_{t}$. This evidently has an anchoring effect
on \emph{`other factor'} $z_{t}$ at the beginning of the sample. In the
model, it acts like a normalisation, as it implies that the natural rate is
driven solely by trend growth $g_{t}$ at sample start. Although $z_{t}$
follows a (zero mean) random walk, so that an initialisation at zero seems
appealing from an econometric perspective, this initialisation has an
important impact on the economic interpretation of $z_{t}$ that should be
more openly discussed if one is to view \emph{`other factor'} $z_{t}$ as a
factor relating to structural changes in an economy. Due to its large impact
on the downward trend in the estimates of the natural rate, understanding
exactly what $z_{t}$ captures and how the zero initialisation affects these
estimates is crucial from a policy perspective.

One final point that needs to be raised relates to \cites{holston.etal:2017}
preference for reporting filtered estimates of the latent states, as opposed
to smoothed ones. It is well know that the mean squared error (MSE) of the
filtered states will in general be larger than the MSE\ of the smoothed
states (see the discussion on page 151 in \cite{harvey:1989}). This is not
surprising, as the smoothed estimates use the full sample --- and therefore
more information --- to estimate the latent states, leading to more
efficient estimates. Moreover, reporting filtered estimates \textit{%
`precludes'} the use of a diffuse prior for the $I(1)$ state vector, since
it generates extreme volatility in the filtered estimates of the states at
the beginning of the sample period. This is not the case with the smoothed
estimates. The large variability in the filtered states is particulary
visible from the three quantities of interest, ie., the estimates of $%
r_{t}^{\ast }$, $g_{t}$ and $z_{t}$, and less so from the output gap (cycle)
estimates.

While it is frequently claimed that the filtered states are \emph{`real time'%
} estimates, and are thus more relevant for policy analysis, one can see
that this cannot be a valid argument in the given context. Not only are the
parameter estimates of the model (ie., the estimates of $\boldsymbol{\theta }%
_{3}$ in \ref{eq:theta3}) based on full sample information, the GDP and PCE\
inflation data that go into the model are also not real time data, that is,
data that were available to policy makers at time $t<T$. Reporting filtered
(one-sided) estimates of the states as in \cite{holston.etal:2017} or as on
the FRBNY website where updates are provided is undesirable from an
estimator efficiency perspective.

\section{Conclusion \label{sec:conclusion}}

\cites{holston.etal:2017} implementation of \cites{stock.watson:1998} Median
Unbiased Estimation (MUE) in Stages 1 and 2 of their procedure to estimate
the natural rate of interest from a larger structural model is unsound. I
show algebraically that their procedure cannot recover the ratios of
interest $\lambda _{g}=\sigma _{g}/\sigma _{y^{\ast }}$ and $\lambda
_{z}=a_{r}\sigma _{z}/\sigma _{\tilde{y}}$ needed for the estimation of the
full structural model of interest. \cites{holston.etal:2017} implementation
of MUE\ in Stage 2 of their procedure is particularly problematic, because
it is based on an \emph{`unnecessarily'} misspecified model as well as an
incorrect MUE\ procedure that spuriously amplifies their estimate of $%
\lambda _{z}$. This has a direct and consequential effect on the severity of
the downward trending behaviour of \emph{`other factor'} $z_{t}$ and thereby
the magnitude of the estimate of the natural rate.

Correcting their Stage 2 model and the implementation of MUE\ leads to a
substantially smaller estimate of $\lambda _{z}$ of close to zero, and
an elimination of the downward trending influence of \emph{`other factor'} $%
z_{t}$ on the natural rate of interest. The correction that is applied is
quantitatively important. It shows that the estimate of $\lambda _{z}$ based
on the correctly specified Stage 2 model is statistically highly
insignificant. The resulting filtered estimates of $z_{t}$ are very close to
zero for the entire sample period, highlighting the lack of evidence of
\textit{`}\emph{other factor'} $z_{t}$ being important for the determination
of the natural rate in this model. Obtaining an accurate estimate of trend
growth for the measurement of the natural rate is therefore imperative. To
provide other benchmark estimates of trend growth, I construct various
simple alternative $g_{t}$ estimates and compare those to the estimate from
\cite{holston.etal:2017}. I find the latter one to be too small,
particularly in the immediate aftermath of the global financial crisis.

Lastly, I\ discuss various other issues with \cites{holston.etal:2017} model
that make it unsuitable for policy analysis. For instance, %
\cites{holston.etal:2017} estimates of the natural rate, trend growth, `%
\emph{other factor}' $z_{t}$ and the output gap are extremely sensitive to
the starting date of the sample used to estimate the model. Using data
beginning in 1972:Q1 (or 1967:Q1) leads to negative estimates of the natural
rate as is the case for their Euro Area estimates. These negative estimates
are again driven purely by the exaggerated downward trending behaviour of
`\emph{other factor}' $z_{t}$. The 1972:Q1 date was chosen to match the
sample used in the estimation of the Euro Area model. Only the Euro Area
estimates of the natural rate turn negative in 2013, and only the Euro Area
sample starts in 1972:Q1 (the others start in 1961:Q1). The fact that it is
possible to generate such negative estimates of the natural rate from
\cites{holston.etal:2017} model for the U.S. as well by simply adjusting the
start of the estimation period suggests that the model is far from robust,
and therefore inappropriate for use in policy analysis.

Also, any Kalman Filtered (or Smoothed) estimates of the state vector will
be a function of the observable variables that enter into the model. If the
central bank controlled nominal interest rate is one of these observables, a
confounding effect between $r_{t}^{\ast }$ and $i_{t}$ will arise, because
any central bank induced change in the policy rate $i_{t}$ is mechanically
transferred to the natural rate via the estimate of the state vector $%
\skew{0}\boldsymbol{\hat{\xi}}_{t|t}$. This makes it impossible to answer
\textit{`causal'} questions regarding the relationship between $r_{t}^{\ast
} $ and $i_{t}$, as one responds as a direct consequence to changes in the
other.

\bigskip

\setlength{\oddsidemargin}{-5.4mm}
\bibliographystyle{LongBibStyleFile}
\bibliography{DanielsBibTexLibrary}

\ifthenelse{\equal{1}{\figs}}{
\IfFileExists{\figtabsname}{\def\plots{0}\cleardoublepage\newpage
\BAP\input{\figtabsname}\EAP}{}}{}

\ifthenelse{\equal{1}{\appndx}}{ \cleardoublepage\newpage
\changepage{0mm}{0mm}{0mm}{0mm}{0mm}{0mm}{0mm}{0mm}{5mm}
\setcounter{page}{1} \setcounter{figure}{0} \setcounter{table}{0}
\setcounter{equation}{0} \setcounter{footnote}{0} \setcounter{section}{0}
\renewcommand{\thepage}{A-\,{\arabic{page}}}
\renewcommand{\thefigure}{A.{\arabic{figure}}}
\renewcommand{\thetable}{A.{\arabic{table}}}
\renewcommand{\theequation}{A.{\arabic{equation}}}
\renewcommand{\thesection}{A.{\arabic{subsection}}} \setstretch{1.234}
\IfFileExists{\appndixname}{\def\filein{0}\cleardoublepage\newpage
\BAP\input{\appndixname}\EAP}{}}{}

\cleardoublepage\newpage

\ifthenelse{\equal{1}{\Rcode}}{
\changepage{3mm}{0mm}{0mm}{0mm}{0mm}{-3mm}{0mm}{0mm}{0mm}
\let\Osubsection\subsection
\renewcommand{\section}[1]{\stepcounter{section}
\Osubsection*{A.\arabic{section}.~~{#1}}
\addcontentsline{toc}{subsection}{\currentname}}
\section{R-Code Snippets \label{Rcode}}

\small

This sections shows various parts of the R-Code that is provided by \cite{holston.etal:2017}
in the zip file from \url{https://www.newyorkfed.org/medialibrary/media/research/economists/williams/data/HLWCode.zip}.
Below the code next to each of the headers, the name of the R-file is
listed from which the code is displayed.

\multilstinputlisting{./HLW_Code_4tex/}{rstar.stage3.R}{1}{123}{R:stage3}

\cleardoublepage\newpage

\multilstinputlisting{./HLW_Code_4tex/}{calculate.covariance.R}{1}{31}{R:covar}
\null \vfill

\cleardoublepage\newpage

\multilstinputlisting{./HLW_Code_4tex/}{unpack.parameters.stage1.R}{1}{45}{R:unpack}
\null \vfill

\cleardoublepage\newpage

\multilstinputlisting{./HLW_Code_4tex/}{kalman.states.wrapper.R}{1}{45}{R:wrapper}
\null \vfill

\cleardoublepage\newpage

\multilstinputlisting{./HLW_Code_4tex/}{rstar.stage2.R}{1}{116}{R:rstarS2}
\null \vfill

\cleardoublepage\newpage

\multilstinputlisting{./HLW_Code_4tex/}{median.unbiased.estimator.stage2.R}{1}{60}{R:MUE2}
\null \vfill

\cleardoublepage\newpage

}{}

\end{document}

%% file: _abstract.5e.tex
                      


\citeauthor*{holston.etal:2017}' (\citeyear{holston.etal:2017}) estimates of
the natural rate of interest are driven by the downward trending behaviour
of `\emph{other factor}' $z_{t}$. I show that their implementation of 
\cites{stock.watson:1998}
Median Unbiased Estimation (MUE) to determine the size of the $\lambda _{z}$
parameter which drives this downward trend in $z_{t}$ is unsound. It cannot
recover the ratio of interest $\lambda _{z}=a_{r}\sigma _{z}/\sigma _{\tilde{%
y}}$ from MUE required for the estimation of the full structural model. This
failure is due to an \emph{`unnecessary'} misspecification in 
\cites{holston.etal:2017}
formulation of the Stage 2 model. More importantly, their implementation of
MUE on this misspecified Stage 2 model spuriously amplifies the point
estimate of $\lambda _{z}$. Using a simulation experiment, I show that their
procedure generates excessively large estimates of $\lambda _{z}$ when
applied to data generated from a model where the true $\lambda _{z}$ is
equal to zero. Correcting the misspecification in their Stage 2 model and
the implementation of MUE leads to a substantially smaller $\lambda _{z}$
estimate, and with this, a more subdued downward trending influence of \emph{%
`other factor}' $z_{t}$ on the natural rate. Moreover, the $\lambda _{z}$
point estimate is statistically highly insignificant, suggesting that there
is no role for \emph{`other factor}' $z_{t}$ in this model. I also discuss
various other estimation issues that arise in 
\cites{holston.etal:2017}
model of the natural rate that make it unsuitable for policy analysis.


\keywrds{Natural rate of interest, Median Unbiased Estimation, Kalman
Filter, spurious relations, misspecified econometric models.}\newline
\jelclss{C32, E43, E52, O40.}\newline


%% file: _figstabs.5e.tex
\renewcommand{\lbfnt}{\fontsize{9pt}{0pt}\selectfont} 
\renewcommand{\sefnt}{\fontsize{8pt}{0pt}\selectfont} 
\renewcommand{\pvfnt}{\fontsize{5pt}{0pt}\selectfont} 
\newcommand{\capfnt}{\fontsize{10.5pt}{13pt}\selectfont} 
\renewcommand{\captionfont}{\capfnt\sf} 
\changepage{33mm}{10mm}{-5mm}{-5mm}{00mm}{-4mm}{-11mm}{0mm}{-4.575mm}         
\setcounter{table}{0} \setcounter{figure}{1}
\renewcommand{\thefigure}{\arabic{figure}}%
\renewcommand{\thetable}{\arabic{table}}  %
\sisetup{detect-all}
\pagestyle{empty}

\vspace*{\figspace}
\section*{{Figures and Tables}}\addcontentsline{toc}{section}{Figures and Tables}
\vspace*{05mm}

\begin{figure}[h!]
\centering
\includegraphics[width=1\textwidth,rotate=00,trim={0 0 0 0},clip]{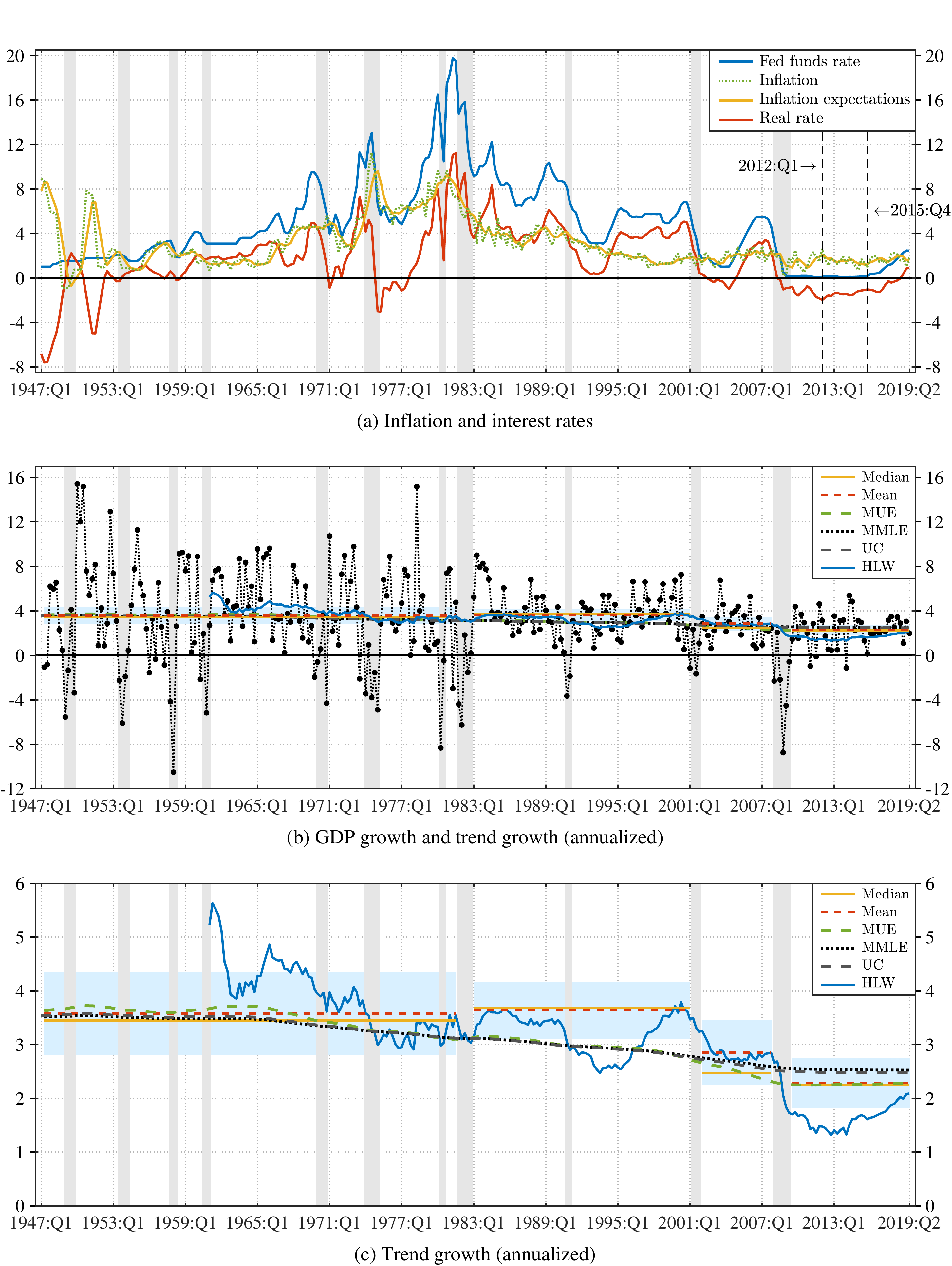} %
\vspace{-06mm}
\caption{Inflation, interest rates, and GDP growth (annualized) from 1947:Q1 to 2019:Q2.}
\label{fig:HLW_factors}
\end{figure}

\cleardoublepage\newpage 

\BT[p!]\caption{Summary statistics of GDP growth over various sub-periods and expansion periods only}    %
\centering\vspace*{-2mm}\renewcommand{\arraystretch}{1.1}\renewcommand                                   %
\tabcolsep{7pt}\fontsize{11pt}{13pt}\selectfont                                                          %
\newcolumntype{N}{S[table-format = 2.5,round-precision = 4]}                                             %
\newcolumntype{K}{S[table-format = 5.2,round-precision = 0]}                                             %
\begin{tabular*}{.97\columnwidth}{p{45mm}NNNKNN}                                                         %
\topline                                                                                                 %
{\hsp[10] Time period}   &                                                                               %
{Mean}    &                                                                                              %
{Median}  &                                                                                              %
{Stdev}    &                                                                                             %
{$T$}      &                                                                                             %
{Stderr}   &                                                                                             %
{HAC-Stderr}  \\ \midrule                                                                                %
{\hsp[3] 1947:Q2 $-$ 1981:Q3} &   3.57460623  &   3.45005454  &   4.62614839   &     138    &     0.39380390  &   0.48408048  \\
{\hsp[3] 1983:Q1 $-$ 2001:Q1} &   3.64188790  &   3.68619125  &   2.29561133   &      73    &     0.26868098  &   0.40155454  \\
{\hsp[3] 1947:Q2 $-$ 2001:Q1} &   3.46739214  &   3.48094368  &   4.04057864   &     216    &     0.27492655  &   0.35302270  \\ \cmidrule(ll){1-7}
{\hsp[3] 1947:Q2 $-$ 1948:Q4} &   2.79515476  &   2.28504979  &   3.40717233   &       7    &     1.28779009  &   1.25301772  \\
{\hsp[3] 1950:Q1 $-$ 1953:Q2} &   7.34709885  &   7.11936691  &   4.92757075   &      14    &     1.31694868  &   1.70601756  \\
{\hsp[3] 1954:Q3 $-$ 1957:Q3} &   3.93896808  &   3.90045256  &   3.65372953   &      13    &     1.01336224  &   1.29851867  \\
{\hsp[3] 1958:Q3 $-$ 1960:Q2} &   5.38582377  &   8.25098474  &   4.77957417   &       8    &     1.68983465  &   1.76820793  \\
{\hsp[3] 1961:Q2 $-$ 1969:Q4} &   4.78091472  &   4.34276189  &   2.98499125   &      35    &     0.50455561  &   0.61041410  \\
{\hsp[3] 1971:Q1 $-$ 1973:Q4} &   4.96221193  &   4.05503386  &   3.82343827   &      12    &     1.10373156  &   0.90409085  \\
{\hsp[3] 1975:Q2 $-$ 1980:Q1} &   4.17954338  &   2.94228977  &   3.66368599   &      20    &     0.81922509  &   0.74844011  \\
{\hsp[3] 1980:Q4 $-$ 1981:Q3} &   4.23481003  &   6.07631882  &   4.98886112   &       4    &     2.49443056  &   1.96043173  \\
{\hsp[3] 1983:Q1 $-$ 1990:Q3} &   4.17109994  &   3.80992734  &   2.21941490   &      31    &     0.39861868  &   0.63002180  \\
{\hsp[3] 1991:Q2 $-$ 2001:Q1} &   3.55219201  &   3.62908024  &   1.88730051   &      40    &     0.29840841  &   0.30730490  \\ \cmidrule(ll){1-7}
{\hsp[3] 2002:Q1 $-$ 2007:Q4} &   2.85460101  &   2.47095951  &   1.49946085   &      24    &     0.30607617  &   0.33238451  \\
{\hsp[3] 2009:Q3 $-$ 2019:Q2} &   2.28637245  &   2.25280082  &   1.47396679   &      40    &     0.23305461  &   0.19459307  \\
\bottomrule
\end{tabular*}
\tabnotes[-2.5mm][.965\columnwidth][-2.0mm]{This table reports estimates of
trend growth computed as the \textit{`average'} of annualized GDP\ growth
computed over various sub-periods and expansion periods only. Columns 2 to 5
report means, medians, standard deviations (Stdev) and sample sizes ($T$)
for the different sub-periods that are listed in column 1. The last two
columns provide simple (Stderr) and HAC\ robust (HAC-Stderr) standard errors
of the sample mean. The first three rows show time periods that include
recession as well as expansion periods over which GDP growth was larger on
average and/or more volatile than the last two rows (excluding the global
financial crisis recession period). The ten rows in the middle provide
summary statistics from 1947:Q2 to 2001:Q1 for expansion periods only.
} \label{tab:sumstatGDP}%
\ET

\cleardoublepage\newpage 

\BT[p!]%
\caption{Replicated results of Tables 4 and 5 in
\cite{stock.watson:1998}}\centering\vspace*{-2mm} \renewcommand{%
\arraystretch}{1.1}\renewcommand\tabcolsep{7pt}\fontsize{11pt}{13pt}%
\selectfont \newcolumntype{K}{S[table-format = 4.5,round-precision = 4]} %
\newcolumntype{L}{S[table-format = 2.7,round-precision = 4]} %
\newcolumntype{U}{S[table-format = 1.2,round-precision = 2]} %
\newcolumntype{N}{S[table-format = 3.6,round-precision = 4]}
\begin{tabular*}{\columnwidth}{p{30mm}NLKp{10mm}Up{1mm}Kp{8mm}U}
\topline
 Test& \multicolumn{1}{c}{Statistic} & \multicolumn{1}{c}{\hsp[-4]$p-$value} & \multicolumn{1}{c}{\hsp[4]$\lambda$} & \multicolumn{2}{c}{90\% CI} & &\multicolumn{1}{c}{\hsp[4]$\sigma_{\Delta \beta}$} & \multicolumn{2}{c}{90\% CI} \\
\midrule
    $L$   & 0.209398 & 0.25  & 4.055867 & [0, & \llap{19.36]} & & 0.130303  & [0, & \llap{0.62]} \\
    MW    & 1.158779 & 0.285 & 3.433543 & [0, & \llap{18.76]} & & 0.11031   & [0, & \llap{0.60]} \\
    EW    & 0.682116 & 0.325 & 3.071203 & [0, & \llap{17.01]} & & 0.098669  & [0, & \llap{0.54]} \\
    QLR   & 3.310513 & 0.48  & 0.778615 & [0, & \llap{13.26]} & & 0.025015  & [0, & \llap{0.42]} \\[1mm]
\end{tabular*}
\renewcommand{\arraystretch}{1.1}\renewcommand\tabcolsep{7pt} %
\fontsize{11pt}{13pt}\selectfont
\newcolumntype{K}{S[table-format = 3.4,round-precision = 4]} %
\newcolumntype{L}{S[table-format = 2.0,round-precision = 2]} %
\newcolumntype{U}{S[table-format = 1.2,round-precision = 2]} %
\newcolumntype{N}{S[table-format = 4.8,round-precision = 8]}
\begin{tabular*}{\columnwidth}{p{30mm}NNNNN}
\topline
Parameter                          & \multicolumn{1}{r}{MPLE\hsp[4]}       & \multicolumn{1}{r}{MMLE\hsp[3]}       & \multicolumn{1}{r}{MUE(0.13)}        & \multicolumn{1}{r}{MUE(0.62)}        & \multicolumn{1}{r}{SW.GAUSS} \\
\midrule
    $\sigma_{\Delta\beta}$         &    0                                    &    0.044400981501                          &    {0.13~~~~~~}                &    {0.62~~~~~~}                &    {0.13~~~~~~}       \\
    $\sigma_{\varepsilon}$         &    3.851994804846                       &    3.858594227694                          &    3.846619226998              &    3.782106578788              &    3.846619168993   \\
    AR(1)                          &    0.337083211459                       &    0.340252340453                          &    0.335014533140              &    0.315444720280              &    0.335014539465   \\
    AR(2)                          &    0.128903279894                       &    0.130746074778                          &    0.127423133110              &    0.120156417873              &    0.127423091443   \\
    AR(3)                          &   -0.009173836441                       &   -0.007251079335                          &   -0.010170600327              &   -0.014889876797              &   -0.010170523754   \\
    AR(4)                          &   -0.085644420982                       &   -0.082478616377                          &   -0.086802971766              &   -0.091560659988              &   -0.086802976773   \\
    $\beta_{00}$                   &    1.795899355603                       &    {~~~~~~---}                             &    2.440999263153              &    2.671500070276              &    2.440999404153   \\
\cmidrule(lr){1-6}
{Log-likelihood}                   & -539.772747031207                       & -547.480464499946                          & -540.692677059246              & -544.907181136766              & -540.692677059247   \\
\bottomrule
\end{tabular*}
\tabnotes[-3mm][.99\columnwidth][-1.0mm]{This table reports replication
results that correspond to Tables 4 and 5 in \cite{stock.watson:1998} on
page 354. The top part of the table shows the 4 different structural break
test statistics together with their $p-$values in the first two columns,
followed by the corresponding MUE estimates of $\lambda $ with 90\% CIs in
square brackets. The last two columns show the implied $\sigma _{\Delta
\beta }$ estimate computed from $T^{-1}{\lambda}\times {\sigma}_{\varepsilon
}/{a}(1)$ and 90\% CIs in square brackets. The first two columns of the
bottom part of the table report results from Maximum Likelihood based
estimation, where MPLE estimates the initial value of the state vector
$\beta_{00}$, while MMLE uses a diffuse prior for the initial value of the
state vector with mean zero and the variance set to $10^6$. Columns under
the heading MUE(0.13) and MUE(0.62) show Median Unbiased Estimates when
$\sigma_{\Delta \beta}$ is held fixed at 0.13, respectively, 0.62, which
correspond to the estimate of $\sigma_{\Delta \beta}$ when $\lambda$ is
computed using \cites{nyblom:1989} $L$ test (and its upper $90\%$ CI). The
last column under the heading SW.GAUSS lists the corresponding MUE(0.13)
estimates obtained from running \cites{stock.watson:1998} GAUSS code. The
row Log-likelihood displays the value of the log-likelihood at the reported
parameter estimates.
The Matlab file \texttt{SW1998\_MUE\_replication.m} replicates these results.}%
\label{tab:sw98_T4}\ET

\cleardoublepage\newpage 

\begin{figure}[p!]
\centering
\includegraphics[width=1\textwidth,rotate=00,trim={0 129mm 0
81mm},clip]{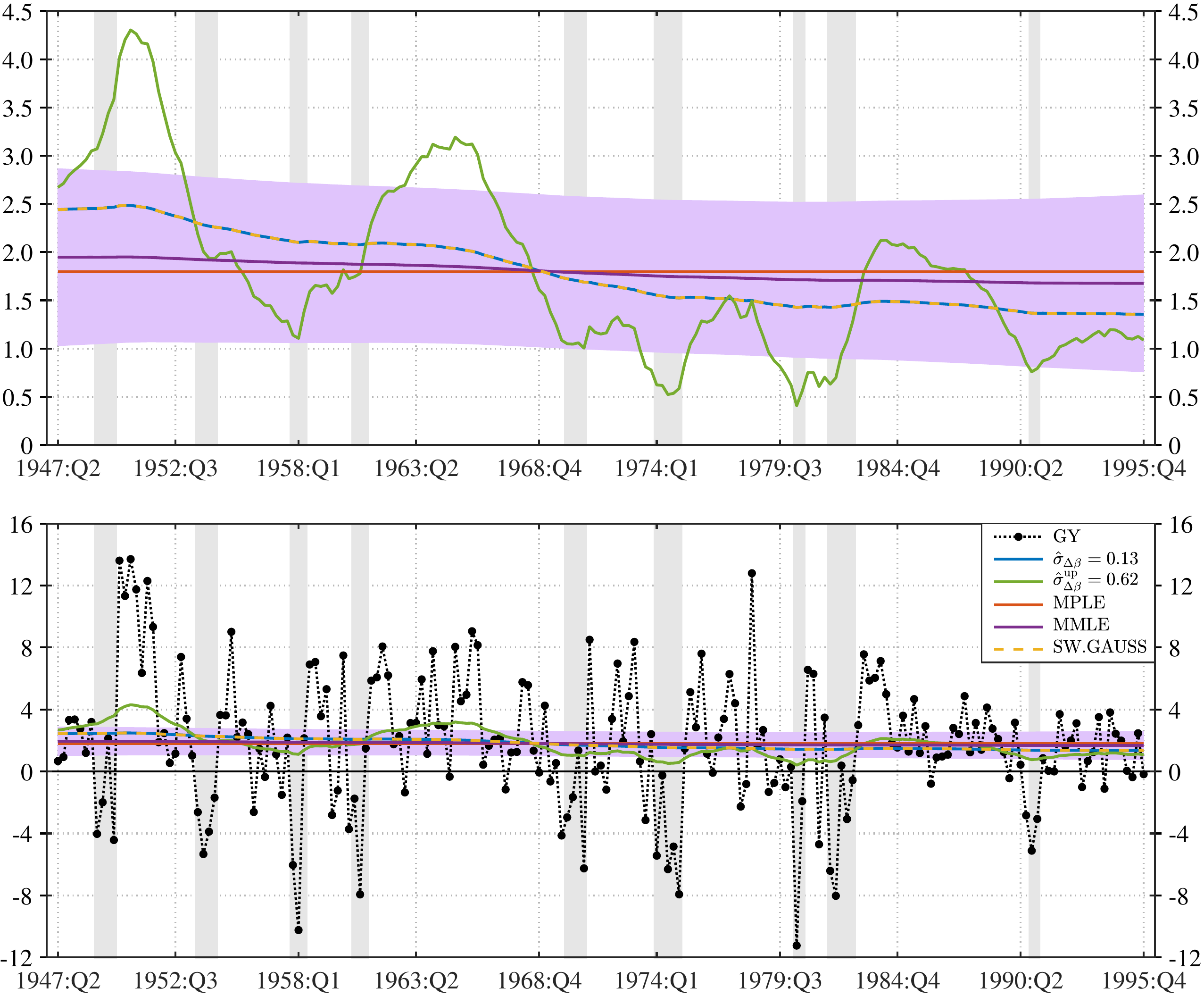} %
\caption{Smoothed trend growth estimates of US real GDP per capita.}
\label{fig:sw98_F4}
\end{figure}

\cleardoublepage\newpage 

\BT[p!]%
\caption{Broader replication results of Tables 4 and 5 in
\cite{stock.watson:1998} using per capita real GDP data from the Federal
Reserve
Economic Data database (FRED2)}\centering\vspace*{-2mm} \renewcommand{%
\arraystretch}{1.1}\renewcommand\tabcolsep{7pt}\fontsize{11pt}{13pt}%
\selectfont \newcolumntype{K}{S[table-format = 4.5,round-precision = 4]} %
\newcolumntype{L}{S[table-format = 2.7,round-precision = 4]} %
\newcolumntype{U}{S[table-format = 1.2,round-precision = 2]} %
\newcolumntype{N}{S[table-format = 3.6,round-precision = 4]}
\begin{tabular*}{\columnwidth}{p{30mm}NLKp{10mm}Up{1mm}Kp{8mm}U}
\topline
 Test& \multicolumn{1}{c}{Statistic} & \multicolumn{1}{c}{\hsp[-4]$p-$value} & \multicolumn{1}{c}{\hsp[4]$\lambda$} & \multicolumn{2}{c}{90\% CI} & &\multicolumn{1}{c}{\hsp[4]$\sigma_{\Delta \beta}$} & \multicolumn{2}{c}{90\% CI} \\
\midrule
    $L$   & 0.046701 & 0.895 & 0.000000 & [0, & \llap{4.099]} & & 0.000000  & [0, &  ~~~\llap{0.1092]} \\
    MW    & 0.251436 & 0.890 & 0.000000 & [0, & \llap{4.296]} & & 0.000000  & [0, &  ~~~\llap{0.1145]} \\
    EW    & 0.132064 & 0.900 & 0.000000 & [0, & \llap{3.910]} & & 0.000000  & [0, &  ~~~\llap{0.1042]} \\
    QLR   & 0.883403 & 0.980 & 0.000000 & [0, & \llap{0.000]} & & 0.000000  & [0, & ~~~~\llap{0.0000]} \\[1mm]
\end{tabular*}
\renewcommand{\arraystretch}{1.1}\renewcommand\tabcolsep{7pt} %
\fontsize{11pt}{13pt}\selectfont
\newcolumntype{K}{S[table-format = 3.4,round-precision = 4]} %
\newcolumntype{L}{S[table-format = 2.0,round-precision = 2]} %
\newcolumntype{U}{S[table-format = 1.2,round-precision = 2]} %
\newcolumntype{N}{S[table-format = 4.9,round-precision = 8]} %
\begin{tabular*}{\columnwidth}{p{44mm}NNNN}
\topline
    Parameter
& \multicolumn{1}{r}{MPLE\hsp[6]}
& \multicolumn{1}{r}{MMLE\hsp[5]}
& \multicolumn{1}{r}{MUE($\sigma_{\Delta \beta}^L$)\hsp[2]}
& \multicolumn{1}{r}{MUE($\mathrm{CI}-\sigma_{\Delta \beta}^L$)\hsp[-2]} \\
\midrule
    $\sigma_{\Delta\beta}$         &    0            &    0                 &    0          &      0.10926099   \\
    $\sigma_{\varepsilon}$         &    3.86603366   &    3.87619022     &    3.86603367 &      3.87574722   \\
    AR(1)                          &    0.31646541   &    0.32120674     &    0.31646541 &      0.32138794   \\
    AR(2)                          &    0.14652905   &    0.14903845     &    0.14652905 &      0.14924197   \\
    AR(3)                          &   -0.11122061   &   -0.10873408     &   -0.11122061 &     -0.10846721   \\
    AR(4)                          &   -0.09512645   &   -0.09050024     &   -0.09512645 &     -0.08983094   \\
    $\beta_{00}$                   &    2.12011198   &    {~~~~---}      &    2.12011200 &      2.07784473   \\ \cmidrule(ll){1-5}
{Log-likelihood}                   & -540.49919714   & -548.38308851     & -540.49919714 &   -541.89394940   \\
\bottomrule
\end{tabular*}
\tabnotes[-3mm][.99\columnwidth][-1.0mm]{This table reports replication
results that correspond to Tables 4 and 5 in \cite{stock.watson:1998} on
page 354, but now using real GDP per capita data (2012 chained dollars)
obtained from the Federal Reserve Economic Data database (FRED2) with series
ID: A939RX0Q048SBEA. The top part of the table shows the 4 different
structural break test statistics together with their $p-$values in the first
two columns, followed by the corresponding MUE estimates of $\lambda $ with
90\% CIs in square brackets. The last two columns show the implied $\sigma
_{\Delta \beta }$ estimate computed from $T^{-1}{\lambda}\times
{\sigma}_{\varepsilon }/{a}(1)$ and 90\% CIs in square brackets. The first
two columns of the bottom part of the table report results from Maximum
Likelihood based estimation, where MPLE estimates the initial value of the
state vector $\beta_{00}$, while MMLE uses a diffuse prior for the initial
value of the state vector with mean zero and the variance set to $10^6$.
Columns under the heading MUE($\sigma_{\Delta \beta}^L$) and
MUE($\mathrm{CI}-\sigma_{\Delta \beta}^L$) show Median Unbiased Estimates
when $\sigma_{\Delta \beta}$ is held fixed again at \cites{nyblom:1989} $L$
test statistic based structural break estimate, respectively, when the upper
$90\%$ CI value is used. The row Log-likelihood displays the value of the
log-likelihood at the reported parameter estimates. The sample period is the
same as in \cite{stock.watson:1998}, that is, from 1947:Q2 to 1995:Q4.
The Matlab file \texttt{estimate\_percapita\_trend\_growth\_v1.m} replicates these results.}%
\label{tab:sw98_T4_2}\ET

\cleardoublepage\newpage 

\BT[p!]\caption{Stage 1 parameter estimates}\centering\vspace*{-2mm}%
\renewcommand{\arraystretch}{1.1}\renewcommand\tabcolsep{7pt}%
\fontsize{11pt}{13pt}\selectfont%
\newcolumntype{N}{S[table-format =
3.6,round-precision = 6]}
\begin{tabular*}{1\columnwidth}{p{30mm}NNNNp{0mm}NN}
\topline
\multirow{2}{*}{\hsp[4]$\boldsymbol{\theta }_{1}$} & \multicolumn{4}{c}{HLW Prior}     & & \multicolumn{2}{c}{Diffuse Prior}                    \\ \cmidrule(rr){2-5} \cmidrule(rr){7-8}
                                &    {\hsp[0.5]HLW.R-File\hsp[-1.5]}    & {\hsp[0.5]$b_{y}\geq 0.025$\hsp[-1.5]} & {\hsp[.5]Alt.Init.Vals\hsp[-1.5]}    &   {\hsp[3]$b_{y}$ Free}    & &{\hsp[0.5]$b_{y}\geq 0.025$\hsp[-1.5]}&{\hsp[3]$b_{y}$ Free} \\
\midrule
$\hsp[3]a_{y,1}$                &    1.51706947457   &     1.51706921391   &    1.55766702528  &    1.45969739764    & &     1.64644427 &    1.56782968    \\
$\hsp[3]a_{y,2}$                &   -0.52880389083   &    -0.52880365096   &   -0.62244312676  &   -0.46382830448    & &    -0.67273238 &   -0.57782969    \\
$\hsp[3]b_{\pi }$               &    0.71249409600   &     0.71249401280   &    0.66995668421  &    0.72908945897    & &     0.71787120 &    0.73306306    \\
$\hsp[3]b_{y}$                  &    0.02500000000   &     0.02500000000   &    0.09718496555  &    0.00574089531    & &     0.02500000 &   -0.00094678    \\
$\hsp[3]g$                      &    0.77639673232   &     0.77639643600   &    0.74377517665  &    0.77947223337    & &     0.70948301 &    0.60465544    \\
$\hsp[3]\sigma _{\tilde{y}}$    &    0.53494302228   &     0.53494313648   &    0.40538024505  &    0.61719013907    & &     0.40636026 &    0.47407731    \\
$\hsp[3]\sigma _{\pi}$          &    0.80773582307   &     0.80773566453   &    0.79068291968  &    0.81180107613    & &     0.80917055 &    0.81302356    \\
$\hsp[3]\sigma _{y^{\ast }}$    &    0.51191069710   &     0.51191040251   &    0.61768886637  &    0.41897679640    & &     0.57474708 &    0.53075566    \\
\cmidrule(lr){1-8}
{Log-likelihood}                & -531.87471383407   &  -531.87471383414   & -531.45144619596  & -531.05106629477    & &  -536.98033619 & -535.95961197    \\
\bottomrule
\end{tabular*}%
\tabnotes[-3mm][.99\columnwidth][-1.0mm]{This table reports replication
results for the Stage 1 model parameter vector $\boldsymbol{\theta }_{1}$ of
\cite{holston.etal:2017}. The table is split in two blocks. The left block
(under the heading HLW Prior) reports estimation results of the Stage 1
model using the initialisation of \cite{holston.etal:2017} for the state
vector $\boldsymbol{\xi }_{t}$, where $\boldsymbol{\xi
}_{00}=[806.45,805.29,804.12]$ and $\mathbf{P}_{00}$ as defined in
\ref{eq:P00S1c}. The right block (under the heading Diffuse Prior) uses a
diffuse prior for $\boldsymbol{\xi }_{t}$ with
$\mathbf{P}_{00}=10^{6}\times\mathbf{I}_{3}$, where $\mathbf{I}_{3}$ is a 3
dimensional identity matrix. In the left block, 4 sets of results are
reported. The first column (HLW.R-File) reports estimates obtained by
running \cites{holston.etal:2017} R-Code for the Stage 1 model. The second
column ($b_{y}\geq 0.025$) shows estimation results using
\cites{holston.etal:2017} initial values for parameter vector
$\boldsymbol{\theta }_{1}$ in the optimisation routine, together with the
lower bound restriction $b_{y}\geq 0.025$. \fnref{FN:initVals} describes how
these initial values were found. The third column (Alt.Init.Vals) shows
estimates when alternative initial values for $\boldsymbol{\theta }_{1}$ are
used, with the $b_{y}\geq 0.025$ restriction still in place. The fourth
column ($b_{y}$ Free)\ reports estimates when the restriction on $b_{y}$ is
removed. The right column block displays estimates of $\boldsymbol{\theta
}_{1}$ with and without the restriction on $b_{y}$ being imposed, but with a
diffuse prior on the state vector. The last row (Log-likelihood) reports the
value of the log-likelihood function at these parameter estimates.
The Matlab file \texttt{Stage1\_replication.m} computes these results.}\label%
{tab:Stage1}\ET

\clearpage\newpage 

\BT[p!]%
\caption{Stage 1 MUE results of $\lambda_g$ for various
$\skew{0}\boldsymbol{\hat{\theta}}_{1}$ and structural break tests} %
\centering\vspace*{-2mm} \renewcommand{\arraystretch}{1.1}%
\renewcommand\tabcolsep{7pt} \fontsize{11pt}{13pt}\selectfont
\newcolumntype{N}{S[table-format =
2.8,round-precision = 7]}
\begin{tabular*}{1\columnwidth}{p{22mm}NNNNp{0mm}NS}
\topline
\multirow{2}{*}{\hsp[2]${\lambda}_{g}$}
&  \multicolumn{4}{c}{HLW Prior}
&& \multicolumn{2}{c}{Diffuse Prior} \\ \cmidrule(lr){2-5} \cmidrule(lr){7-8}
              & {HLW.R-File}         & {$b_{y}\geq 0.025$}   & {Alt.Init.Vals}    & {$b_y$ Free} & & {$b_{y}\geq 0.025$} & {$b_y$ free}  \\
\midrule
{$L$}         &  {---}              & 0.073287980348           &   0.094199118618  &   0.032862425776            & & 0.047520315055   &   0               \\
{MW}          &  0.06518061332957   & 0.065180695002           &   0.089453277795  &   0.031465385660            & & 0.041827386386   &   0               \\
{EW}          &  0.05386903777145   & 0.053869107878           &   0.080675754152  &   0.025383515899            & & 0.042378997103   &   0               \\
{QLR}         &  0.04938177758158   & 0.049381833434           &   0.079201458499  &   0.019428919676            & & 0.041187690190   &   0               \\
\bottomrule
\end{tabular*}\label{tab:Stage1_lambda_g}
\tabnotes[-3mm][.99\columnwidth][-1.0mm]{This table reports Stage 1
estimates of the ratio $\lambda _{g}=\sigma_{g}/\sigma _{y^{\ast }}$ which
is equal to \cites{stock.watson:1998} MUE $\lambda/T$ for the various estimates
of $\boldsymbol{{\theta}}_{1}$ reported in \autoref{tab:Stage1} and the four
different structural break tests. The table is split into left and right
column blocks as in \autoref{tab:Stage1}. Under the heading HLW.R-File,
estimates of $\lambda _{g}$ obtained from running \cites{holston.etal:2017}
R-Code are reported for reference. These are computed for the MW, EW and QLR
structural break tests only. The remaining columns report the replicated
$\lambda _{g}$ from the various $\boldsymbol{{\theta}}_{1}$ estimates from
\autoref{tab:Stage1}. } \ET

\cleardoublepage\newpage 

\BT[p!]%
\caption{Stage 1 MUE results of $\lambda_g$ after AR(1) filtering $\Delta
\hat{y}^\ast_{t|T}$ as in \cite{stock.watson:1998}}
\centering\vspace*{-2mm} \renewcommand{%
\arraystretch}{1.1}\renewcommand\tabcolsep{7pt}\fontsize{11pt}{13pt}\selectfont
\newcolumntype{K}{S[table-format = 4.5,round-precision = 4]} %
\newcolumntype{L}{S[table-format = 2.7,round-precision = 4]} %
\newcolumntype{U}{S[table-format = 1.2,round-precision = 2]} %
\newcolumntype{N}{S[table-format = 3.6,round-precision = 4]}
\begin{tabular*}{.85\columnwidth}{p{25mm}NLKp{10mm}Up{0mm}K}
\topline
 Test& \multicolumn{1}{c}{Statistic} & \multicolumn{1}{c}{\hsp[-4]$p-$value} & \multicolumn{1}{c}{\hsp[2]$\lambda$} & \multicolumn{2}{c}{90\% CI} & &\multicolumn{1}{c}{\hsp[4]$\lambda_g=\frac{\lambda}{T}$}\\
\midrule
    $L$   &  2.28146146 & 0.00500000 & 20.38331071 & [4.36, & {\hsp[-4]80.00]} & & 0.09059249  \\
    MW    & 15.35436143 & 0.00500000 & 20.58395335 & [4.47, & {\hsp[-4]80.00]} & & 0.09148424  \\
    EW    &  8.45806593 & 0.00500000 & 15.99034241 & [3.53, & {\hsp[-4]52.81]} & & 0.07106819  \\
    QLR   & 20.75957973 & 0.00500000 & 14.81635270 & [3.14, & {\hsp[-4]48.48]} & & 0.06585046  \\[1mm]
\bottomrule
\end{tabular*}
\tabnotes[-3mm][.84\columnwidth][-2.0mm]{This table reports
\cites{stock.watson:1998} MUE estimation results after the constructed
$\Delta \hat{y}^\ast_{t|T}$ variable was AR(1) filtered to remove the serial
correlation. The first two columns report the 4 different structural break
test statistics together with the corresponding $p-$values, followed by the
implied MUE estimates of $\lambda $ with 90\% CIs in square brackets. The
last column lists \cites{holston.etal:2017} $\lambda_g=\frac{\lambda}{T}$
to facilitate the comparison to the results listed under column one in
\autoref{tab:Stage1_lambda_g}.}%
\label{tab:stage1_MUE_AR1}\ET

\cleardoublepage\newpage 

\BT[p!]%
\caption{MUE estimates of the transformed Stage 1 model using an
AR(4) model for $u_t$} \centering\vspace*{-2mm}\renewcommand{%
\arraystretch}{1.1} \renewcommand\tabcolsep{7pt}\fontsize{11pt}{13pt}%
\selectfont\newcolumntype{K}{S[table-format = 3.6,round-precision = 6]} %
\newcolumntype{L}{S[table-format = 2.7,round-precision = 4]} %
\newcolumntype{U}{S[table-format = 1.2,round-precision = 2]} %
\newcolumntype{N}{S[table-format = 3.6,round-precision = 4]}
\begin{tabular*}{1\columnwidth}{p{25mm}NLKp{10mm}Up{1mm}Kp{8mm}U}
\topline
 Test& \multicolumn{1}{c}{Statistic}     &
 \multicolumn{1}{c}{\hsp[-4]$p-$value}     &
 \multicolumn{1}{c}{\hsp[4]$\lambda$}     &
 \multicolumn{2}{c}{90\% CI} & &
 \multicolumn{1}{c}{\hsp[4]$\sigma_{g}$} &
 \multicolumn{2}{c}{90\% CI} \\
\midrule
    $L$   & 0.316151035998  & 0.120000000000  & 5.914618545073 & [0, & \llap{23.95]} & & 0.154213440892  & [0, & \llap{0.62]} \\
    MW    & 1.787457322584  & 0.145000000000  & 5.650430931635 & [0, & \llap{23.88]} & & 0.147325206157  & [0, & \llap{0.62]} \\
    EW    & 1.066331073356  & 0.180000000000  & 4.883718810124 & [0, & \llap{20.97]} & & 0.127334514698  & [0, & \llap{0.54]} \\
    QLR   & 4.602919591679  & 0.285000000000  & 3.511961413462 & [0, & \llap{17.65]} & & 0.091568314968  & [0, & \llap{0.46]} \\[1mm]
\end{tabular*}\renewcommand{\arraystretch}{1.1}\renewcommand\tabcolsep{7pt} %
\fontsize{11pt}{13pt}\selectfont%
\newcolumntype{N}{S[table-format =
4.10,round-precision = 8]}
\begin{tabular*}{1\columnwidth}{p{38mm}NNNN}
\topline
Parameter                   & \multicolumn{1}{c}{MPLE} & \multicolumn{1}{c}{MMLE} & \multicolumn{1}{c}{$\mathrm{MUE}(\lambda_{\mathrm{EW}})$}  & \multicolumn{1}{c}{$\mathrm{MUE}(\lambda_{\mathrm{EW}}^{\mathrm{Up}})$}         \\
\midrule
    $\sigma_{g}$            &      0                              &    0.10621860661               &     0.12733451470               &    0.54678210579 \\
    $\sigma_{\varepsilon}$  &      2.99782489961                  &    2.98030098731               &     2.97346405372               &    2.90800214575 \\
    AR(1)                   &      0.28603147365                  &    0.27433173122               &     0.26988229275               &    0.24291125758 \\
    AR(2)                   &      0.16828174224                  &    0.16079307466               &     0.15789805142               &    0.14866123650 \\
    AR(3)                   &     -0.02046076069                  &   -0.02734561640               &    -0.02996690605               &   -0.03106235496 \\
    AR(4)                   &      0.06570210187                  &    0.05750551407               &     0.05423838150               &    0.06119692006 \\
    $g_{00}$                &      3.02198580916                  &    {---}                       &     4.09740641791               &    5.17204700003 \\
\cmidrule(ll){1-5}
    Log-likelihood          &   -566.39181042995                  & -573.64230971420               & -566.57435245187                & -570.81021839246 \\
\bottomrule
\end{tabular*}
\tabnotes[-3mm][.99\columnwidth][-1.0mm]{This table reports MUE estimation
results of the transformed (expressed in local level model form) Stage 1
model, using an AR(4) process for $u_{t}$. The top part of the table shows
the 4 different structural break test statistics together with their
$p-$values in the first two columns, followed by the corresponding MUE
estimates of $\lambda $ with 90\% CIs in square brackets. The last two
columns show the implied $\sigma _{g}$ estimate computed from
$T^{-1}{\lambda}\times {\sigma}_{\varepsilon }/{a}(1)$ and 90\%
CIs in square brackets. The first two columns of the bottom part of the
table report results from Maximum Likelihood based estimation, where MPLE
estimates the initial value of the state vector $g_{00}$, while MMLE uses a
diffuse prior for the initial value of the state vector with mean zero and
the variance set to $10^{6}$. Columns under the heading
MUE(${\lambda}_{\mathrm{EW}}$) and
MUE(${\lambda}_{\mathrm{EW}}^{\mathrm{UP}}$) show Median Unbiased
Estimates when $\sigma _{g}$ is held fixed at its MUE point estimate and
upper $90\%$ CI, respectively, from the EW structural break test. The row
Log-likelihood displays the value of the log-likelihood at the reported
parameter estimates. The Matlab file
\texttt{Stage1\_local\_level\_model\_SW98\_MUE\_Clark\_UC.m} replicates these results.}%
\label{tab:MUE_S1}\ET 

\cleardoublepage\newpage 

\BT[p!]\caption{Parameter estimates of \cites{clark:1987} UC model}%
\centering\vspace*{-2mm}\renewcommand{\arraystretch}{1.1}\renewcommand%
\tabcolsep{7pt}\fontsize{11pt}{13pt}\selectfont%
\newcolumntype{N}{S[table-format =
4.8,round-precision = 8]}
\newcolumntype{K}{S[table-format =
4.10,round-precision = 8]}
\begin{tabular*}{1\columnwidth}{p{38mm}NNp{7mm}NK}
\topline
{Parameter}                               & {\hsp[4]Clark's UC0}  & {\hsp[5]Std.error} & &    {\hsp[5]Clark's UC}   &  {\hsp[2]Std.error} \\
\midrule
$a_{y,1}$                                 &    1.6688617339     &  0.1094874136 & &    1.2954481785  &   0.2353595457       \\
$a_{y,2}$                                 &   -0.7242805140     &  0.1124274886 & &   -0.5674869068  &   0.2168834975       \\
$\sigma _{y^{\ast}}$                      &    0.5898417486     &  0.0584209144 & &    1.1575382576  &   0.2250901380       \\
$\sigma _{g}$                             &    0.0463214922     &  0.0227693495 & &    0.0321901826  &   0.0222178828       \\
$\sigma _{\tilde{y}}$                     &    0.3462603749     &  0.0972702766 & &    0.8095072197  &   0.3646114322       \\
$\mathrm{Corr}(\mathring{\varepsilon}_{t}^{\tilde{y}},\varepsilon _{t}^{y^{\ast }})$ & 0 &   {~~~~---}     & &   -0.9426313454  &   0.0971454061       \\
\cmidrule(lr){1-6}
{Log-likelihood}                          & -270.0007183929     &     {~~~~---}     & & -269.8750406078  &        {~~~~---}           \\
\bottomrule
\end{tabular*}
\tabnotes[-3mm][.99\columnwidth][-1.0mm]{This table reports parameter
estimates of \cites{clark:1987} UC model. Two sets of results are reported.
In the left part of \autoref{tab:clarkddUC}, parameter estimates and
standard errors (Std.errors) from Clark's UC0\ model which assumes
$\mathrm{Corr}(\mathring{\varepsilon}_{t}^{\tilde{y}},\varepsilon
_{t}^{y^{\ast }})=0$ are reported. In the right part, parameter estimates
and standard errors for Clark's correlated UC\ model are shown, where
$\mathrm{Corr}(\mathring{\varepsilon}_{t}^{\tilde{y}},\varepsilon
_{t}^{y^{\ast }})$ is explicitly estimated. Standard errors are computed
from the inverse of the Hessian matrix of the log-likelihood. I use a
diffuse prior for the $I(1)$ part of the state vector, with the variance set
to $10^{6}$. The stationary part of the state vector is initialized at its
unconditional mean and variance. I do not estimate the initial value of the
state vector. This is analogous to MMLE in \cite{stock.watson:1998}. The
Matlab file \texttt{Stage1\_local\_level\_model\_SW98\_MUE\_Clark\_UC.m}
replicates
these results.} \label{tab:clarkddUC}\ET 

\cleardoublepage\newpage 

\begin{figure}[p!]
\centering
\includegraphics[width=1.01\textwidth,trim={0 00mm 0
50mm},clip]{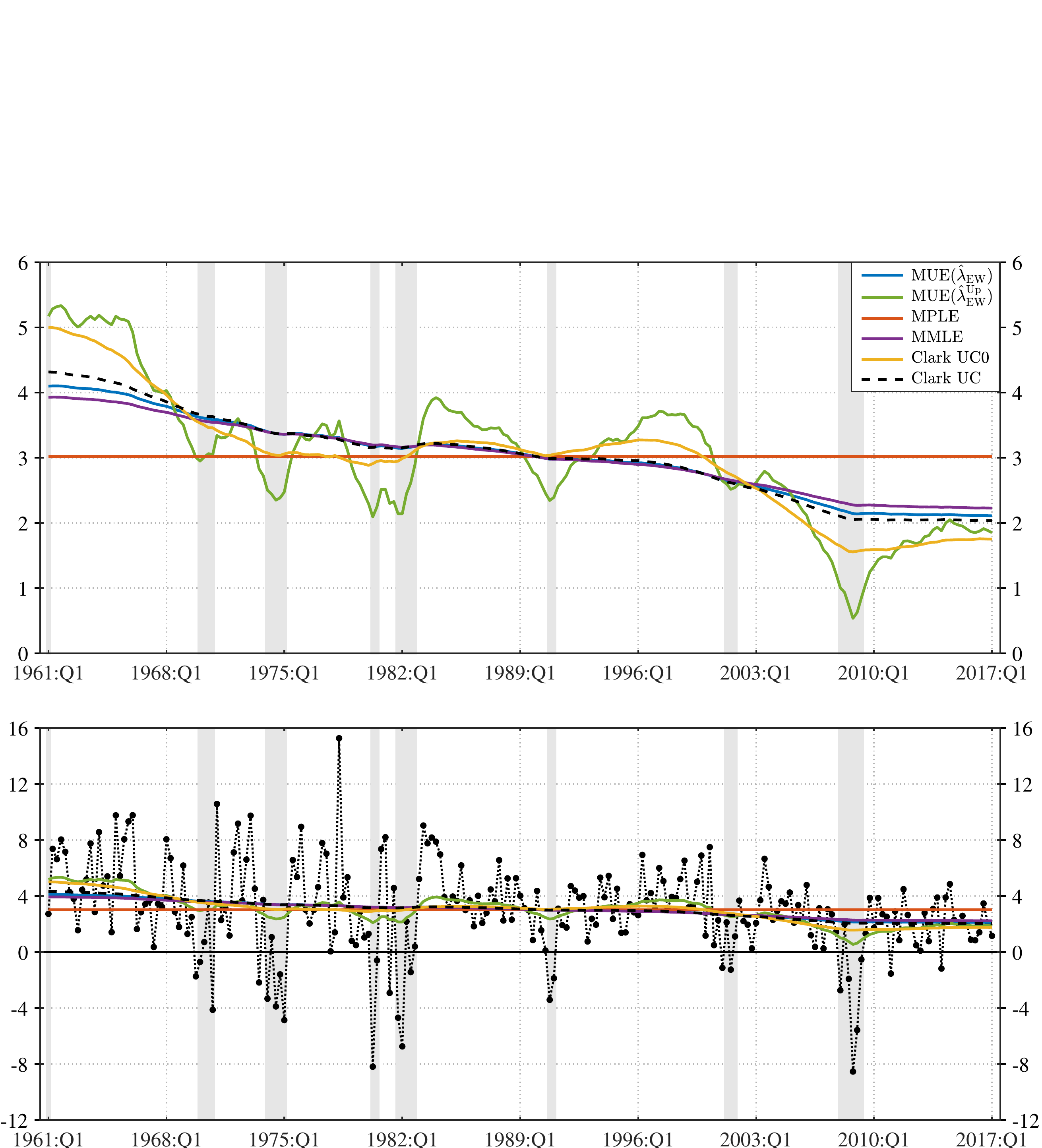}
\caption{Smoothed trend growth estimates from the modified Stage 1 model.}
\label{fig:MUE_S1}
\end{figure}

\cleardoublepage\newpage 

\BT[p!]
\caption{Stage 2 parameter estimates}\centering\vspace*{-2mm}%
\renewcommand{\arraystretch}{1.1}\renewcommand\tabcolsep{7pt}%
\fontsize{11pt}{13pt}\selectfont%
\newcolumntype{N}{S[table-format = 5.8,round-precision = 7]}
\newcolumntype{U}{S[table-format = 5.7,round-precision = 7]}
\newcolumntype{L}{S[table-format = 4.7,round-precision = 7]}
\begin{tabular*}{.93\columnwidth}{p{38mm}NNNN}
\topline
\hsp[5]$\boldsymbol{\theta }_{2}$
& {\hsp[0]HLW.R-File\hsp[-4.0]}
& {\hsp[6]Replicated\hsp[1.5]}
& {\hsp[4]MLE($\sigma_g$) \hsp[-3]}
& {\hsp[3]MLE($\sigma_g$).$\mathcal{M}_0$\hsp[-2]}\\
\midrule
$\hsp[3]a_{y,1}  $                  &     1.5139908988  &     1.5139908874   &     1.4735944910  &      1.4947610514         \\
$\hsp[3]a_{y,2}  $                  &    -0.5709338972  &    -0.5709338892   &    -0.5321668125  &     -0.5531450715         \\
$\hsp[3]a_{r}    $                  &    -0.0736646657  &    -0.0736646651   &    -0.0831539459  &     -0.0755562707         \\
$\hsp[3]a_{0}    $                  &    -0.2630693940  &    -0.2630693778   &    -0.2548597292  &       {~~~~~---}          \\
$\hsp[3]a_{g}    $                  &     0.6078665929  &     0.6078665690   &     0.6277123919  &       {~~~~~---}          \\
$\hsp[3]b_{\pi } $                  &     0.6627428265  &     0.6627428246   &     0.6655286128  &      0.6692918543         \\
$\hsp[3]b_{y}    $                  &     0.0844720258  &     0.0844720318   &     0.0819057776  &      0.0802934388         \\
$\hsp[3]\sigma _{\tilde{y}}$        &     0.3582701455  &     0.3582701554   &     0.3636497583  &      0.3742315512         \\
$\hsp[3]\sigma _{\pi }     $        &     0.7872279651  &     0.7872279652   &     0.7881905740  &      0.7895136932         \\
$\hsp[3]\sigma _{y^{\ast }}$        &     0.5665698145  &     0.5665698109   &     0.5534536560  &      0.5526272640         \\
$\hsp[3]\sigma _{g}$ {(implied)}    &    (0.0305205)    &    (0.0305205)     &     0.0437060828  &      0.0448689280         \\
$\hsp[3]\lambda_g  $ {(implied)}    &     0.0538690378  &     0.0538690378   &    (0.0789697)    &     (0.0811920)           \\  \cmidrule(lr){1-5}
{Log-likelihood}                    &  -513.5709576473  &  -513.5709576473   &  -513.2849624670  &   -514.1458025902         \\
\bottomrule
\end{tabular*}%
\tabnotes[-3mm][.925\columnwidth][-1.0mm]{This table reports replication
results for the Stage 2 model parameter vector $\boldsymbol{\theta }_{2}$ of
\cite{holston.etal:2017}. The first column (HLW.R-File) reports estimates
obtained by running \cites{holston.etal:2017} R-Code for the Stage 2 model.
The second column
(Replicated) shows the replicated results using the same set-up as in %
\cites{holston.etal:2017}. The third column (MLE($\sigma _{g}$)) reports
estimates when $\sigma _{g}$ is freely estimated by MLE together with the
other parameters of the Stage 2 model, rather than imposing the ratio $%
\lambda _{g}=\sigma _{g}/\sigma _{y^{\ast }}=0.0538690378$ obtained from
Stage 1. The last column (MLE($\sigma _{g}$)$.\mathcal{M}_{0}$) provides
estimates of the \emph{"correctly specified"} Stage 2 model in
\ref{S2full0}, with $\sigma _{g} $ again estimated directly by MLE. Values
in round brackets give the implied
$\sigma _{g}$ or $\lambda _{g}$ values when either $\lambda _{g}$ is fixed or when $%
\sigma _{g}$ is estimated. The last row (Log-likelihood) reports the value
of the log-likelihood function at these parameter estimates. The Matlab file
\texttt{Stage2\_replication.m} replicates these results.}\label%
{tab:Stage2} \ET

\cleardoublepage\newpage 

\begin{figure}[p!]
\centering
\includegraphics[width=1\textwidth,trim={0 0 0 0},clip]{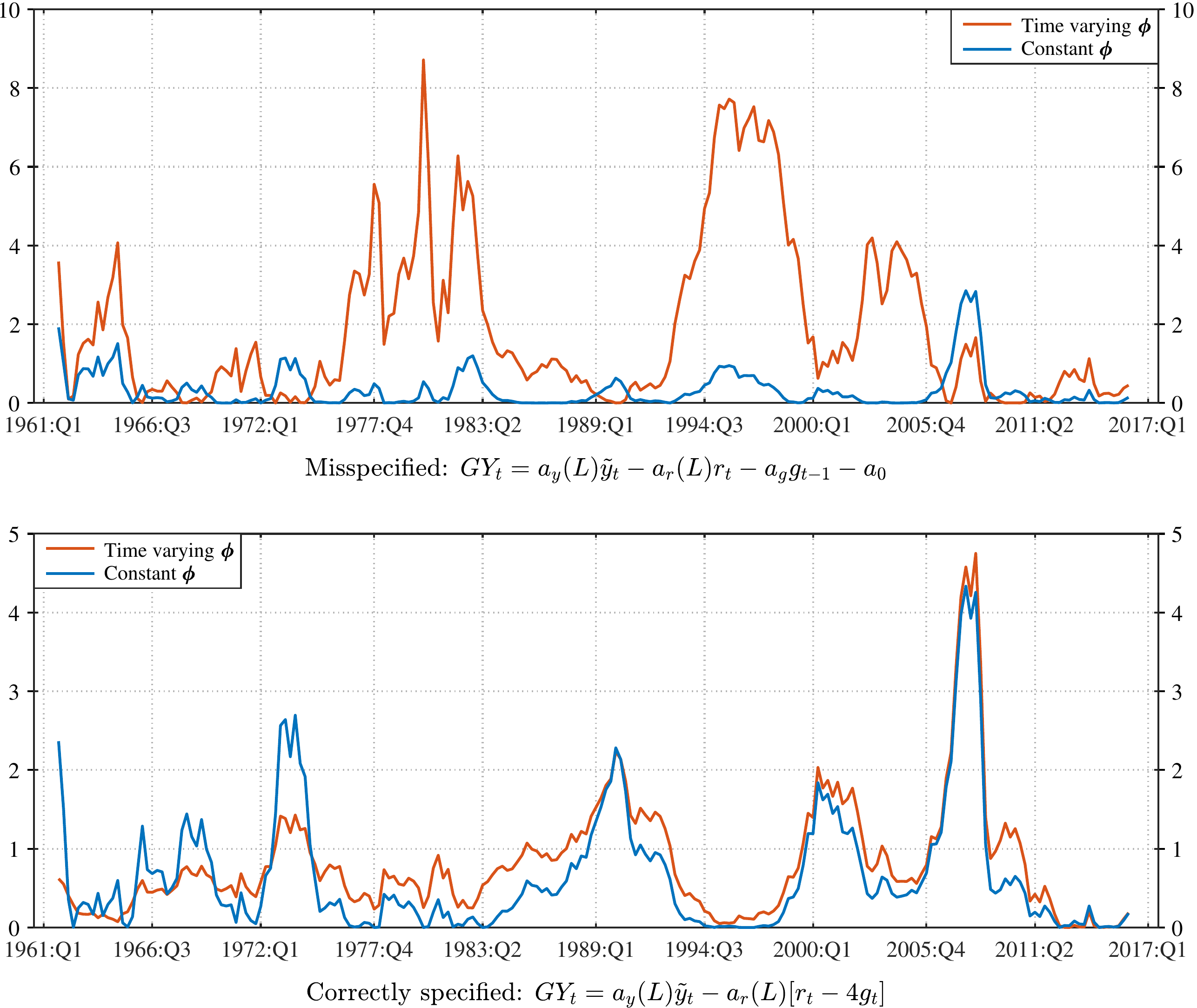}\vspace{-0mm}
\caption{Sequence of $\{F(\tau )\}_{\tau =\tau _{0}}^{\tau _{1}}$ statistics on the dummy
variable coefficients $\{\hat{\zeta}_{1}(\tau )\}_{\tau =\tau _{0}}^{\tau
_{1}}$ used in the construction of the structural break test statistics.}
\label{fig:seqaF}
\end{figure}

\cleardoublepage\newpage 

\BST[p!]%
\BAW[5]%
\caption{Stage 2 MUE results of $\lambda_z$ with corresponding 90\%
confidence intervals, structural break test statistics and $p-$values} %
\centering\vspace*{-2mm}\renewcommand{%
\arraystretch}{1.15}\renewcommand\tabcolsep{7pt} %
\fontsize{10pt}{12pt}\selectfont %
\newcolumntype{N}{S[table-format = 1.6,round-precision = 6]} %
\newcolumntype{K}{S[table-format = 1.5,round-precision = 6]} %
\newcolumntype{Q}{S[table-format = 1.4,round-precision = 6]} %
\begin{tabular*}{.94\columnwidth}{p{7mm}NNNNNNNp{0mm}NNNNNN}
\topline
\multirow{2}{*}{\hsp[2]$\lambda_{z}$}
&     \multicolumn{7}{c}{\emph{`Time varying} $\boldsymbol{\phi }$\textit{'}}
&&    \multicolumn{6}{c}{\emph{`Constant} $\boldsymbol{\phi }$\textit{'}} \\ \cmidrule(rr){2-8} \cmidrule(rr){10-15}

& {HLW.R-File} & {\hsp[-2] Replicated \hsp[-2]} & {[90\% CI]} & {\hsp[-2]
MLE($\sigma_g$) \hsp[-2]} & {[90\% CI]} & {\hsp[-2]
MLE($\sigma_g$).$\mathcal{M}_0$ \hsp[-2]} & {[90\% CI]} & {} & {\hsp[-2]
Replicated \hsp[-2]} & {[90\% CI]} & {\hsp[-2] MLE($\sigma_g$) \hsp[-2]} &
{[90\% CI]} & {\hsp[-2] MLE($\sigma_g$).$\mathcal{M}_0$ \hsp[-2]} & {[90\%
CI]}
\\
\midrule
        $L$  & {---}                & 0          & {[0, 0.02]}  & 0           & {[0, 0.00]}  & 0          & {[0, 0.05]}  &&  0          & {[0, 0.02]}   & 0           & {[0, 0.00]} &  0           & {[0, 0.05]} \\
        MW   & 0.0249690419675479   & 0.02496905 & {[0, 0.11]}  & 0.03299656  & {[0, 0.14]}  & 0.00892013 & {[0, 0.07]}  &&  0          & {[0, 0.03]}   & 0           & {[0, 0.02]} &  0           & {[0, 0.06]} \\
        EW   & 0.0302172209051429   & 0.03021723 & {[0, 0.11]}  & 0.03379822  & {[0, 0.12]}  & 0.00779609 & {[0, 0.06]}  &&  0          & {[0, 0.03]}   & 0           & {[0, 0.02]} &  0.00075430  & {[0, 0.06]} \\
        QLR  & 0.0342646997381782   & 0.03426471 & {[0, 0.12]}  & 0.03894233  & {[0, 0.14]}  & 0.01719852 & {[0, 0.08]}  &&  0          & {[0, 0.05]}   & 0           & {[0, 0.04]} &  0.01470321  & {[0, 0.07]} \\
\midrule
\multirow{1}{*}{} & \multicolumn{14}{c}{Structural break test statistics ($p-$values in parenthesis)\hsp[0]}     \\
\midrule
        $L$  & {---}                & 0.05085088 & {(0.8700)}   &  0.03808119 & {(0.9400)}   & 0.10830144 & {(0.5450)}   &&  0.05085088 & {(0.8700)}   & 0.03808119   & {(0.9400)}   &  0.10830144  & {(0.5450)} \\
        MW   & 1.8705612948815300   & 1.87056176 & {(0.1300)}   &  2.68314033 & {(0.0600)}   & 0.80746921 & {(0.4300)}   &&  0.33010788 & {(0.8100)}   & 0.25322628   & {(0.8900)}   &  0.65167084  & {(0.5250)} \\
        EW   & 1.6930140233544000   & 1.69301457 & {(0.0800)}   &  2.12052740 & {(0.0450)}   & 0.50616478 & {(0.4300)}   &&  0.20293551 & {(0.7850)}   & 0.14831394   & {(0.8750)}   &  0.43448586  & {(0.4900)} \\
        QLR  & 8.7144611146321900   & 8.71446298 & {(0.0450)}   & 10.30303534 & {(0.0250)}   & 4.75129241 & {(0.2700)}   &&  2.85143418 & {(0.5700)}   & 2.20271913   & {(0.7150)}   &  4.33470147  & {(0.3150)} \\
\bottomrule
\end{tabular*}\label{tab:Stage2_lambda_z}
\tabnotes[-2.5mm][.935\columnwidth][-0.5mm]{This table reports the Stage 2
estimates of $\lambda _{z}$ for the different
$\boldsymbol{\theta }_{2}$ estimates corresponding to the \emph{%
"misspecified"} and \emph{"correctly specified"} Stage 2 models reported in %
\autoref{tab:Stage2}. The table is split into two column blocks, showing the
results for the \emph{`Time varying} $\boldsymbol{\phi }$\textit{'} and
\emph{`Constant} $\boldsymbol{\phi }$\textit{'} scenarios in the left and
right blocks, respectively. In the bottom half of the table, the four
different structural break test statistics for the considered models are
shown. The results under the heading `HLW.R-File' show the $\lambda _{z}$
estimates obtained from running \cites{holston.etal:2017} R-Code for the
Stage 2 model as reference values. The second column `Replicated' shows my
replicated results. Under the heading `MLE($\sigma _{g}$)', results for the
\emph{"misspecified}" Stage 2 model are shown with $\sigma _{g}$ estimated
directly by MLE rather
than from the first stage estimate of $\lambda _{g}$. Under the heading `MLE(%
$\sigma _{g}$).$\mathcal{M}_{0}$', results for the \emph{"correctly
specified"} Stage 2 model are reported where $\sigma _{g}$ is again
estimated by MLE. The values in square brackets in the top half of the table
report 90\% confidence intervals for $\lambda _{z}$ computed from %
\cites{stock.watson:1998} tabulated values provided in their GAUSS\ files.
These were divided by sample size $T$ to make them comparable to $\lambda
_{z}$. In the bottom panel, $p-$values of the various structural break tests
are reported in round brackets. These were also extracted from %
\cites{stock.watson:1998} GAUSS\ files.}%
\EAW \EST%

\cleardoublepage\newpage 


\BT[p!]%
\caption{Summary statistics of the $\lambda_z$ estimates obtained from applying
\cites{holston.etal:2017} Stage 2 MUE procedure to simulated data} %
\centering\vspace*{-2mm}\renewcommand{%
\arraystretch}{1.1}\renewcommand\tabcolsep{7pt} %
\fontsize{11pt}{13pt}\selectfont %
\newcolumntype{N}{S[table-format = 3.8,round-precision = 6]} %
\newcolumntype{K}{S[table-format = 5.9,round-precision = 6]} %
\newcolumntype{Q}{S[table-format = 1.0,round-precision = 6]} %
\begin{tabular*}{1\columnwidth}{p{48mm}NNp{3mm}NN}
\topline
\multirow{2}{*}{\hsp[1] Summary Statistics}
&     \multicolumn{2}{c}{DGPs when $\boldsymbol{\theta }_{2}$ held fixed at $\skew{0}\boldsymbol{\hat{\theta}}_{2}$~~~} &
&     \multicolumn{2}{c}{DGPs when $\boldsymbol{\theta}_{2}$ is re-estimated~~~}
\\ \cmidrule(rr){2-3} \cmidrule(rr){5-6}

& {$r_t^{\ast}=4g_{t}$}
& {$r_t^{\ast}=4g_{t}+z_{t}$} &
& {$r_t^{\ast}=4g_{t}$}
& {$r_t^{\ast}=4g_{t}+z_{t}$} \\ \midrule
Minimum                                 & 0          & 0          && 0           &  0           \\  
Maximum                                 & 0.10121984 & 0.09642681 && 0.11688594  &  0.11644479  \\  
Standard deviation                      & 0.01624549 & 0.01658156 && 0.01851192  &  0.01964657  \\  
Mean                                    & 0.02884249 & 0.03072566 && 0.02510333  &  0.02746184  \\  
Median                                  & 0.02839441 & 0.02960898 && 0.02221494  &  0.02511532  \\  
$\mathrm{Pr}(\hat\lambda^s_z> 0.030217)$& 0.4570     & 0.4900     && 0.3390      &  0.3930      \\  
\bottomrule                                                                                         
\end{tabular*}\label{tab:Stage2_lambda_z_o}                                                         
\tabnotes[-3mm][.994\columnwidth][-1.5mm]{This table reports summary statistics                     
of the $\lambda _{z}$ estimates that                                                                
one obtains from implementing \cites{holston.etal:2017} Stage 2 MUE                                 
procedure on artificial data that was simulated from two different data
generating processes (DGPs). The first DGP simulates data from the full
structural model in \ref{eq:hlw} under the parameter estimates of \cite%
{holston.etal:2017}, but where the natural rate is determined solely by
trend growth. That is, in the output gap equation in \ref{IS}, $r_{t}^{\ast
}=4g_{t}$. The second DGP simulates data from the full model of \cite%
{holston.etal:2017} where $r_{t}^{\ast }=4g_{t}+z_{t}$. The summary
statistics that are reported are the minimum, maximum, standard deviation,
mean, median, as well as the
empirical frequency of observing a value larger than the estimate of $%
0.030217$ obtained by \cite{holston.etal:2017}, denoted by $\Pr (\hat{\lambda%
}_{z}^{s}>0.030217)$. The table shows four different estimates, grouped in 2
block pairs. The left block under the heading `DGPs when $\boldsymbol{\theta
}_{2}$ is held fixed' shows the simulation results for the two DGPs when the
Stage 2 parameter vector $\boldsymbol{\theta }_{2}$ is held fixed at the
Stage 2 estimates and is not re-estimated on the simulated data. The right block under the heading
`DGPs when $\boldsymbol{\theta }_{2}$ is re-estimated' shows the simulation
results when $\boldsymbol{\theta }_{2}$ is re-estimated for each simulated
series. Simulations are performed on a sample size equivalent to the empirical data, with
$1000$ repetitions. }
\ET %

\cleardoublepage\newpage 

\begin{figure}[p!]
\centering
\subfigure[Stage 2 parameters held fixed at $\skew{0}\boldsymbol{\hat{\theta}}_{2}$ from column 1 of \autoref{tab:Stage2}]{\includegraphics[width=.75\textwidth]{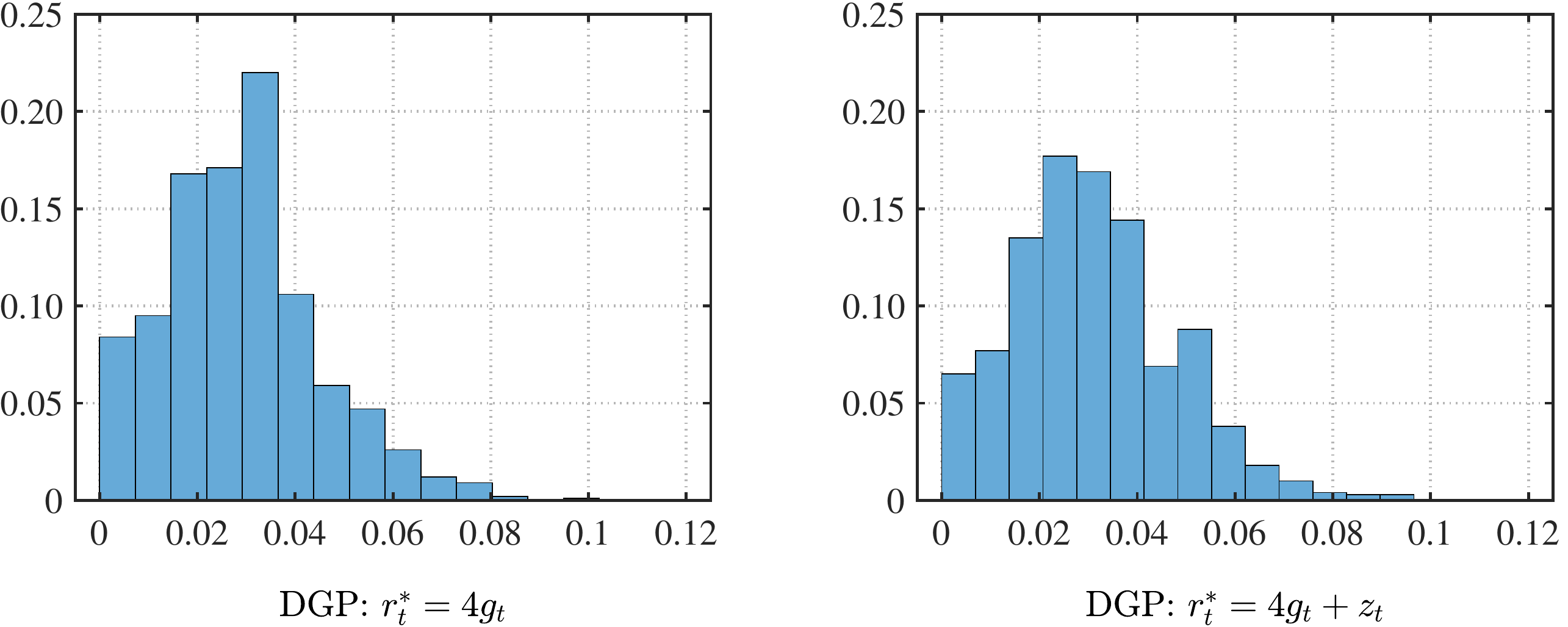}}\vspace{4mm}
\subfigure[Stage 2 parameters re-estimated on each simulated series]{\includegraphics[width=.75\textwidth]{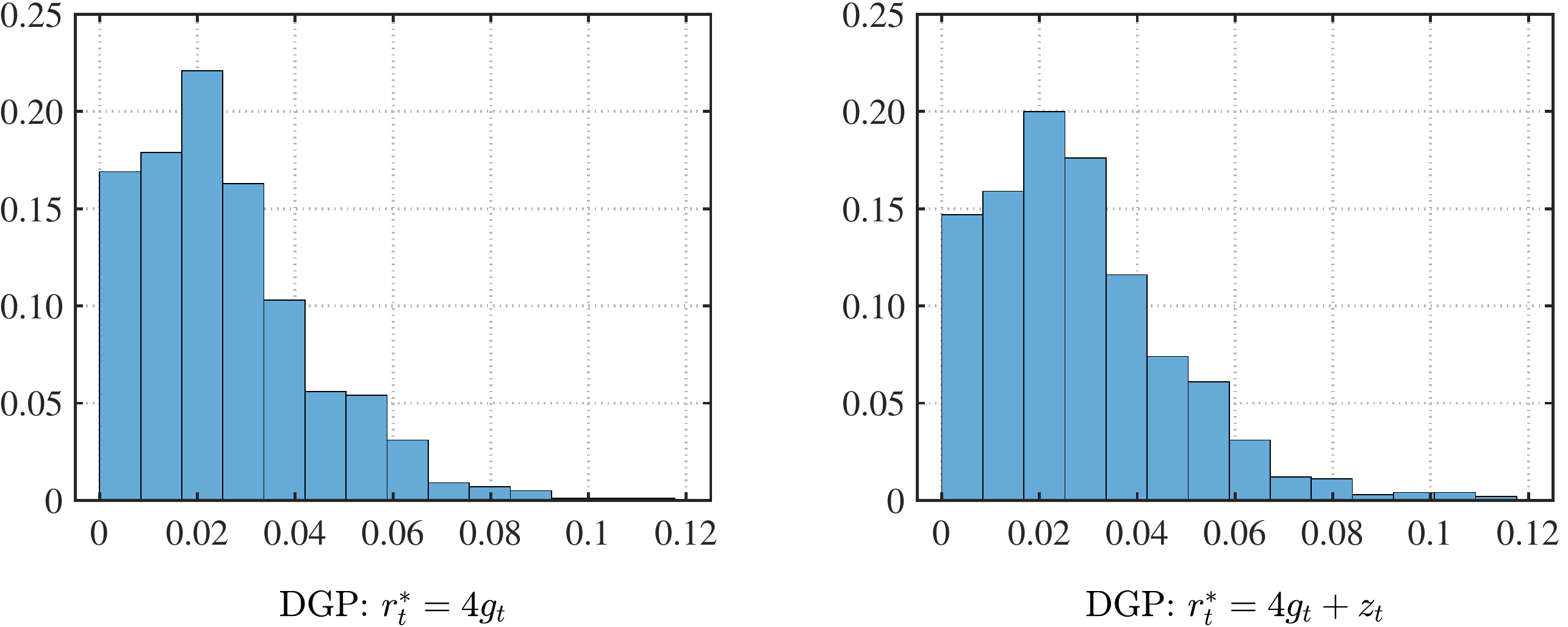}}
\caption{Histograms of the estimated $\left\{ \hat\lambda_z^s\right\}_{s=1}^{S}$
sequence corresponding to the summary statistics shown in \autoref{tab:Stage2_lambda_z_o}.
On the left and right columns, histograms for the two different DGPs are shown. To top two
histograms show the results when $\boldsymbol{\theta
}_{2}$ is held fixed in the simulations and is not re-estimated, while the bottom plots
show the results when $\boldsymbol{\theta}_{2}$ is re-estimated on each simulated series
that is generated.}
\label{fig:S2Lam_z_sim}
\end{figure}

\cleardoublepage\newpage 

\BT[p!]
\caption{Stage 3 parameter estimates}\centering\vspace*{-2mm}%
\renewcommand{\arraystretch}{1.1}\renewcommand\tabcolsep{7pt}%
\fontsize{11pt}{13pt}\selectfont%
\newcolumntype{N}{S[table-format = 4.8,round-precision = 8]}
\newcolumntype{U}{S[table-format = 4.8,round-precision = 8]}
\newcolumntype{L}{S[table-format = 4.8,round-precision = 8]}
\begin{tabular*}{1\columnwidth}{p{27mm}NNNNN}
\topline
\hsp[5]$\boldsymbol{\theta }_{3}$
& {\hsp[2]HLW.R-File\hsp[-4]}
& {\hsp[2]Replicated\hsp[-4]}
& {\hsp[2]MLE($\sigma_g|\lambda_z^{\mathrm{HLW}})$\hsp[-3]}
& {\hsp[3]MLE($\sigma_g|\lambda_z^{\mathcal{M}_0})$\hsp[-4]}
& {\hsp[2]MLE($\sigma_g,\sigma_z)$\hsp[-4]}\\
\midrule
$\hsp[3]a_{y,1}  $                  &     1.5295724886 &    1.5295724693 &    1.4944246197  &    1.4956671184 &    1.4956614728     \\
$\hsp[3]a_{y,2}  $                  &    -0.5875641518 &   -0.5875641351 &   -0.5537026759  &   -0.5544894229 &   -0.5544821188     \\
$\hsp[3]a_{r}    $                  &    -0.0711956862 &   -0.0711956881 &   -0.0794159824  &   -0.0752549575 &   -0.0752523989     \\
$\hsp[3]b_{\pi } $                  &     0.6682070533 &    0.6682070526 &    0.6712819687  &    0.6691946828 &    0.6691999284     \\
$\hsp[3]b_{y}    $                  &     0.0789577832 &    0.0789577841 &    0.0759360415  &    0.0805490066 &    0.0805471559     \\
$\hsp[3]\sigma _{\tilde{y}}$        &     0.3534684542 &    0.3534684662 &    0.3604311438  &    0.3738137616 &    0.3738293536     \\
$\hsp[3]\sigma _{\pi }     $        &     0.7891948659 &    0.7891948667 &    0.7902998169  &    0.7894892114 &    0.7894909384     \\
$\hsp[3]\sigma _{y^{\ast }}$        &     0.5724192458 &    0.5724192433 &    0.5591574254  &    0.5529381760 &    0.5529301787     \\
$\hsp[3]\sigma _{g}$ {(implied)}    &    (0.03083567)  &   (0.03083567)  &    0.0458385200  &    0.0449745020 &    0.0449741366     \\
$\hsp[3]\sigma _{z}$ {(implied)}    &    (0.15002080)  &   (0.15002080)  &   (0.13714150)   &   (0.00374682)  &    0.0000000051     \\
$\hsp[3]\lambda_g  $ {(implied)}    &     0.0538690379 &    0.0538690379 &   (0.08197784)   &   (0.08133730)  &   (0.08133782)      \\
$\hsp[3]\lambda_z  $ {(implied)}    &     0.0302172209 &    0.0302172209 &    0.0302172209  &    0.0007542990 &   (0.00000000)      \\  \cmidrule(ll){1-6}
{Log-likelihood}                    &  -515.1447052780 & -515.1447059855 & -514.8307054362  & -514.2898742606 & -514.2895896936     \\
\bottomrule
\end{tabular*}\label{tab:Stage3}%
\tabnotes[-3mm][.99\columnwidth][-1.0mm]{This table reports replication
results for the Stage 3 model parameter vector $\boldsymbol{\theta }_{3}$ of
\cite{holston.etal:2017}. The first column (HLW.R-File) reports estimates
obtained by running \cites{holston.etal:2017} R-Code for the Stage 3 model.
The second column (Replicated) shows the replicated results using the same set-up as in %
\cites{holston.etal:2017}. The third column
(MLE($\sigma_g|\lambda_z^{\mathrm{HLW}})$) reports estimates when $\sigma
_{g}$ is directly estimated by MLE together with the other parameters of the
Stage 3 model, while $\lambda _{z}$ is held fixed at
$\lambda _{z}^{\mathrm{HLW}}=0.030217$ obtained from %
\cites{holston.etal:2017} \emph{"misspecified"} Stage 2 procedure. In the
forth column (MLE($\sigma_g|\lambda_z^{\mathcal{M}_0})$), $\sigma _{g}$ is
again estimated directly by MLE together with the other parameters of the
Stage 3 model, but with $\lambda _{z}$ now fixed at $\lambda
_{z}^{\mathcal{M}_{0}}=0.000754$ obtained from the \emph{"correctly
specified"} Stage 2 model in \ref{S2full0}. The last column
(MLE($\sigma_g,\sigma_g)$) shows estimates when all parameters are computed
by MLE. Values in round brackets give the implied $\{\sigma _{g}, \sigma
_{z}\}$ or $\{\lambda _{g},\lambda _{z}\}$ values when either is fixed or
estimated. The last row (Log-likelihood) reports the value of the
log-likelihood function at these parameter estimates. The Matlab file
\texttt{Stage3\_replication.m} replicates these results.} \ET

\cleardoublepage\newpage 

\begin{figure}[p!]
\centering 
\includegraphics[width=1\textwidth,trim={0 0 0 0},clip]{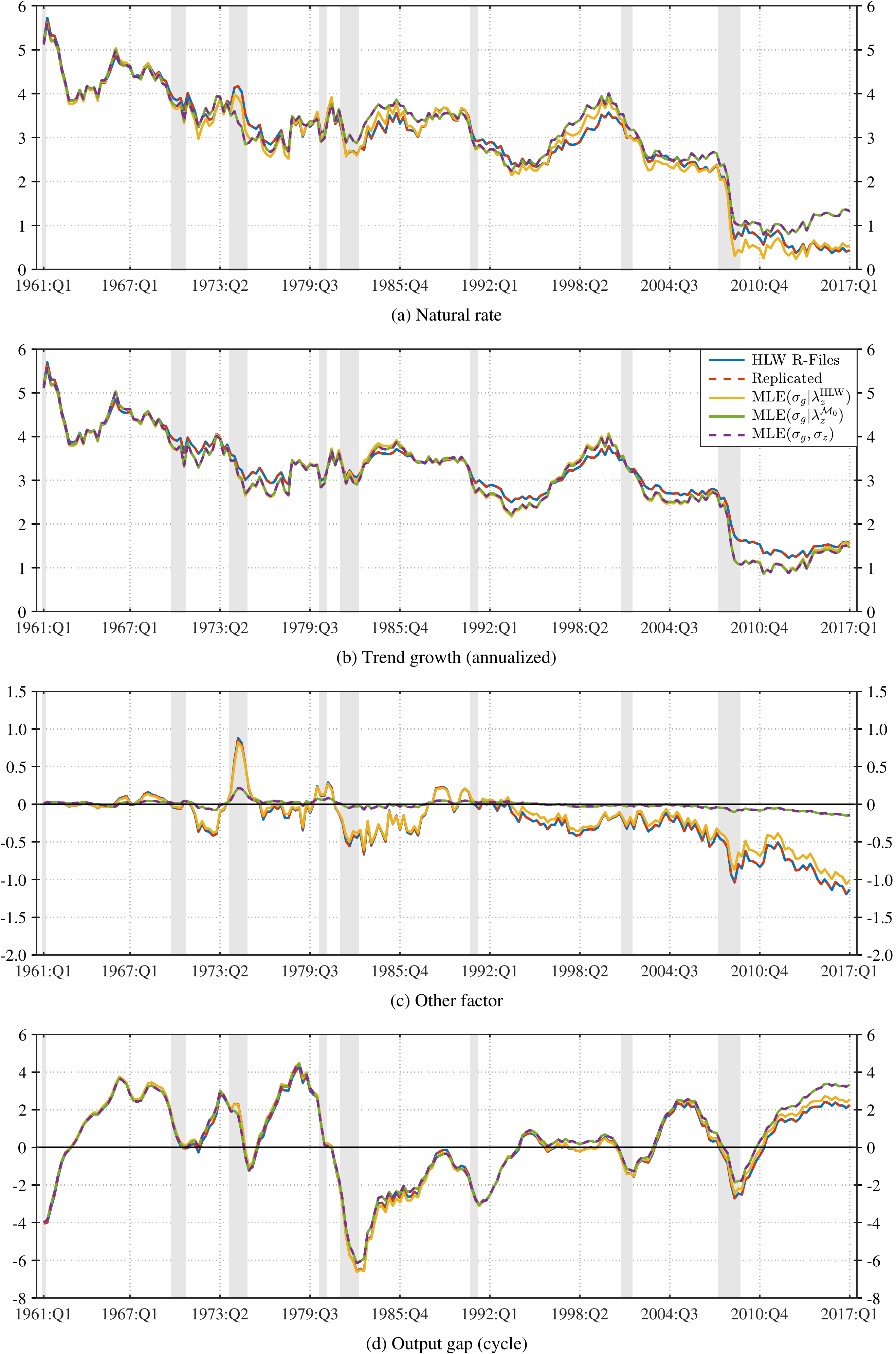}
\vspace{-3mm}
\caption{Filtered estimates of the natural rate $r^{\ast}_t$,
annualized trend growth $g_t$, \emph{`other factor'} $z_t$, and the output gap
(cycle) variable $\tilde{y}_t$.}
\label{fig:2017KF}
\end{figure}

\cleardoublepage\newpage 

\begin{figure}[p!]
\centering 
\includegraphics[width=1\textwidth,trim={0 0 0 0},clip]{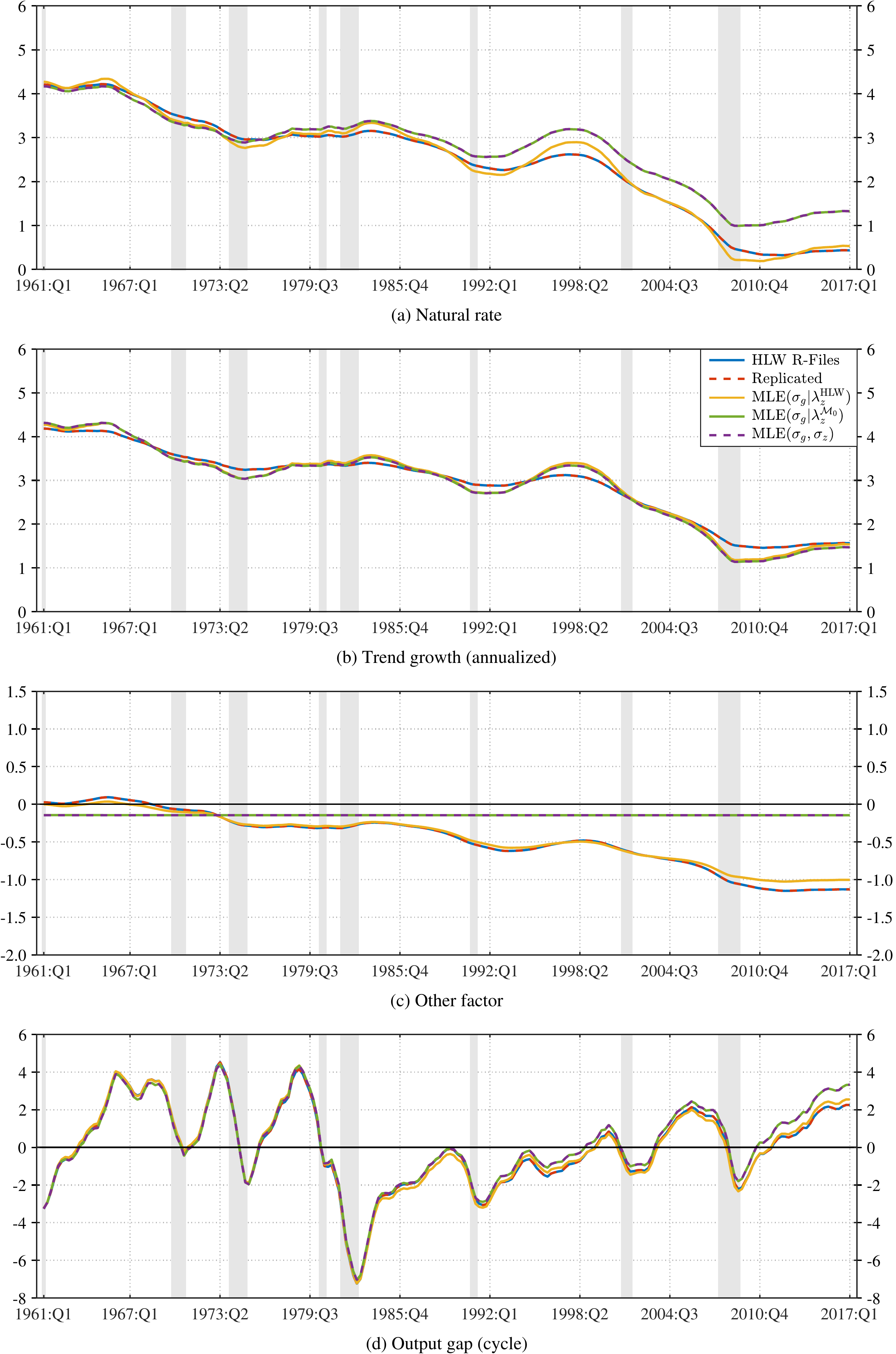}
\vspace{-3mm}
\caption{Smoothed estimates of the natural rate $r^{\ast}_t$,
annualized trend growth $g_t$, \emph{`other factor'} $z_t$, and the output gap
(cycle) variable $\tilde{y}_t$.}
\label{fig:2017KS}
\end{figure}

\cleardoublepage\newpage 

\begin{figure}[p!]
\centering \vspace{-3mm}
\includegraphics[width=1\textwidth,trim={0 0 0 0},clip]{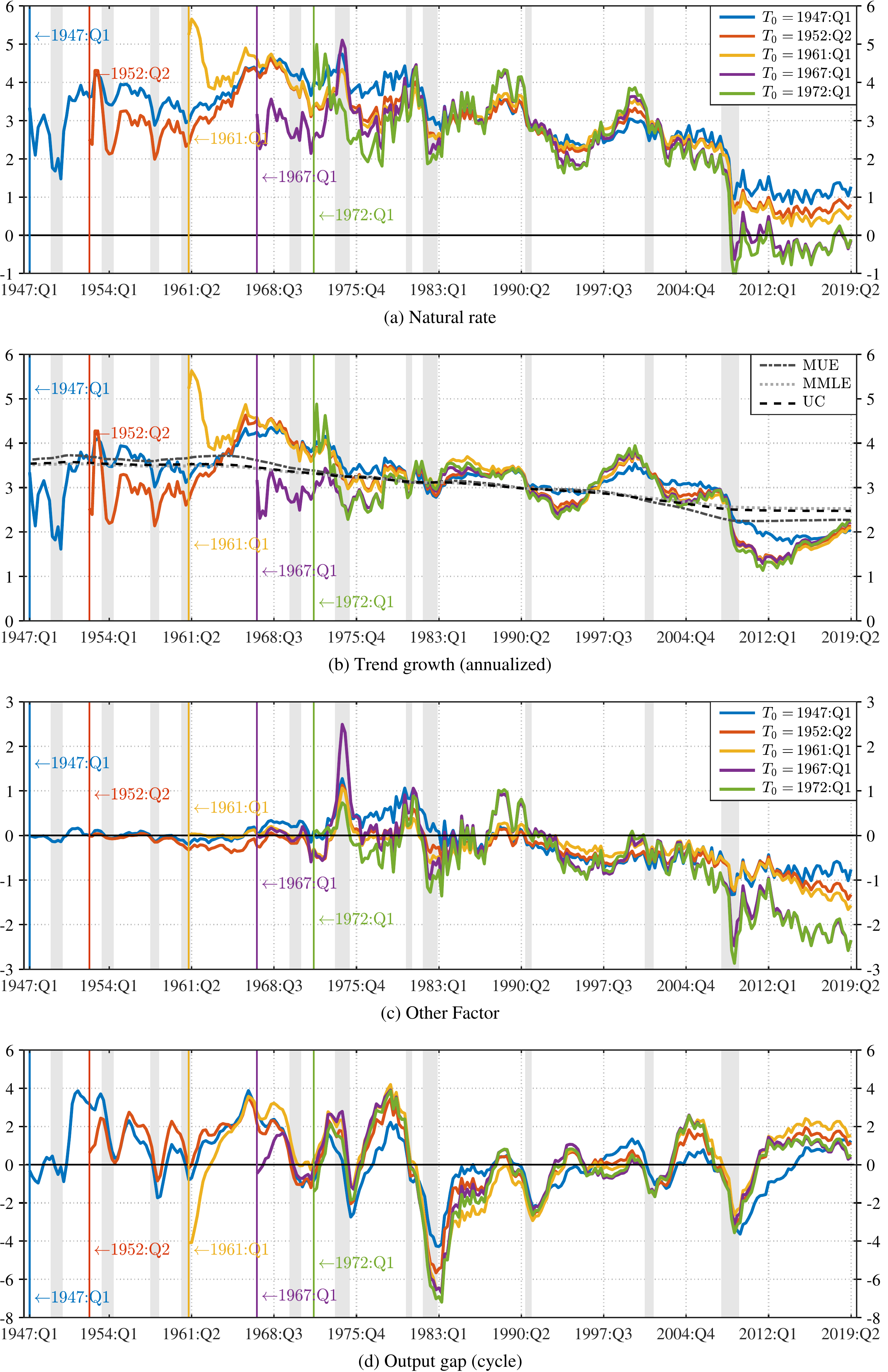}
\vspace{-3mm}
\caption{Filtered estimates of annualized trend growth $g_t$, \emph{`other factor'} $z_t$
and the natural rate $r^{\ast}_t$ based on different starting dates}
\label{fig:T0KF}
\end{figure}

\cleardoublepage\newpage 

\begin{figure}[p!]
\centering \vspace{-3mm}
\includegraphics[width=1\textwidth,trim={0 0 0 0},clip]{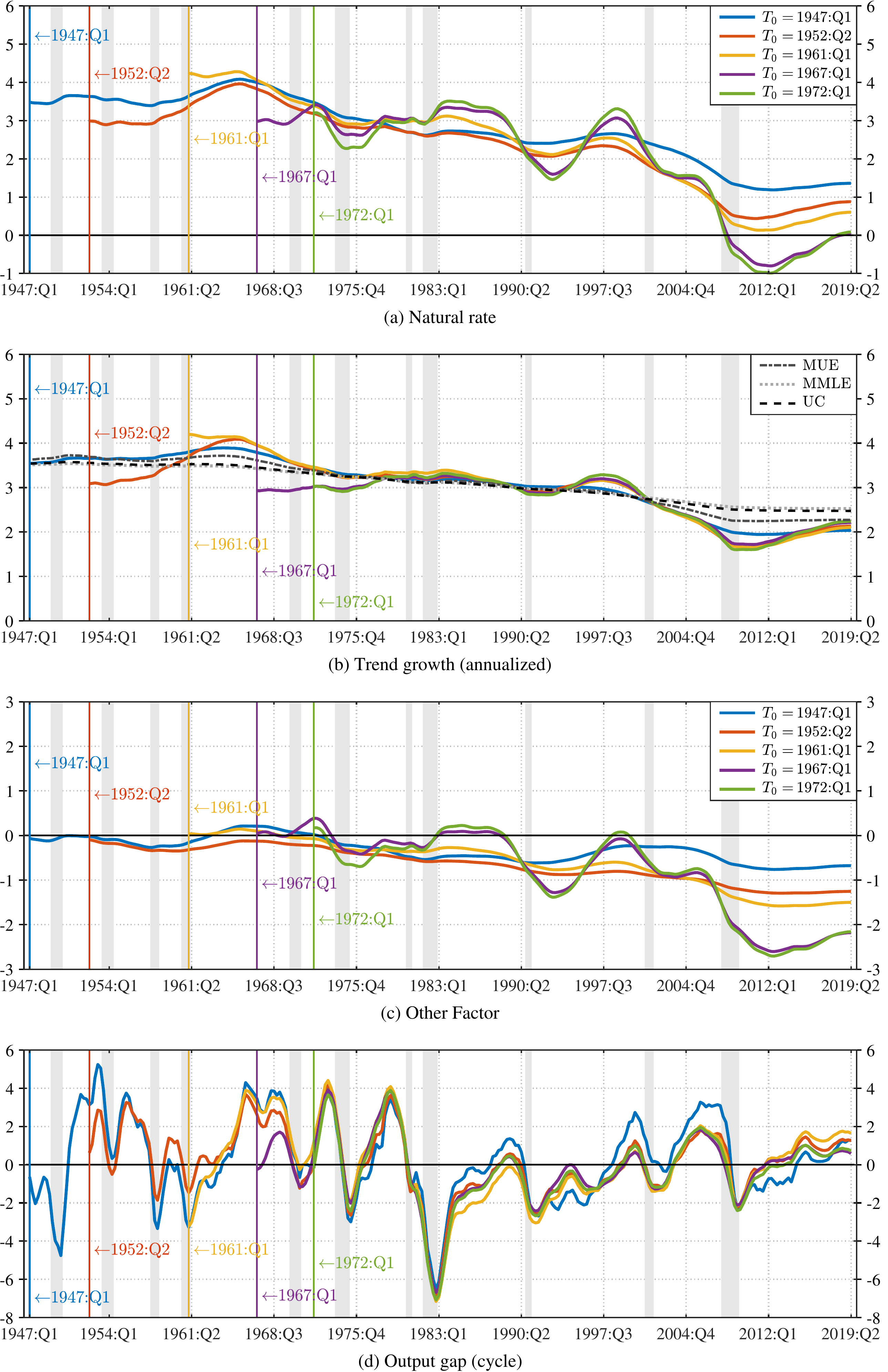}
\vspace{-3mm}
\caption{Smoothed estimates of annualized trend growth $g_t$, \emph{`other factor'} $z_t$
and the natural rate $r^{\ast}_t$ based on different starting dates}
\label{fig:T0KS}
\end{figure}

\cleardoublepage\newpage 

\newpage\cleardoublepage

\ifthenelse{\equal{1}{\plots}}{

%% file: _appendix.5e.tex



\small%

\section*{\hfil Appendix \label{appendix} \hfil}

\addcontentsline{toc}{section}{\currentname}


\let\Osubsection\subsection
\renewcommand{\section}[1]{\stepcounter{section}
\Osubsection*{A.\arabic{section}.~~{#1}}
\addcontentsline{toc}{subsection}{\currentname}}%
\let\Osubsubsection\subsubsection
\renewcommand{\subsection}[1]{\stepcounter{subsection}
\Osubsubsection*{A.\arabic{section}.\arabic{subsection}.~~{#1}}
\addcontentsline{toc}{subsubsection}{\currentname}}%


This appendix provides additional information on the \cite{holston.etal:2017}
model, their estimation procedure as well as snippets of R-Code. Matrix
details regarding the three stages of their procedure are taken from the
file \texttt{HLW\_Code\_Guide.pdf} which is contained in the \texttt{%
HLW\_Code.zip} file available from John Williams' website at the Federal
Reserve Bank of New York: %
\url{https://www.newyorkfed.org/medialibrary/media/research/economists/williams/data/HLWCode.zip}%
.

The state-space model notation is:%
\begin{equation}
\begin{array}{l}
\mathbf{y}_{t}=\mathbf{Ax}_{t}+\mathbf{H}\boldsymbol{\xi }_{t}+\boldsymbol{%
\nu }_{t} \\
\boldsymbol{\xi }_{t}=\mathbf{F}\boldsymbol{\xi }_{t-1}+\underbrace{\mathbf{S%
}\boldsymbol{\varepsilon }_{t}}_{\boldsymbol{\epsilon }_{t}}%
\end{array}%
\text{, \ \ where }%
\begin{bmatrix}
\boldsymbol{\nu }_{t} \\
\boldsymbol{\varepsilon }_{t}%
\end{bmatrix}%
\sim \mathsf{MNorm}\left(
\begin{bmatrix}
\boldsymbol{0} \\
\boldsymbol{0}%
\end{bmatrix}%
,%
\begin{bmatrix}
\mathbf{R} & \boldsymbol{0} \\
\boldsymbol{0} & \mathbf{W}%
\end{bmatrix}%
\right) ,
\end{equation}%
where $\mathbf{S}\boldsymbol{\varepsilon }_{t}=\boldsymbol{\epsilon }_{t}$,
so that $\mathrm{Var}(\mathbf{S}\boldsymbol{\varepsilon }_{t})=\mathrm{Var}(%
\boldsymbol{\epsilon }_{t})=\mathbf{SWS}^{\prime }=\mathbf{Q}$, with $%
\boldsymbol{\epsilon }_{t}$ and $\mathbf{Q}$ being the notation used in the
online appendix of \cite{holston.etal:2017} for the state vector's
disturbance term and its variance-covariance matrix.

\section{Stage 1 Model \label{sec:AS1}}

The first Stage model is defined by the following system matrices:\bsq\label%
{AS1_M}
\begin{align}
\mathbf{y}_{t}& =[y_{t},~\pi _{t}]^{\prime }  \label{AS1:y} \\
\mathbf{x}_{t}& =[y_{t-1},~y_{t-2},~\pi _{t-1},~\pi _{t-2,4}]^{\prime }
\label{AS1:x} \\
\boldsymbol{\xi }_{t}& =[y_{t}^{\ast },~y_{t-1}^{\ast },~y_{t-2}^{\ast
}]^{\prime },  \label{AS1:xi}
\end{align}%
\esq\vsp[-5]
\begin{equation*}
\mathbf{A}=%
\begin{bmatrix}
a_{y,1} & a_{y,2} & 0 & 0 \\
b_{y} & 0 & b_{\pi } & (1-b_{\pi })%
\end{bmatrix}%
,~\mathbf{H}=%
\begin{bmatrix}
1 & -a_{y,1} & -a_{y,2} \\
0 & -b_{y} & 0%
\end{bmatrix}%
,~\mathbf{F}=%
\begin{bmatrix}
1 & 0 & 0 \\
1 & 0 & 0 \\
0 & 1 & 0%
\end{bmatrix}%
,~\mathbf{S}=%
\begin{bmatrix}
1 \\
0 \\
0%
\end{bmatrix}%
.
\end{equation*}%
From this, the measurement relations are:%
\begin{align}
\mathbf{y}_{t}& =\mathbf{Ax}_{t}+\mathbf{H}\boldsymbol{\xi }_{t}+\boldsymbol{%
\nu }_{t}  \notag \\
\begin{bmatrix}
y_{t} \\
\pi _{t}%
\end{bmatrix}%
& =%
\begin{bmatrix}
a_{y,1} & a_{y,2} & 0 & 0 \\
b_{y} & 0 & b_{\pi } & (1-b_{\pi })%
\end{bmatrix}%
\begin{bmatrix}
y_{t-1} \\
y_{t-2} \\
\pi _{t-1} \\
\pi _{t-2,4}%
\end{bmatrix}%
+%
\begin{bmatrix}
1 & -a_{y,1} & -a_{y,2} \\
0 & -b_{y} & 0%
\end{bmatrix}%
\begin{bmatrix}
y_{t}^{\ast } \\
y_{t-1}^{\ast } \\
y_{t-2}^{\ast }%
\end{bmatrix}%
+%
\begin{bmatrix}
\varepsilon _{t}^{\tilde{y}} \\
\varepsilon _{t}^{\pi }%
\end{bmatrix}
\label{AS1:m}
\end{align}%
with the corresponding state equations being:%
\begin{align}
\boldsymbol{\xi }_{t}& =\mathbf{F}\boldsymbol{\xi }_{t-1}+\mathbf{S}%
\boldsymbol{\varepsilon }_{t}  \notag \\
\begin{bmatrix}
y_{t}^{\ast } \\
y_{t-1}^{\ast } \\
y_{t-2}^{\ast }%
\end{bmatrix}%
& =%
\begin{bmatrix}
1 & 0 & 0 \\
1 & 0 & 0 \\
0 & 1 & 0%
\end{bmatrix}%
\begin{bmatrix}
y_{t-1}^{\ast } \\
y_{t-2}^{\ast } \\
y_{t-3}^{\ast }%
\end{bmatrix}%
+%
\begin{bmatrix}
1 \\
0 \\
0%
\end{bmatrix}%
\begin{bmatrix}
\varepsilon _{t}^{y^{\ast }}%
\end{bmatrix}%
.  \label{AS1:s}
\end{align}

Expanding \ref{AS1:m} and \ref{AS1:s} yields:\bsq\label{AS1_0}%
\begin{align*}
y_{t}& =y_{t}^{\ast }+a_{y,1}(y_{t-1}-y_{t-1}^{\ast
})+a_{y,2}(y_{t-2}-y_{t-2}^{\ast })+\varepsilon _{t}^{\tilde{y}} \\
\pi _{t}& =b_{y}(y_{t-1}-y_{t-1}^{\ast })+b_{\pi }\pi _{t-1}+\left( 1-b_{\pi
}\right) \pi _{t-2,4}+\varepsilon _{t}^{\pi }
\end{align*}%
and%
\begin{align*}
y_{t}^{\ast }& =y_{t-1}^{\ast }+\varepsilon _{t}^{y^{\ast }} \\
y_{t-1}^{\ast }& =y_{t-1}^{\ast } \\
y_{t-2}^{\ast }& =y_{t-2}^{\ast },
\end{align*}%
\esq respectively, for the measurement and state equations. Defining output $%
y_{t}$ as trend plus cycle, and ignoring the identities, yields then the
following relations for the Stage 1 model:\bsq\label{AS1}%
\begin{align}
y_{t}& =y_{t}^{\ast }+\tilde{y}_{t}  \label{AS1:a} \\
\pi _{t}& =b_{\pi }\pi _{t-1}+\left( 1-b_{\pi }\right) \pi _{t-2,4}+b_{y}%
\tilde{y}_{t-1}+\varepsilon _{t}^{\pi }  \label{AS1:b} \\
\tilde{y}_{t}& =a_{y,1}\tilde{y}_{t-1}+a_{y,2}\tilde{y}_{t-2}+\varepsilon
_{t}^{\tilde{y}}  \label{AS1:c} \\
y_{t}^{\ast }& =y_{t-1}^{\ast }+\varepsilon _{t}^{y^{\ast }}.  \label{AS1:d}
\end{align}%
\esq If we disregard the inflation equation \ref{AS1:b} for now, the
decomposition of output into trend and cycle can be recognized as the
standard Unobserved Component (UC) model of \cite{harvey:1985}, \cite%
{clark:1987}, \cite{kuttner:1994}, \cite{morley.etal:2003} and others. \cite%
{holston.etal:2017} write on page S64: "\dots \textit{we follow Kuttner
(1994) and apply the Kalman filter to estimate the natural rate of output,
omitting the real rate gap term from Eq. (4)} [our Equation \ref{AS1:c}]
\textit{and assuming that the trend growth rate, }$g$\textit{, is constant.}"

One key difference is, nevertheless, that no drift term is included in the
trend specification in \ref{AS1:d}, so that $y_{t}^{\ast }$ follows a random
walk \emph{without} drift. Evidently, this cannot match the upward sloping
pattern in the GDP series. The way that \cite{holston.etal:2017} deal with
this mismatch is by `\textit{detrending'} output $y_{t}$ in the estimation.
This is implemented by re-placing $\{y_{t-j}\}_{j=0}^{2}$ in $\mathbf{y}_{t}$
and $\mathbf{x}_{t}$ in \ref{AS1_M} by $(y_{t}-gt)$, where $g$ is a
parameter (and not a trend growth state variable) to be estimated, and $t$
is a linear time trend defined as $t=[1,\ldots ,T]^{\prime }$. This is
hidden away from the reader and is not described in the documentation in
either text or equation form. Only from the listing of the vector of
parameters to be estimated by MLE, referred to as $\boldsymbol{\theta }_{1}$
in the middle of page 10 in the documentation, does it become evident that
an additional parameter --- confusingly labelled as $g$ --- is included in
the estimation. That is, the vector of Stage 1 parameters to be estimated is
defined as:
\begin{equation}
\boldsymbol{\theta }_{1}=[a_{y,1},~a_{y,2},~b_{\pi },~b_{y},~g,~\sigma _{%
\tilde{y}},~\sigma _{\pi },~\sigma _{y^{\ast }}]^{\prime }.  \label{AS1:t1}
\end{equation}

Note that the parameter $g$ in $\boldsymbol{\theta }_{1}$ is not found in
any of the system matrices that describe the Stage 1 model on page 10 of the
documentation. This gives the impression that it is a typographical error in
the documentation, rather than a parameter that is added to the model in the
estimation. However, from their R-Code file \texttt{%
unpack.parameters.stage1.R}, which is reproduced in \coderef{R:unpack}{3},
one can see that part of the unpacking routine, which is later called by the
log-likelihood estimation function, `\textit{detrends'} the data (see the
highlighted lines 29 to 31 in \coderef{R:unpack}{3}, where \texttt{$\ast $
parameter[5]} refers to parameter $g$ in $\boldsymbol{\theta }_{1}$). Due to
the linear time trend removal in the estimation stage, it has to be added
back to the Kalman Filter and Smoother extracted trends $y_{t}^{\ast }$,
which is is done in \texttt{kalman.states.wrapper.R }(see the highlighted
lines 29 to 30 in \coderef{R:wrapper}{4}, where the if statement: \texttt{if
(stage == 1) \{} on line 28 of this file ensures that this is only done for
the Stage 1 model). The actual equation for the trend term $y_{t}^{\ast }$
is thus:%
\begin{align}
y_{t}^{\ast }& =g+y_{t-1}^{\ast }+\varepsilon _{t}^{y^{\ast }}
\label{s1:y*1} \\
& =y_{0}^{\ast }+gt+\sum_{s=1}^{t}\varepsilon _{s}^{y^{\ast }},
\label{s1:y*2}
\end{align}%
where $g$ is an intercept term that captures \textit{constant} trend growth,
and $y_{0}^{\ast }$ is the initial condition of the state vector set to
806.45 from the HP filter output as discussed in \fnref{fn:1}. Why \cite%
{holston.etal:2017} prefer to use this way of dealing with the drift term
rather than simply adding an intercept term to the state equation in \ref%
{AS1:s} is not clear, and not discussed anywhere.

In the estimation of the Stage 1 model, the state vector $\boldsymbol{\xi }%
_{t}$ is initialized using the same procedure as outlined in \ref{eq:P00S1a}
and \fnref{fn:1} with the numerical value of $\boldsymbol{\xi }_{00}$ and $%
\mathbf{P}_{00}$ set at:%
\begin{align}
\boldsymbol{\xi }_{00}& =[806.4455,~805.2851,~804.1248]  \label{AS1:xi00} \\
\mathbf{P}_{00}& =%
\begin{bmatrix}
0.4711 & 0.2 & 0.0 \\
0.2 & 0.2 & 0.0 \\
0.0 & 0.0 & 0.2%
\end{bmatrix}%
.  \label{AS1:P00}
\end{align}

\section{Stage 2 Model\label{sec:AS2}}

The second Stage model of \cite{holston.etal:2017} is defined by the
following model matrices:%
\begin{align}
\mathbf{y}_{t}& =[y_{t},~\pi _{t}]^{\prime }  \label{AS2:y} \\
\mathbf{x}_{t}& =[y_{t-1},~y_{t-2},~r_{t-1},~r_{t-2},~\pi _{t-1},~\pi
_{t-2,4},~1]^{\prime }  \label{AS2:x} \\
\boldsymbol{\xi }_{t}& =[y_{t}^{\ast },~y_{t-1}^{\ast },~y_{t-2}^{\ast
},~g_{t-1}]^{\prime }  \label{AS2:xi}
\end{align}%
\vsp[-11]
\begin{align*}
\mathbf{A}& =%
\begin{bmatrix}
a_{y,1} & a_{y,2} & \frac{a_{r}}{2} & \frac{a_{r}}{2} & 0 & 0 & a_{0} \\
b_{y} & 0 & 0 & 0 & b_{\pi } & (1-b_{\pi }) & 0%
\end{bmatrix}%
,~\mathbf{H}=%
\begin{bmatrix}
1 & -a_{y,1} & -a_{y,2} & a_{g} \\
0 & -b_{y} & 0 & 0%
\end{bmatrix}%
, \\
& \\
\mathbf{F}& =%
\begin{bmatrix}
1 & 0 & 0 & 1 \\
1 & 0 & 0 & 0 \\
0 & 1 & 0 & 0 \\
0 & 0 & 0 & 1%
\end{bmatrix}%
,~\mathbf{S}=%
\begin{bmatrix}
1 & 0 \\
0 & 0 \\
0 & 0 \\
0 & 1%
\end{bmatrix}%
.
\end{align*}%
The measurement and state relations are given by:%
\begin{align}
\mathbf{y}_{t}& =\mathbf{Ax}_{t}+\mathbf{H}\boldsymbol{\xi }_{t}+\boldsymbol{%
\nu }_{t}  \notag \\
\begin{bmatrix}
y_{t} \\
\pi _{t}%
\end{bmatrix}%
& =%
\begin{bmatrix}
a_{y,1} & a_{y,2} & \frac{a_{r}}{2} & \frac{a_{r}}{2} & 0 & 0 & a_{0} \\
b_{y} & 0 & 0 & 0 & b_{\pi } & (1-b_{\pi }) & 0%
\end{bmatrix}%
\begin{bmatrix}
y_{t-1} \\
y_{t-2} \\
r_{t-1} \\
r_{t-2} \\
\pi _{t-1} \\
\pi _{t-2,4} \\
1%
\end{bmatrix}%
+%
\begin{bmatrix}
1 & -a_{y,1} & -a_{y,2} & a_{g} \\
0 & -b_{y} & 0 & 0%
\end{bmatrix}%
\begin{bmatrix}
y_{t}^{\ast } \\
y_{t-1}^{\ast } \\
y_{t-2}^{\ast } \\
g_{t-1}%
\end{bmatrix}%
+%
\begin{bmatrix}
\varepsilon _{t}^{\tilde{y}} \\
\varepsilon _{t}^{\pi }%
\end{bmatrix}
\label{AS2:m}
\end{align}%
and%
\begin{align}
\boldsymbol{\xi }_{t}& =\mathbf{F}\boldsymbol{\xi }_{t-1}+\mathbf{S}%
\boldsymbol{\varepsilon }_{t}  \notag \\
\begin{bmatrix}
y_{t}^{\ast } \\
y_{t-1}^{\ast } \\
y_{t-2}^{\ast } \\
g_{t-1}%
\end{bmatrix}%
& =%
\begin{bmatrix}
1 & 0 & 0 & 1 \\
1 & 0 & 0 & 0 \\
0 & 1 & 0 & 0 \\
0 & 0 & 0 & 1%
\end{bmatrix}%
\begin{bmatrix}
y_{t-1}^{\ast } \\
y_{t-2}^{\ast } \\
y_{t-3}^{\ast } \\
g_{t-2}%
\end{bmatrix}%
+%
\begin{bmatrix}
1 & 0 \\
0 & 0 \\
0 & 0 \\
0 & 1%
\end{bmatrix}%
\begin{bmatrix}
\varepsilon _{t}^{y^{\ast }} \\
\varepsilon _{t-1}^{g}%
\end{bmatrix}%
.  \label{AS2:s}
\end{align}%
Note that $\sigma _{g}^{2}$ in $\mathrm{Var}(\boldsymbol{\varepsilon }_{t})=%
\mathbf{W}=\mathrm{\mathrm{diag}}([\sigma _{y^{\ast }}^{2},~\sigma
_{g}^{2}]) $ is replaced by $(\hat{\lambda}_{g}\sigma _{y^{\ast }})^{2}$
where $\hat{\lambda}_{g}$ is the estimate from the first Stage, so that we
obtain: \vsp[-8]

\begin{align}
\mathrm{Var}(\mathbf{S}\boldsymbol{\varepsilon }_{t})& =\mathbf{SWS}^{\prime
}  \notag \\
& =%
\begin{bmatrix}
1 & 0 \\
0 & 0 \\
0 & 0 \\
0 & 1%
\end{bmatrix}%
\begin{bmatrix}
\sigma _{y^{\ast }}^{2} & 0 \\
0 & (\hat{\lambda}_{g}\sigma _{y^{\ast }})^{2}%
\end{bmatrix}%
\begin{bmatrix}
1 & 0 \\
0 & 0 \\
0 & 0 \\
0 & 1%
\end{bmatrix}%
^{\prime }  \notag \\[0.08in]
\mathbf{Q}& =%
\begin{bmatrix}
\sigma _{y^{\ast }}^{2} & 0 & 0 & 0 \\
0 & 0 & 0 & 0 \\
0 & 0 & 0 & 0 \\
0 & 0 & 0 & (\hat{\lambda}_{g}\sigma _{y^{\ast }})^{2}%
\end{bmatrix}%
,  \label{S2Q}
\end{align}%
which is then used in the Kalman Filter routine and ML to estimate the Stage
2 model parameters.

Expanding the relations in \ref{AS2:m} and \ref{AS2:s} leads to the
measurement:\bsq\label{AS2m}%
\begin{align}
y_{t}& =y_{t}^{\ast }+a_{y,1}(y_{t-1}-y_{t-1}^{\ast
})+a_{y,2}(y_{t-2}-y_{t-2}^{\ast })+\tfrac{a_{r}}{2}%
(r_{t-1}+r_{t-2})+a_{0}+a_{g}g_{t-1}+\varepsilon _{t}^{\tilde{y}} \\
\pi _{t}& =b_{y}(y_{t-1}-y_{t-1}^{\ast })+b_{\pi }\pi _{t-1}+\left( 1-b_{\pi
}\right) \pi _{t-2,4}+\varepsilon _{t}^{\pi }
\end{align}%
\esq and corresponding state relations\bsq%
\begin{align}
y_{t}^{\ast }& =y_{t-1}^{\ast }+g_{t-2}+\varepsilon _{t}^{y^{\ast }} \\
y_{t-1}^{\ast }& =y_{t-1}^{\ast } \\
y_{t-2}^{\ast }& =y_{t-2}^{\ast } \\
g_{t-1}& =g_{t-2}+\varepsilon _{t-1}^{g}.
\end{align}%
\esq Defining output $y_{t}$ as before as trend plus cycle, dropping
identities, simplifying and rewriting gives the following Stage 2 system
relations:\bsq\label{AS2}%
\begin{align}
y_{t}& =y_{t}^{\ast }+\tilde{y}_{t} \\
\pi _{t}& =b_{\pi }\pi _{t-1}+\left( 1-b_{\pi }\right) \pi _{t-2,4}+b_{y}%
\tilde{y}_{t-1}+\varepsilon _{t}^{\pi } \\
a_{y}(L)\tilde{y}_{t}& =a_{0}+\tfrac{a_{r}}{2}%
(r_{t-1}+r_{t-2})+a_{g}g_{t-1}+\varepsilon _{t}^{\tilde{y}} \\
y_{t}^{\ast }& =y_{t-1}^{\ast }+g_{t-2}+\varepsilon _{t}^{y^{\ast }}
\label{ystar_mis} \\
g_{t-1}& =g_{t-2}+\varepsilon _{t-1}^{g},
\end{align}%
\esq where the corresponding vector of parameters to be estimated by MLE is:%
\begin{equation}
\boldsymbol{\theta }_{2}=[a_{y,1},~a_{y,2},~a_{r},~a_{0},~a_{g},~b_{\pi
},~b_{y},~\sigma _{\tilde{y}},~\sigma _{\pi },~\sigma _{y^{\ast }}]^{\prime
}.  \label{eq:theta2}
\end{equation}%
The state vector $\boldsymbol{\xi }_{t}$ in the estimation of the Stage 2
model is initialized using the procedure outlined in \ref{eq:P00S1a} and %
\fnref{fn:1}, with the numerical value of $\boldsymbol{\xi }_{00}$ and $%
\mathbf{P}_{00}$ set at:%
\begin{align}
\boldsymbol{\xi }_{00}& =[806.4455,~805.2851,~804.1248,~1.1604]
\label{AS2:xi00} \\
\mathbf{P}_{00}& =%
\begin{bmatrix}
0.7185 & 0.2 & 0.0 & 0.2 \\
0.2 & 0.2 & 0.0 & 0.0 \\
0.2 & 0.2 & 0.0 & 0.0 \\
0.2 & 0.0 & 0.2 & 0.2009%
\end{bmatrix}%
.  \label{AS2:P00}
\end{align}

Notice from the trend specification in \ref{ystar_mis} that $g_{t-2}$
instead of $g_{t-1}$ is included in the equation. This is not a
typographical error, but rather a `\emph{feature}' of the Stage 2 model
specification of \cite{holston.etal:2017}, and is not obvious until the
Stage 2 model relations are written out as above in equations \ref{AS2:m} to %
\ref{AS2}.\ I use the selection matrix $\mathbf{S}$ to derive what the
variance-covariance matrix of $\mathbf{S}\boldsymbol{\varepsilon }_{t}$,
that is, $\mathrm{Var}(\mathbf{S}\boldsymbol{\varepsilon }_{t})=\mathrm{Var}(%
\boldsymbol{\epsilon }_{t})=\mathbf{SWS}^{\prime }=\mathbf{Q}$, should look
like. \cite{holston.etal:2017} only report the $\mathbf{Q}$ matrix in their
online appendix included in the R-Code zip file (see page 10, lower half of
the page in Section 7.4).

In the Stage 3 model, \cite{holston.etal:2017} use a `\emph{trick}' to
arrive at the correct trend specification for $y_{t}^{\ast }$ by including
both, the $\varepsilon _{t-1}^{g}$ as well as the $\varepsilon
_{t}^{y_{t}^{\ast }}$ error terms in the equation for $y_{t}^{\ast }$ (see %
\ref{eq:g_trick} below). This can also be seen from the $\mathbf{Q}$ matrix
on page 11 in Section 7.5 of their online appendix or \ref{Q3} below, which
now includes off-diagonal terms in the Stage 3 model.

\subsection{Getting the correct Stage 2 Model from the Stage 3 Model \label%
{sec:AS21}}

We can apply this same `\emph{trick}' for the Stage 2 model, by taking the
Stage 3 model state-space form and deleting the row, respectively, column
entries of the $\mathbf{F}$, $\mathbf{H}$, and $\mathbf{S}$ matrices to make
them conformable with the required Stage 2 model. The state and measurement
equations of the correct Stage 2 model then look as follows:%
\begin{align}
\mathbf{y}_{t}& =\mathbf{Ax}_{t}+\mathbf{H}\boldsymbol{\xi }_{t}+\boldsymbol{%
\nu }_{t}  \notag \\
\begin{bmatrix}
y_{t} \\
\pi _{t}%
\end{bmatrix}%
& =%
\begin{bmatrix}
a_{y,1} & a_{y,2} & \frac{a_{r}}{2} & \frac{a_{r}}{2} & 0 & 0 \\
b_{y} & 0 & 0 & 0 & b_{\pi } & (1-b_{\pi })%
\end{bmatrix}%
\begin{bmatrix}
y_{t-1} \\
y_{t-2} \\
r_{t-1} \\
r_{t-2} \\
\pi _{t-1} \\
\pi _{t-2,4}%
\end{bmatrix}%
+%
\begin{bmatrix}
1 & -a_{y,1} & -a_{y,2} & -\frac{a_{r}}{2} & -\frac{a_{r}}{2} \\
0 & -b_{y} & 0 & 0 & 0%
\end{bmatrix}%
\begin{bmatrix}
y_{t}^{\ast } \\
y_{t-1}^{\ast } \\
y_{t-2}^{\ast } \\
g_{t-1} \\
g_{t-2}%
\end{bmatrix}%
+%
\begin{bmatrix}
\varepsilon _{t}^{\tilde{y}} \\
\varepsilon _{t}^{\pi }%
\end{bmatrix}%
\end{align}%
\begin{align*}
\boldsymbol{\xi }_{t}& =\mathbf{F}\boldsymbol{\xi }_{t-1}+\mathbf{S}%
\boldsymbol{\varepsilon }_{t} \\
\begin{bmatrix}
y_{t}^{\ast } \\
y_{t-1}^{\ast } \\
y_{t-2}^{\ast } \\
g_{t-1} \\
g_{t-2}%
\end{bmatrix}%
& =%
\begin{bmatrix}
1 & 0 & 0 & 1 & 0 \\
1 & 0 & 0 & 0 & 0 \\
0 & 1 & 0 & 0 & 0 \\
0 & 0 & 0 & 1 & 0 \\
0 & 0 & 0 & 1 & 0%
\end{bmatrix}%
\begin{bmatrix}
y_{t-1}^{\ast } \\
y_{t-2}^{\ast } \\
y_{t-3}^{\ast } \\
g_{t-2} \\
g_{t-3}%
\end{bmatrix}%
+%
\begin{bmatrix}
1 & 1 \\
0 & 0 \\
0 & 0 \\
0 & 1 \\
0 & 0%
\end{bmatrix}%
\begin{bmatrix}
\varepsilon _{t}^{y^{\ast }} \\
\varepsilon _{t-1}^{g}%
\end{bmatrix}%
, \\
\intxt{which, upon expanding and dropping of identities, yields:}y_{t}&
=y_{t}^{\ast }+\tilde{y}_{t} \\
\pi _{t}& =b_{\pi }\pi _{t-1}+\left( 1-b_{\pi }\right) \pi _{t-2,4}+b_{y}%
\tilde{y}_{t-1}+\varepsilon _{t}^{\pi } \\
a_{y}(L)\tilde{y}_{t}& =\tfrac{a_{r}}{2}(r_{t-1}-g_{t-1})+\tfrac{a_{r}}{2}%
(r_{t-2}-g_{t-2})+\varepsilon _{t}^{\tilde{y}} \\
y_{t}^{\ast }& =y_{t-1}^{\ast }+\overbrace{g_{t-2}+\varepsilon _{t-1}^{g}}%
^{g_{t-1}}+\varepsilon _{t}^{y^{\ast }} \\
g_{t-1}& =g_{t-2}+\varepsilon _{t-1}^{g}.
\end{align*}%
These last relations correspond to \ref{S2full0}, with $\varepsilon _{t}^{%
\tilde{y}}$ being the counterpart to $\mathring{\varepsilon}_{t}^{\tilde{y}%
}=-a_{r}(L)z_{t}+\varepsilon _{t}^{\tilde{y}}$ if we take the full Stage 3
model as the true model.

Using the Stage 3 state-space form and simply adjusting it as shown above
yields the correct Stage 2 equations for trend $y_{t}^{\ast }$ and the
output gap $\tilde{y}_{t}$. With this form of the state-space model, it is
also clear that the variance-covariance matrix $\mathbf{Q}=\mathrm{Var}(%
\mathbf{S}\boldsymbol{\varepsilon }_{t})$ will be:%
\begin{align}
\mathbf{Q}& =\mathbf{SWS}^{\prime }  \notag \\
& =%
\begin{bmatrix}
1 & 1 \\
0 & 0 \\
0 & 0 \\
0 & 1 \\
0 & 0%
\end{bmatrix}%
\begin{bmatrix}
\sigma _{y^{\ast }}^{2} & 0 \\
0 & (\lambda _{g}\sigma _{y^{\ast }})^{2}%
\end{bmatrix}%
\begin{bmatrix}
1 & 1 \\
0 & 0 \\
0 & 0 \\
0 & 1 \\
0 & 0%
\end{bmatrix}%
^{\prime }  \notag \\
& =%
\begin{bmatrix}
\sigma _{y^{\ast }}^{2}+(\hat{\lambda}_{g}\sigma _{y^{\ast }})^{2} & 0 & 0 &
(\hat{\lambda}_{g}\sigma _{y^{\ast }})^{2} & 0 \\
0 & 0 & 0 & 0 & 0 \\
0 & 0 & 0 & 0 & 0 \\
(\hat{\lambda}_{g}\sigma _{y^{\ast }})^{2} & 0 & 0 & (\hat{\lambda}%
_{g}\sigma _{y^{\ast }})^{2} & 0 \\
0 & 0 & 0 & 0 & 0%
\end{bmatrix}%
,  \label{S2Qcorrect}
\end{align}%
where $(\hat{\lambda}_{g}\sigma _{y^{\ast }})^{2}$ again replaces $\sigma
_{g}^{2}$, as before.\ Since the\textbf{\ }$\mathbf{Q}$ matrix in \cite%
{holston.etal:2017} takes the form of \ref{S2Q} and not \textrm{\ref%
{S2Qcorrect}}, we can see that this `\emph{trick}' of rewriting the trend
growth equation as in the Stage 3 model specification was not applied to the
Stage 2 model. Given that the correct Stage 2 model is easily obtained from
the full Stage 3 model specification, it is not clear why the Stage 2 model
is defined incorrectly as in \ref{eq:stag2}.

\section{Stage 3 Model\label{sec:AS3}}

The third and final Stage model is defined as follows:%
\begin{align}
\mathbf{y}_{t}& =[y_{t},~\pi _{t}]^{\prime }  \label{AS3:y} \\
\mathbf{x}_{t}& =[y_{t-1},~y_{t-2},~r_{t-1},~r_{t-2},~\pi _{t-1},~\pi
_{t-2,4}]^{\prime }  \label{AS3:x} \\
\boldsymbol{\xi }_{t}& =[y_{t}^{\ast },~y_{t-1}^{\ast },~y_{t-2}^{\ast
},~g_{t-1},~g_{t-2},~z_{t-1},~z_{t-2}]^{\prime }  \label{AS3:xi}
\end{align}%
\vsp[-11]
\begin{eqnarray*}
\mathbf{A} &=&%
\begin{bmatrix}
a_{y,1} & a_{y,2} & \frac{a_{r}}{2} & \frac{a_{r}}{2} & 0 & 0 \\
b_{y} & 0 & 0 & 0 & b_{\pi } & (1-b_{\pi })%
\end{bmatrix}%
,~\mathbf{H}=%
\begin{bmatrix}
1 & -a_{y,1} & -a_{y,2} & -\frac{a_{r}}{2} & -\frac{a_{r}}{2} & -\frac{a_{r}%
}{2} & -\frac{a_{r}}{2} \\
0 & -b_{y} & 0 & 0 & 0 & 0 & 0%
\end{bmatrix}%
, \\[5mm]
\mathbf{F} &=&%
\begin{bmatrix}
1 & 0 & 0 & 1 & 0 & 0 & 0 \\
1 & 0 & 0 & 0 & 0 & 0 & 0 \\
0 & 1 & 0 & 0 & 0 & 0 & 0 \\
0 & 0 & 0 & 1 & 0 & 0 & 0 \\
0 & 0 & 0 & 1 & 0 & 0 & 0 \\
0 & 0 & 0 & 0 & 0 & 1 & 0 \\
0 & 0 & 0 & 0 & 0 & 1 & 0%
\end{bmatrix}%
,~\mathbf{S}=%
\begin{bmatrix}
1 & 1 & 0 \\
0 & 0 & 0 \\
0 & 0 & 0 \\
0 & 1 & 0 \\
0 & 0 & 0 \\
0 & 0 & 1 \\
0 & 0 & 0%
\end{bmatrix}%
.
\end{eqnarray*}%
The measurement and state relations are:\vsp[-11]\BAW[9]
\begin{align}
\mathbf{y}_{t}& =\mathbf{Ax}_{t}+\mathbf{H}\boldsymbol{\xi }_{t}+\boldsymbol{%
\nu }_{t}  \notag \\
\begin{bmatrix}
y_{t} \\
\pi _{t}%
\end{bmatrix}%
& =%
\begin{bmatrix}
a_{y,1} & a_{y,2} & \frac{a_{r}}{2} & \frac{a_{r}}{2} & 0 & 0 \\
b_{y} & 0 & 0 & 0 & b_{\pi } & (1-b_{\pi })%
\end{bmatrix}%
\begin{bmatrix}
y_{t-1} \\
y_{t-2} \\
r_{t-1} \\
r_{t-2} \\
\pi _{t-1} \\
\pi _{t-2,4}%
\end{bmatrix}%
+%
\begin{bmatrix}
1 & -a_{y,1} & -a_{y,2} & -\frac{a_{r}}{2} & -\frac{a_{r}}{2} & -\frac{a_{r}%
}{2} & -\frac{a_{r}}{2} \\
0 & -b_{y} & 0 & 0 & 0 & 0 & 0%
\end{bmatrix}%
\begin{bmatrix}
y_{t}^{\ast } \\
y_{t-1}^{\ast } \\
y_{t-2}^{\ast } \\
g_{t-1} \\
g_{t-2} \\
z_{t-1} \\
z_{t-2}%
\end{bmatrix}%
+%
\begin{bmatrix}
\varepsilon _{t}^{\tilde{y}} \\
\varepsilon _{t}^{\pi }%
\end{bmatrix}
\label{AS3:m}
\end{align}%
\vsp[-15]\EAW and%
\begin{align}
\boldsymbol{\xi }_{t}& =\mathbf{F}\boldsymbol{\xi }_{t-1}+\mathbf{S}%
\boldsymbol{\varepsilon }_{t}  \notag \\
\begin{bmatrix}
y_{t}^{\ast } \\
y_{t-1}^{\ast } \\
y_{t-2}^{\ast } \\
g_{t-1} \\
g_{t-2} \\
z_{t-1} \\
z_{t-2}%
\end{bmatrix}%
& =%
\begin{bmatrix}
1 & 0 & 0 & 1 & 0 & 0 & 0 \\
1 & 0 & 0 & 0 & 0 & 0 & 0 \\
0 & 1 & 0 & 0 & 0 & 0 & 0 \\
0 & 0 & 0 & 1 & 0 & 0 & 0 \\
0 & 0 & 0 & 1 & 0 & 0 & 0 \\
0 & 0 & 0 & 0 & 0 & 1 & 0 \\
0 & 0 & 0 & 0 & 0 & 1 & 0%
\end{bmatrix}%
\begin{bmatrix}
y_{t-1}^{\ast } \\
y_{t-2}^{\ast } \\
y_{t-3}^{\ast } \\
g_{t-2} \\
g_{t-3} \\
z_{t-2} \\
z_{t-3}%
\end{bmatrix}%
+%
\begin{bmatrix}
1 & 1 & 0 \\
0 & 0 & 0 \\
0 & 0 & 0 \\
0 & 1 & 0 \\
0 & 0 & 0 \\
0 & 0 & 1 \\
0 & 0 & 0%
\end{bmatrix}%
\begin{bmatrix}
\varepsilon _{t}^{y_{t}^{\ast }} \\
\varepsilon _{t-1}^{g} \\
\varepsilon _{t-1}^{z}%
\end{bmatrix}%
.  \label{AS3:s}
\end{align}%
In the Stage 3 model, \cite{holston.etal:2017} replace $\sigma _{g}^{2}$ and
$\sigma _{z}^{2}$ in $\mathrm{Var}(\boldsymbol{\varepsilon }_{t})=\mathbf{W}=%
\mathrm{\mathrm{diag}}([\sigma _{y^{\ast }}^{2},~\sigma _{g}^{2},~\sigma
_{z}^{2}])$ with $(\hat{\lambda}_{g}\sigma _{y^{\ast }})^{2}$ and $(\hat{%
\lambda}_{z}\sigma _{\tilde{y}}/a_{r})^{2}$, respectively, from the two
previous estimation steps, so that:\vsp[-4]%
\begin{align}
\mathrm{Var}(\mathbf{S}\boldsymbol{\varepsilon }_{t})& =\mathbf{SWS}^{\prime
}  \notag \\
& =%
\begin{bmatrix}
1 & 1 & 0 \\
0 & 0 & 0 \\
0 & 0 & 0 \\
0 & 1 & 0 \\
0 & 0 & 0 \\
0 & 0 & 1 \\
0 & 0 & 0%
\end{bmatrix}%
\begin{bmatrix}
\sigma _{y^{\ast }}^{2} & 0 & 0 \\
0 & (\hat{\lambda}_{g}\sigma _{y^{\ast }})^{2} & 0 \\
0 & 0 & (\hat{\lambda}_{z}\sigma _{\tilde{y}}/a_{r})^{2}%
\end{bmatrix}%
\begin{bmatrix}
1 & 1 & 0 \\
0 & 0 & 0 \\
0 & 0 & 0 \\
0 & 1 & 0 \\
0 & 0 & 0 \\
0 & 0 & 1 \\
0 & 0 & 0%
\end{bmatrix}%
^{\prime }  \notag \\[0.08in]
\mathbf{\mathbf{Q}}& =%
\begin{bmatrix}
\sigma _{y^{\ast }}^{2}+(\hat{\lambda}_{g}\sigma _{y^{\ast }})^{2} & 0 & 0 &
(\hat{\lambda}_{g}\sigma _{y^{\ast }})^{2} & 0 & 0 & 0 \\
0 & 0 & 0 & 0 & 0 & 0 & 0 \\
0 & 0 & 0 & 0 & 0 & 0 & 0 \\
(\hat{\lambda}_{g}\sigma _{y^{\ast }})^{2} & 0 & 0 & (\hat{\lambda}%
_{g}\sigma _{y^{\ast }})^{2} & 0 & 0 & 0 \\
0 & 0 & 0 & 0 & 0 & 0 & 0 \\
0 & 0 & 0 & 0 & 0 & (\hat{\lambda}_{z}\sigma _{\tilde{y}}/a_{r})^{2} & 0 \\
0 & 0 & 0 & 0 & 0 & 0 & 0%
\end{bmatrix}%
,  \label{Q3}
\end{align}%
which enters the Kalman Filter routine and ML estimation of the final Stage
3 parameters.

Expanding the relations in \ref{AS3:m} and \ref{AS3:s} leads to the
following measurement:%
\begin{align*}
y_{t}& =y_{t}^{\ast }+a_{y,1}(y_{t-1}-y_{t-1}^{\ast
})+a_{y,2}(y_{t-2}-y_{t-2}^{\ast })+\tfrac{a_{r}}{2}%
(r_{t-1}-g_{t-1}-z_{t-1})+\tfrac{a_{r}}{2}(r_{t-2}-g_{t-2}-z_{t-2})+%
\varepsilon _{t}^{\tilde{y}} \\
\pi _{t}& =b_{y}(y_{t-1}-y_{t-1}^{\ast })+b_{\pi }\pi _{t-1}+\left( 1-b_{\pi
}\right) \pi _{t-2,4}+\varepsilon _{t}^{\pi }
\end{align*}%
and corresponding state relations%
\begin{align}
y_{t}^{\ast }& =y_{t-1}^{\ast }+\overbrace{g_{t-2}+\varepsilon _{t-1}^{g}}%
^{g_{t-1}}+\varepsilon _{t}^{y_{t}^{\ast }}  \label{eq:g_trick} \\
y_{t-1}^{\ast }& =y_{t-1}^{\ast }  \notag \\
y_{t-2}^{\ast }& =y_{t-2}^{\ast }  \notag \\
g_{t-1}& =g_{t-2}+\varepsilon _{t-1}^{g}  \notag \\
g_{t-2}& =g_{t-2}  \notag \\
z_{t-1}& =z_{t-2}+\varepsilon _{t-1}^{z}  \notag \\
z_{t-2}& =z_{t-2}.  \notag
\end{align}%
\bigskip Defining output $y_{t}$ once again as trend plus cycle, dropping
identities and simplifying gives the following system of Stage 3 relations:%
\bsq\label{AS3}%
\begin{align}
y_{t}& =y_{t}^{\ast }+\tilde{y}_{t} \\
\pi _{t}& =b_{\pi }\pi _{t-1}+\left( 1-b_{\pi }\right) \pi _{t-2,4}+b_{y}%
\tilde{y}_{t-1}+\varepsilon _{t}^{\pi } \\
a_{y}(L)\tilde{y}_{t}& =\tfrac{a_{r}}{2}(r_{t-1}-g_{t-1}-z_{t-1})+\tfrac{%
a_{r}}{2}(r_{t-2}-g_{t-2}-z_{t-2})+\varepsilon _{t}^{\tilde{y}} \\
y_{t}^{\ast }& =y_{t-1}^{\ast }+g_{t-1}+\varepsilon _{t}^{y^{\ast }} \\
g_{t-1}& =g_{t-2}+\varepsilon _{t-1}^{g} \\
z_{t-1}& =z_{t-2}+\varepsilon _{t-1}^{z},
\end{align}%
\esq with the corresponding vector of Stage 3 model parameters to be
estimated by MLE being:%
\begin{equation}
\boldsymbol{\theta }_{3}=[a_{y,1},~a_{y,2},~a_{r},~b_{\pi },~b_{y},~\sigma _{%
\tilde{y}},~\sigma _{\pi },~\sigma _{y^{\ast }}]^{\prime }.
\end{equation}%
For the Stage 3 model, the variance of the state vector $\boldsymbol{\xi }%
_{t}$ is initialized once more as outlined in \ref{eq:P00S1a} and %
\fnref{fn:1}, with the numerical value of $\boldsymbol{\xi }_{00}$ and $%
\mathbf{P}_{00}$ being:%
\begin{align}
\boldsymbol{\xi }_{00}& =[806.4455,~805.2851,~804.1248,~1.1604,~1.1603,~0,~0]
\label{AS3:xi00} \\
\mathbf{P}_{00}& =%
\begin{bmatrix}
0.7272 & 0.2 & 0 & 0.2009 & 0.2 & 0 & 0 \\
0.2 & 0.2 & 0 & 0 & 0 & 0 & 0 \\
0 & 0 & 0.2 & 0 & 0 & 0 & 0 \\
0.2009 & 0 & 0 & 0.2009 & 0.2 & 0 & 0 \\
0.2 & 0 & 0 & 0.2 & 0.2 & 0 & 0 \\
0 & 0 & 0 & 0 & 0 & 0.2227 & 0.2 \\
0 & 0 & 0 & 0 & 0 & 0.2 & 0.2%
\end{bmatrix}%
.  \label{AS3:P00}
\end{align}

\smallskip

\section{Additional simulation results\label{sec:AS4}}

As an additional experiment, I simulate entirely unrelated univariate time
series processes as inputs into the $\mathcal{Y}_{t}$ and $\boldsymbol{%
\mathcal{X}}_{t}$ vector series needed for the structural break regressions
in \ref{eqS2regs}. As before, the simulated inputs that are required are the
cycle variable $\tilde{y}_{t}$, trend growth $g_{t}$ as well as the real
rate $r_{t}$. To avoid having to use the observed exogenous interest rate
series that makes up the real rate via the relation $r_{t}=i_{t}-\pi
_{t}^{e} $ ($\pi _{t}^{e}$ is expected inflation as defined in \ref{pi}) as
it will be function of $r_{t}^{\ast }$ and hence $g_{t}$ and $z_{t}$, I fit
a low order ARMA process to $r_{t}$. I then use the coefficients from this
estimated ARMA\ model to generate a simulated sequence of $T$ observations
from the real interest rate. I follow the same strategy to generate a
simulated series for $\tilde{y}_{t}$. Note that I do not simply use the
AR(2) model structure for the cycle series $\tilde{y}_{t}$ as is implied by
the left hand side of \ref{S2:ytilde} together with the $a_{y,1}$ and $%
a_{y,2}$ estimates from the Stage 2 model in the simulation. The reason for
this is that the empirical $\hat{\tilde{y}}_{t|T}$ series that \cite%
{holston.etal:2017} use in their procedure is the Kalman Smoother based
estimate of $\tilde{y}_{t}$ which portrays a more complicated
autocorrelation pattern than an AR(2) process. In order to match the
autocorrelation pattern of the $\hat{\tilde{y}}_{t|T}$ series as closely as
possible, I fitted the best (low order) ARMA\ process to $\hat{\tilde{y}}%
_{t|T}$, and used those coefficients to generate the simulated cycle series.

For the $g_{t}$ series, I use three different simulation scenarios. First, I
replace the trend growth estimate in $\boldsymbol{\mathcal{X}}_{t}$ by the
Kalman Smoother estimate of $g_{t}$ denoted by $\hat{g}_{t-1|T}$ above. This
is the same series that \cite{holston.etal:2017}. Second, I simulate $%
g_{t-1} $ from a pure random walk (RW) process with the standard deviation
of the error term set equal to $\hat{\sigma}_{g}=0.0305205$, the implied
estimate reported in column 1 of \ref{tab:Stage2}. Third, I simulate a
simple (Gaussian) white noise (WN) for $g_{t-1}$. And last, I fit a low
order ARMA\ process to the first difference of $\hat{g}_{t-1|T}$. The "\emph{%
empirical}" $\hat{g}_{t-1|T}$ series is very persistent and its dynamics are
not sufficiently captured by a pure RW. I therefore use the coefficients
from a fitted ARMA\ model to $\Delta \hat{g}_{t-1|T}$ to simulate the first
difference process $\Delta g_{t-1}$, and then construct the $g_{t-1}$ as the
cumulative sum of\ $\Delta \hat{g}_{t-1|T}$ in $\boldsymbol{\mathcal{X}}_{t}$%
. All simulation scenarios are based on 1000 repetitions of sample size $T$
and the \textrm{EW} structural break test.

\begin{figure}[h]
\centering
\includegraphics[width=.825\textwidth]{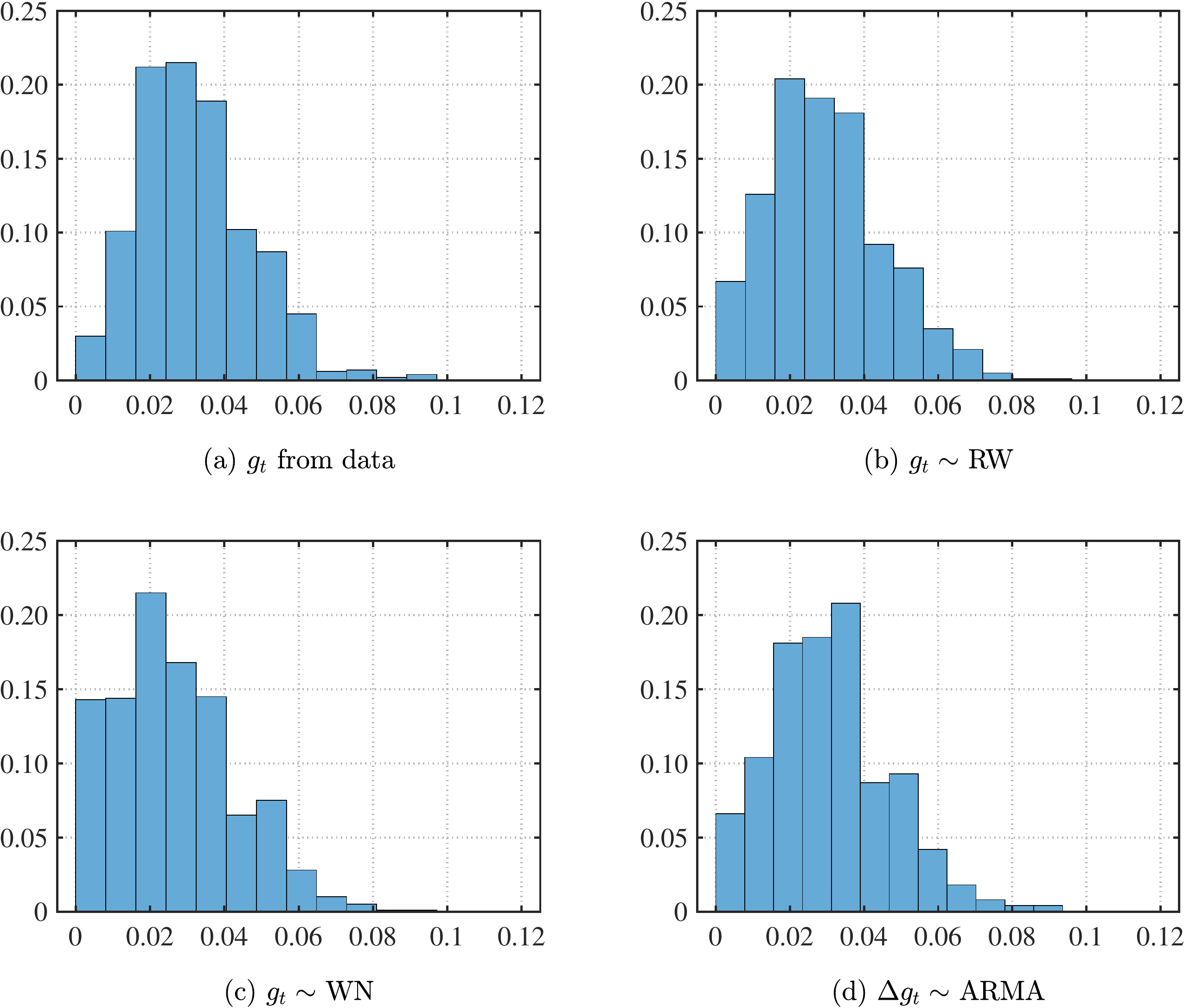} \vspace{-0mm}
\caption{Histograms of the estimated $\{ \hat{\protect\lambda}%
_{z}^{s}\} _{s=1}^{S}$ sequence corresponding to the summary
statistics shown in \autoref{tab:MUE2_Sim_extra}}
\label{fig:MUE2_Sim_extra}
\end{figure}

In \autoref{tab:MUE2_Sim_extra}, summary statistics of the $\lambda _{z}$
estimates obtained from implementing \cites{holston.etal:2017} MUE procedure
in Stage 2 are shown. The summary statistics are means and medians, as well
as empirical probabilities of observing an $\lambda _{z}$ estimate computed
from the simulated data being larger than the Stage 2 estimate of $0.030217$
from \cite{holston.etal:2017}. In \autoref{fig:MUE2_Sim_extra}, I show
histogram plots corresponding to the summary statistics of the $\lambda _{z}$
estimates computed from the simulated data. These are shown as supplementary
information to complement the summary statistics in \autoref%
{tab:MUE2_Sim_extra} and to avoid concerns related to unusual simulation
patterns.

\BT[h!]%
\caption{Summary statistics of $\lambda_z$ estimates of the Stage 2
MUE procedure applied to data simulated from unrelated univariate ARMA
processes} \centering\vspace*{-2mm}\renewcommand{\arraystretch}{1.1}%
\renewcommand\tabcolsep{7pt} \fontsize{11pt}{13pt}\selectfont%
\newcolumntype{N}{S[table-format =
3.8,round-precision = 6]}
\newcolumntype{K}{S[table-format =
5.9,round-precision = 6]}
\newcolumntype{Q}{S[table-format =
1.0,round-precision = 6]}
\begin{tabular*}{1\columnwidth}{p{40mm}NNNN}
\topline Summary Statistic & {$g_{t-1}=\hat{g}_{t-1|T}$} & {$g_{t-1}\sim
\mathrm{RW}$} & {~~~$g_{t-1}\sim \mathrm{WN}$~~~} & {\hsp[3]$\Delta
g_{t-1}\sim \mathrm{ARMA}$\hsp[3]} \\ \midrule
Minimum                              & 0           &  0           &  0           &  0             \\
Maximum                              & 0.09701863  &  0.09591360  &  0.09678872  &  0.09334030    \\
Standard deviation                   & 0.01523990  &  0.01585832  &  0.01680254  &  0.01633487    \\
Mean                                 & 0.03179750  &  0.02970785  &  0.02611747  &  0.03044940    \\
Median                               & 0.03016459  &  0.02864655  &  0.02425446  &  0.02943455    \\
$\mathrm{Pr}(\lambda^s_z> 0.030217)$ & 0.49800000  &  0.45600000  &  0.38400000  &  0.48200000    \\
\bottomrule
\end{tabular*}\label{tab:MUE2_Sim_extra}
\tabnotes[-3mm][.995\columnwidth][-1.25mm]{This table reports summary
statistics of the Stage 2 estimates of $\lambda_{z}$ that one obtains when
applying \cites{holston.etal:2017} MUE procedure to simulated data without
the $z_t$ process. The summary statistics that are reported are the minimum,
maximum, standard deviation, mean, median, as well as the empirical
frequency of observing a value larger than the estimate of $0.030217$
obtained by \cite{holston.etal:2017}, denoted by $\Pr
(\hat{\lambda}_{z}^{s}>0.030217)$. The columns show the estimates for the
four different data generating processes for trend growth $g_t$. The first
column reports results when the Kalman Smoothed estimate $\hat{g}_{t-1|T}$
is used for $g_{t-1}$. The second and third columns show estimates when
$g_{t-1}$ is generated as pure random walk (RW) or (Gaussian) white noise
(WN) process. The last column reports results when $g_{t-1}$ is computed as
the cumulative sum of $\Delta g_{t-1}$, which is simulated from the
coefficients obtained from a low order ARMA process fitted to
$\Delta\hat{g}_{t-1|T}$.The cycle and real rate series are also constructed
by first finding the best fitting low order ARMA processes to the individual
series and then simulating from fitted coefficients.} \ET

Looking over the results in \autoref{tab:MUE2_Sim_extra} and histograms in %
\autoref{fig:MUE2_Sim_extra}, it is clear that there are many instances
where the estimates of $\lambda _{z}$ from the simulated data are not only
non-zero, but rather sizeable, being larger than the estimate of $\lambda
_{z}=0.030217$ that \cite{holston.etal:2017} compute from the empirical
data. Note that there is no $z_{t}$ process simulated, yet with \cite%
{holston.etal:2017} Stage 2 MUE procedure one can recover an estimate that
is at least as large as the empirical one around 40 to 50 percent of the
time, depending on how $g_{t}$ is simulated. This simulation exercise thus
highlights how spurious \cites{holston.etal:2017} MUE procedure to estimate $%
\lambda _{z}$ is. As the downward trend in the $z_{t}$ process drives the
movement in the natural rate, where the severity of the downward trend is
related to the magnitude of $\sigma _{z}$, which is through $\lambda _{z}$, %
\cites{holston.etal:2017} estimates of the natural rate are likely to be
downward biased.

\section{Additional figures and tables\label{sec:AS_FT}}

This section presents additional figures and tables to complement the
results reported in the main text. Some of these results
are based on an expanded sample period using data that ends in 2019:Q2.

\cleardoublepage\newpage

\begin{figure}[p!]
\centering
\includegraphics[width=1\textwidth,trim={0 0 0 0},clip]
{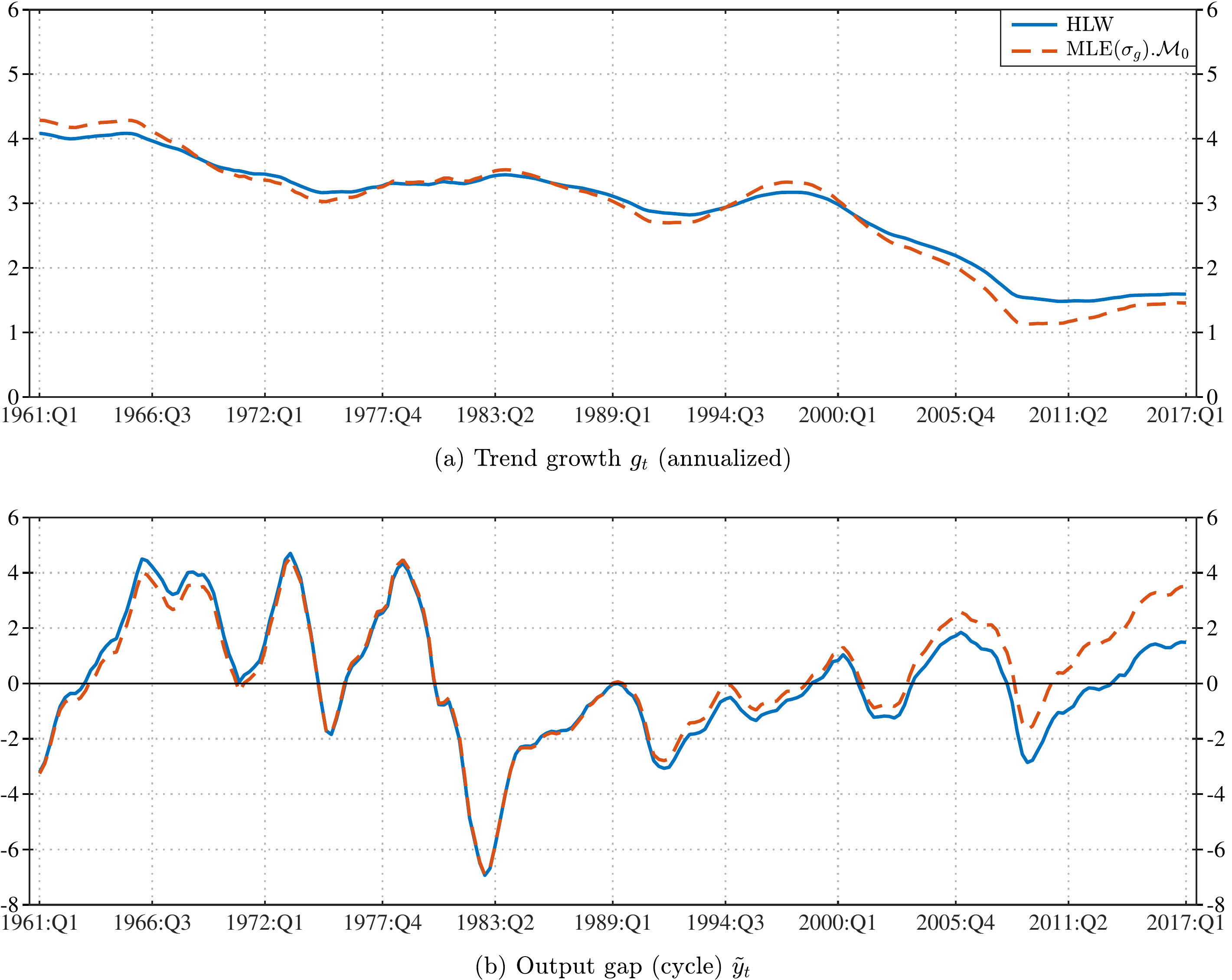}\vspace{-0mm}
\caption{Kalman smoothed estimates of (annualized) trend growth $g_t$
and output gap (cycle) $\tilde{y}_t$ from \cites{holston.etal:2017}
\emph{`misspecified'} Stage 2 model (HLW blue solid line) and the \emph{`correctly specified'} Stage 2 model
(MLE$(\sigma _{g}).\mathcal{M}_{0}$ red dashed lined). These are used as inputs into the structural break
dummy variable regression in \ref{eqS2regs}.}
\label{fig:MUE_comp_input}
\end{figure}

\cleardoublepage\newpage

\begin{figure}[p!]
\centering
\includegraphics[width=1\textwidth,trim={0 0 0 0},clip]
{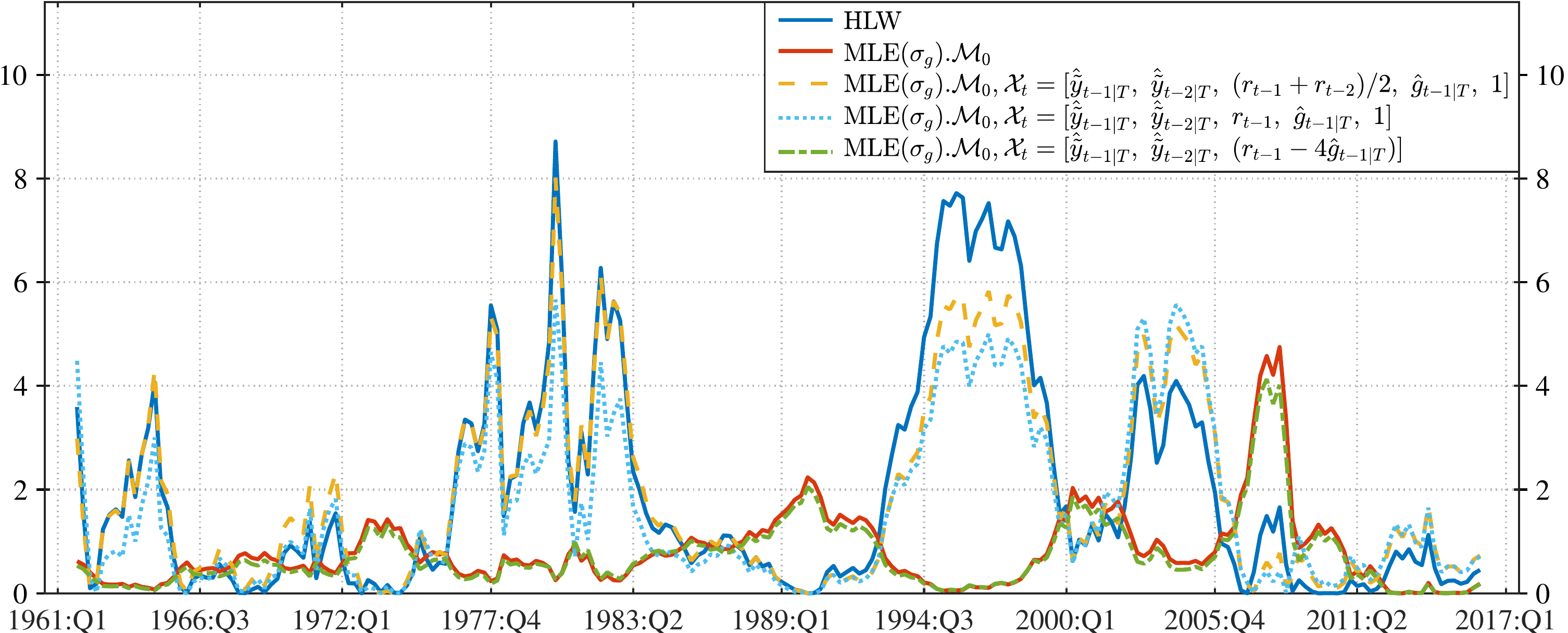}\vspace{-0mm}
\caption{Sequences of $\{F(\protect\tau )\}_{\protect\tau =%
\protect\tau _{0}}^{\protect\tau _{1}}$ statistics from the structural break
dummy variable regressions in \ref{eqS2regs} for the different scenarios that are considered.}
\label{fig:MUE_comp}
\end{figure}

\cleardoublepage\newpage

\begin{figure}[p!]
\centering
\includegraphics[width=1\textwidth,trim={0 0 0
0},clip]{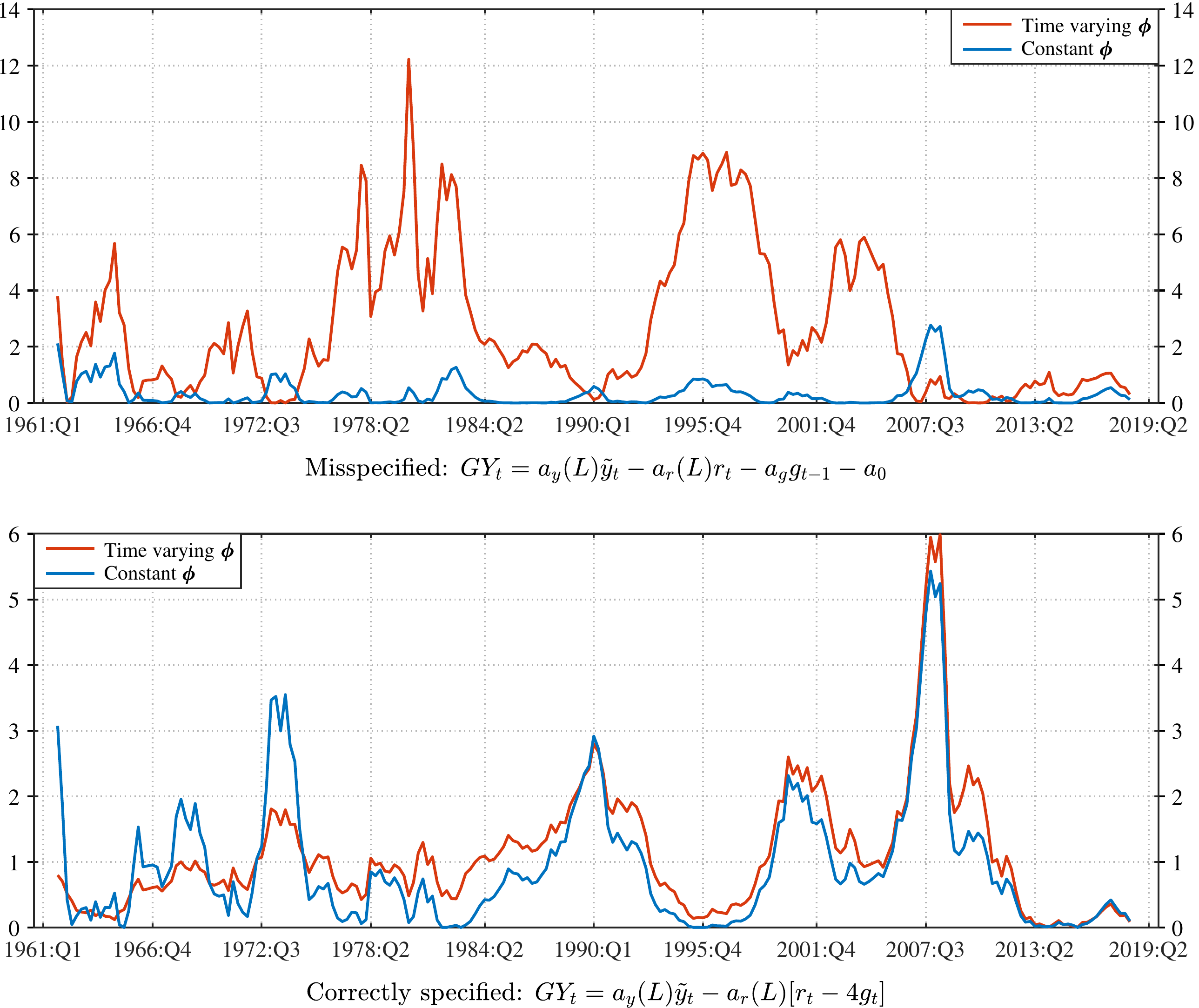}\vspace{-0mm}
\caption{Sequence of $\{F(\protect\tau )\}_{\protect\tau =%
\protect\tau _{0}}^{\protect\tau _{1}}$ statistics on the dummy variable
coefficients $\{\hat{\protect\zeta}_{1}(\protect\tau )\}_{\protect\tau =%
\protect\tau _{0}}^{\protect\tau _{1}}$ used in the construction of the
structural break test statistics.}
\label{fig:seqaF_2019Q2}
\end{figure}

\cleardoublepage\newpage

\BST[p!]%
\BAW[7]%
\caption{Stage 2 MUE results of $\lambda_z$ with corresponding 90\%
confidence intervals, structural break test statistics and $p-$values using data up to 2019:Q2} %
\centering\vspace*{-2mm}\renewcommand{%
\arraystretch}{1.15}\renewcommand\tabcolsep{7pt} %
\fontsize{10pt}{12pt}\selectfont %
\newcolumntype{N}{S[table-format = 1.6,round-precision = 6]} %
\newcolumntype{K}{S[table-format = 1.5,round-precision = 6]} %
\newcolumntype{Q}{S[table-format = 1.4,round-precision = 6]} %
\begin{tabular*}{1.07\columnwidth}{p{7mm}NNNNNNNp{0mm}NNNNNN}
\topline
\multirow{2}{*}{\hsp[2]$\lambda_{z}$}
&     \multicolumn{7}{c}{Time varying $\boldsymbol{{\phi}}$}
&&    \multicolumn{6}{c}{Constant $\boldsymbol{{\phi}}$} \\ \cmidrule(rr){2-8} \cmidrule(rr){10-15}

& {HLW.R-File} & {\hsp[-2] Replicated \hsp[-2]} & {[90\% CI]} & {\hsp[-2]
MLE($\sigma_g$) \hsp[-2]} & {[90\% CI]} & {\hsp[-2]
MLE($\sigma_g$).$\mathcal{M}_0$ \hsp[-2]} & {[90\% CI]} & {} & {\hsp[-2]
Replicated \hsp[-2]} & {[90\% CI]} & {\hsp[-2] MLE($\sigma_g$) \hsp[-2]} &
{[90\% CI]} & {\hsp[-2] MLE($\sigma_g$).$\mathcal{M}_0$ \hsp[-2]} & {[90\%
CI]}
\\
\midrule
        $L$  & {---}                 &   0          & {[0, 0.02]}  & 0           & {[0, 0.00]}  & 0.01159724 & {[0, 0.07]}  &&  0          & {[0, 0.02]}   & 0           & {[0, 0.00]} & 0.01159724   & {[0, 0.05]} \\
        MW   & 0.03169923666015      &   0.03169924 & {[0, 0.14]}  & 0.03925465  & {[0, 0.17]}  & 0.01462208 & {[0, 0.08]}  &&  0          & {[0, 0.03]}   & 0           & {[0, 0.02]} & 0.01104361   & {[0, 0.06]} \\
        EW   & 0.03520151475188      &   0.03520152 & {[0, 0.13]}  & 0.04041613  & {[0, 0.14]}  & 0.01477308 & {[0, 0.07]}  &&  0          & {[0, 0.03]}   & 0           & {[0, 0.02]} & 0.01227187  & {[0, 0.06]} \\
        QLR  & 0.04402871075924      &   0.04402873 & {[0, 0.15]}  & 0.04800876  & {[0, 0.16]}  & 0.02265371 & {[0, 0.09]}  &&  0          & {[0, 0.05]}   & 0           & {[0, 0.04]} & 0.02006166  & {[0, 0.07]} \\
\midrule
\multirow{1}{*}{} & \multicolumn{14}{c}{Structural break test statistics ($p-$values in parenthesis)\hsp[0]}     \\
\midrule
        $L$  & {---}                &   0.04985073 & {(0.8750)}   &  0.03736874 & {(0.9450)}   & 0.15984012 & {(0.3600)}   &&  0.04985073 & {(0.8750)}   & 0.03736874   & {(0.9450)}   &  0.15984012  & {(0.3600)} \\
        MW   & 2.67857637252549     &   2.67857737 & {(0.0600)}   &  3.79606247 & {(0.0200)}   & 1.10732323 & {(0.3050)}   &&  0.33027268 & {(0.8100)}   & 0.25484507   & {(0.8850)}   &  0.92809881  & {(0.3750)} \\
        EW   & 2.48662072082676     &   2.48662194 & {(0.0300)}   &  3.14005922 & {(0.0100)}   & 0.73638858 & {(0.3000)}   &&  0.20088486 & {(0.7900)}   & 0.15008242   & {(0.8750)}   &  0.64238445  & {(0.3450)} \\
        QLR  & 12.22151692530160    &  12.22152141 & {(0.0100)}   & 13.64872048 & {(0.0050)}   & 5.98786206 & {(0.1600)}   &&  2.76790912 & {(0.5900)}   & 2.17481293   & {(0.7250)}   &  5.43201383  & {(0.2000)} \\
\bottomrule
\end{tabular*}\label{Atab:Stage2_lambda_z_2019}
\tabnotes[-2.5mm][1.06\columnwidth][-0.5mm]{This table reports the Stage 2
estimates of $\lambda _{z}$ for the different
$\boldsymbol{\theta }_{2}$ estimates corresponding to the \emph{%
"misspecified"} and \emph{"correctly specified"} Stage 2 models reported in %
\autoref{tab:Stage2} using data updated to 2019:Q2. The table is split into
two column blocks, showing the
results for the \emph{"time varying} $\boldsymbol{\phi }$" and \emph{%
"constant} $\boldsymbol{\phi }$" scenarios in the left and right blocks,
respectively. In the bottom half of the table, the four different structural
break test statistics for the considered models are shown. The results under
the heading `HLW.R-File' show the $\lambda _{z}$ estimates obtained from
running \cites{holston.etal:2017} R-Code for the Stage 2 model as reference
values. The second column `Replicated' shows my replicated results. Under
the heading `MLE($\sigma _{g}$)', results for the \emph{"misspecified}"
Stage 2 model are shown with $\sigma _{g}$ estimated directly by MLE rather
than from the first stage estimate of $\lambda _{g}$. Under the heading `MLE(%
$\sigma _{g}$).$\mathcal{M}_{0}$', results for the \emph{"correctly
specified"} Stage 2 model are reported where $\sigma _{g}$ is again
estimated by MLE. The values in square brackets in the top half of the table
report 90\% confidence intervals for $\lambda _{z}$ computed from %
\cites{stock.watson:1998} tabulated values provided in their GAUSS\ files.
These were divided by sample size $T$ to make them comparable to $\lambda
_{z}$. In the bottom panel, $p-$values of the various structural break tests
are reported in round brackets. These were also extracted from %
\cites{stock.watson:1998} GAUSS\ files.}%
\EAW \EST%

\cleardoublepage\newpage

\BT[p!] \caption{Stage 3 parameter estimates using data up to
2019:Q2}\centering\vspace*{-2mm}\renewcommand{\arraystretch}{1.1}%
\renewcommand\tabcolsep{7pt}\fontsize{11pt}{13pt}\selectfont%
\newcolumntype{N}{S[table-format = 4.8,round-precision = 8]} %
\newcolumntype{U}{S[table-format = 4.8,round-precision = 8]} %
\newcolumntype{L}{S[table-format = 4.8,round-precision = 8]} \BAW[7]
\begin{tabular*}{1.085\columnwidth}{p{27mm}NNNNN}
\topline
\hsp[5]$\boldsymbol{\theta }_{3}$
& {\hsp[2]HLW.R-File\hsp[-4]}
& {\hsp[2]Replicated\hsp[-4]}
& {\hsp[2]MLE($\sigma_g|\lambda_z^{\mathrm{HLW}})$\hsp[-3]}
& {\hsp[3]MLE($\sigma_g|\lambda_z^{\mathcal{M}_0})$\hsp[-4]}
& {\hsp[2]MLE($\sigma_g,\sigma_z)$\hsp[-4]}\\
\midrule
$\hsp[3]a_{y,1}  $                  &      1.5387645830 &    1.5387645934 &    1.5108322346 &    1.5165911490 &    1.5166962002    \\
$\hsp[3]a_{y,2}  $                  &     -0.5970026371 &   -0.5970026497 &   -0.5705368374 &   -0.5763753973 &   -0.5764593051    \\
$\hsp[3]a_{r}    $                  &     -0.0685404334 &   -0.0685404321 &   -0.0756111341 &   -0.0702967130 &   -0.0700067459    \\
$\hsp[3]b_{\pi } $                  &      0.6733154496 &    0.6733154483 &    0.6763890042 &    0.6746345037 &    0.6748341057    \\
$\hsp[3]b_{y}    $                  &      0.0775545028 &    0.0775545062 &    0.0745405463 &    0.0788842702 &    0.0788524299    \\
$\hsp[3]\sigma _{\tilde{y}}$        &      0.3359069291 &    0.3359069170 &    0.3359838098 &    0.3474767025 &    0.3482610569    \\
$\hsp[3]\sigma _{\pi }     $        &      0.7881255370 &    0.7881255365 &    0.7892181385 &    0.7885495236 &    0.7886203640    \\
$\hsp[3]\sigma _{y^{\ast }}$        &      0.5757731876 &    0.5757731957 &    0.5678951962 &    0.5635923645 &    0.5632780334    \\
$\hsp[3]\sigma _{g}$ {(implied)}    &      (0.03082331) &    (0.03082331) &    0.0451784946 &    0.0438616880 &    0.0437898193    \\
$\hsp[3]\sigma _{z}$ {(implied)}    &      (0.17251762) &    (0.17251762) &    0.1564206011 &    0.0606598161 &    0.0525049366    \\
$\hsp[3]\lambda_g  $ {(implied)}    &      0.0535337714 &    0.0535337714 &    (0.07955428) &    (0.07782519) &    (0.07774103)    \\
$\hsp[3]\lambda_z  $ {(implied)}    &      0.0352015148 &    0.0352015148 &    (0.03520151) &    (0.01227186) &    (0.01055443)    \\  \cmidrule(ll){1-6}
{Log-likelihood}                    &   -533.3698452415 & -533.3698455030 & -533.1654750148 & -532.8287486005 & -532.8263754097    \\
\bottomrule
\end{tabular*}\label{Atab:S3_2019}%
\tabnotes[-2.6mm][1.08\columnwidth][-4.0mm]{This table reports replication
results for the Stage 3 model parameter vector $\boldsymbol{\theta }_{3}$ of
\cite{holston.etal:2017}. The first column (HLW.R-File) reports estimates
obtained by running \cites{holston.etal:2017} R-Code for the Stage 3 model.
The second column (Replicated) shows the replicated results using the same set-up as in \cites{holston.etal:2017}. The third column
(MLE($\sigma_g|\lambda_z^{\mathrm{HLW}})$) reports estimates when $\sigma
_{g}$ is directly estimated by MLE together with the other parameters of the
Stage 3 model, while $\lambda _{z}$ is held fixed at
$\lambda _{z}^{\mathrm{HLW}}=0.035202$ obtained from \cites{holston.etal:2017} \emph{"misspecified"} Stage 2 procedure. In the
forth column (MLE($\sigma_g|\lambda_z^{\mathcal{M}_0})$), $\sigma _{g}$ is
again estimated directly by MLE together with the other parameters of the
Stage 3 model, but with $\lambda _{z}$ now fixed at $\lambda
_{z}^{\mathcal{M}_{0}}=0.012272$ obtained from the \emph{"correctly
specified"} Stage 2 model in \ref{S2full0}. The last column
(MLE($\sigma_g,\sigma_g)$) shows estimates when all parameters are computed
by MLE. Values in round brackets give the implied $\{\sigma _{g}, \sigma
_{z}\}$ or $\{\lambda _{g},\lambda _{z}\}$ values when either is fixed or
estimated. The last row (Log-likelihood) reports the value of the
log-likelihood function at these parameter estimates. The Matlab file
\texttt{Stage3\_replication.m} replicates these results.} \EAW \ET

\cleardoublepage\newpage

\begin{figure}[p!]
\BAW[10]\centering\vspace{-15mm}
\includegraphics[width=1.05\textwidth,trim={0 0 0
0},clip]{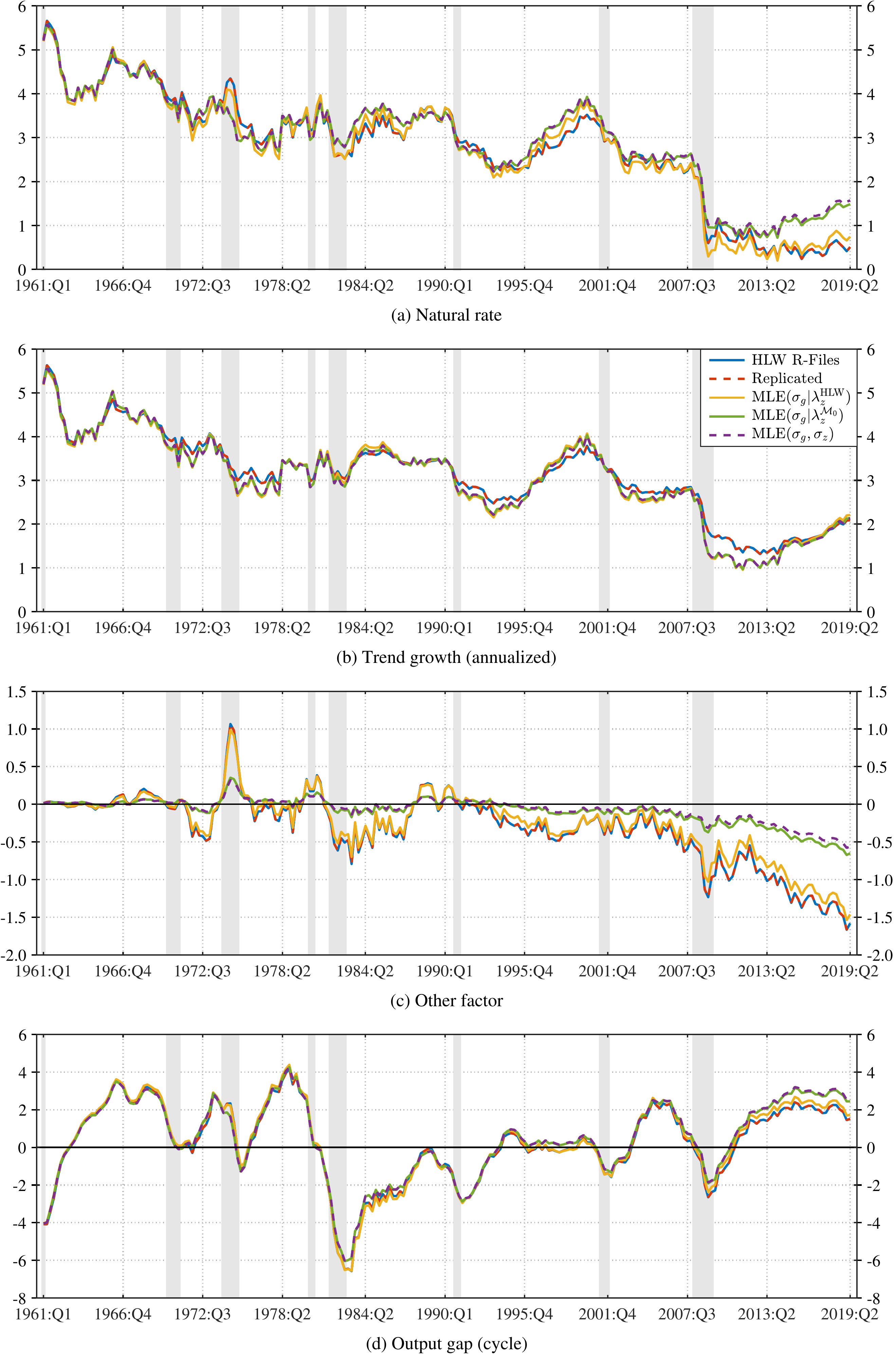} \vspace{-3mm} \EAW
\caption{Filtered estimates of the natural rate $r^{\ast}_t$, annualized
trend growth $g_t$, \emph{`other factor'} $z_t$, and the output gap (cycle)
variable $\tilde{y}_t$ up to 2019:Q2.}
\label{Afig:2019KF}
\end{figure}

\cleardoublepage\newpage

\begin{figure}[p!]
\BAW[10]\centering\vspace{-15mm}
\includegraphics[width=1.05\textwidth,trim={0 0 0
0},clip]{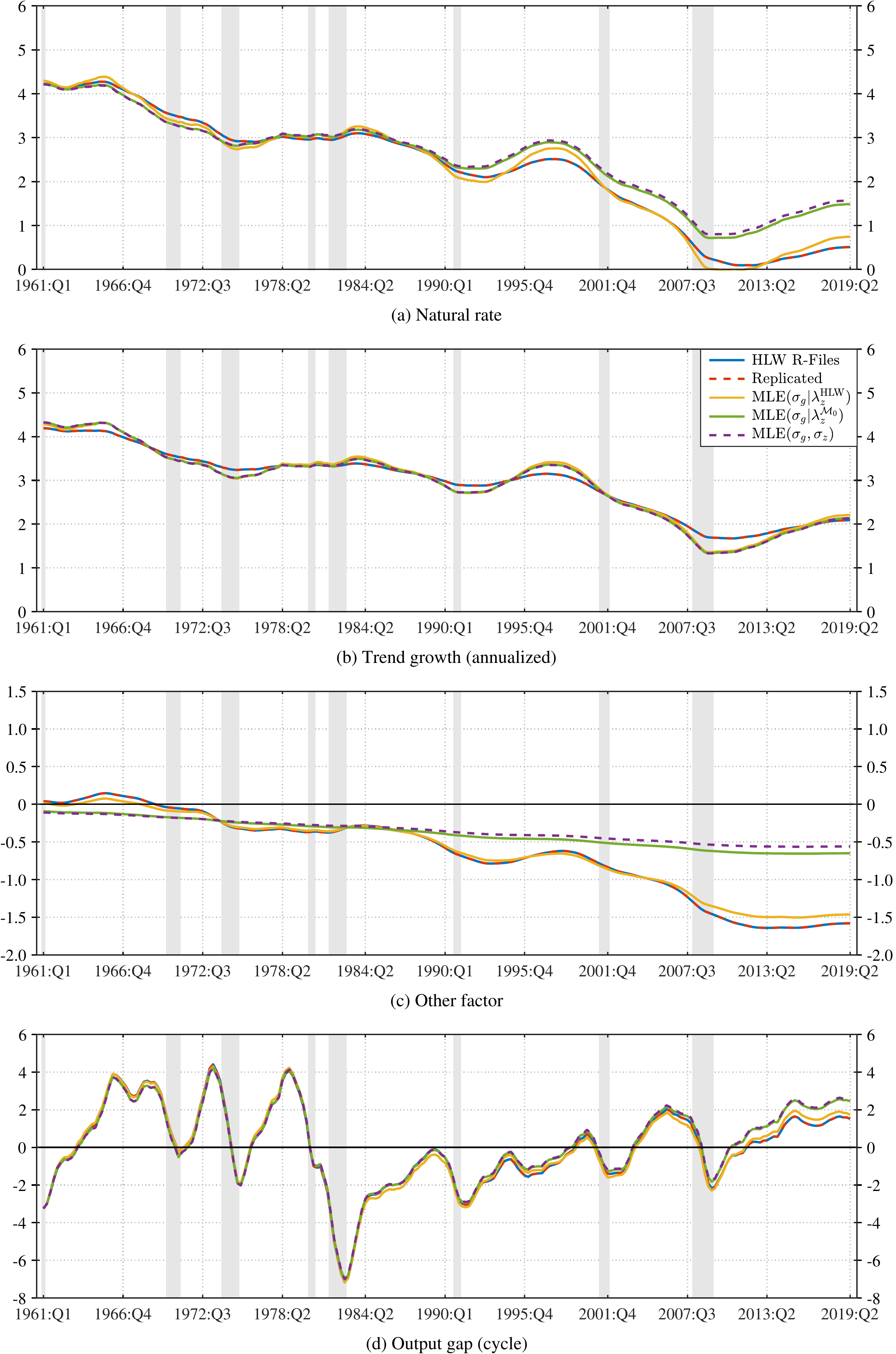} \vspace{-3mm} \EAW
\caption{Smoothed estimates of the natural rate $r^{\ast}_t$, annualized
trend growth $g_t$, \emph{`other factor'} $z_t$, and the output gap (cycle)
variable $\tilde{y}_t$ up to 2019:Q2.}
\label{Afig:2019KS}
\end{figure}

\cleardoublepage\newpage

\vspace*{-13mm}

\begin{figure}[H] 
\centering
\includegraphics[width=.96\textwidth,trim={0 0 0 0},clip]{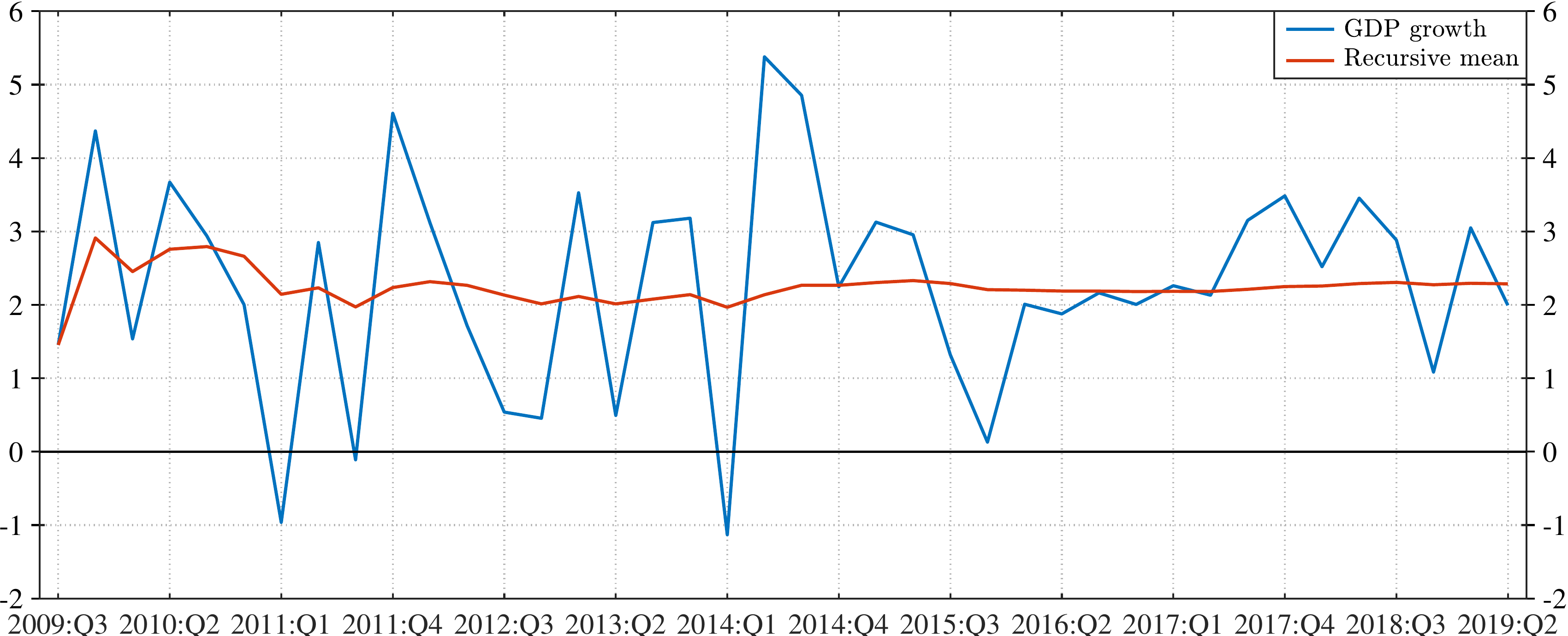} \vspace{0mm}
\caption{GDP growth and recursively estimated mean of GDP growth from 2009:Q3 to 2019:Q2.}
\label{Afig:rec_mean}%
\end{figure}

\vspace*{1mm}

\begin{figure}[H]
\centering
\includegraphics[width=\textwidth,trim={0 0 0 0},clip]{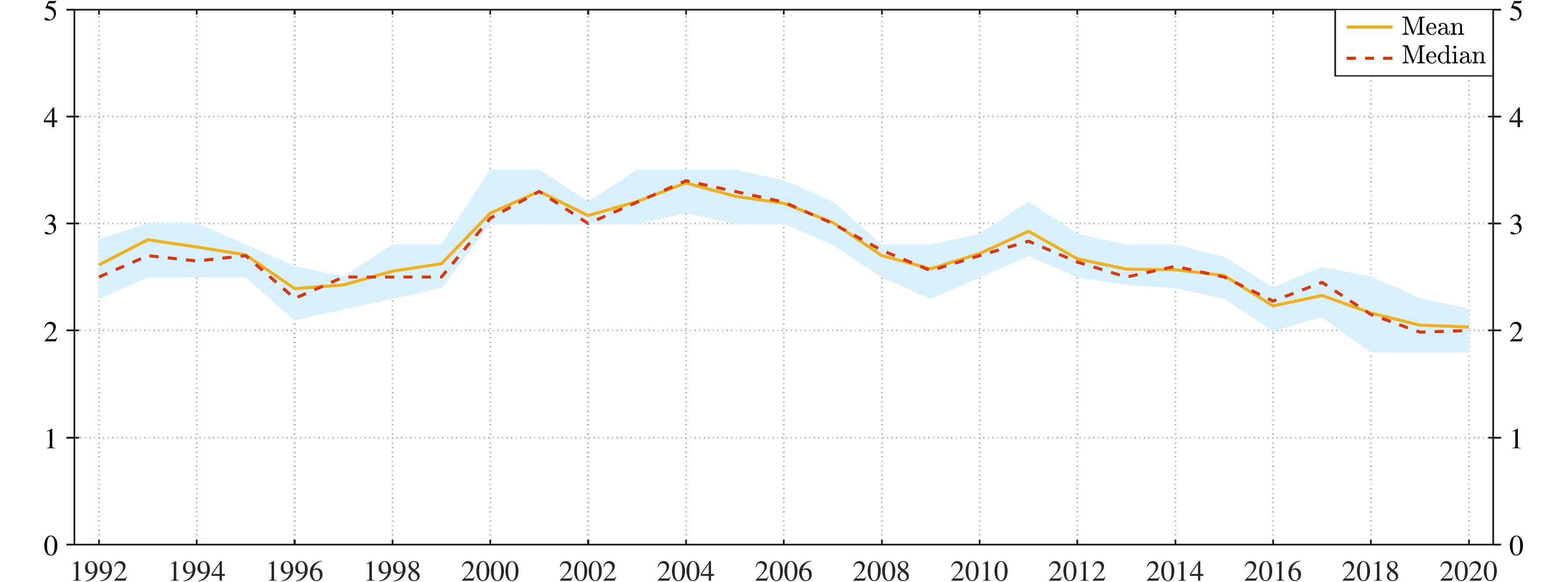}\vspace{-0mm}
\caption{Mean and Median 10 year (real) GDP growth forecasts from the Survey of
Professional forecasters (SPF) obtained from
\url{https://www.philadelphiafed.org/research-and-data/real-time-center/survey-of-professional-forecasters/data-files/rgdp10}.
The blue shaded region marks the $25^{th}$ to $75^{th}$ percentile
region of the cross-section of forecaster.}
\label{Afig:SPF_GDP_growth} %
\end{figure}

\vspace*{2mm}

\begin{figure}[H]
\centering
\includegraphics[width=.99\textwidth,trim={0 0 0 0},clip]{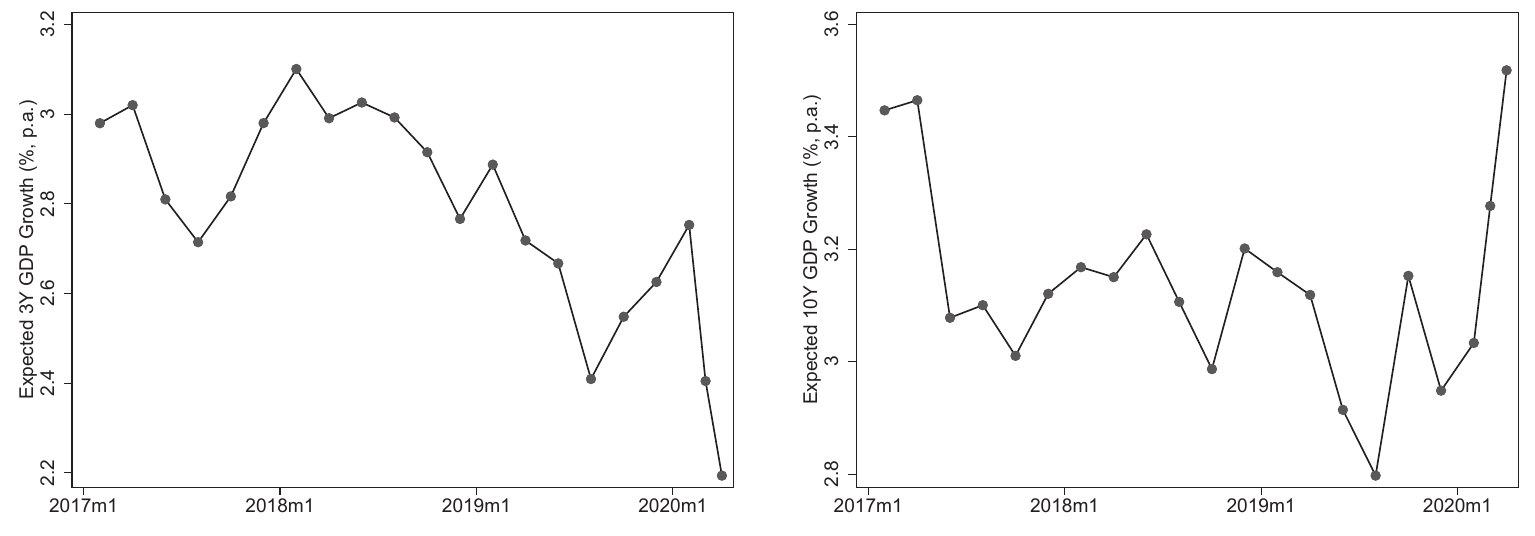} \vspace{0mm}
\caption{GMSU-Vanguard survey based expected 3 year and 10 year (real) GDP growth from February 2017
to April 2020, taken from Figure II on page 5 in \cite{giglio.etal:2020} (see the appendix in
\cite{giglio.etal:2020} for more details on the design of the client/investor survey).}
\label{Afig:giglio_GDP_growth} %
\end{figure}

\clearpage